\documentclass[aps,amsmath,amssymb,preprint, 12pt]{revtex4-1}
\usepackage{amsmath, latexsym, color, graphicx, tabularx, booktabs, enumerate}
\usepackage{multirow}
\usepackage[normalem]{ulem} 
\usepackage{caption}
\usepackage{epstopdf}
\usepackage{mathtools}
\usepackage{soul,xcolor}
\setstcolor{red}
\begin{document}
\title{First-principles study of mechanical and electronic properties of bent monolayer transition metal dichalcogenides}
\author{Niraj K. Nepal$^1$}
\author{Liping Yu$^2$}
\author{Qimin Yan$^1$}
\author{Adrienn Ruzsinszky$^1$}

\affiliation{[1] Department of Physics, Temple University, Philadelphia, Pennsylvania 19122, United States}

\affiliation{[2] Department of Physics and Astronomy, University of Maine, Orono, Maine, United States}

\begin{abstract}
	The mechanical and electronic properties of transition metal dichalcogenide (TMD) monolayers corresponding to transition groups IV, VI, and X are explored under mechanical bending from first principles calculations using the strongly constrained and appropriately normed (SCAN) meta-GGA (MGGA). SCAN provides an accurate description of the phase stability of the TMD monolayers. Our calculated lattice parameters and other structure parameters agree well with experiment. We find that bending stiffness (or flexural rigidity) increases as the transition metal group goes from IV to X to VI, with the exception of PdTe$_2$. Variation in mechanical properties (local strain, physical thickness) and electronic properties (local charge density, band structure) with bending curvature is discussed. The local strain profile of these TMD monolayers under mechanical bending is highly non-uniform. The mechanical bending tunes not only the thickness of the TMD monolayers, but also the local charge distribution as well as the band structures, adding more functionalization options to these materials.
\end{abstract}
\maketitle

\section{Introduction}
Layered transition metal dichalcogenides (TMD) offer a wide variety of physical and chemical properties from metal to insulator  \cite{WY69, XLSC13, KT16} and are extensively studied  \cite{TAN14, JS12, CSELLZ13, YYR17}. An increasing interest and recent progress towards these materials led to a variety of improved applications such as sensors, energy storage, photonics, optoelectronics, and spintronics  \cite{CHS12, APH14, WKKCS12, WKKCS12}. In particular, atomically thin monolayer TMDs have attracted most of the attention due to the unique mechanical and electronic properties related to their high flexibility  \cite{HPMS13, CSBAPJ14, LZZZT16}. A large scope of flexible electronics has been realized via applications such as flexible displays  \cite{JLK09, ZWWSPJ06, R16, WS09}, wearable sensors  \cite{SC14, KGLR12, WW13}, and electronic skins  \cite{STM13, PKV14, HCTTB13}. Each TMD (TX$_2$) monolayer consisting of 3 atomic layers (X-T-X stacking) can undergo bending deformation, possessing higher flexural rigidity than graphene (D$_{MoS_2}$ $\sim$ 7-8 D$_{Graphene}$   \cite{AB17}). The bending behavior (curvature effect) of 2D TMD monolayers, especially of MoS$_2$, has been studied both theoretically  \cite{JQPR13, XC16} and experimentally  \cite{ZDLHSR15, CSBAPJ14}. For 2D materials such as MoS$_2$, the bending can induce localization or delocalization in the electronic charge distribution. This change in the charge distribution results in changes in electronic properties such as the Fermi level, effective mass, and band gap  \cite{YRP16}. However, the bending behavior of other TMD monolayers is largely unexplored at least from first-principles. Quantitatively, the resistance of a material against bending is characterized by the bending stiffness. The bending stiffness or flexural rigidity of the TMD monolayers can be estimated using first-principles as in Refs.~\onlinecite{RBM92, AB04, JQPR13}. Most of the earlier studies used nanotubes of different radii created by rolling an infinitely extended sheet to estimate the bending stiffness of 2D monolayers  \cite{RBM92, AB04, HWH06}. However, such a scheme has several limitations. (1) It does not mimic the edges present in the monolayer. (2) The nanoribbons unfolded from differently sized nanotubes have different widths which contribute to different quantum confinement effects along with the curvature effect. By utilizing the bending scheme similar to the bending of a thin plate, we restore the edges as well as fix the width of the nanoribbon, thereby eliminating the quantum confinement effect resulting from difference in width between various configurations of nanoribbons from flat to bent ones. However, the edge effects due to their finite width may not be completely eliminated.\\
 
Here we report a comprehensive first-principles study of the structural, mechanical, and electronic properties of flat and bent monolayer TMD compounds, i.e., TX$_2$ (T = transition metal, X = chalcogen atom). As in Ref.~\onlinecite{WY69}, we represent each TMD (TX$_2$) with its transition metal group. For example, d$^0$ for group IV, d$^2$ for group VI, and d$^6$ for group X. Their layer structures have been observed in experiment: group IV (T = Ti, Zr or Hf; X = S, Se or Te) and group X (PdTe$_2$ and PtX$_2$) TMDs prefer the 1T phase, while group VI TMDs crystallize in the 1H (T = Mo or W; X = S, Se) as well as the distorted T (1T$'$) phase (WTe$_2$)  \cite{WY69}. We first investigate the relative stability of a monolayer in three different phases (1H, 1T, 1T$'$). The mechanical and electronic properties have been studied only for those most stable phases. The organization of the rest of the paper is as follows. The computational details are presented in Sec. II. Section III presents our results, followed by some discussion and conclusions in Sec. IV.\\


\section{Computational Details}

\begin{figure}[h!]
	\includegraphics[scale=0.5]{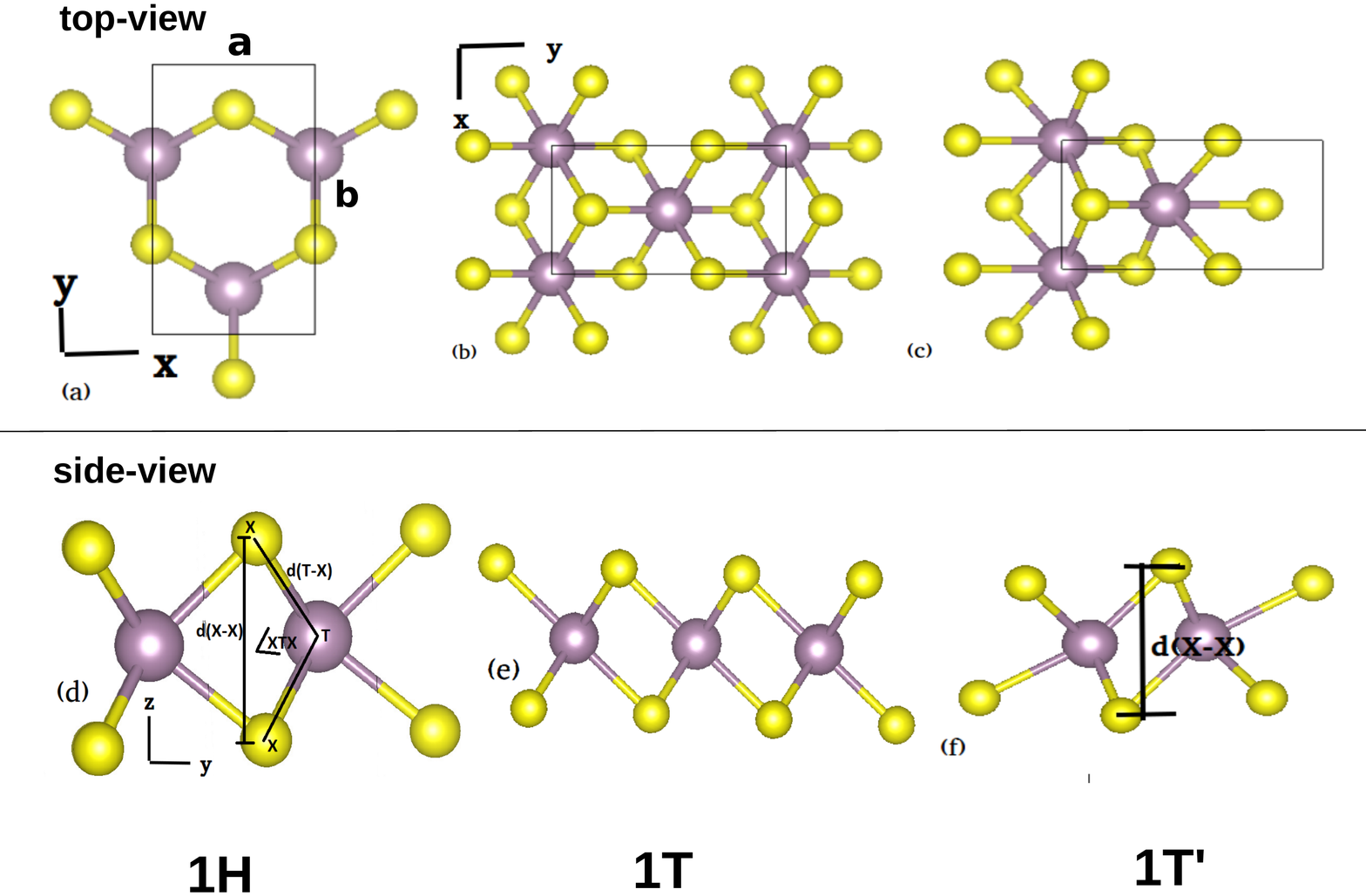}

	\caption{Rectangular unit-cells of types 1H, 1T, and 1T$'$ (WTe$_2$) used in the calculations. The first row represents the top view (a-c) while second (d-f) corresponds to the side view; d(T-X) is metal-chalcogen distance, $\angle$XTX is an angle made by two d(T-X) sides, and d(X-X) (or d$_{X-X}$) is the distance between the outer and inner layer of flat monolayer bulk TMDs.}
	\label{fig:struc}
\end{figure}
The ground state calculations were performed using  the Vienna ab initio simulation package (VASP)  \cite{VASP} with projected augmented wave (PAW)  \cite{B94} pseudo-potentials (PS)  \cite{KJ99} as implemented in the VASP code  \cite{K01}, modified to include the kinetic energy density required for meta-GGA (MGGA) calculations. We used pseudo-potentials recommended in VASP for all elements except for tungsten (W), where we used a pseudopotential such that the valence electron configuration includes 6s$^1$5d$^5$ electrons. The exchange-correlation energy was approximated using the strongly constrained and appropriately normed (SCAN) MGGA  \cite{SRP15}. It can describe an intermediate range of dispersion via the kinetic energy density and is proven to deliver sufficiently accurate ground state properties for diversely bonded systems  \cite{NBR18, SRZ16, SSP18, BLBRSB17}, as compared to local density approximation (LDA) and the generalized gradient approximation (GGA) of Perdew, Burke, and Ernzerhof (PBE). The unit-cell calculations for all pristine TMD monolayers were carried out using a rectangular supercell consisting of two MX$_2$-units with three different configurations 1H, 1T, and 1T$'$-WTe$_2$ to determine the most stable ground state. We used the energy cutoff of 550 eV and 24 $\times$ 16 $\times$ 1 and 16 $\times$ 24 $\times$ 1 Gamma-centered Monkhorst-Pack k-meshes  \cite{MP76} to sample the Brillouin zone. Periodic boundary conditions were applied along the in-plane direction, while a vacuum of about 20 $\AA$ was inserted along the out-of-plane direction. The geometry optimization of the mono-layer unit-cell was achieved by converging all the forces and energies within 0.005 eV/$\AA$ and 10$^{-6}$ eV respectively. To estimate the bending stiffness, we relaxed our nano-ribbons having a width of 3-4 nm ({\color{blue} Supplementary Table S1}) with forces less than 0.01 eV/$\AA$, using an energy cutoff of 450 eV. The Brillouin zone was sampled using Gamma-centered Monkhorst-Pack k-meshes of 8 $\times$ 1 $\times$ 1 and 1 $\times$ 8 $\times$ 1.\\

\begin{figure}
   \includegraphics[scale = 0.45]{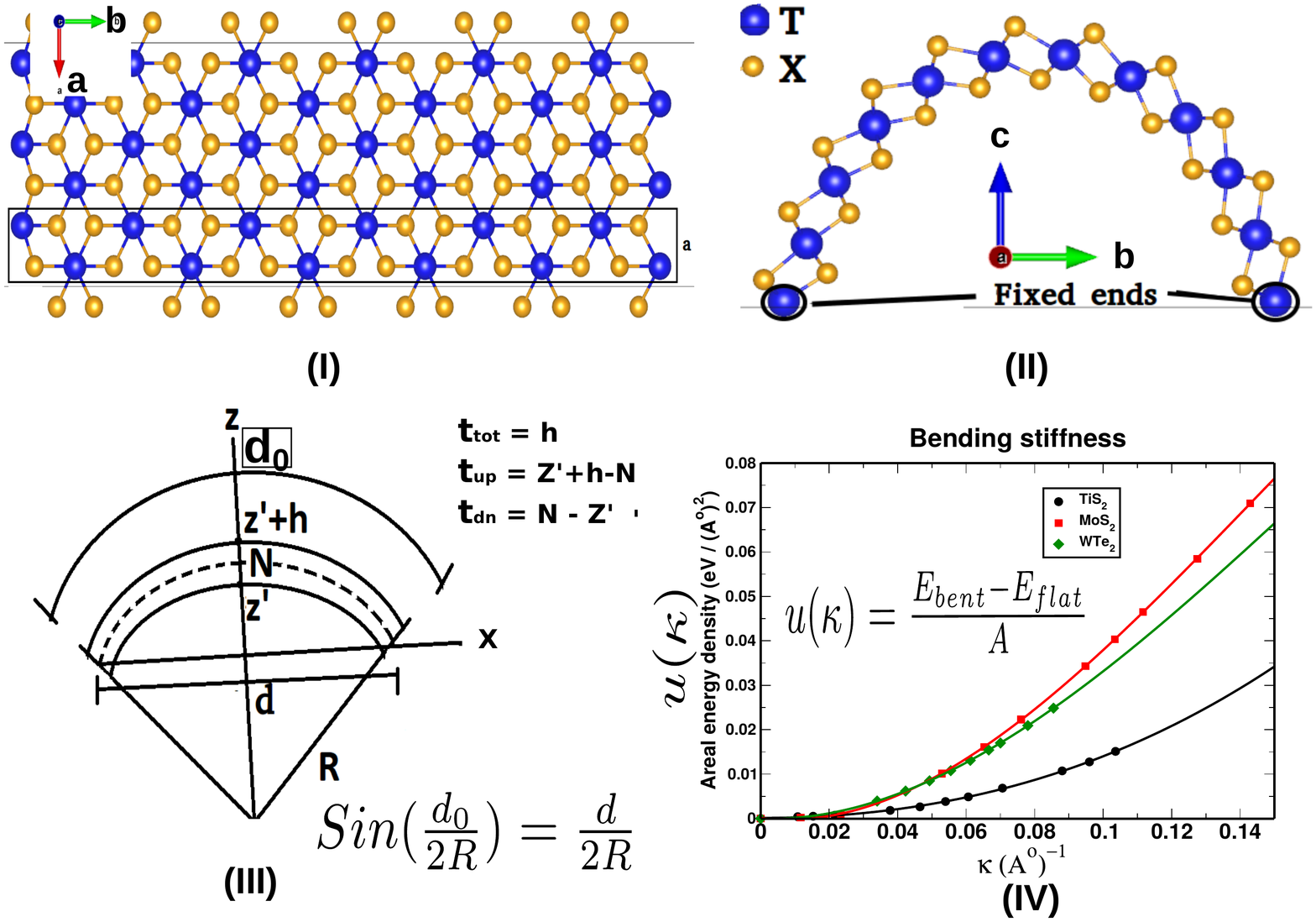}
\caption{(I) A nanoribbon (enclosed by rectangle) is taken to simulate an extended sheet of 1T monolayer; a is the lattice constant with the ribbon extended along the \textit{a}-axis and a vacuum of 20 Angstroms is inserted along \textit{b-} and \textit{c-} axes ({\color{blue} Supplementary Figure S4}); bent sample of 1T nano-ribbon; (III) a schematic bending of a thin plate. d0, d, and R are the length of a thin plate before bending, length after bending, and radius of curvature respectively. N is the neutral surface denoted by a dashed line. t$_{tot}$, t$_{up}$, and t$_{dn}$ are the physical thicknesses of the bent nano-ribbon, assuming that the middle layer coincides with the neutral surface (N); (IV) areal bending energy density vs bending curvature curve to estimate the bending stiffness. E$_{bent}$, E$_{flat}$, and A are the total energy of bent nanoribbon, total energy of flat nanoribbon, and cross-sectional area of flat nanoribbon (length * width) respectively.}
\label{fig:ribbon}
\end{figure}

To estimate the in-plane stiffness, we applied strain along one direction (say the x-direction) and relaxed the system along the lateral direction (i.e., the y-direction) or vice versa (See Figure~\ref{fig:struc}). An in-plane stiffness then can be estimated using
\begin{equation}
 Y_{2D} =\frac{1}{A_0} \frac{\partial^2 E_s}{\partial \epsilon^2},
 \label{Y2D}
\end{equation}
  where E$_s$ $=$ E($\epsilon = s$) - E($\epsilon = 0$) is the strain energy, $\epsilon = \frac{\textnormal{Change in length ($\Delta l$)}} {\textnormal{equilibrium length ($l_0$)}} $ is the linear strain, and A$_0$ is an equilibrium area of an unstrained supercell. We also applied a 5\% axial strain and relaxed the rectangular supercell in the transverse direction to estimate the lateral strain and hence found the Poisson's ratio. We first relaxed the flat ribbon using various edge schemes. The choices of edges are mainly due to either relaxation of the flat nanoribbon or to satisfy the condition, areal bending energy density u($\kappa $)\hspace{1mm}$=$ $\frac{E_{bent} - E_{flat}}{Area (A)}$  $\rightarrow$ 0 as the bending curvature $\kappa$ \hspace{1mm}$=$ $\frac{1}{\textnormal{radius of curvature (R)}}$ $\rightarrow 0$ (Figure~\ref{fig:ribbon} (IV)). We have taken stoichiometric (n(X):n(T)$=$ 2:1) nano-ribbons ({\color{blue} Supporting Figure S4}) for most of the calculations in which TiTe$_2$, MoTe$_2$-1T$'$, and WX$_2$ (X = S, Se, or Te) were stabilized using hydrogen passivated edges whereas others were relaxed without hydrogen passivation. We also relaxed TiSe$_2$, HfS$_2$, PdTe$_2$, and PtSe$_2$ nano-ribbons in symmetric configuration (Figure~\ref{fig:ribbon} II). Finally, the bent structures of different bending curvatures were created by relaxing the ribbons along the infinite length direction, while keeping the transition atoms fixed at the opposite end, and applying strain along the width direction. A 20 $\AA$ of vacuum was introduced along the y- and z- direction to eliminate an interaction between the system and its image ({\color{blue} Supplementary Figure S4}). The areal bending energy density (u($\kappa$)) vs bending curvature ($\kappa$) curve were fitted with a cubic polynomial to capture the non-linear behavior (Figure~\ref{fig:ribbon} (IV)). The quadratic coefficient of the cubic fitting was utilized to estimate the bending stiffness, 
  \begin{equation}
  S_b = \frac{\partial^2 u(\kappa)}{\partial \kappa^2} |_{\kappa = 0}.
\end{equation}

\section{Results}
\subsection{Relative Stability}
\begin{figure}[htbp!]
	
	\includegraphics[height=3.5in, width=4.5in]{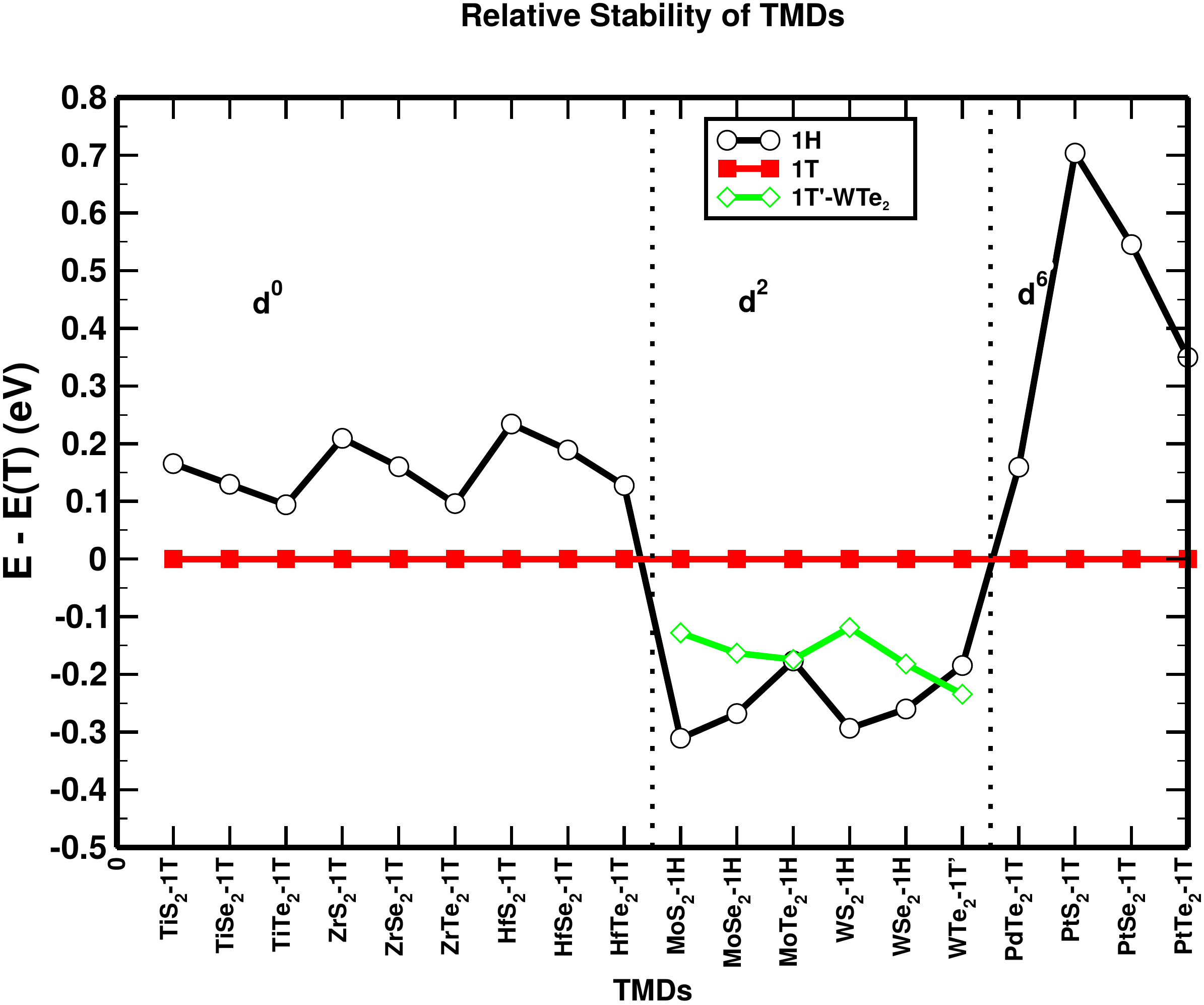}
	\captionsetup{font=scriptsize}
	\caption{Stability (relative to the 1T phase) from SCAN calculations for TMDs between the 3 experimentally observed phases 1H, 1T, and 1T$'$-WTe$_2$. The \textit{x}-axis represents the TMD with a phase corresponding to the minimum ground state (GS) energy, and the relative GS energies per atom of the TMDs of any phase with respect to corresponding GS of 1T phase are presented on the \textit{y}-axis. The straight line parallel to the \textit{x}-axis passing through the origin represents the GS energies of 1T phases. SCAN correctly predicts the ground state for these compounds. Also, MoTe$_2$ seems to be iso-energetic between 1H and 1T$'$-WTe$_2$ phases.}
	\label{fig:stability}
\end{figure}

Experimentally, it is largely known which phase is preferred in the bulk layer structure. However, the relative stability of their monolayer structures remained elusive. We have performed relative stability analysis of monolayer TX$_2$ among 3 different phases, namely 1H, 1T, and 1T$'$-WTe$_2$, to test the predictive power of SCAN. Energies of TMDs in different phases relative to the 1T phase are presented in Figure~\ref{fig:stability}. Among two different phases, 1H and 1T, group (IV) and (X) TMD monolayers prefer the 1T phase. We could not find a distorted phase (1T$'$) for these TMD monolayers. In addition to the 1H and 1T phase, group (VI) TMDs MoTe$_2$ and WTe$_2$ also crystallize in the distorted (1T$'$) form. Our relative stability analysis shows that TX$_2$ with X$=$S or Se prefers the 1H phase, while it depends on the transition metal for X$=$Te, consistent with the experimental predictions  \cite{WY69}. WTe$_2$ prefers the 1T$'$ phase while the cohesive energies of 1H and 1T$'$ phases of MoTe$_2$ are almost identical (favoring the 1H phase by 5 meV), leading to an easy modulation between 2 phases  \cite{WNGJDY17}. Satisfying 17 exact known constraints, SCAN accurately captures the necessary interactions present in these TMD monolayers and predicts the correct ground state structure.

\subsection{Structural properties}
\begin{figure}[htbp!]	
	\includegraphics[height=3in, width=5in]{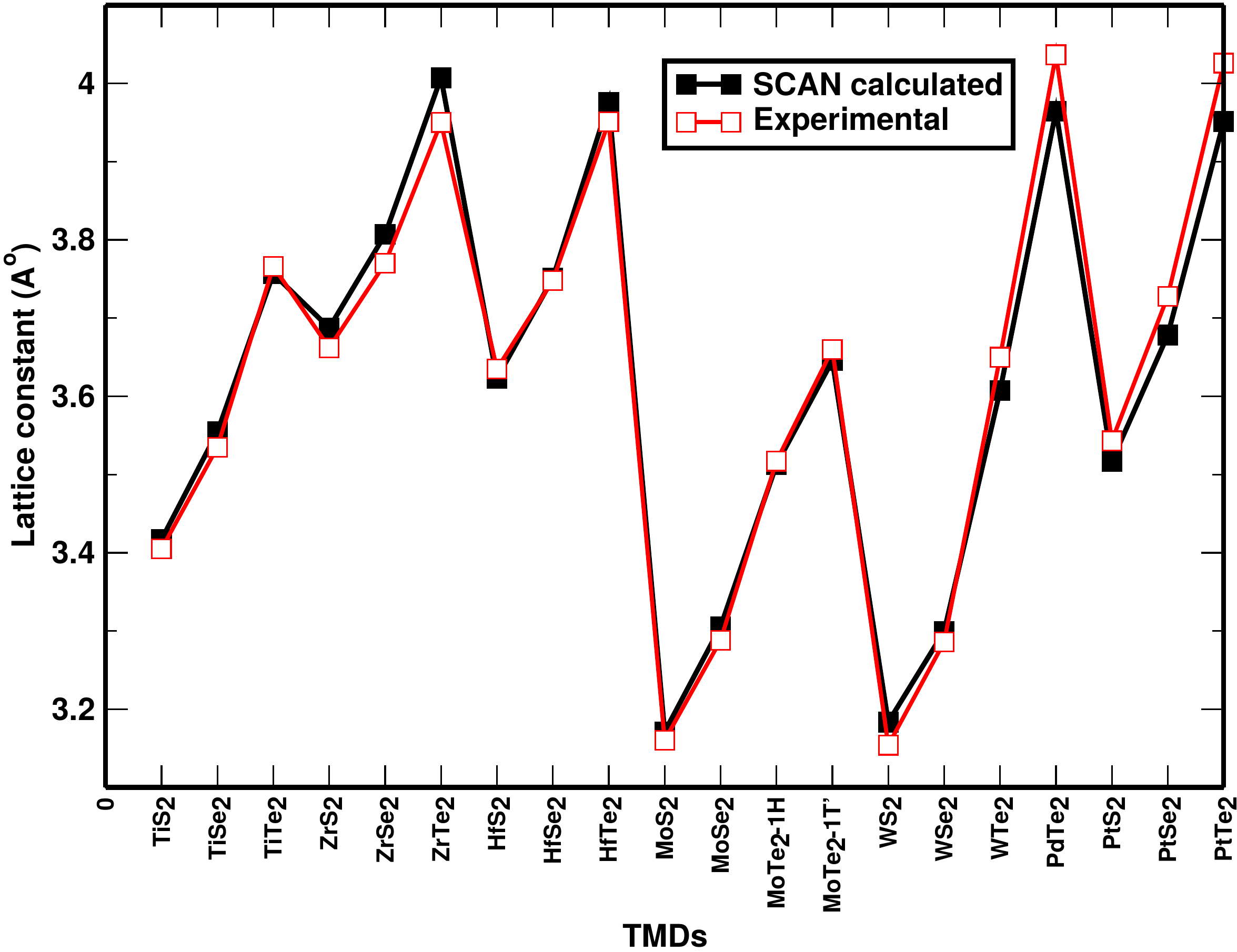}
	\captionsetup{font=scriptsize}
	\caption{Comparision of the SCAN-calculated in-plane lattice constants of various TMD mono-layers in the ground state with respect to the bulk lattice constants available in the literature  \cite{WY69, AMTTRD16, LWBR73}.}
	\label{fig:lattice}
\end{figure}

Comparison has been made for the estimated in-plane lattice constant of monolayers with the experimental bulk results in Figure \ref{fig:lattice}. The lattice constants are in good agreement with the experimental results with a mean absolute error (MAE) and a mean absolute percentage error (MAPE) of 0.03 $\AA$ and 0.7\% respectively. The results for the structural parameters related to the monolayer bulk are in good agreement with reference values  \cite{CHS12}. The structural parameters related to the lattice constant such as d$_{T-X}$, d$_{X-X}$, and $\theta_{X-T-X}$ increase from S to Se to Te. The decreasing cohesive energies from S to Se to Te make them more loosely bound, thereby increasing the lattice parameters.\\

\subsection{In-plane stiffness and Poisson's ratio}
The strength of a material is crucial for a device's performance and its durability. As a measure of the strength, we computed an in-plane stiffness or 2D Young's modulus (Eq.~\ref{Y2D}) of the most stable ground state and tabulated it in Table~\ref{tab:eff-t}. Similar to the cohesive energy, the in-plane stiffness decreases from S to Se to Te, indicating a softening of TMD monolayers from S to Te under an application of linear strain. The estimated 2D in-plane stiffness of MoS$_2$ is 141.59 N/m, which is in close agreement with the experimental value of 180 $\pm$ 60 N/m \cite{BBK11}. \\

Under Poisson's effect, materials tend to expand (or contract) in a direction perpendicular to the axis of compression (or expansion). It can be measured using Poisson's ratio $\nu_{ij} = -\frac{d\epsilon_j}{d\epsilon_i}$, where $d\epsilon_j$ and $d\epsilon_i$ are transverse and axial strains respectively. The in-plane (-$\frac{d\epsilon_y}{d\epsilon_x}$ or $-\frac{d\epsilon_x}{d\epsilon_y}$) and an out of plane Poisson's ratio (-$\frac{d\epsilon_z}{d\epsilon_x}$) are also calculated and tabulated. The in-plane Poisson's ratio is different than that of the out of plane Poisson's ratio for 1T compounds. For example, PtS$_2$ has $\nu_{xy} = 0.29$ and $\nu_{xz} = 0.58$. However, the Poisson's ratio of 1H monolayers is almost isotropic ($\nu_{xy} \approx \nu_{xz}$ ).

\subsection{Mechanical bending}
The primary focus of this study is to understand the response of the TMD monolayers to mechanical bending. We have calculated the bending stiffness and studied the change in various physical and electronic properties due to bending. Since previous studies  \cite{YRP16, ZDLHSR15} showed that the bending stiffness is independent of the type of the armchair or zigzag edges (chiral), we only utilized armchair-edge nanoribbons for the 1H structures. The bending stiffness of 20 TMDs are compared and tabulated in Table~\ref{tab:eff-t}. Unlike the in-plane stiffness, the overall bending stiffness increases from S to Se to Te (Table~\ref{tab:eff-t}), indicating a hardening of the nanoribbons from S to Se to Te. The d$^0$ compounds, especially S and Se, along with the PdTe$_2$ have lower ($<$ 3 eV) bending stiffness. The lower flexural rigidity of these compounds can result in enormous changes in their local strain as well as the charge density profile under mechanical bending. The 1H compounds have higher bending stiffness, possessing higher flexural rigidity against mechanical bending. The estimated bending stiffness of 12.29 eV for MoS$_2$ agrees with the experimental values of 6.62-13.24 eV  \cite{CSBAPJ14} as well as 10-16 eV \cite{ZDLHSR15}. To explore the trend of mechanical strengths with respect to transition metal, one can look into the d-band filling of valence electrons. The filling of the d band increases from transition metal group IV ($\sim$ sparsely-filled) to VI ($\sim$ half-filled) to X ($\sim$ completely filled) within the same row in periodic table. Both quantities Y$_{2D}$ and S$_b$ increase as the number of valence d electrons increases until the shell becomes nearly half-filled. To facilitate the claim further, we have estimated the in-plane stiffness and bending stiffness of 1H-NbS$_2$ and 1H-TaS$_2$ corresponding to group V (d$^1$) transition metals. The in-plane stiffness of NbS$_2$ and TaS$_2$ were found to be 95.74 N/m and 115.04 N/m respectively. In addition, the bending stiffness was obtained as 4.87 eV and 6.43 eV respectively for NbS$_2$ and TaS$_2$. Comparing TMDs (TX$_2$) having the same chalcogen atom, we can see the trend d$^0$ $<$ d$^1$ $<$ d$^2$ for both stiffness. However, there is a decrement in both Y$_{2D}$ and S$_b$ while going from half-filled (d$^2$) to nearly completely filled (d$^6$) d-band transition metal. Moreover, the large bending stiffness of group VI compounds decreases on changing phase from 1H to distorted 1T phase, for instance, 1H to 1T$^\prime$ transformation in MoTe$_2$.\\

\noindent We utilized 
\begin{align}
t_{eff} = \sqrt{12S_b/Y_{2D}} \\
\shortintertext{and}
Y_{3D} = Y_{2D}/t_{eff}
 \end{align}
 to estimate the effective thickness as well as the 3D Young's modulus. An effective thickness is the combination of d$_{X-X}$ distance and the total effective decay length of electron density into the vacuum. Experimentally, it is difficult to define the total effective decay length of the electronic charge distribution. Therefore, it is a common practice to take a range from the d$_{X-X}$ distance to the inter layer metal-metal distance within the bulk structure as the effective thickness, which gives the range for both in-plane stiffness and bending stiffness \cite{CSBAPJ14, ZDLHSR15}. Using equation 3, one can estimate a reasonable value for the effective thickness for a wide range of TMDs. However, the computed effective thicknesses t$_{eff}$ of certain TX$_2$ (T=Ti, Zr, Hf; X=S, Se) are less than their d$_{X-X}$ distance (Figure 1), which means that bending is much easier than stretching. Similar underestimation was found for the effective thickness of a carbon monolayer estimated by various methods  \cite{YS97, W04, KGB01, Z00}. Utilizing eq. (3), Yakobson et al.  \cite{YS97}, Wang  \cite{W04}, and Yu et al.\cite{YRP16} estimated the effective thickness of the carbon monolayer to be around 0.7-0.9 $\AA$, which is much less than 3.4 $\AA$, the normal spacing between sheets in graphite. Such huge underestimation indicates the possible breakdown of the expression (3) to estimate the effective thickness in the case of atomically thin carbon layer \cite{W04}. The 3D Young's modulus (eq.4) allows us to compare the strength between various 2D and 3D materials, for instance, MoS$_2$ against steel. Similar to 2D in-plane stiffness, the 3D Young's modulus of TMD monolayers decreases from S to Se to Te. Due to the larger underestimation of the effective thickness, there is a huge overestimation in the 3D Young's modulus of group IV compounds with sulfur as the chalcogen atom.  With that in mind, one can conclude that MoS$_2$ as well as WS$_2$ have large 3D Young's moduli of 347.03 GPa and 351.02 GPa respectively, agreeing with the experimental value of 270$\pm$100 GPa \cite{BBK11} for MoS$_2$.\\

\subsection{Effect of bending on physical properties}
\textbf{I. Local Strain} \\

Local strain ($\epsilon$ $=$ $\frac{\delta - \delta_{flat}}{\delta_{flat}}$) projected on the y-z plane (see b-c plane in Figure~\ref{fig:ribbon} (II)) of different TMD nano-ribbons corresponding to the bending curvature around 0.09 $\AA^{-1}$ are presented in {\color{blue}Supplementary Figure S1}. The inner layer gets contracted while the outer layer gets expanded, and this is consistent with the elastic theory of bending of a thin plate  \cite{LL86}. The expansion of the outer layer is close to the contraction of the inner layer for 1T compounds, while the expansion dominates the contraction in the case of 1H compounds ({\color{blue}Supplementary Figure S1}). The middle metal layer is expanded up to 2\% in the case of 1T while it is 5-10\% for 1H, indicating that the middle layer is closer to the neutral axis for 1T than that of the 1H compounds. For 1T$'$ compounds (MoTe$_2$ and WTe$_2$), the outer layer is expanded more as compared to the contraction of the inner layer with a distortion represented by the zigzag structure in the strain profile ({\color{blue} Supplementary Figure S1}).\\

\begin{figure}[h!]
    \includegraphics[scale=0.4]{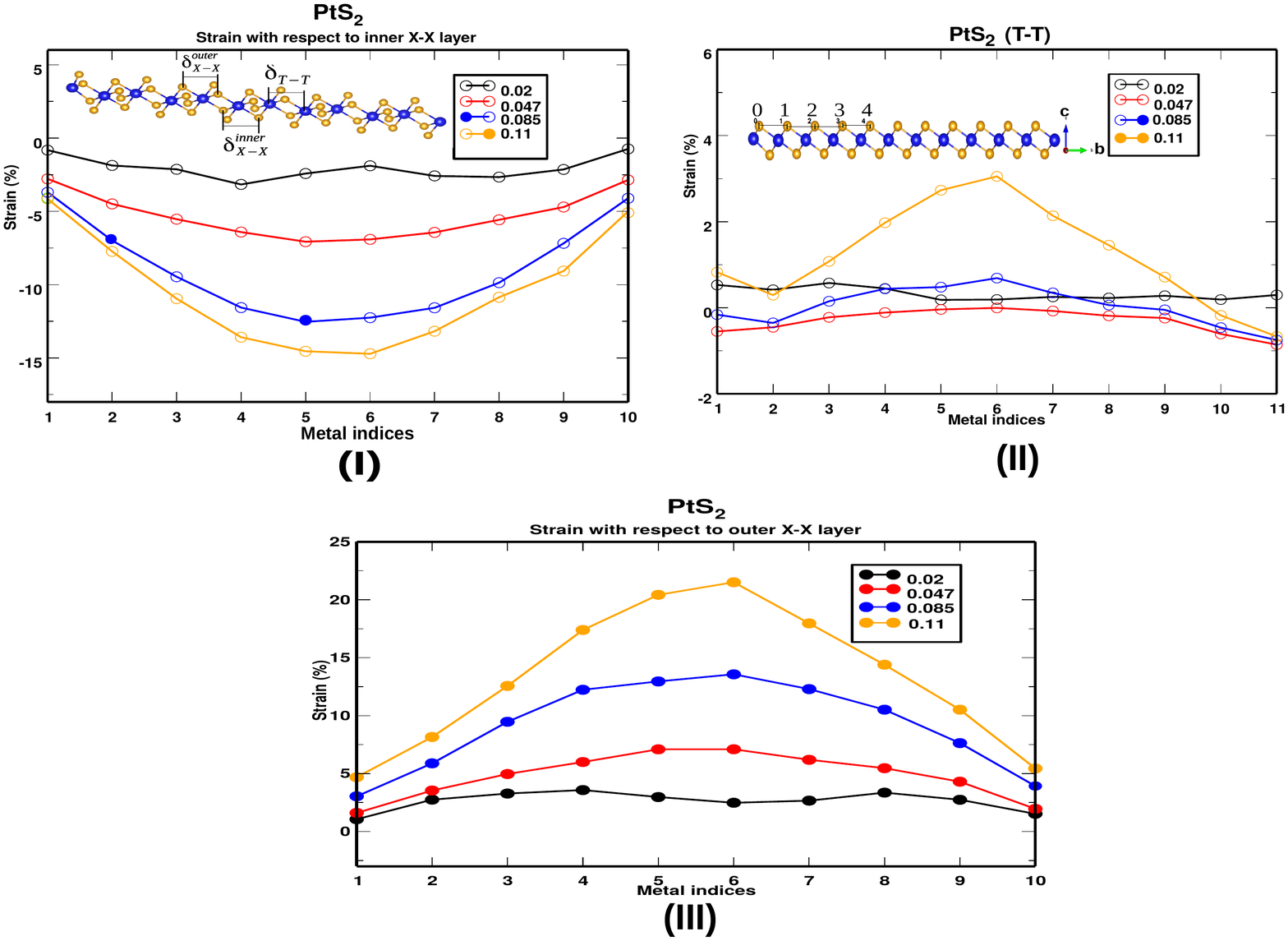}
	\caption{Local strain ($\epsilon$ $=$ $\frac{\delta - \delta_{flat}}{\delta_{flat}}$) with respect to the inner chalcogen-chalcogen ($\epsilon^{inner}_{X-X}$), metal-metal ($\epsilon_{T-T}$), and outer chalcogen-chalcogen distance ($\epsilon^{outer}_{X-X}$)  projected in the y-z plane for PtS$_2$. Strain at metal indices ``i" (see 2$^{nd}$ subfigure) is calculated with respect to the distance between two metals at indices i-1 and i where i = 1, 2, ...10 (or 11)}
		\label{fig:strain-PtS}
\end{figure}

To study the effect of bending on the local strain profiles, we compare the local strain profiles of the PtS$_2$ nano-ribbon projected on y-z plane, as shown in Figure~\ref{fig:strain-PtS}. The inner layer is contracted while the outer layer gets expanded. This effect increases upon increasing the bending curvature. For PtS$_2$, the middle layer is expanded within 2-3\%, while the expansion is 16-20\% for the inner and the outer layer. Such large local strain can induce a highly non-uniform local potential and hence affect the charge distribution. Both lattice expansion in the outer layer and the lattice contraction in the inner layer could be applicable in tuning adsorption (binding distance and energy) of the 2D materials, similar to the linear strain modulated adsorption properties of various semiconducting or metallic surfaces  \cite{KDCF14, CW16, MHN98}. The tensile strain strengthens the hydrogen adsorption in TMD surfaces, while a compressive strain weakens it \cite{CW16}. By utilizing both the concave (compressive strain) and convex (tensile strain) surfaces of a bent monolayers, one can tune the Gibb's free energy of hydrogen adsorption to zero when it is respectively more negative and more positive. \\

\pagebreak

\textbf{II. Physical thickness}\\

\begin{figure}[h!]
	\includegraphics[scale=0.5]{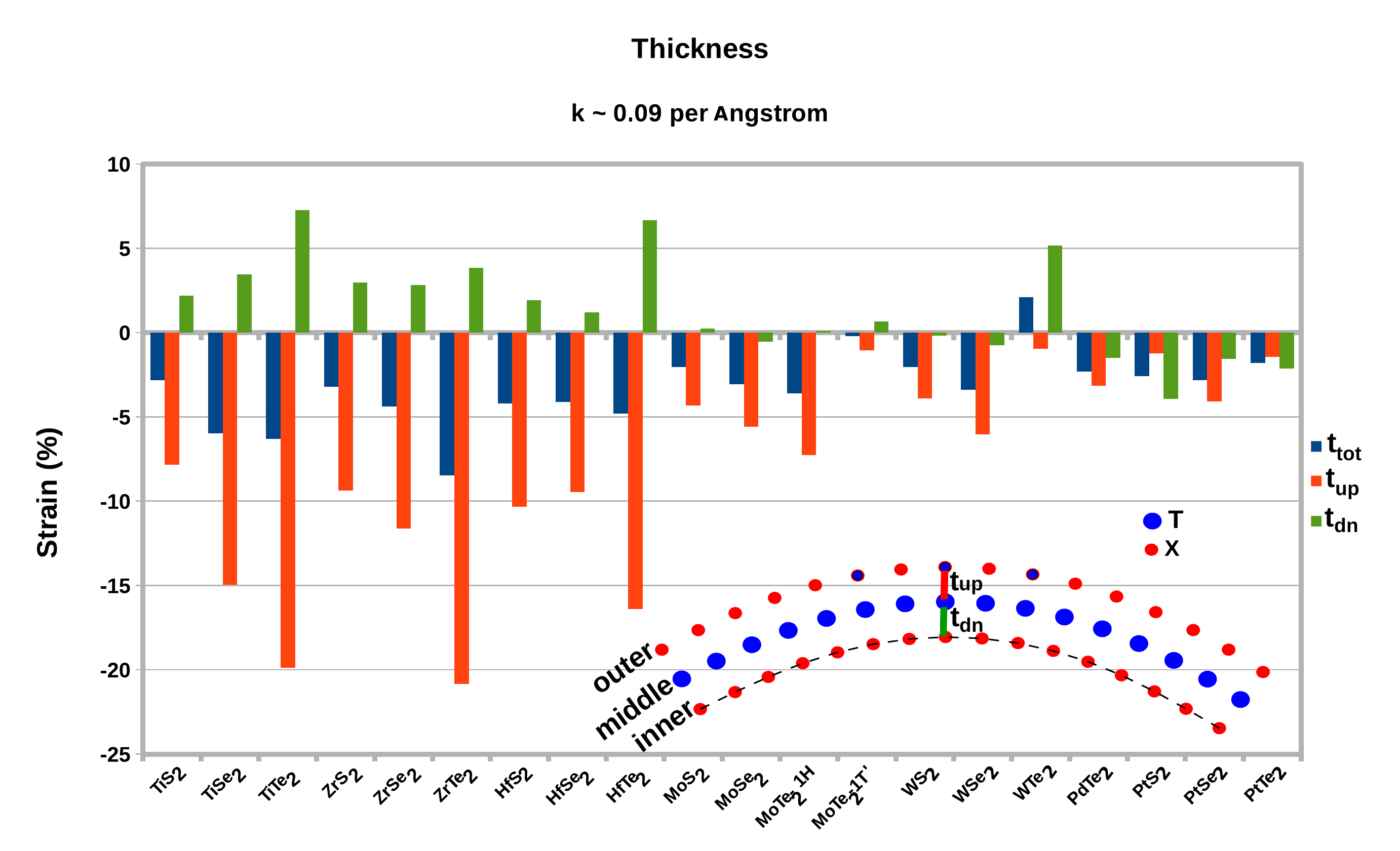}
		\caption{The strain with respect to the physical thickness of the bent nano-ribbon around 0.09 $\AA^{-1}$ for various TMD compounds; t$_{tot}$ (t$_{up}$ + t$_{dn}$, {\color{blue} blue}) is outer-inner layer thickness; t$_{up}$ ({\color{red} red}) and t$_{dn}$ ({\color{green} green}) are measured between outer-middle and middle-inner layers respectively  (see Figure~\ref{fig:ribbon} (III)).}
	\label{fig:compare-t}
\end{figure}
The behavior of different layers within the TMD nano-ribbon under mechanical bending can be understood by looking at the variation of the physical thickness (t$_{tot}$, defined later in this section and shown in Figure~\ref{fig:compare-t}) with respect to bending curvature. Moreover, tuning of the physical thickness can be particularly useful in nano-electronic applications due to an enhancement of the electron confinement in 2D materials with an out-of-plane compression  \cite{BB18, GDKF17}. A percentage change in the thickness (t$_{tot}$, t$_{up}$, or t$_{dn}$) at the middle of various bent nano-ribbons with respect to the unbent ones is presented in {\color{blue}Supplementary Figure S2}. t$_{tot}$ represents an outer-inner chalcogen atom layer thickness at the vertex of a bent ribbon, while t$_{up}$ and t$_{dn}$ correspond to outer-middle and middle-inner layers respectively. We fitted a 6$^{th}$ order polynomial to each layer of the bent nanoribbon to estimate the thickness ({\color{blue}Supplementary Figure S3}). The thickness measured between outer and inner chalcogen layers is described by t$_{tot}$ (t$_{up}$ + t$_{dn}$, {\color{blue} blue}) while t$_{up}$ ({\color{red} red}) and t$_{dn}$ ({\color{green} green}) are measured between the outer-to-middle and middle-to-inner layers respectively (see Figure~\ref{fig:compare-t}). When a thin plate is bent, it undergoes both compression (z' to N, t$_{dn}$) and expansion (N to z'+h, t$_{up}$) with ``N" being the neutral surface  \cite{LL86} (see figure~\ref{fig:ribbon} (III)). As the middle layer does not mimic the neutral surface (N), t$_{up}$ and t$_{dn}$ do not respectively increase and decrease with the bending curvature. For most of the compounds, t$_{up}$ decreases on increasing the bending curvature. On the other hand, t$_{dn}$ slightly increases for d$^0$-1T compounds, but depends on the bending curvature for d$^2$-1H and d$^6$-1T compounds ({\color{blue}Supplementary Figure S2}). For a quantitative comparison among different materials, we plot the thicknesses for various TMDs around the bending curvature of 0.09 $\AA^{-1}$ as shown in Figure~\ref{fig:compare-t}.
Group IV compounds have a lower flexural rigidity, therefore have more of a decrement in the physical thickness (t$_{tot}$) than group VI and X compounds. \\

\subsection{Effect of the bending on electronic properties}
\begin{figure}[h!]
	\includegraphics[scale = 0.46]{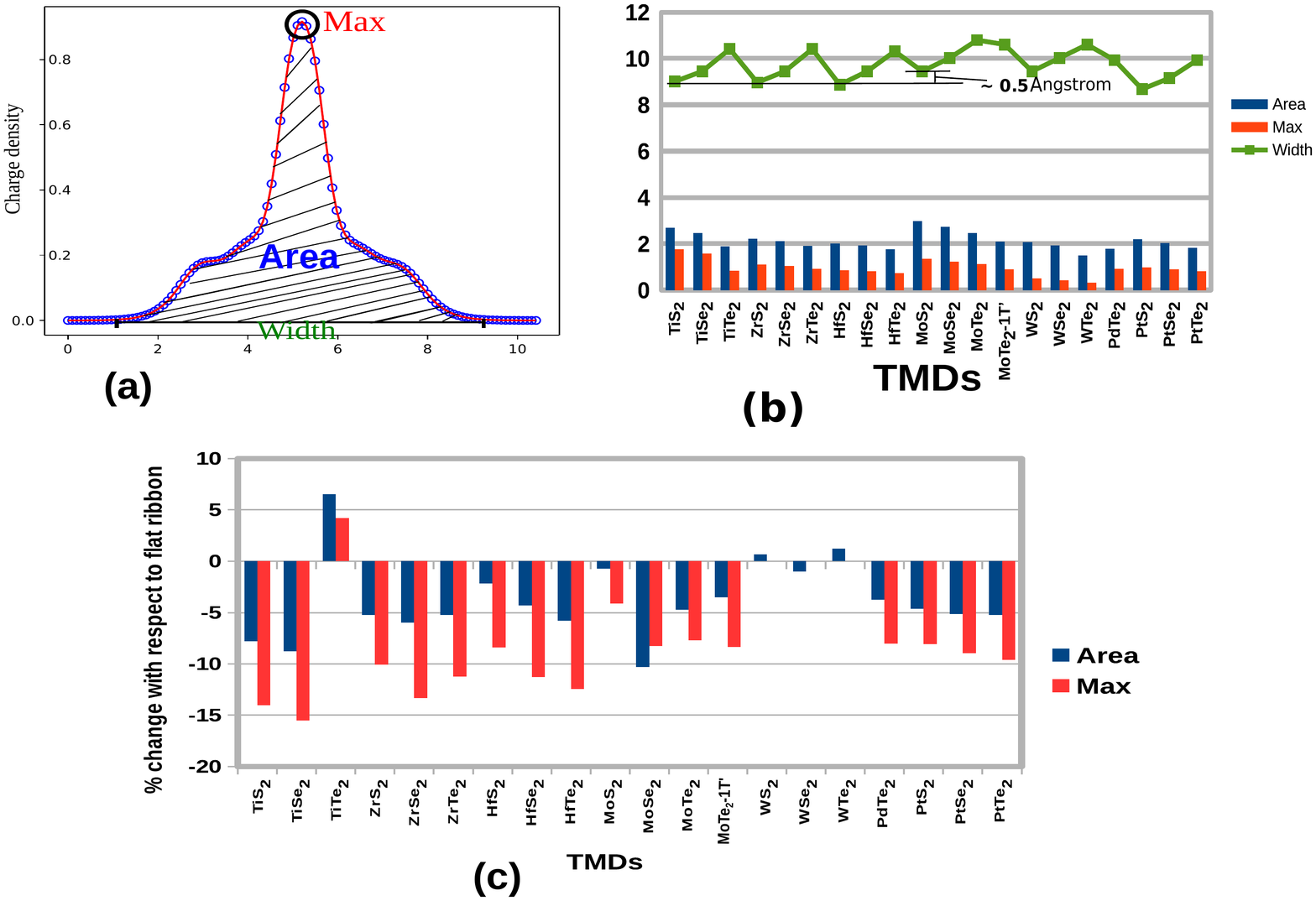}
	\caption{(a) The local charge density along the out of plane (z) direction of the nano-ribbon. (b) Width ($\AA$), max ($e/\AA^3$, e: an electronic charge), and areal density ($e/{\AA^{2}}$) of flat nanoribbon. (c) \% change in an area and the max of the bent nanoribbons having a bending curvature around 0.09 $\AA^{-1}$ with respect to the flat nano-ribbon; result of max. value is not shown for WX$_2$ as it possesses multiple local maxima.}
	\label{fig:local-charge}
\end{figure}

\textbf{I. Local electronic charge density}\\
Along with the change in physical properties, mechanical bending also affects the electronic properties. The local charge density (average over ab-plane, [Figure~\ref{fig:ribbon} (I)]) is computed and plotted against distance along an out-of-plane direction (\textit{c}- axis) [Figure~\ref{fig:ribbon} (II)]. The different nature of the local charge distribution of flat WX$_2$ (X=S, Se, Te) ribbon with two equal local maxima may be related to the different pseudopotential used in the calculation. We choose a narrow window (within 2 black vertical lines) at the middle of a nano-ribbon (for both flat and bent) to study the local charge distribution near the surface-vacuum interface as shown in {\color{blue}Supplementary Figure S4}. We define 3 different quantities Width, Max, and an Area of the local charge density (left) and compared among the flat nano-ribbons of various TMDs (right), as shown in Figure~\ref{fig:local-charge}. The ``Width" represents the distance over which the charge density decays to a smaller non-zero value ($\epsilon < 10^{-4}$) in vacuum ({\color{blue} Supplementary Figure S4}) which also gives a tentative idea about the total effective decay length of electron density. In addition, the areal density ($\int_{0}^{Width}\rho(z)dz$, an area under the curve) represents the average number of electrons per unit area, as shown in Figure~\ref{fig:local-charge}.\\

For the flat nano-ribbons, the width increases whereas Max and the Area decrease as we go from S to Se to Te for a given transition metal. Increasing the width from S to Se to Te indicates an increase in the total effective decay length of electron density, hence the effective thickness. Also, the width corresponding to flat 1H nano-ribbons is shifted upward by atleast 0.5 $\AA$ compared to that of 1T flat nano-ribbons which then contributes to an effective thickness giving larger bending stiffness. Our results suggest that the overall bending stiffness follows the trend of the width of an electron density and hence the effective thickness.  
The variation of the local charge density along an out of plane direction for different TMD nano-ribbons with the bending curvature is presented in {\color{blue}Supplementary Figure S5}. When a nanoribbon is bent, the local charge density shrinks with the bending curvature within an outer layer-vacuum interface while expanding near the inner layer-vacuum interface leaving the total width unaffected. However, both the Max and the Area decrease with increasing bending curvature for most of the TMD compounds except for TiTe$_2$ and WX$_2$. For WX$_2$, the max. value of local maximum closer to the surface-vacuum interface decreases on increasing the bending curvature (Circular region in the {\color{blue}Supplementary Figure S5}) whereas the other local maxima have an opposite trend. To study the effect of bending on the aforementioned local maximum (Max) and areal density (Area) among different materials, we estimate their percentage change with respect to the flat ribbon, as in Figure~\ref{fig:local-charge}. The bending produces noticeable changes in the charge distribution within the surface-vacuum interfaces.\\

\begin{table}[h!]
	\centering
	\captionsetup{font=scriptsize}
	\caption{The ground state properties of TMD mono-layers having 1H or 1T phase: Relaxed in-plane lattice constant, a; Metal-chalcogen and chalcogen-chalcogen distance, $d_{T-X}$ and $d_{X-X}$ respectively (See Fig \ref{fig:struc}); X-T-X angle, $\theta_{X-T-X}$; Cohesive energy per atom, $E_c$; in-plane ($\nu_{in}$) and out-of-plane ($\nu_{out}$) Poisson's ratios; 2D Young's modulus, $Y_{2D}$; Bending stiffness, S$_b$; and Effective thickness, t$_{eff}$. Results for structural parameters of TiX$_2$ (X $=$ S, Se, Te), MoX$_2$, and WX$_2$ are in good agreement with the LDA+U results from reference~\onlinecite{CHS12}. The structure parameters of distorted T compounds,WTe$_2$ and MoTe$_2$ can be estimated from {\color{blue}Supplementary Table S2}. The representations of T$^{4+}$ such as d$^0$, d$^2$, and d$^6$ are taken from Ref.~\onlinecite{WY69}.}
	\resizebox{1.0\textwidth}{!}{%
	\begin{tabular}{lrrrrrrrrrrrr}
		\hline
		\hline
		T$^{4+}$ & TMDs  & \multicolumn{1}{l}{a} & \multicolumn{1}{l}{d$_{T-X}$} & \multicolumn{1}{l}{d$_{X-X}$} & \multicolumn{1}{l}{$\theta_{X-T-X}$} & \multicolumn{1}{l}{E$_c$} & \multicolumn{1}{l}{$\nu_{in}$} & \multicolumn{1}{l}{$\nu_{out}$}& $Y_{2D}$& S$_b$ & t$_{eff}$ & $Y_{3D}$($\frac{Y_{2D}}{t_{eff}}$)\\
		& & \multicolumn{1}{l}{($\AA$)} & \multicolumn{1}{l}{($\AA$)} & \multicolumn{1}{l}{($\AA$)} & \multicolumn{1}{l}{degree} & \multicolumn{1}{l}{(eV/atom)} & & & (N/m)& (eV) & ($\AA$) &  (GPa)\\
		\hline
	 d$^0$	& TiS$_2$ & 3.42 & 2.42 & 2.80 & 90.16 & 6.80 & 0.17 & 0.42  & 85.20  & 2.25 & 2.25 & 378.67\\
	& TiSe$_2$ & 3.55 & 2.55 & 3.04 & 91.76 & 6.17 & 0.23 & 0.43  & 59.74  & 2.86  & 3.03 & 197.72\\
	&	TiTe$_2$ & 3.76 & 2.77 & 3.44 & 94.55 & 5.41 & 0.24 & 0.38  & 44.46  & 3.29  & 3.77 & 117.93\\
	&	ZrS$_2$ & 3.67 & 2.57 & 2.87 & 88.14 &  7.35 & 0.19 & 0.52  & 83.76  & 2.13  & 2.21 & 379.00\\
	 &	ZrSe$_2$ & 3.81 & 2.70 & 3.12 & 90.14 & 6.71 & 0.22 & 0.47  & 71.30  & 2.57  & 2.63 & 271.10\\
		& ZrTe$_2$ & 4.01 & 2.91 & 3.53 & 92.94 & 5.89 & 0.18 & 0.44  & 43.16  & 3.01  & 3.66 & 117.92\\
	  &	HfS$_2$ & 3.62 & 2.53 & 2.85 & 88.65 & 7.35 & 0.19 & 0.52   & 85.78  & 2.82 & 2.51 & 341.75\\
	  &	HfSe$_2$ & 3.75 & 2.66 & 3.09 & 90.37 & 6.67 & 0.21 & 0.47  & 77.75  & 3.64  & 3.00 & 259.17\\
	  &	HfTe$_2$ & 3.98 & 2.88 & 3.47 & 92.58 & 5.80 & 0.15 & 0.41  & 46.77  & 3.92  & 4.01 & 116.63\\
	  \hline
	  d$^2$ &	MoS$_2$ & 3.17 & 2.40 & 3.10 & 80.56 & 7.86 & 0.26 & 0.30   & 141.59 & 12.29  & 4.08 & 347.03\\
	  &	MoSe$_2$ & 3.30 & 2.53 & 3.31 & 81.86 & 7.22 & 0.26 & 0.32  & 114.97 & 14.60 & 4.94 & 232.73\\
	  &	MoTe$_2$-1H & 3.51 & 2.71 & 3.59 & 83.04 & 6.54 & 0.28 & 0.34  & 87.88  & 14.63  &
	  	5.65 & 155.54\\
	  &	MoTe$_2$-1T$'$ & 3.65 & -- & -- & -- & 6.54 & 0.28 & 0.46  & 61.85  & 7.28  &
	  	  	4.75 & 130.21\\
	  &	WS$_2$  & 3.16 & 2.40 & 3.10 & 80.25 & 7.91 & 0.26 & 0.33   & 143.92 & 12.61 &
		4.10 & 351.02\\
	  &	WSe$_2$ & 3.29 & 2.53 & 3.32 & 82.16 & 7.20 & 0.33 & 0.35   & 130.03 & 14.48 &
		4.62 & 281.45\\
	&	WTe$_2$-1T$'$ & 3.61 & -- & -- & -- & 6.49 & 0.35 & 0.60   & 86.79 & 8.96 &
			4.45 & 195.03\\
	
	  	\hline
	
	 d$^6$ &	PdTe$_2$ & 3.96 & 2.67 & 2.73 & 83.91 &  4.07 & 0.32 & 0.64 & 61.82  & 2.78  &
	  2.94 & 210.27\\
	  &	PtS$_2$ & 3.52 & 2.37 & 2.45 & 84.25 & 5.73 & 0.29 & 0.58   & 105.81  & 5.66 &
		3.20 & 330.65\\
	  &	PtSe$_2$ & 3.68 & 2.49 & 2.60 & 84.83 & 5.32 & 0.26 & 0.59  & 87.01 & 6.33  &
		3.74 & 232.65\\
	  &	PtTe$_2$ & 3.95 & 2.66 & 2.74 & 84.15 & 5.07 & 0.26 & 0.57  & 81.41  & 4.58 &
		3.29 & 247.45\\
		\hline
		\hline
	\end{tabular}}%
	\label{tab:eff-t}%
\end{table}%


 \textbf{II. Band structure}\\
 
\begin{figure}[h!]

\includegraphics[scale=0.3]{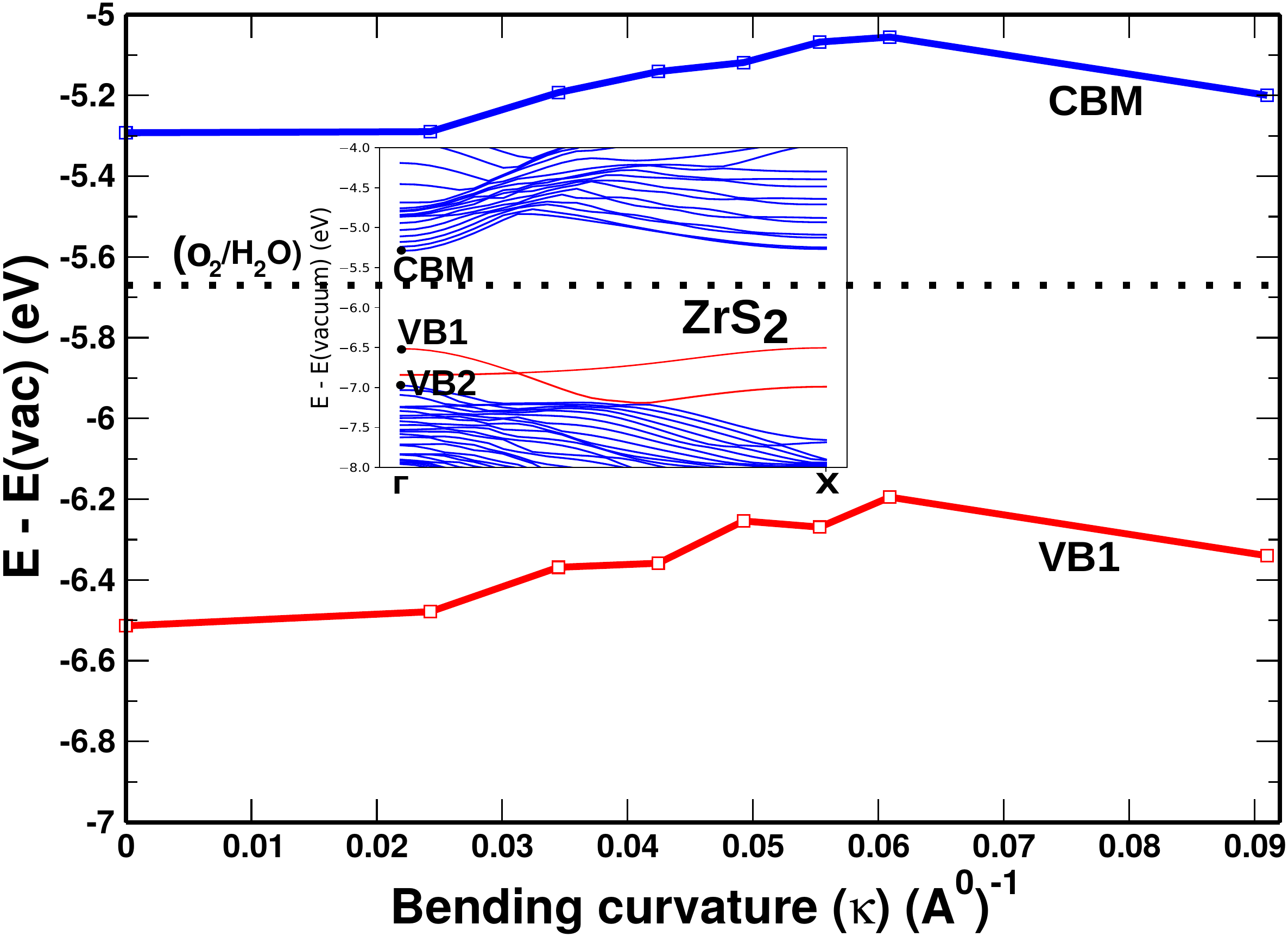}
\includegraphics[scale=0.3]{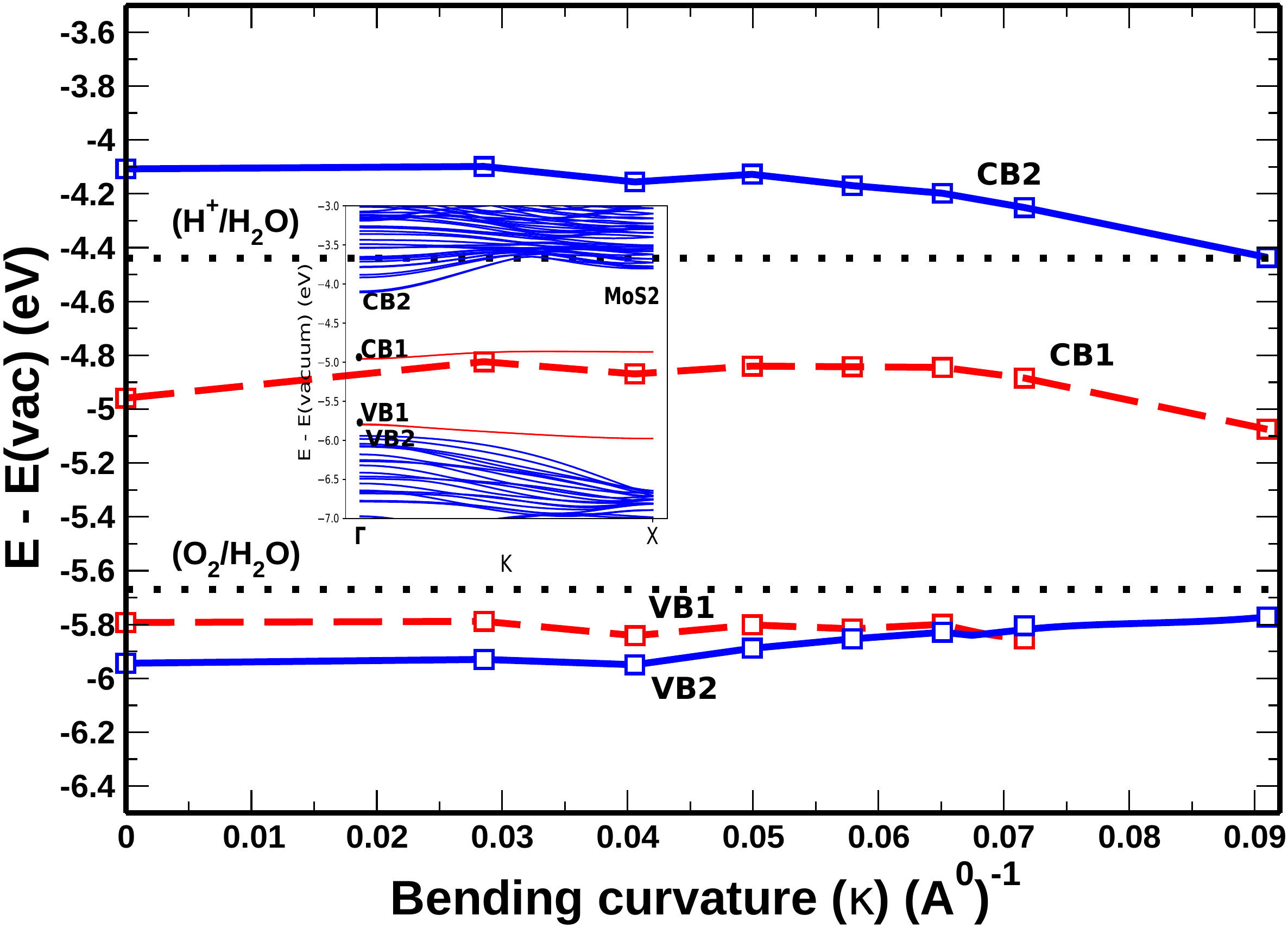}
\caption{Variation of band edges with respect to the bending curvature for ZrS$_2$ (left) and MoS$_2$ (right); CBM and VB1 are the conduction band minimum and edge state VB (valence band) respectively; CB1 (CBM), CB2, VB1 (VBM), and VB2 respectively are edge state CB (conduction band), bulk CB, edge state VB (valence band), and bulk VB. For flat MoS$_2$ ribbon, VB1 represents the VBM while for higher bending curvature ($\kappa$ $=$ $0.09 \AA^{-1}$) VB2 switches to VBM.}
\label{fig:band-semi}
\end{figure}

The band structure plots of group IV, VI, and X TMDs with respect to vacuum with various bending curvatures are shown in {\color{blue}Supplementary Figures S6, S7, and S8} respectively. The dashed lines in the band structure plots indicate the SCAN estimated Fermi energy with respect to vacuum (``-ve" of the work function) while the red bands correspond to in-gap edge states. The edge states are identified by comparing the band structures of the ribbon with that of the monolayer bulk, and are highlighted by red color. The bulk band-gap (E$_g$ (eV)) (excluding edge states) and the work function ($\phi$ (eV)) of our flat nano-ribbons are extracted and tabulated in {\color{blue} Supplementary Table S1}. Out of TMD nano-ribbons considered, ZrX$_2$, HfX$_2$, MoY$_2$, and WX$_2$ (X = S, Se; Y = S, Se, Te) are semiconductors. To study the changes in the band structure of these semiconductors with respect to bending, we utilized the hydrogen passivated edges. A few of the low band-gap semiconductors such as TiY$_2$, TTe$_2$ (T=Zr, Hf) and group (X) indirect band-gap semiconductors (PtX$_2$) become metallic due to the edge states. We did not observe any substantial effect of bending on metallic compounds. An effect of the mechanical bending on the band-gap is of particular interest for semiconductors, due to a wide range of applications in nano-electronics. One each from the 1T and the 1H group, respectively ZrS$_2$ and MoS$_2$, are chosen to study the effect of bending on the band structure as shown in Figure~\ref{fig:band-semi}.\\

The nature of edge states is different for 1T and 1H semiconductors. The 1T nanoribbon has edge states only below the Fermi level while both the edge states above and below the Fermi level are present in the 1H nanoribbon. The horizontal black dashed lines represent water redox potentials with respect to the vacuum level, -4.44 eV for the reduction (H$^+$/H$_2$), and -5.67 eV for the oxidation (O2/H$_2$O) at pH 0  \cite{CAAWSS07}. When the band edges straddle these potentials, materials possess good water splitting properties. The band edges CB2, VB1 (VBM), and VB2 of MoS$_2$ straddle the water redox potentials while only the edge state CB1 stays within the gap. As semilocal DFT functionals underestimate the band gap  \cite{P85}, a correction is always expected at the G$_0$W$_0$ level ({\color{blue} Supplementary Table S1}), which shifts the bands above and below the Fermi level even further up and below respectively  \cite{YRP16}. However, it is known that such correction for localized states (in the case of point defects) is less considerable than that for the delocalized bulk states  \cite{ABP08}.\\

\textbf{(a) Tuning of band edges}\\

The band edges (conduction band minimum (CBM) and valence band maximum (VBM)) of ZrS$_2$ and other 1T semiconductors increase on increasing the bending curvature, while this varies from one band edge to another for MoS$_2$ and other 1H semiconductors. For example, shifting of the band energies with respect to vacuum is negligible for edge states as compared to the bulk ones for MoS$_2$. The shifting of band edges also leads to changing of the Fermi level as well as the band gap ({\color{blue} Supplementary Figure S10)}. For MoS$_2$, VB2 increases while VB1 decreases on increasing the bending curvature and eventually results in the removal of some of the edge states, though, complete elimination might not be possible. Since the mechanical bending shifts the band edges only by a little, the photocatalytic properties of MoS$_2$ and WS$_2$ is preserved even for a larger bending curvature. On the other hand, bending can shift the band edges of 1T semiconductors by a considerable amount for bending curvature up to 0.06 $\AA^{-1}$, but shift downward for higher bending curvature. For example, one can shift the band edges of ZrS$_2$ upward by 0.25 eV when applying the bending curvature of 0.06 $\AA^{-1}$. Moreover, the G$_0$W$_0$ calculated band structure shows that the CBM (-4.58 eV and -4.53 eV respectively) of ZrS$_2$ and HfS$_2$ is slightly lower than the reduction potential (-4.44 eV) while the VBM (-7.15 eV and -6.98 eV) is significantly lower than the water oxidation potential (-5.67 eV)  \cite{ZH13}. Mechanical bending can shift the band edges in the upward direction to straddle the water redox potentials, enhancing the photocatalytic activity. The effect of bending on the band edges of 1H-TSe$_2$ semiconductors is different than that of 1H-TS$_2$ ({\color{blue} Supplementary Figure S9}), especially in the bulk valence band maximum (VB2). The VB2 is almost constant for lower bending curvature for TSe$_2$, while there is an appreciable increase in the case of TS$_2$.\\ 

\pagebreak
\textbf{(b) Charge localization and Conductivity}\\

\begin{figure}[h!]
	\includegraphics[scale=0.5]{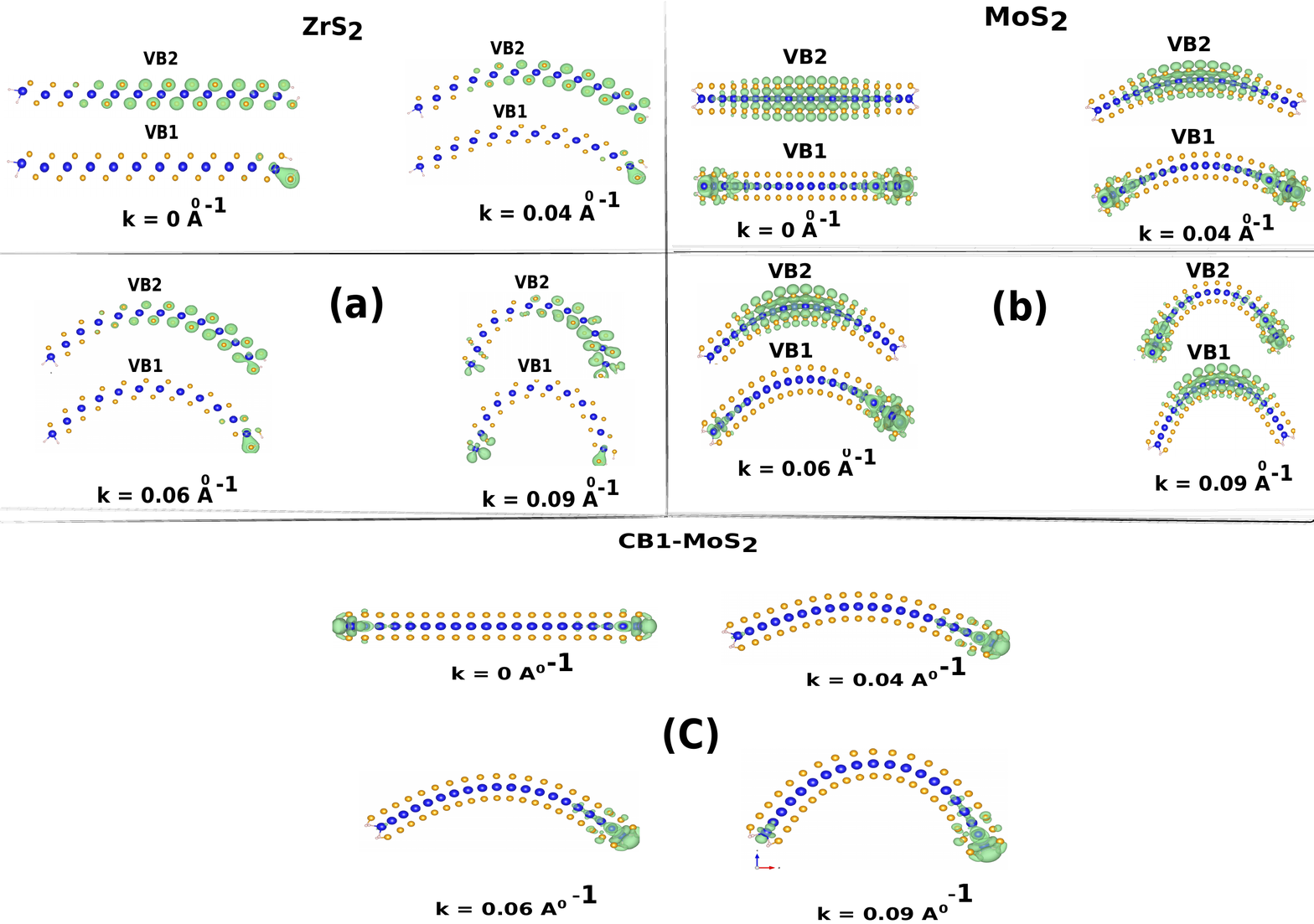}
	\caption{{Variation in the isosurface of partial charge densities at VB1 and VB2 (holes) with respect to the bending curvature; (a) ZrS$_2$; (b) MoS$_2$; (c) Variation in the isosurface of the partial charge densities (donor-like) of MoS$_2$ at CB1 with bending curvatures.}}
	\label{fig:band-charge}
	
\end{figure}

\begin{figure}[h!]
	\includegraphics[scale=0.4]{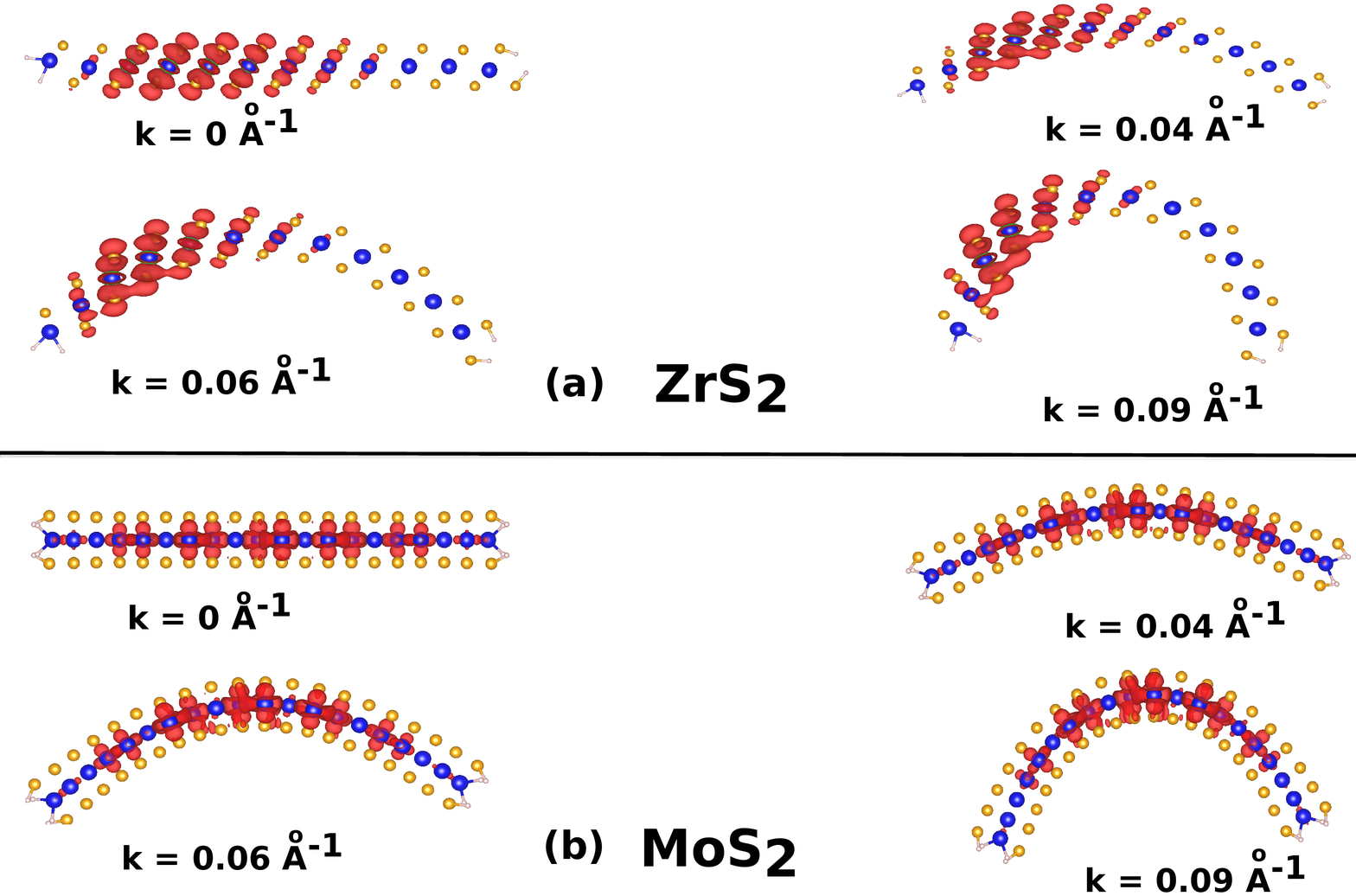}
\caption{Variation in the isosurface of partial charge density (electrons) with respect to the bending curvature at bulk conduction band minimum; (a) CBM for ZrS$_2$; (b) CB2 for MoS$_2$.}
	\label{fig:cbm}
\end{figure}

In this section, we describe the effect of bending on band edges in terms of localization or delocalization of the charge carriers at those band edges. The variation of an isosurface of the partial charge (electrons or holes) density with respect to bending curvature are presented in Figures~\ref{fig:band-charge} and \ref{fig:cbm}. Using the mechanical bending, one can tune the conductivity of TMD monolayers \cite{YRP16}. Before bending, the charge carriers (holes) of ZrS$_2$ at VB2 are delocalized over the whole ribbon width, decreasing in magnitude from S-edge to Zr-edge. The mechanical bending localizes the charges towards the S-edge while depleting along the Zr-edge, reducing the conductivity from one edge to the other. On the other hand, the charge density on top of VB1 does not change much with the bending for lower bending curvatures. However, at $\kappa$ $=$ 0.09 $\AA^{-1}$ some charges accumulate at the Zr-edge, thereby changing the trend of band energy with respect to vacuum (see Figure~\ref{fig:band-semi}). Unlike ZrS$_2$, the charge carriers (holes) of MoS$_2$ at VB2 are delocalized over the whole width, decrease in magnitude from the center of the ribbon to either side of edges symmetrically. With bending, the charge carriers localize at the middle of the ribbon and deplete at the edges, reducing the conductivity due to holes from one edge to the other \cite{YRP16}. At a higher bending curvature beyond $\kappa > 0.065 \AA^{-1}$, edge state VB1 crosses the bulk-VB and becomes VB2 and vice versa. Similar to VB1, CB1 also has the same behavior before and after bending, except it does not cross the CB2. Instead, it is also shifted down as VB1 does.\\
	
Conversely, the charge carriers (electrons) of ZrS$_2$ at the CBM (CB2) decrease in magnitude from Zr-edge to the S-edge. Again, mechanical bending localizes the electrons towards the Zr-edge. On the other hand, the electronic conductivity does not change even for larger bending curvature for MoS$_2$. The electrons are delocalized uniformly over the whole ribbon width which remains unaffected for a wide range of bending curvature. The conductivity of a semiconductor is the sum of conductivity of both electrons and holes. The mechanical bending reduces both types of conductivity in 1T semiconductors, while it only reduces hole-type conductivity in 1H semiconductors.\\

\pagebreak

\textbf{G.  Stability of nano-ribbons and finite width effect}\\ 

\begin{figure}[h!]
 \includegraphics[scale=0.4]{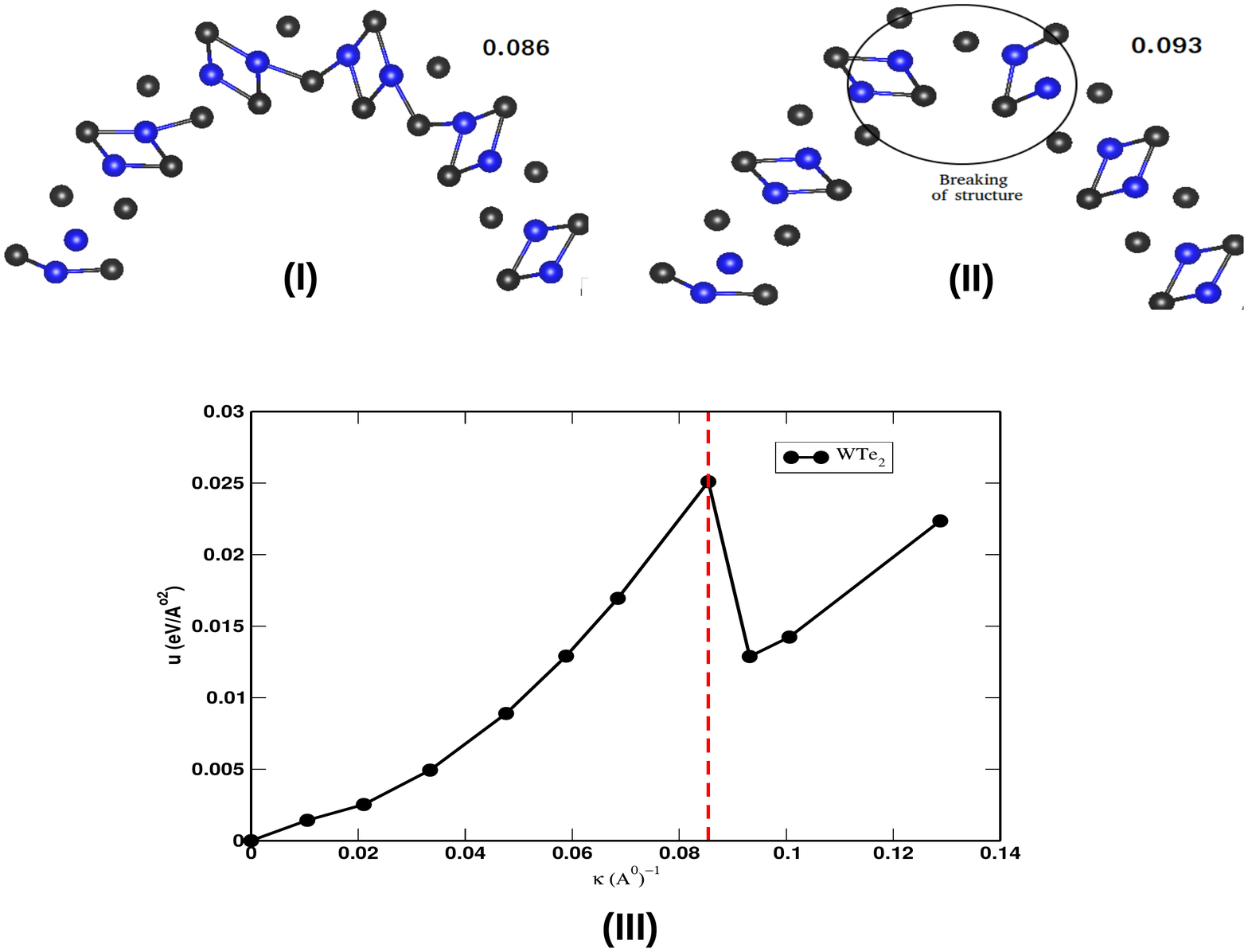}
  \caption{(I)-(II): Structures for 2 different bending curvatures, showing the breaking of the ribbon within the curvature region; The figure on left is for $\kappa$ $=$ 0.086 $\AA^{-1}$ while the one on the right is for $\kappa$ $=$ 0.093 $\AA^{-1}$. (III) An areal bending energy density with respect to bending curvature for WTe$_2$, showing the breaking of structure.}
 \label{fig:WTe}
 \end{figure}

Based on our calculation, we have found that the stability of the flat nano-ribbons also depends on the type of edge. We have taken stoichiometric (n(X):n(T)$=$ 2:1) nano-ribbons ({\color{blue} Supplementary Figure S4}) for most of the calculations. However, we could not relax TiSe$_2$, HfS$_2$, PdTe$_2$, and PtSe$_2$ nano-ribbons in this configuration. We confirm that the instability of these flat ribbons cannot be removed simply by increasing the width of the ribbon. We chose a symmetric edge nano-ribbon by removing 2 dangling X (S, Se or Te) atoms from one of the edges for these compounds (Figure~\ref{fig:ribbon} II).
Our calculation shows that the TMD nano-ribbons are stable against mechanical bending for a wide range of bending curvature, except for WTe$_2$. The bond breaking at the curvature region is observed for $\kappa$ $>$ 0.086 $\AA^{-1}$, as shown in Figure~\ref{fig:WTe}. Upon bending, one of the chalcogen atoms in the curvature region moves towards the middle layer, causing a further separation of the 2 metal atoms, as shown inside the circle, creating a sudden jump, as shown in an areal bending energy density vs curvature plot (See Figure~\ref{fig:WTe} (III)).\\

\noindent We utilized the thin plate bending model in our assessment in which we fix the width between flat and bent nanoribbons. It eliminates the quantum confinement effect present in the nanotube method due to dissimilarity of the width between flat and bent nanoribbons of the different radii of curvatures. However, the edge effects due to the finite width may remain uneliminated. Rafael I. Gonz$\acute{a}$lez et al. \cite{GVRVSKM18}, using classical molecular dynamics simulation, reported that the bending stiffness of MoS$_2$ estimated with a 0.95 nm width nanoribbon is only 46\% of those estimated using a 8 nm width nanoribbon. But, it recovers 88-93\% of bending stiffness when the width increases up to 3-4 nm, leaving the overall trend unaffected. We believe that such an accuracy would be a reasonable tradeoff to the computational complexity that arises while using a larger width. Moreover, we expect that the finite size effect would be less present in our results than in those calculated from MD simulation, as the quantum effects are more properly treated.
 
 \section{conclusion and discussion}
 The 2D materials offer a wide range of electronic properties efficiently applicable in sensors, energy storage, photonics, and optoelectronic devices. The higher flexural rigidity and strain-tunable properties of these compounds make them potential functional materials for future flexible electronics. In this work, we have employed the SCAN functional to explore the physical and mechanical properties of the 2D transition metal dichalcogenide (TMD) monolayers under mechanical bending. SCAN performs reasonably well in predicting the correct ground state phase as well as the geometrical properties. Also, a wide variety of flexural rigidities can be observed while scanning the periodic table for TMDs. The in-plane stiffness decreases from S to Se to Te, while the bending stiffness has the opposite trend. Overall, the bending stiffness also depends on the d band filling in the transition metal. The bending stiffness increases on increasing the filling of the d band from sparsely-filled (d$^0$) to nearly half-filled (d$^2$). However, decrease in bending stiffness is observed on moving from nearly half-filled (d$^2$) to completely-filled (d$^6$) d band. The out-of-plane Poisson's ratios are found to be different from the in-plane Poisson's ratio for 1T and 1T$'$ monolayers, while the difference is negligible in the case of 1H compounds, showing an anisotropic behavior of 1T and 1T$'$ monolayers.\\
 
Despite the extraordinary physical and electronic properties of TMDs, there are still challenges to make use of TMD semiconductors in nanoelectronics. The strong Fermi level pinning and high contact resistance are key bottlenecks in contact-engineering which are mainly due to in-plane, in-gap edge states and do not depend too much on the work function of a contact metal  \cite{KMLCAN17}. Thanks to mechanical bending, tuning of various properties of monolayer TMDs is possible, including band edges, thickness, and local strain. Bending deformation produces highly non-uniform local strain up to 40\% ({\color{blue} Supplementary Figure S1}), which is almost impossible with a linear strain ($\epsilon$). The high out-of-plane compressive strain developed within the layers due to bending reduces the mechanical thickness and makes the materials thinner in the curvature region. Moreover, one can remove strong Fermi-level pinning while using it in contact-engineering. Besides that, the optimal band alignment with the HER redox potential can be achieved for 1T semiconductors ZrS$_2$ and HfS$_2$ under mechanical bending, which are not present in an unbent monolayer. Furthermore, both electron and hole conductivities are affected in 1T semiconductors, while only the hole conductivity is affected in 1H semiconductors \cite{YRP16}. Similar to graphene  \cite{YS97, W04, KGB01, Z00}, the estimated effective thickness of group IV TMDs, especially sulfide and selenide, is underestimated as compared to chalcogen-chalcogen distance (d$_{X-X}$), which is quite puzzling and needs further investigation. 

     
 \section{Acknowledgement}
 We thank Prof. John P. Perdew for useful comments on the manuscript.
 This research was supported as part of the Center for Complex Materials from First Principles (CCM), 
 an Energy Frontier Research Center funded by the U.S. Department of Energy (DOE), Office of Science,
 Basic Energy Sciences (BES), under Award No. DE-SC0012575. Computational support was provided by National Energy Research Scientific Computing Center (NERSC). Some of calculations were carried out on Temple University's HPC resources and thus was supported in part by the National Science Foundation through major research instrumentation grant number 1625061 and by the US Army Research Laboratory under contract number W911NF-16-2-0189.
 

 \begin{center}
 \textbf{Supplementary material\\}
 \end{center}

 \begin{table}[h!]
 	\renewcommand\thetable{S1}
 	\caption{Width of the relaxed flat nanoribbons used in the calculations ($\AA$); The bulk band gap (excluding edge states) of the semiconducting unbent nano-ribbons (E$_g$); Workfunction ($\phi$ (eV)); The G$_0$W$_0$ quasi-particle gap of monolayer semiconductors (E$^{QP}_g$) is shown for comparison.}
 	\begin{tabular}{|l|r|r|r|r|}
 		\hline
 		TMDs & Width ($\AA$) & \multicolumn{1}{l|}{E$_g$ (eV)} & \multicolumn{1}{l|}{$\phi = E_{vacuum} - E_{Fermi}$  (eV)} & E$^{QP}_g$ (eV)\cite{ZH13} \\ \hline
 		TiS$_2$ & 32.04 &-- & 5.236 & \\ \hline
 		TiSe$_2$ & 32.62 &-- & 5.508 & \\ \hline
 		TiTe$_2$ & 34.11 &-- & 5.09 & \\ \hline
 		ZrS$_2$ & 34.64 &1.549 & 5.886 & 2.56 \\ \hline
 		ZrSe$_2$ & 35.55 &1.025 & 5.549 & 1.54\\ \hline
 		ZrTe$_2$ & 37.10 & -- & 5.084 & \\ \hline
 		HfS$_2$ & 34.18 & 1.751 & 5.919 & 2.45 \\ \hline
 		HfSe$_2$ & 35.01 & 1.092 & 5.386 & 1.39\\ \hline
 		HfTe$_2$ & 36.77 & -- & 4.938 & \\ \hline
 		MoS$_2$ & 29.85 & 1.836 & 5.376 & 2.36 \\ \hline
 		MoSe$_2$ & 30.95 & 1.709 & 4.952 & 2.04\\ \hline
 		MoTe$_2$ & 32.62 & 1.349 & 4.631 & 1.54 \\ \hline
 		MoTe$_2$-1T$^\prime$ & 33.72 & -- & 4.795 & \\ \hline
 		WS$_2$ & 29.80 & 2.094 & 5.126 & 2.64 \\ \hline
 		WSe$_2$ & 30.73 & 1.893 & 4.736 & 2.26\\ \hline
 		WTe$_2$ & 33.26 & -- & 4.584 & \\ \hline
 		PdTe$_2$ & 37.07 & -- & 4.4 & \\ \hline
 		PtS$_2$ & 35.36 & -- & 5.482 & \\ \hline
 		PtSe$_2$ & 34.32 & -- & 4.958 & \\ \hline
 		PtTe$_2$ & 37.20 & -- & 4.466 & \\ \hline
 	\end{tabular}
 	\label{tab:band}
 \end{table}

 \begin{table}[h!]
 	\renewcommand\thetable{S2}
 	\centering
 	\caption{The calculated structure parameters for the rectangular 1T$^\prime$ monolayer unit cell of WTe$_2$-type having 2 TX$_2$ units: For MoTe$_2$, a $=$ 3.43439, b $=$ 6.31457 $\AA$. Similarly for WTe$_2$, a $=$ 3.45822, b $=$ 6.24802 $\AA$ (Figure 1 in the main text). The combined-fractional coordinates X, Y, and Z represent the position of the corresponding atom in the same row. Lattice constants can be estimated as $b/\sqrt3$. Chalcogen-chalcogen distances d$_{X-X}$ for the distorted MoTe$_2$ and WTe$_2$ are (2.9486 $\AA$ and 4.0940 $\AA$) and (2.9118 $\AA$ and 4.1386 $\AA$) respectively.}
 	
 	\begin{tabular}{rlrrr}
 		\hline
 		\hline
 		\multicolumn{1}{l}{TMD} & Atom  & \multicolumn{1}{l}{X} & \multicolumn{1}{l}{Y} & \multicolumn{1}{l}{Z}\\
 		\hline
 		& Mo1   & 0.01712 & 0     & 0.50454 \\
 		& Mo2   & 0.37804 & 0.5   & 0.49545 \\
 		\multicolumn{1}{l}{MoTe$_2$} & Te1   & 0.27688 & 0     & 0.39764  \\
 		& Te2   & 0.62769 & 0     & 0.57361\\
 		& Te3   & 0.11724 & 0.5   & 0.60234 \\
 		& Te4   & 0.76664 & 0.5   & 0.42641 \\
 		&       &       &       &      \\
 		& W1    & 0.02032 & 0     & 0.50509\\
 		& W2    & 0.37395 & 0.5   & 0.49487 \\
 		\multicolumn{1}{l}{WTe$_2$} & Te1   & 0.27811 & 0     & 0.39653 \\
 		& Te2   & 0.62559 & 0     & 0.57281 \\
 		& Te3   & 0.11669 & 0.5   & 0.60346 \\
 		& Te4   & 0.76895 & 0.5   & 0.42722\\
 		\hline
 		\hline
 	\end{tabular}%
 	\label{tab:eff-2t}%
 \end{table}%

 \begin{figure}[h!]
 	\renewcommand\thefigure{S1}
 	\includegraphics[scale=0.33]{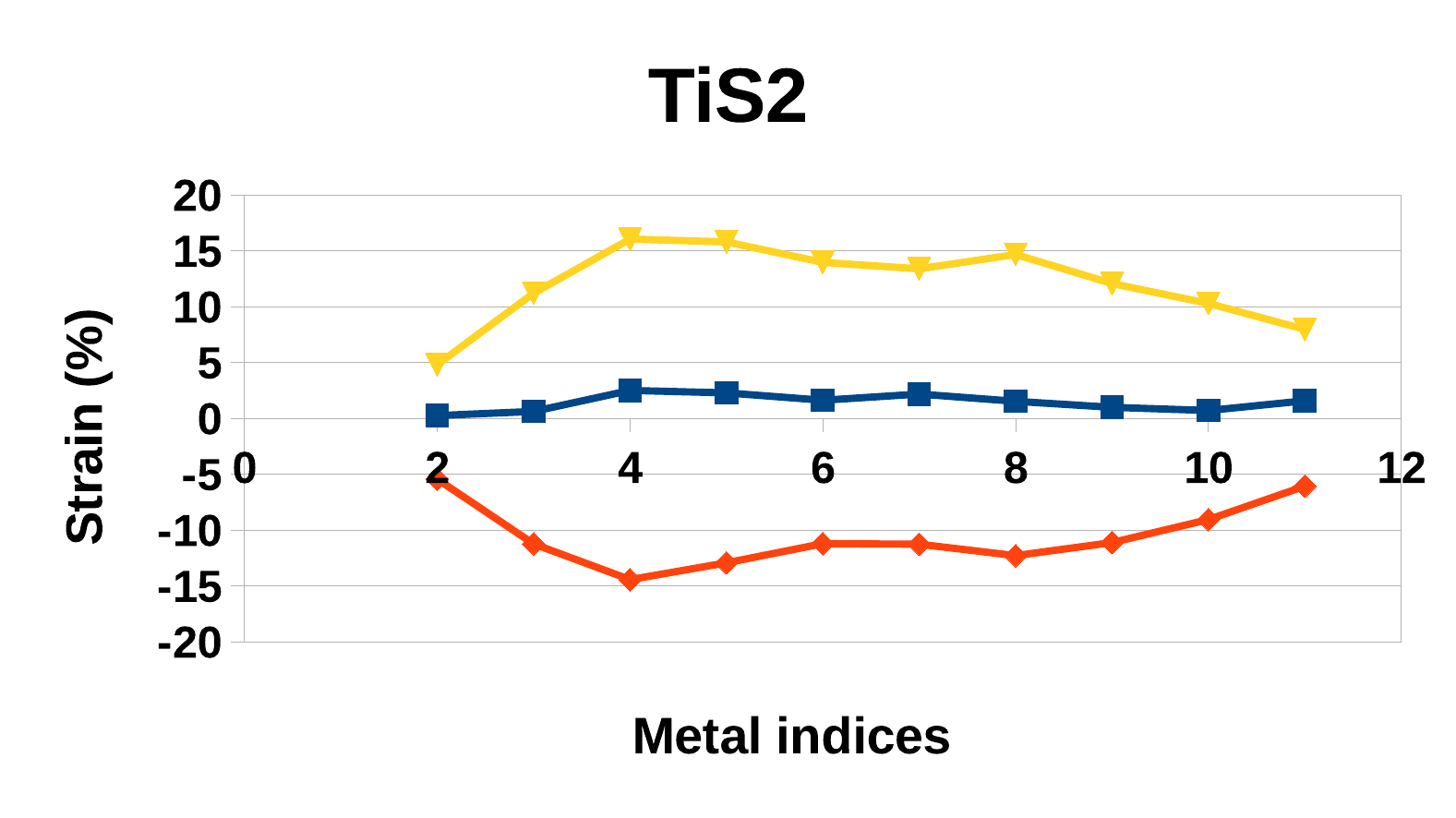}
 	\includegraphics[scale=0.33]{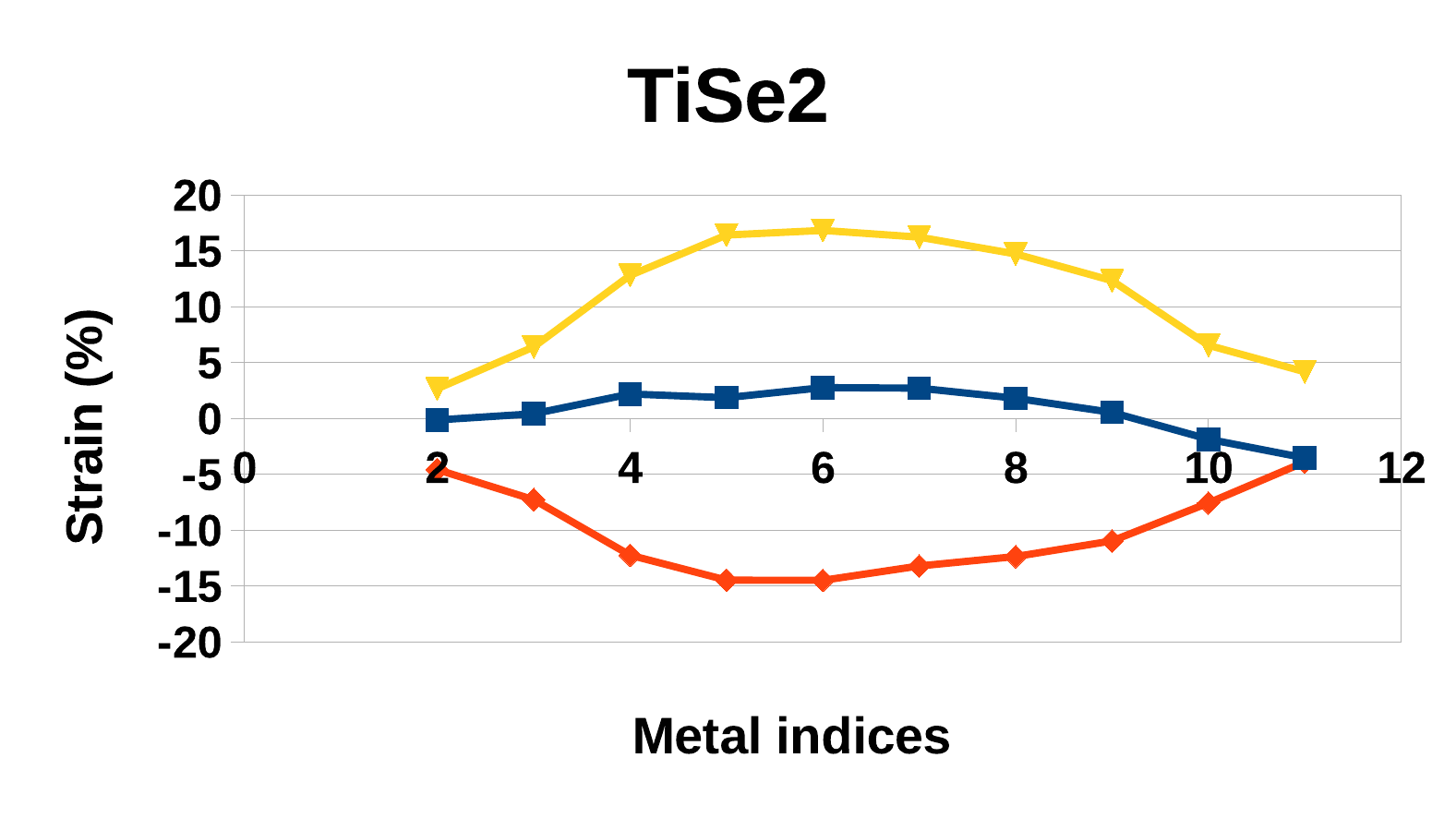}
 	\includegraphics[scale=0.33]{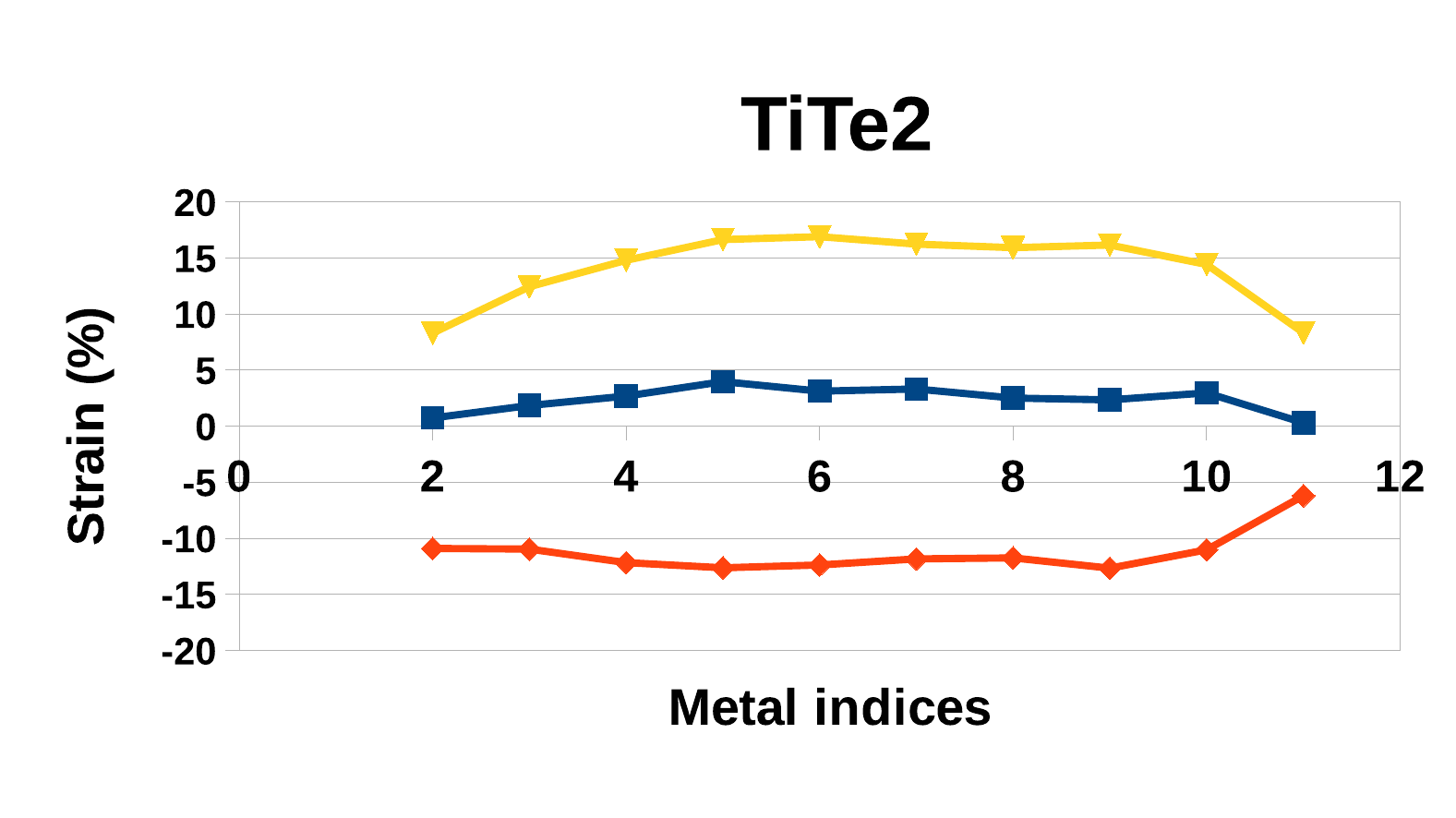}
 	\includegraphics[scale=0.33]{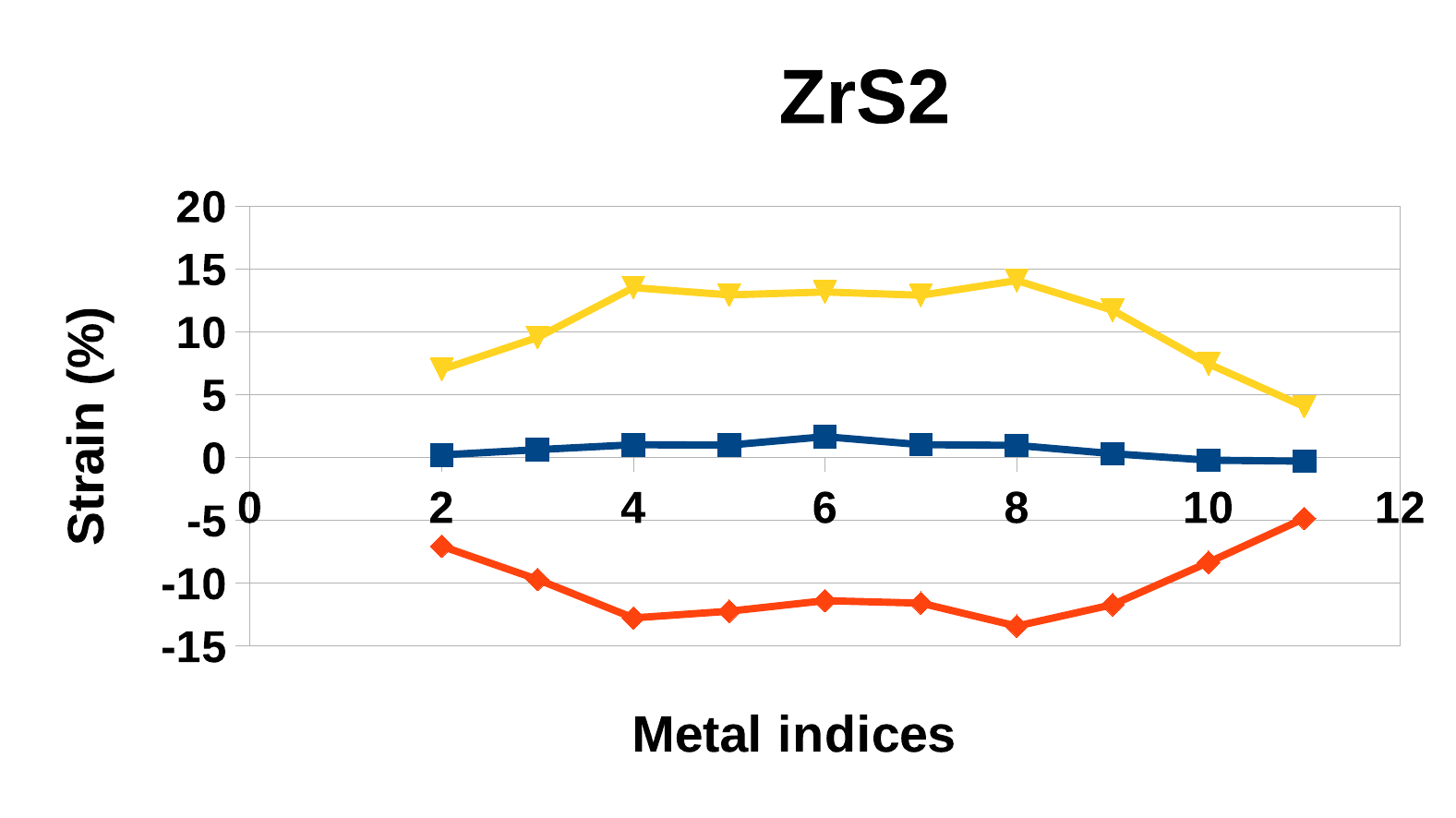}
 	\includegraphics[scale=0.33]{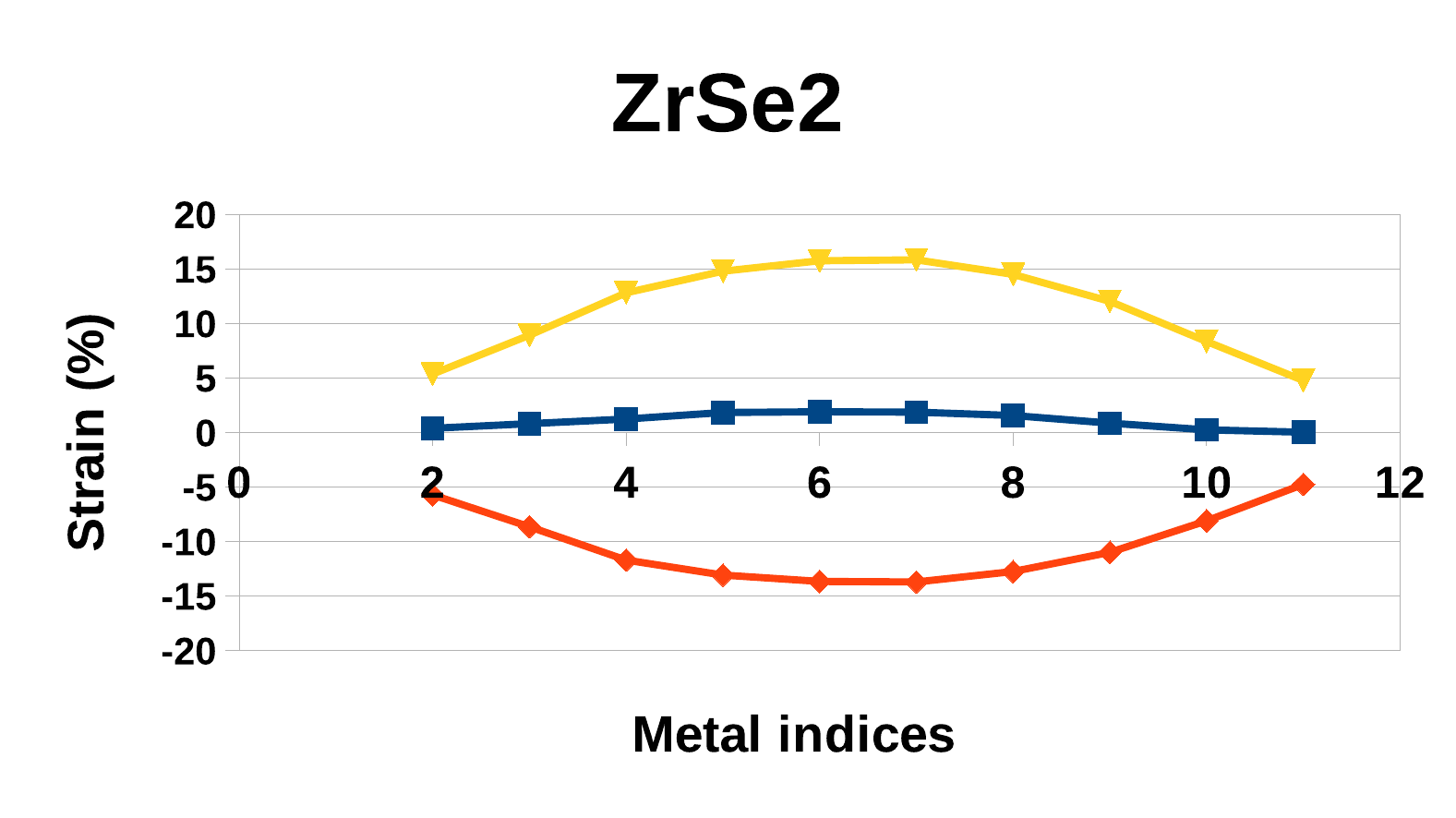}
 	\includegraphics[scale=0.33]{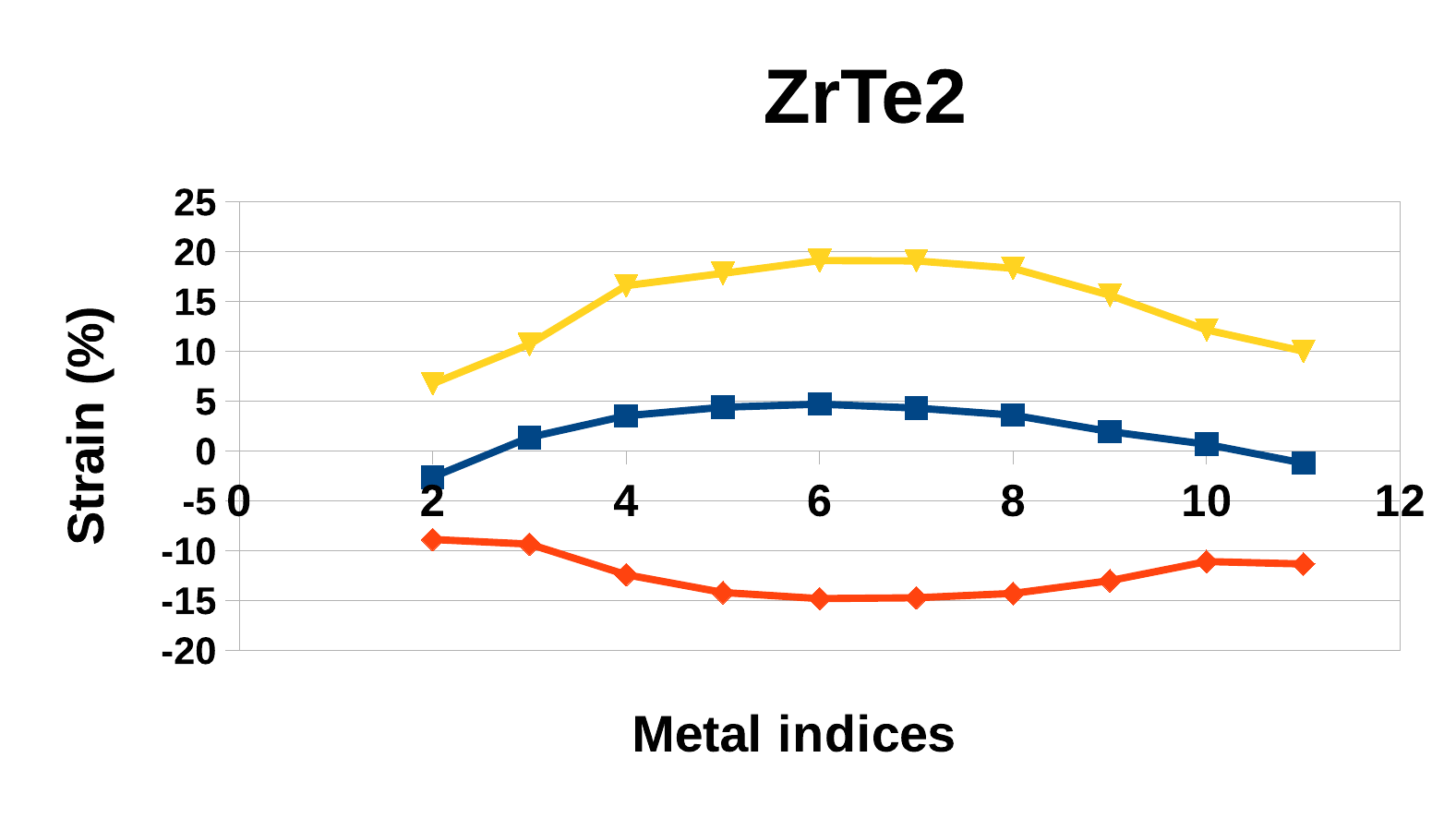}
 	\includegraphics[scale=0.33]{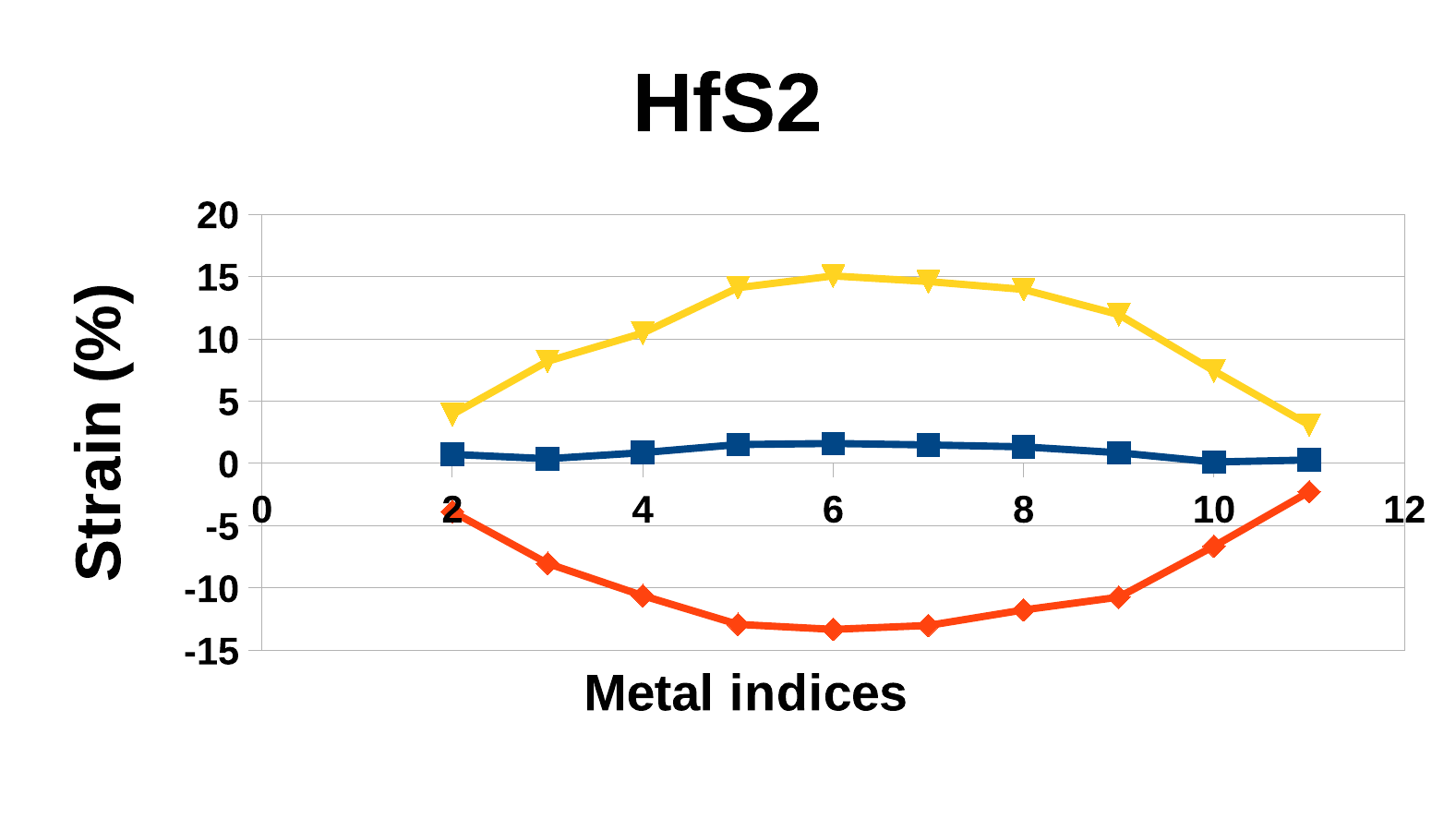}
 	\includegraphics[scale=0.33]{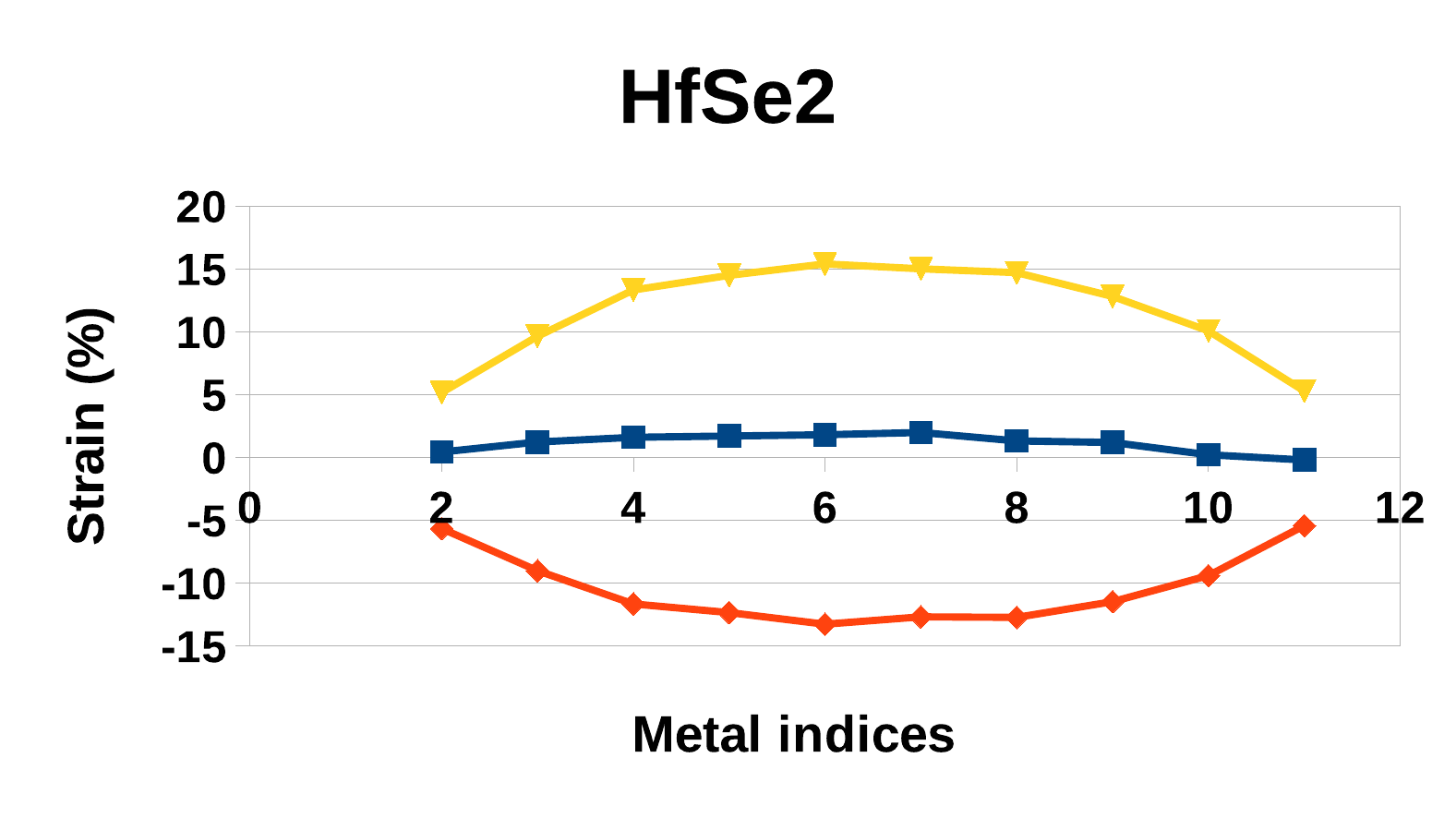}
 	\includegraphics[scale=0.33]{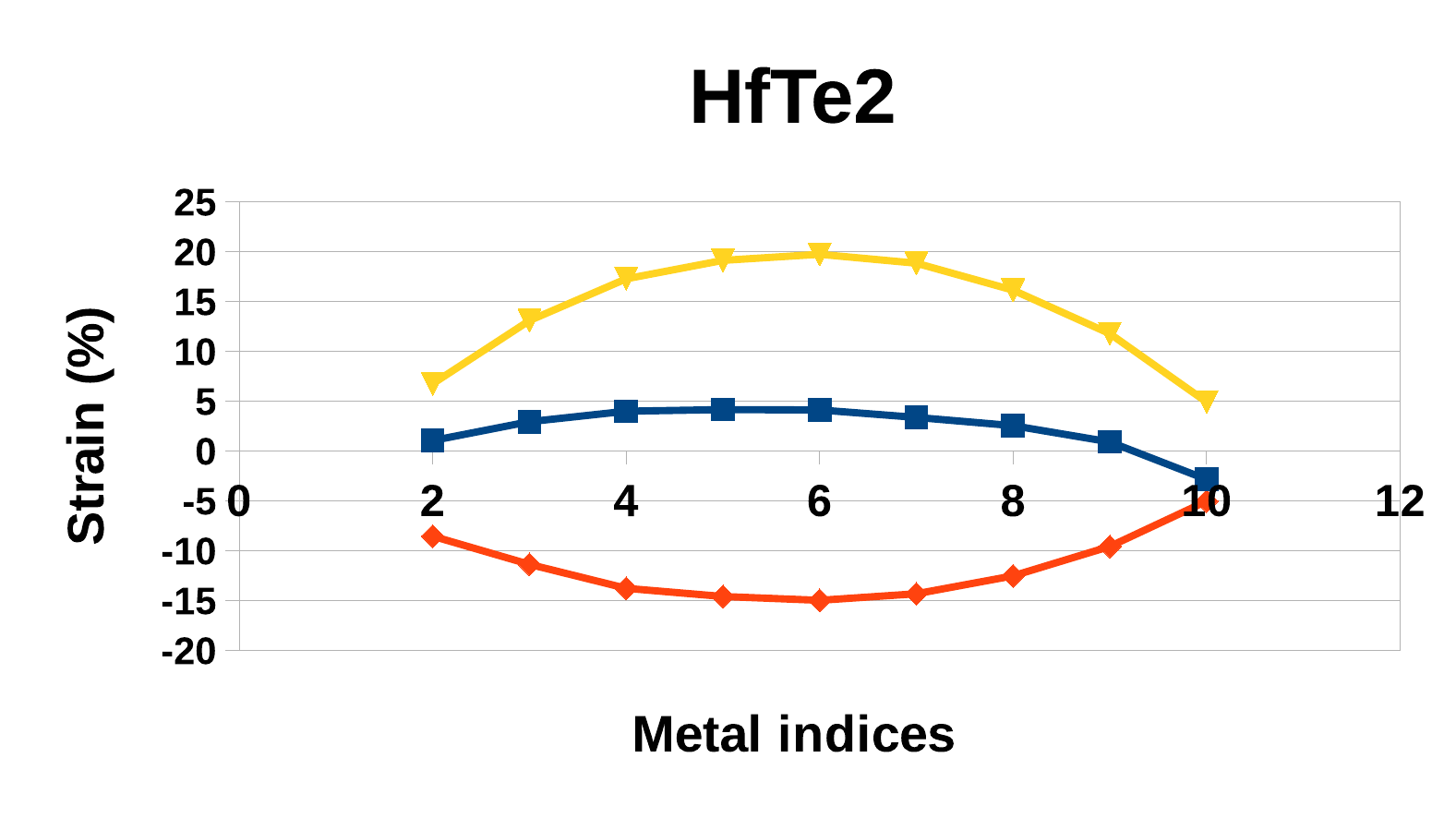}
 	\includegraphics[scale=0.33]{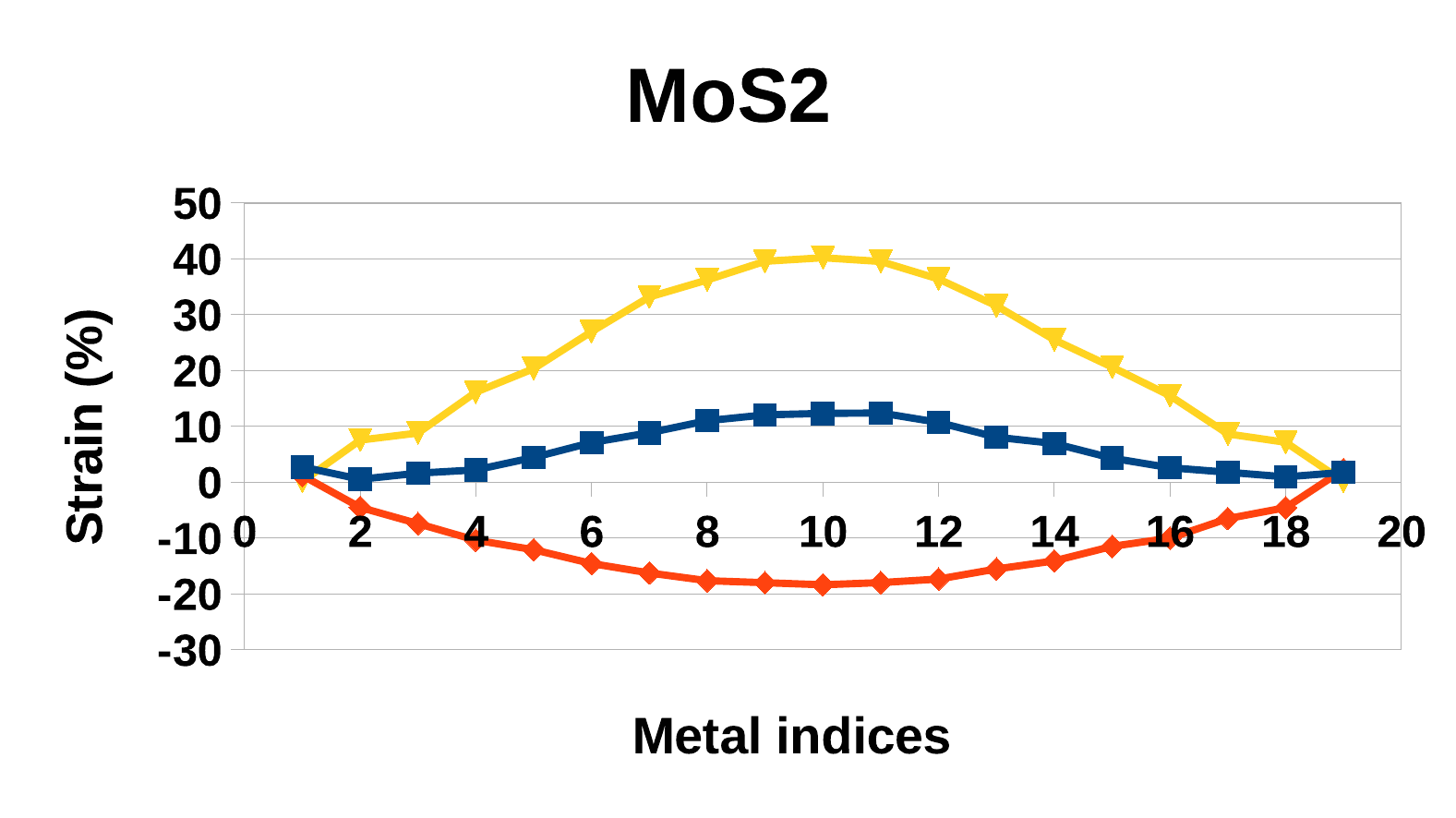}
 	\includegraphics[scale=0.33]{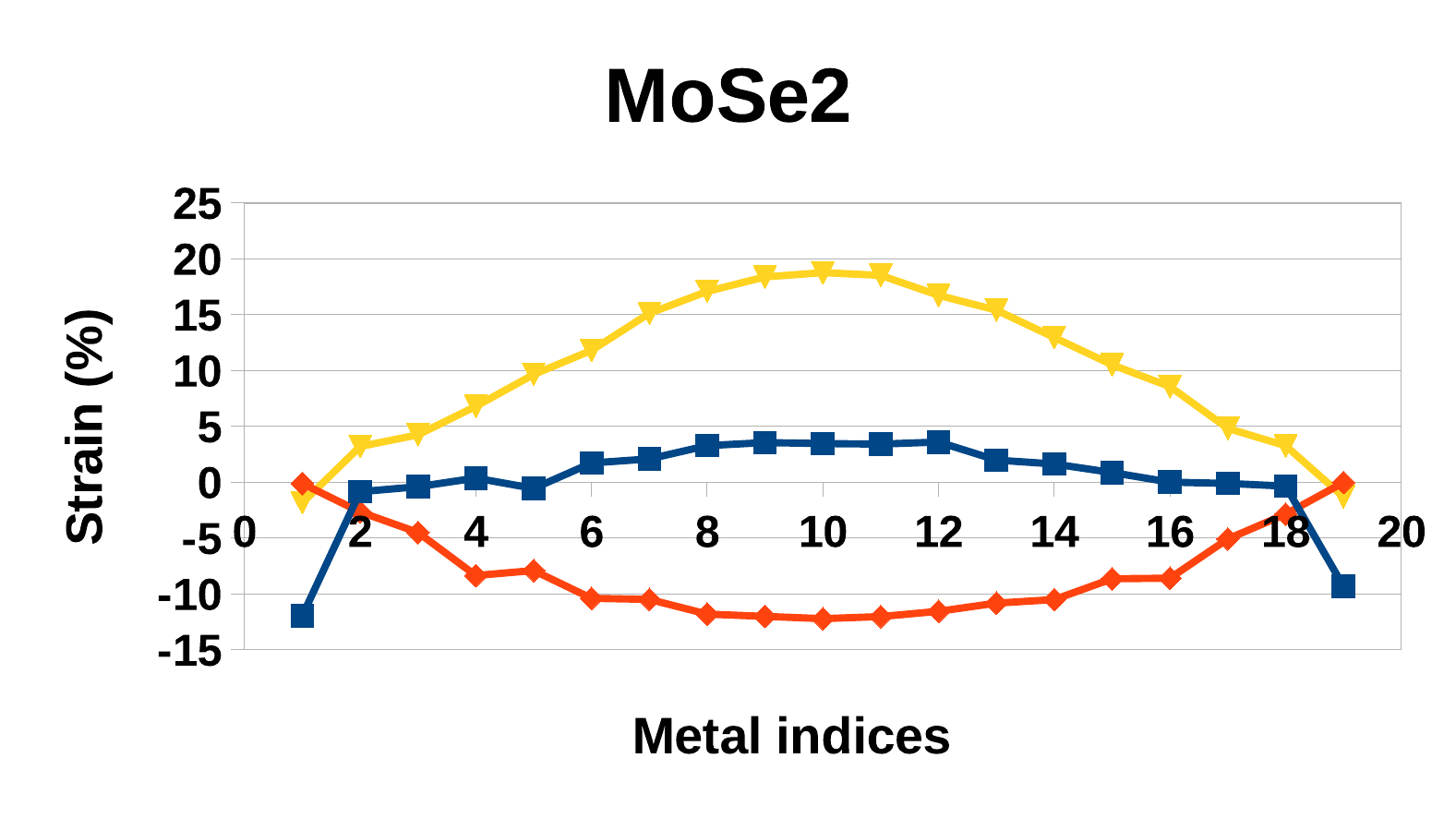}
 	\includegraphics[scale=0.33]{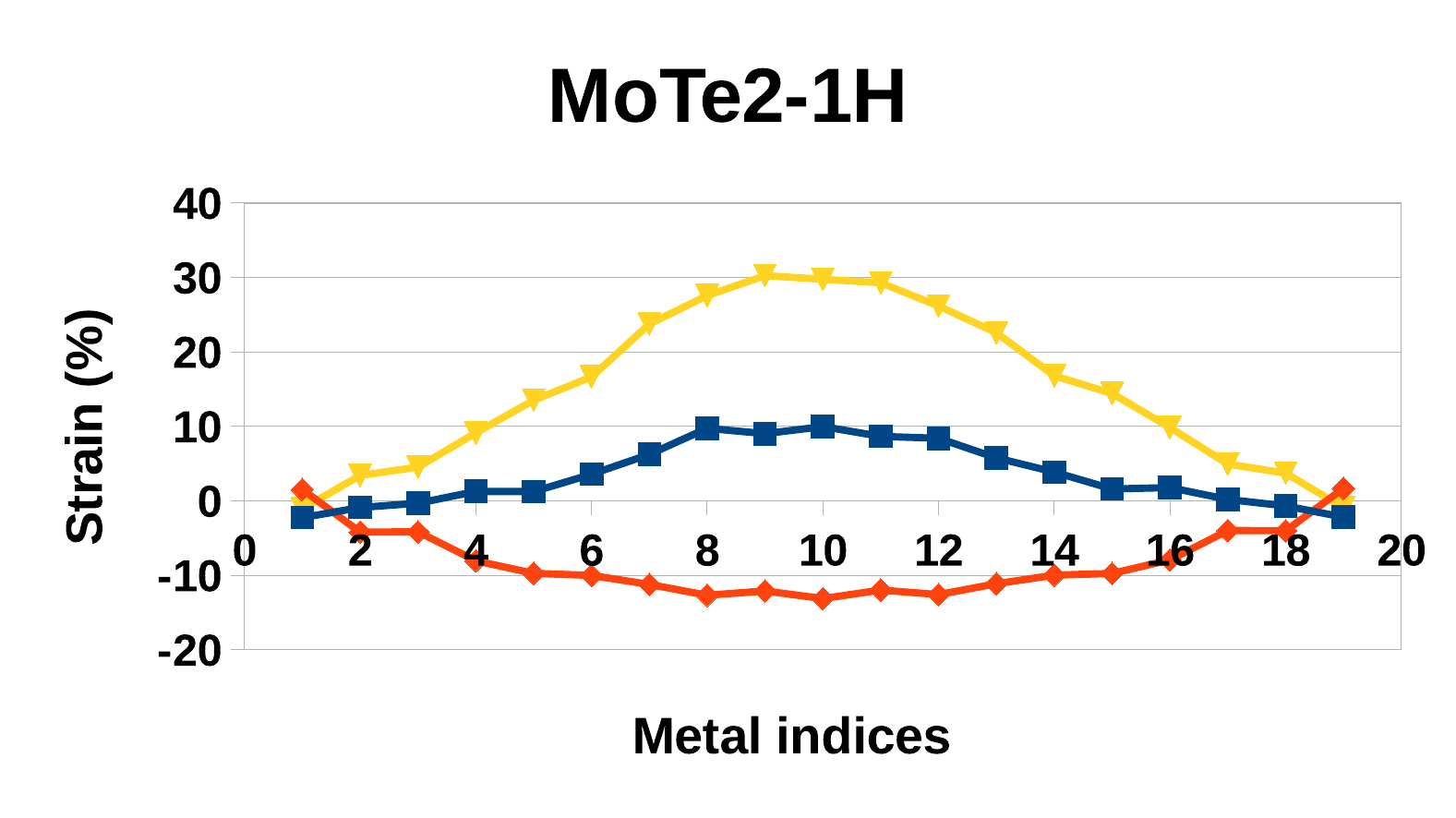}
 	\includegraphics[scale=0.33]{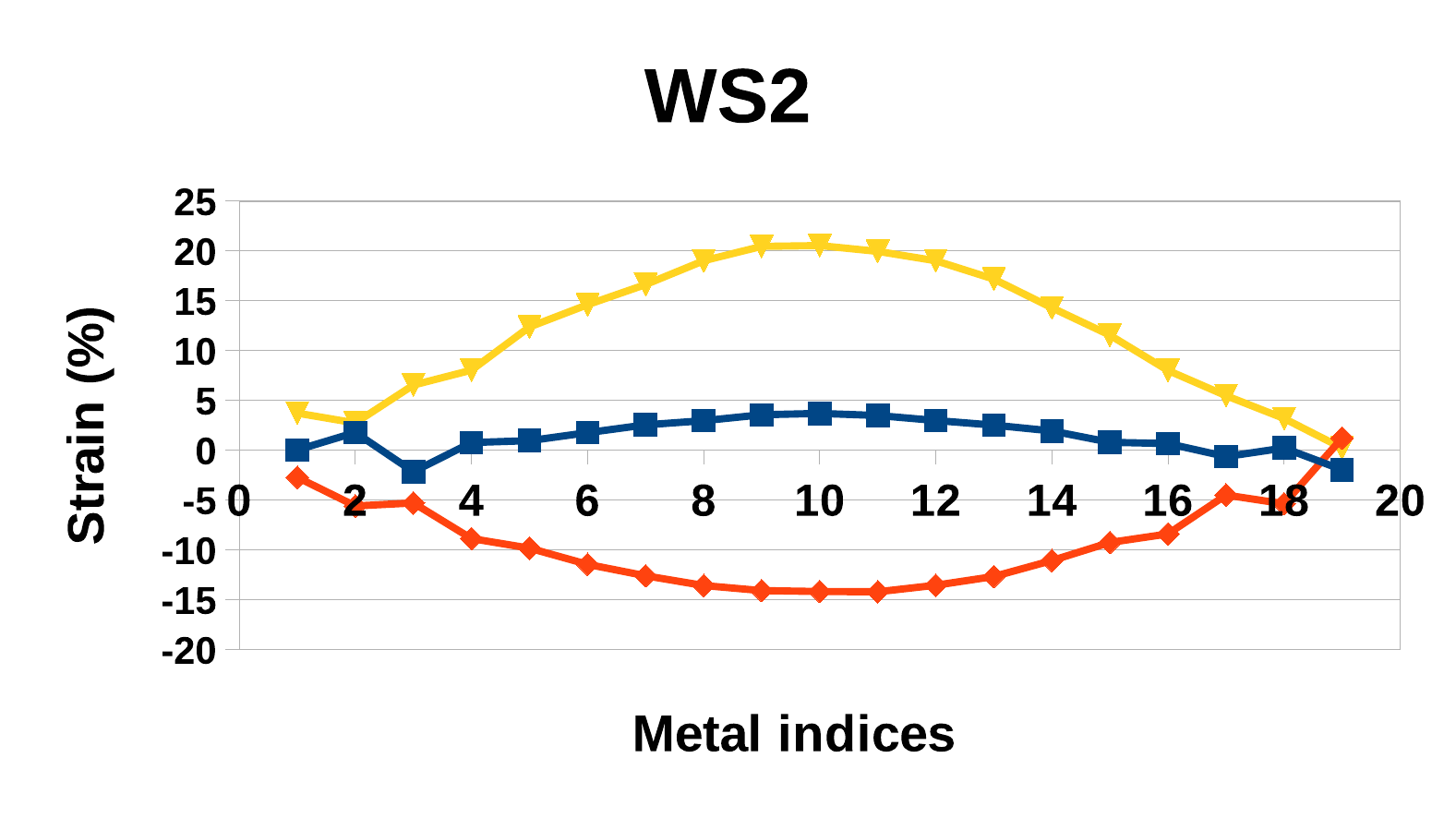}
 	\includegraphics[scale=0.33]{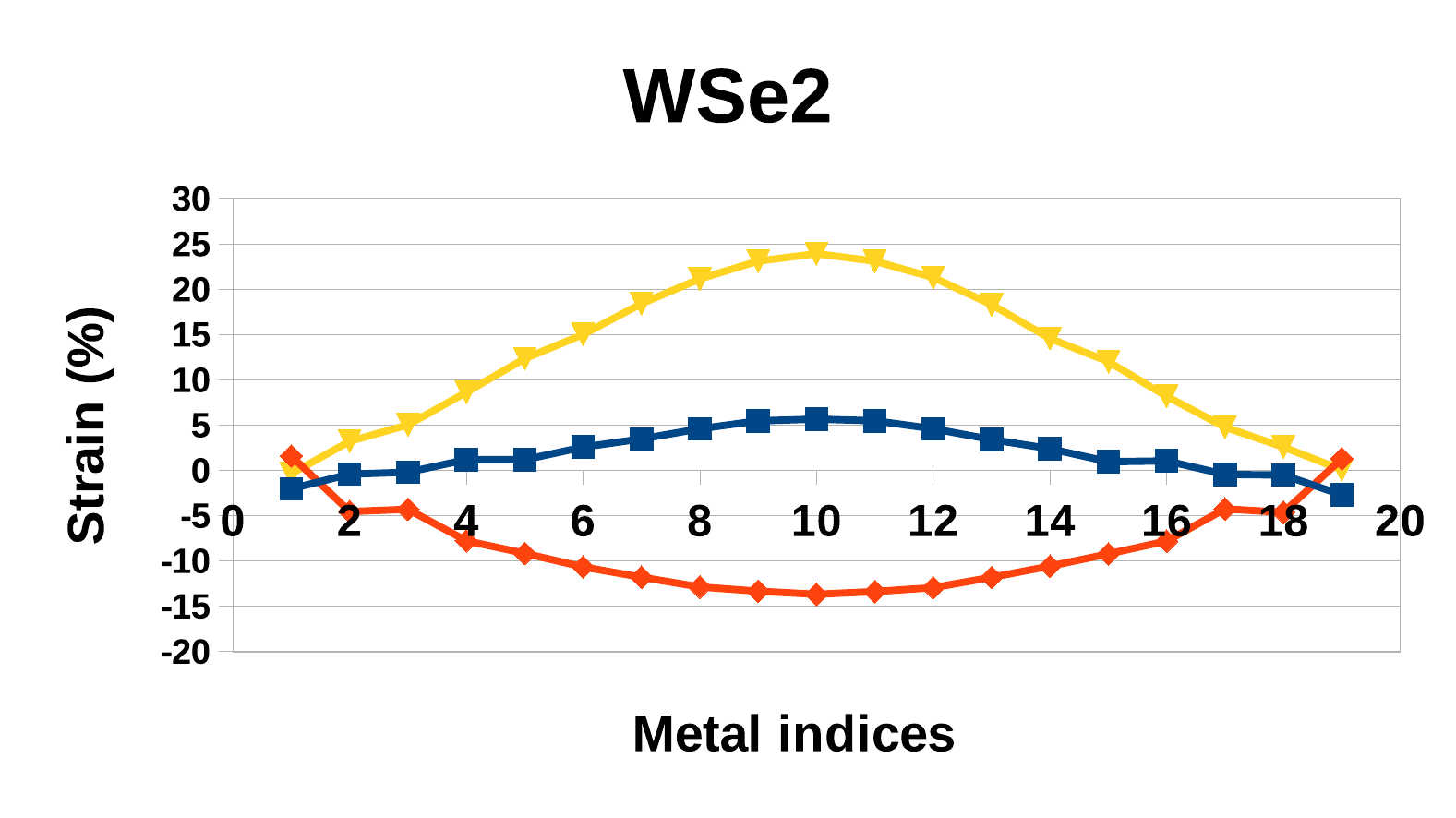}
 	\includegraphics[scale=0.33]{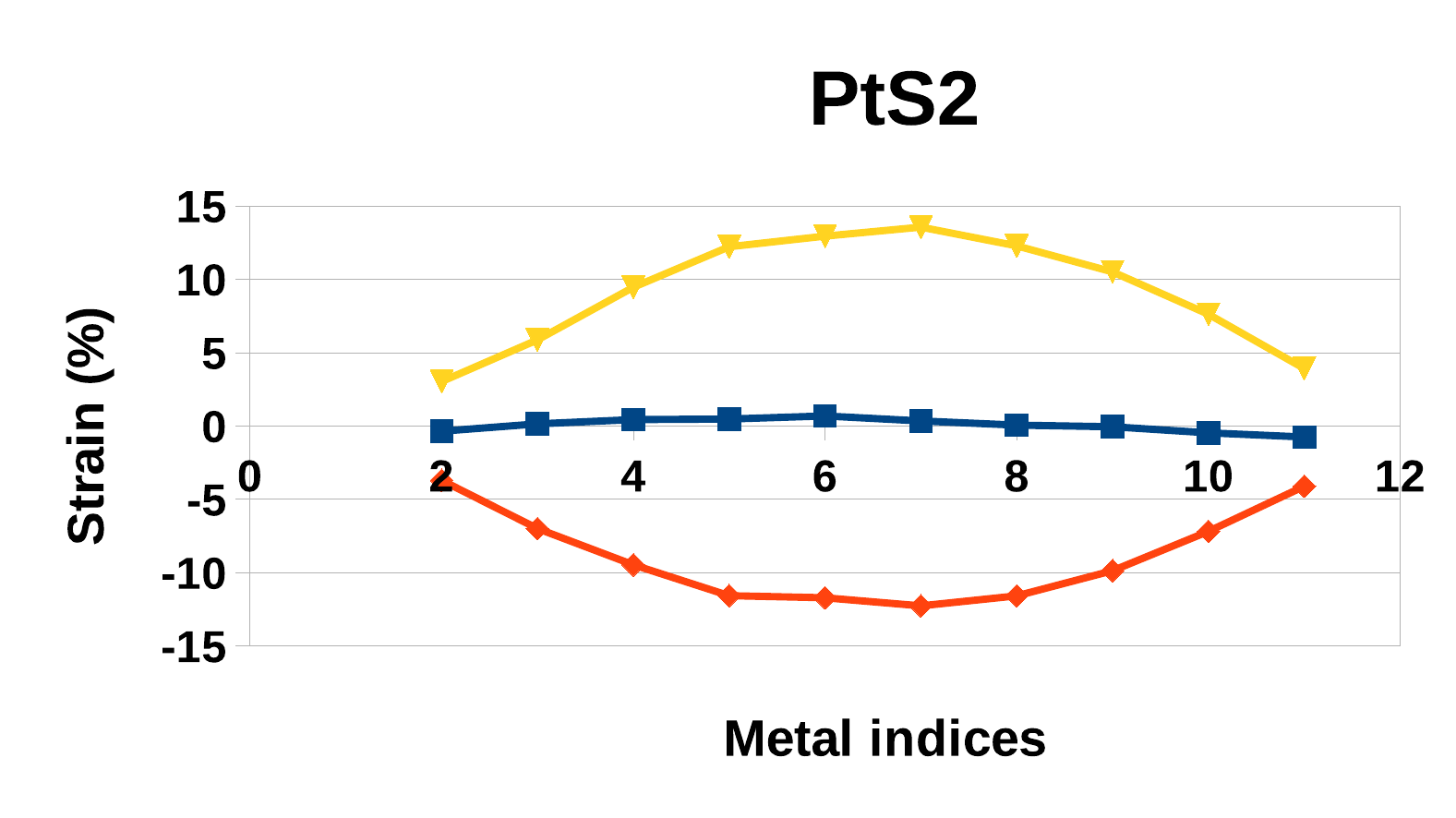}
 	\includegraphics[scale=0.33]{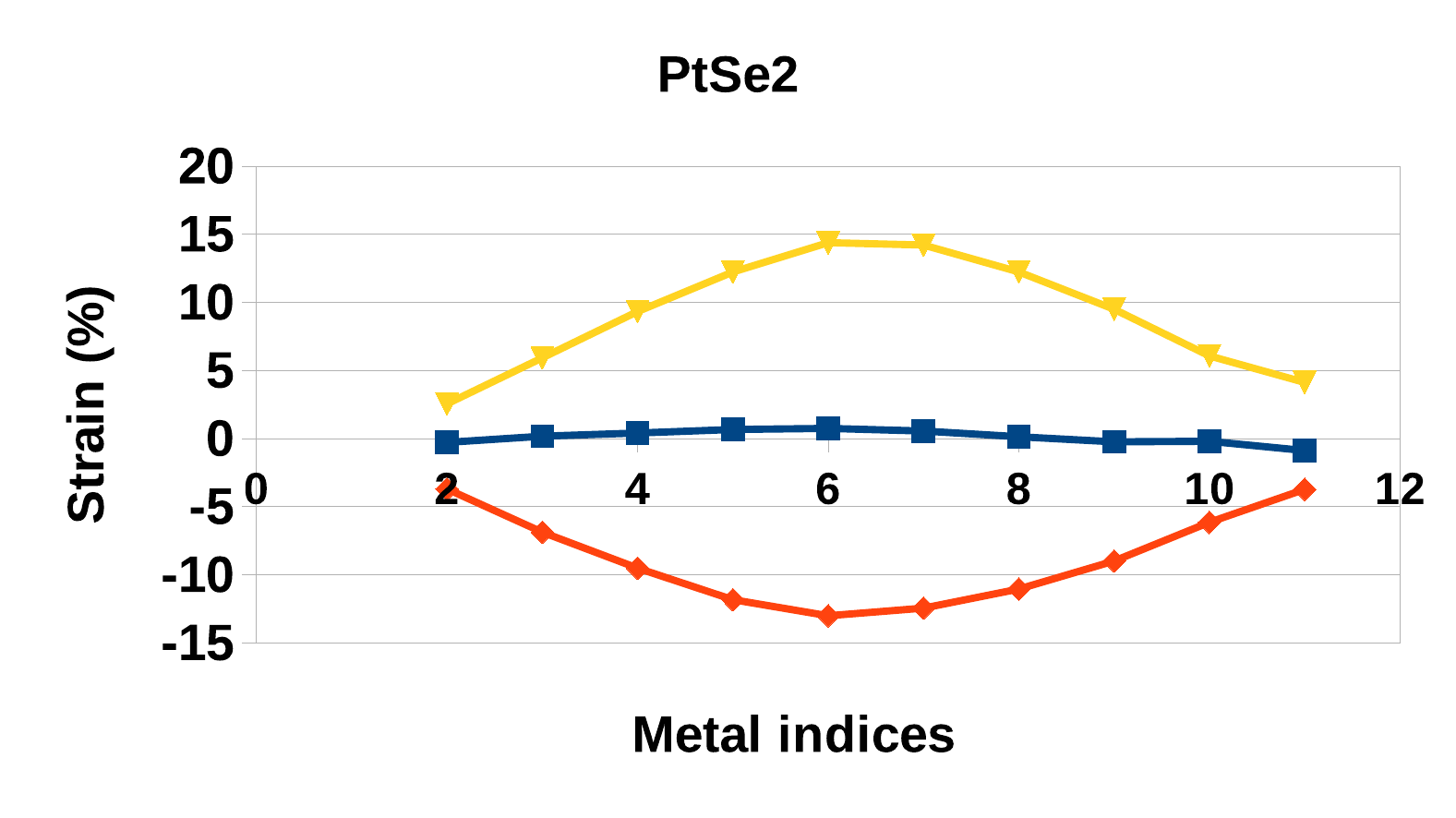}
 	\includegraphics[scale=0.33]{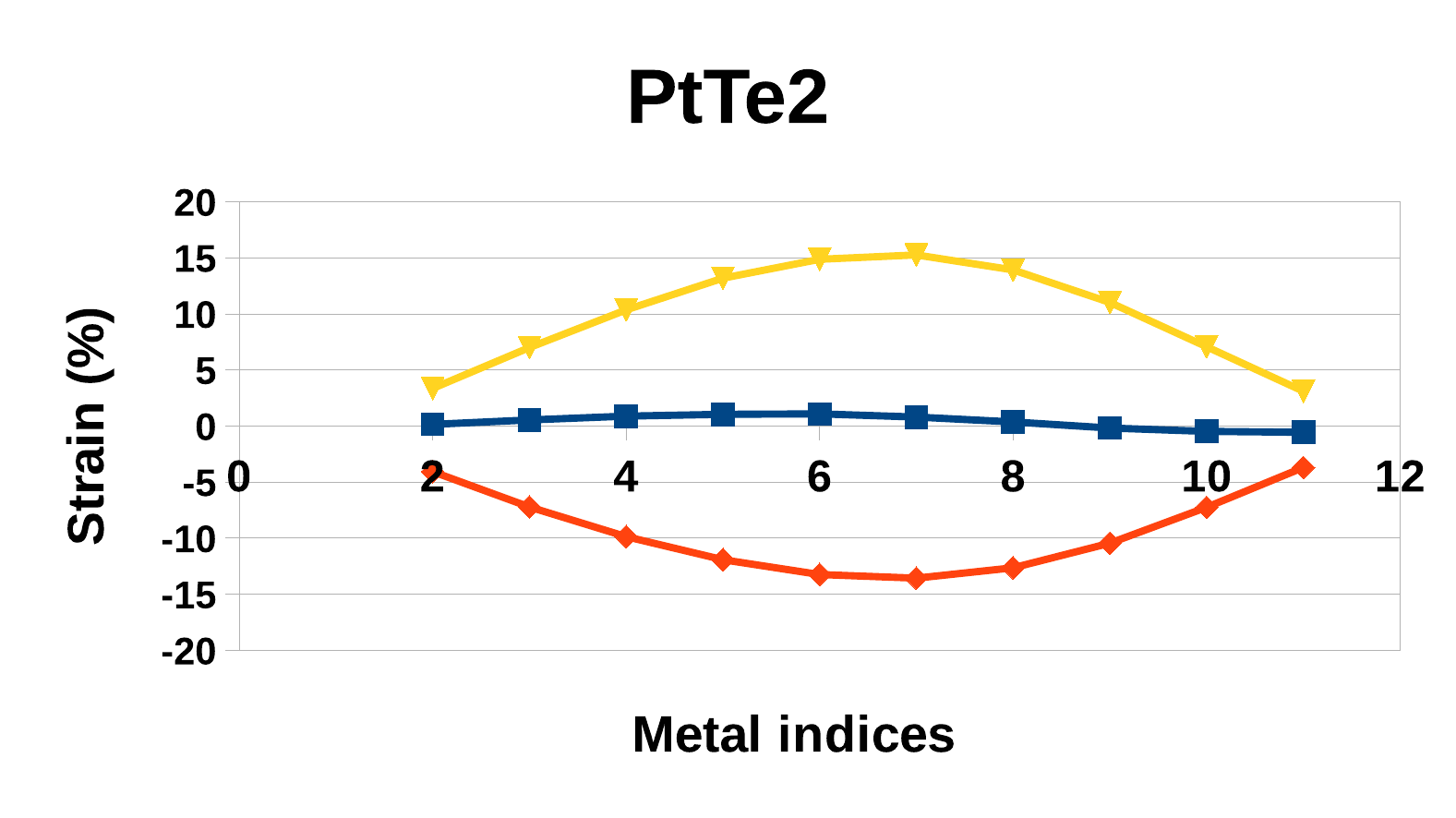}
 	\includegraphics[scale=0.33]{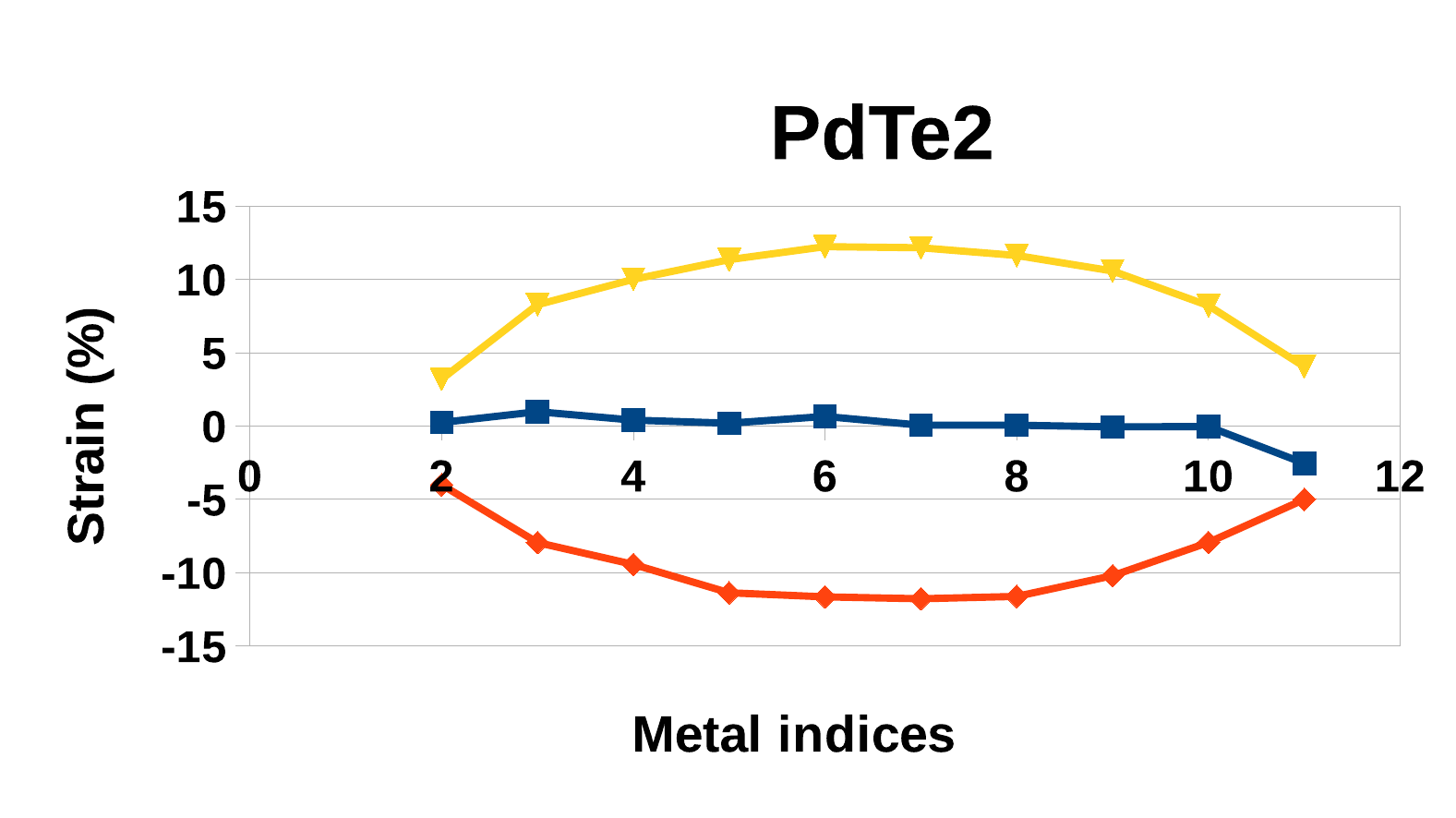}
 	\includegraphics[scale=0.33]{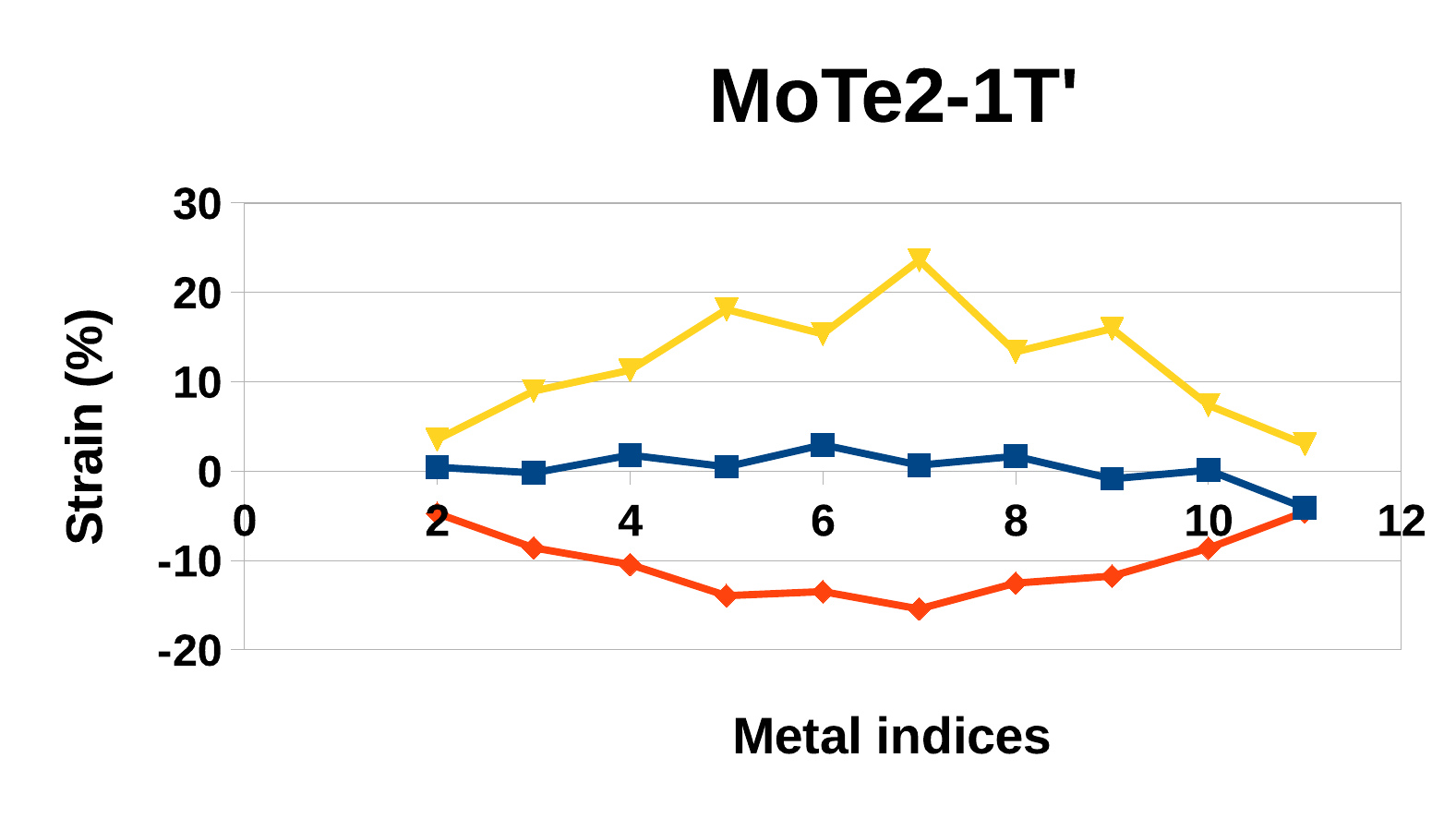}
 	\includegraphics[scale=0.33]{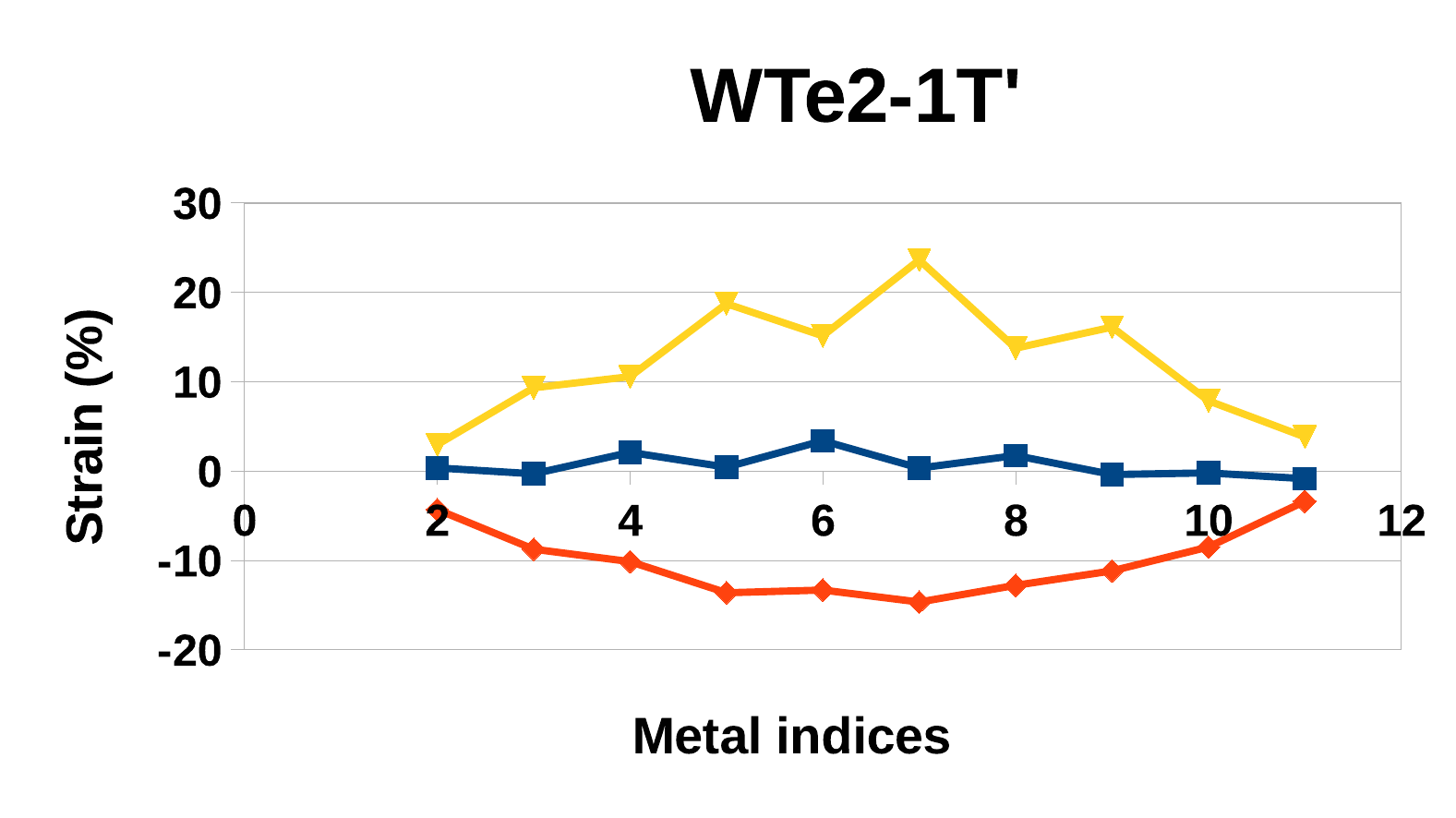}
 	\caption{The local strain projected onto the bending plane (bc) for the bending curvature around 0.09 $\AA^{-1}$. The upper and lower plots correspond to outer and inner chalcogen layers respectively, while the middle one corresponds to the metallic layer.}
 	\label{lab:local-strain}
 	
 \end{figure}
 
 \begin{figure}[h!]
 	\renewcommand\thefigure{S2}
 	\includegraphics[scale=0.33]{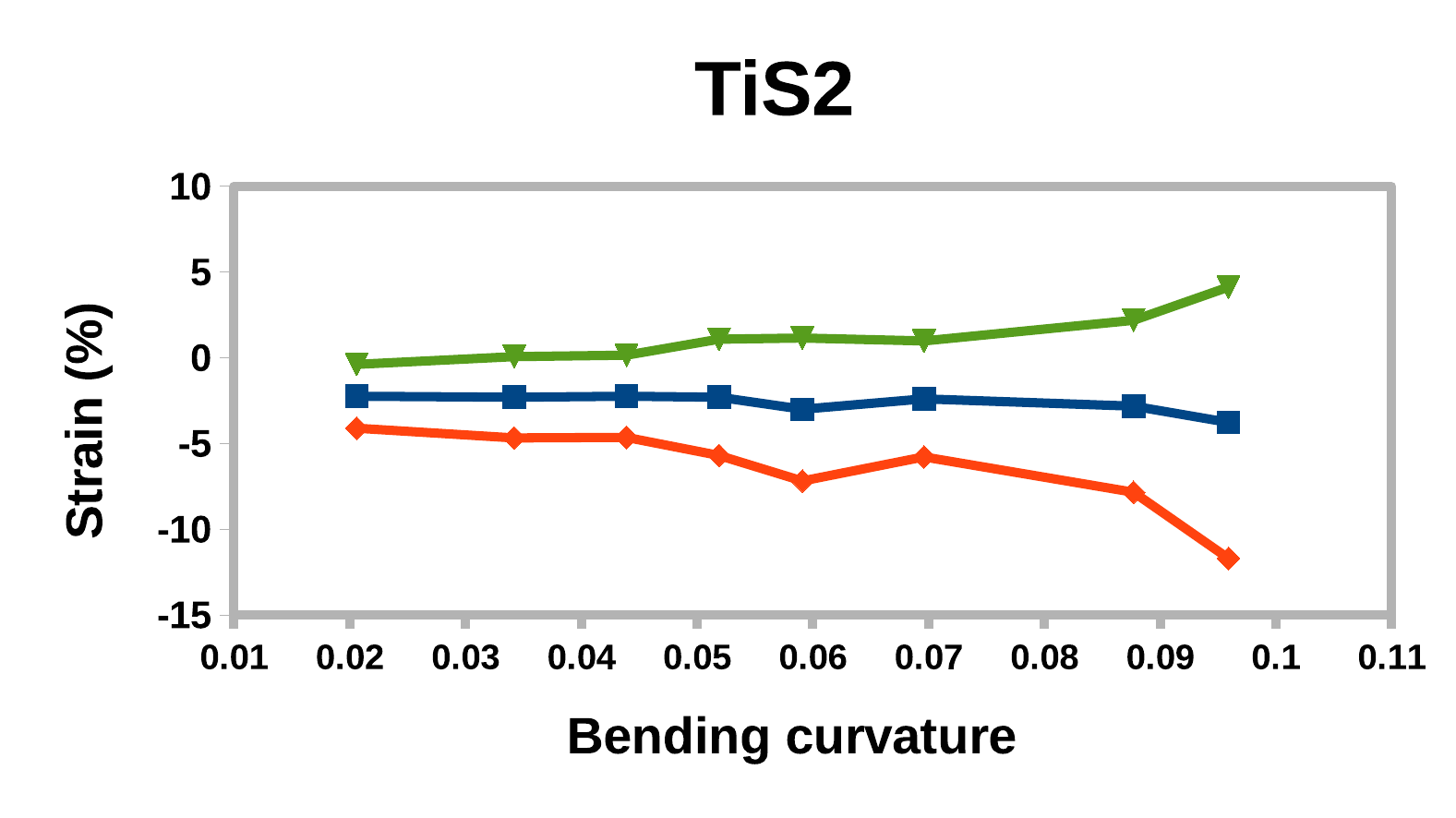}
 	\includegraphics[scale=0.33]{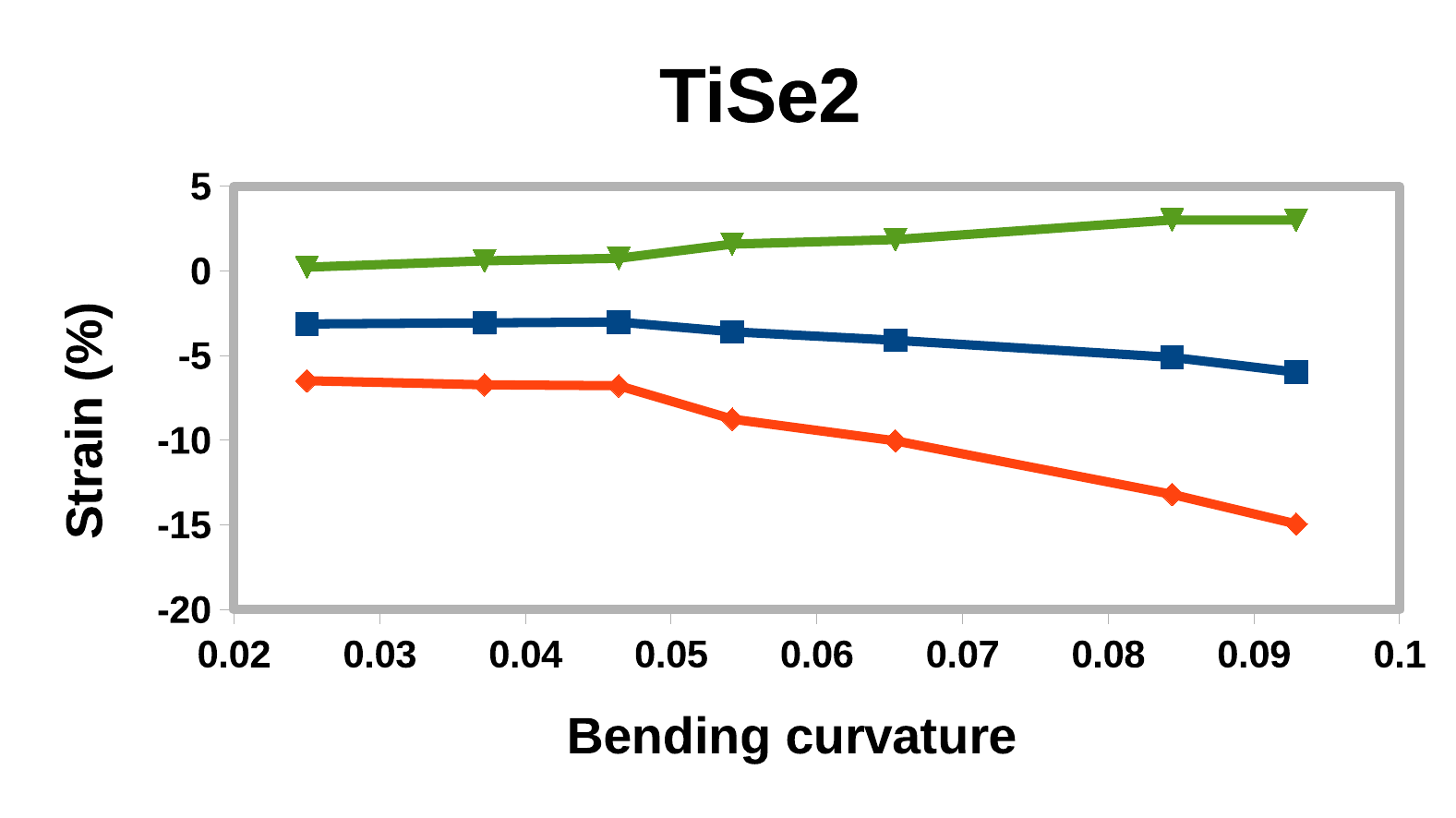}
 	\includegraphics[scale=0.33]{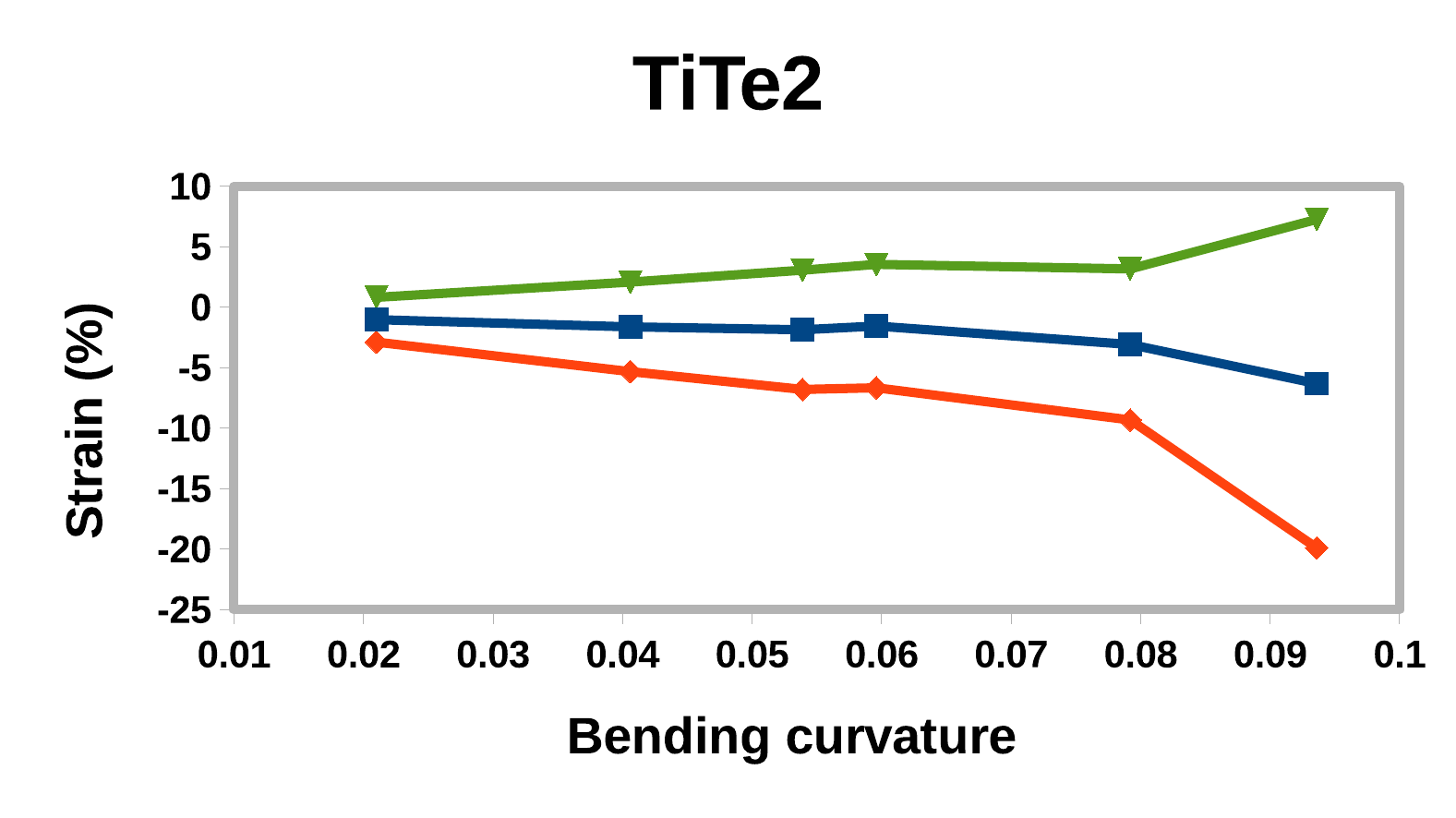}
 	\includegraphics[scale=0.33]{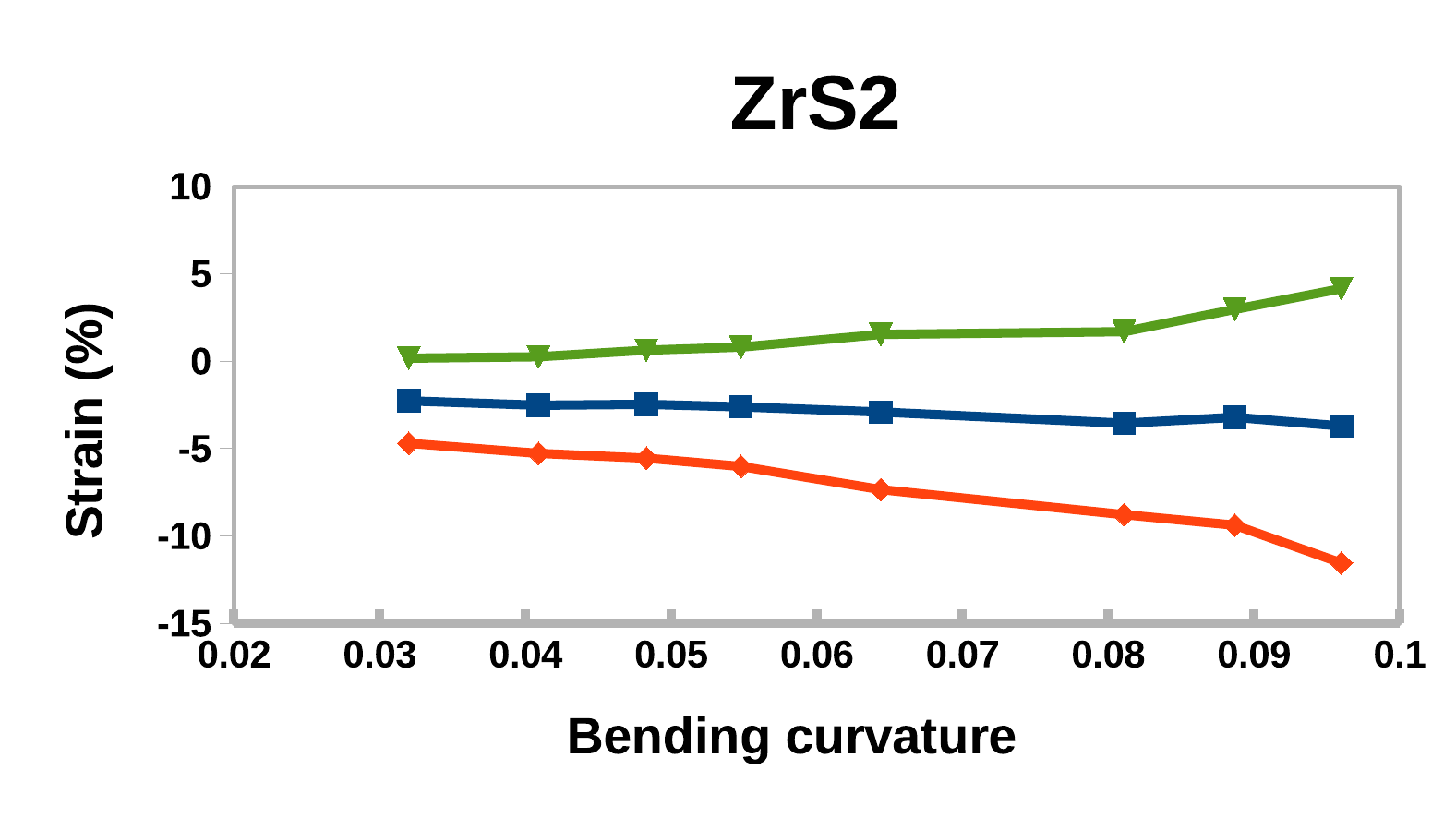}
 	\includegraphics[scale=0.33]{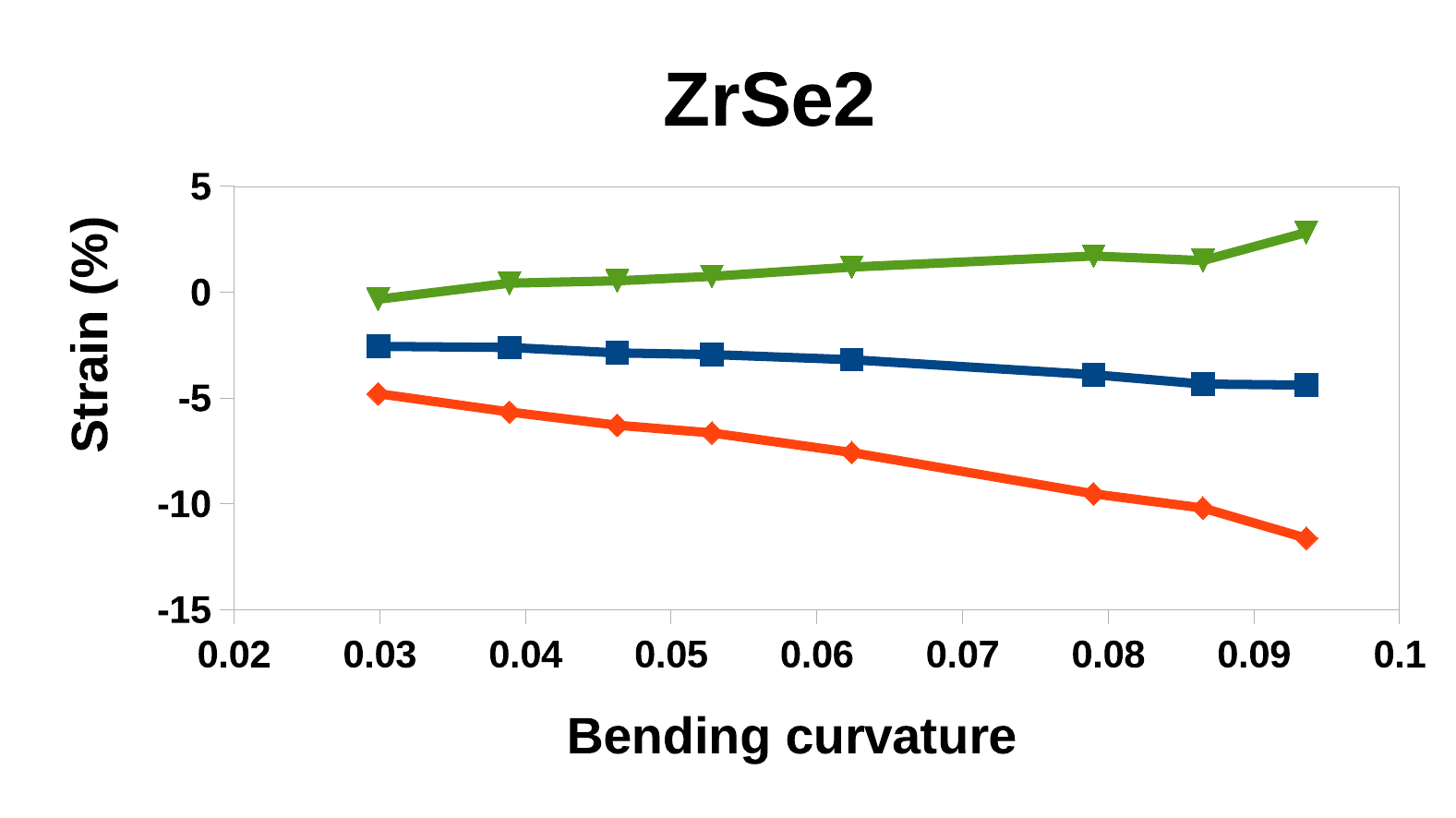}
 	\includegraphics[scale=0.33]{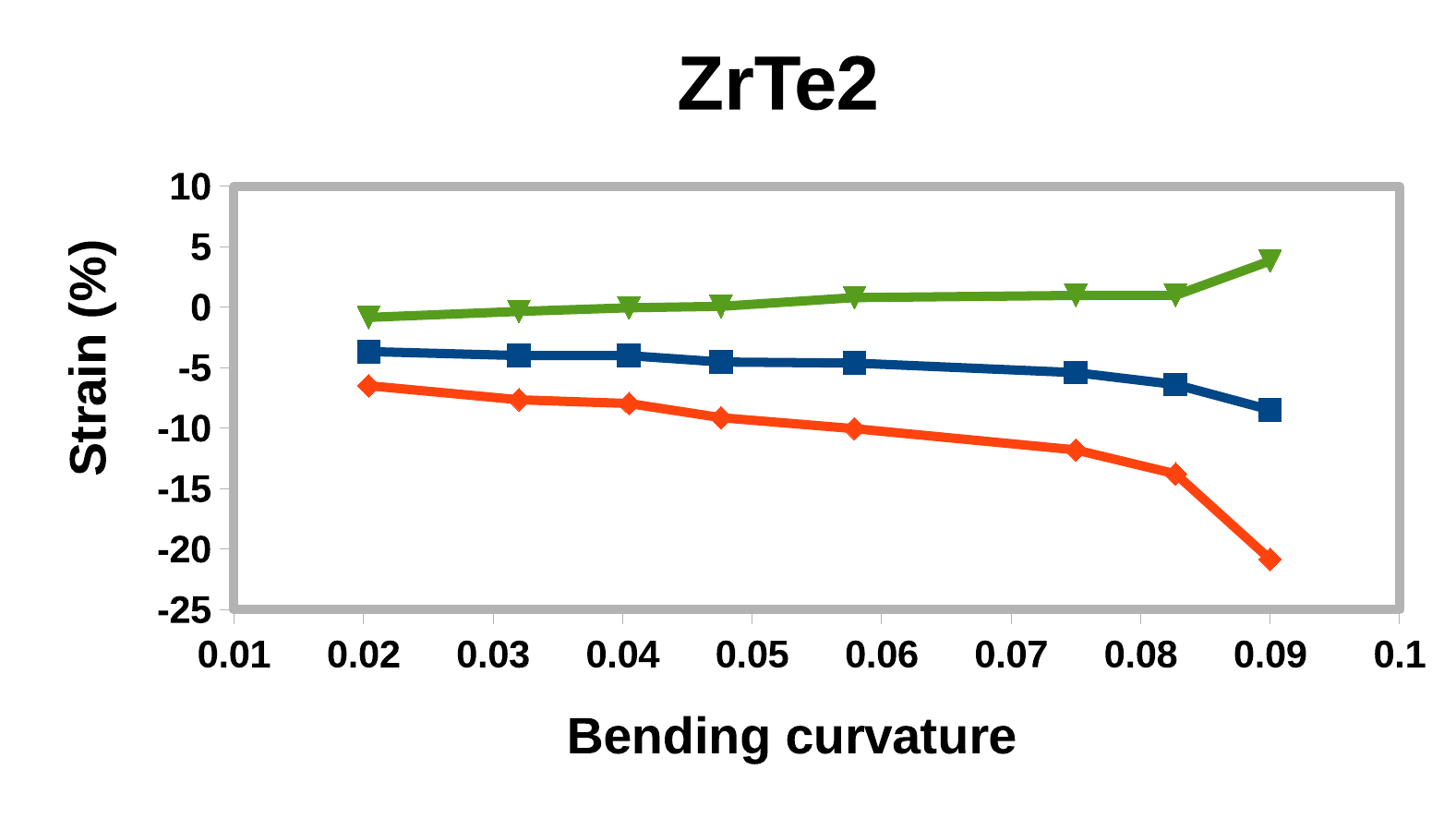}
 	\includegraphics[scale=0.33]{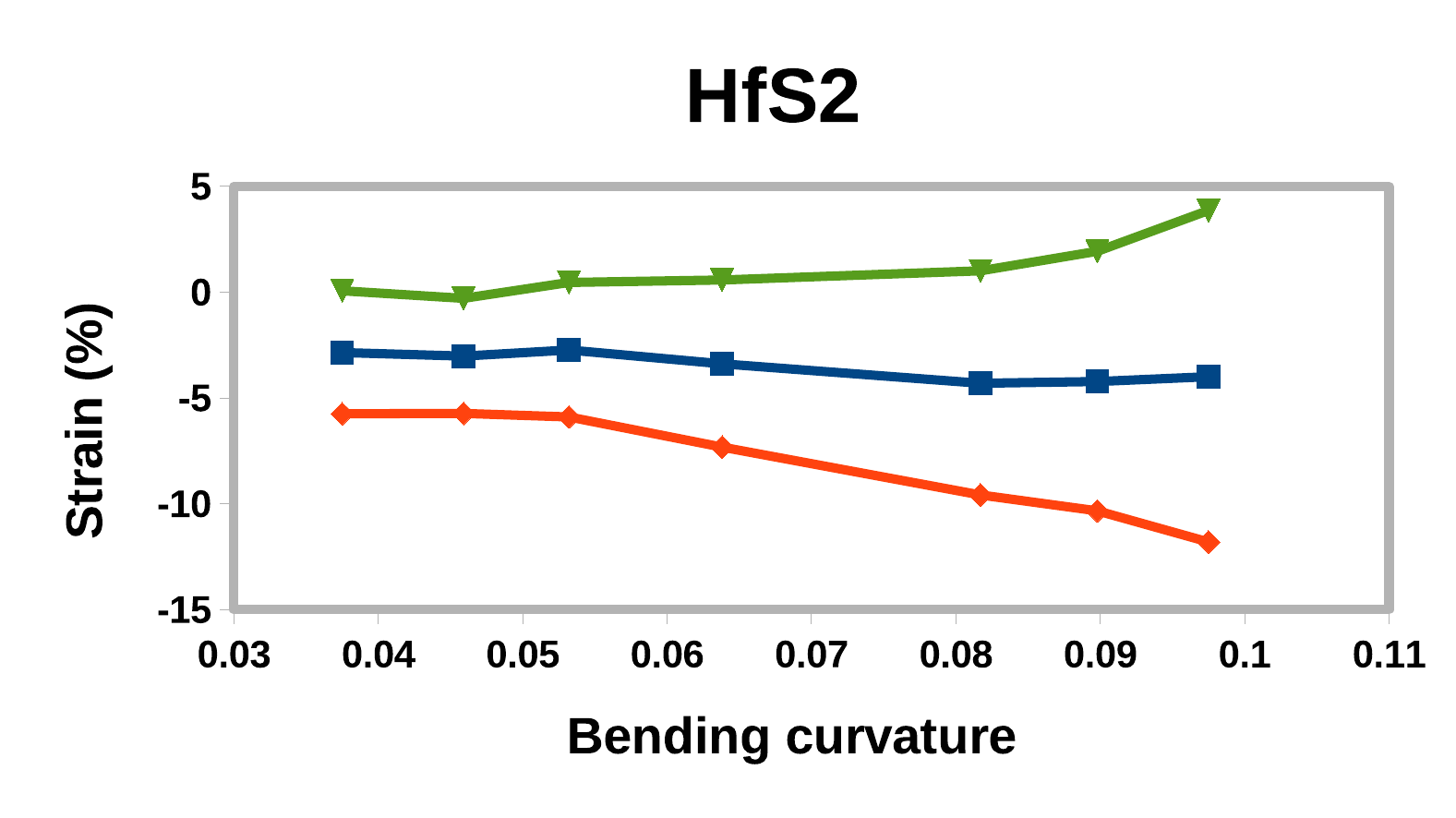}
 	\includegraphics[scale=0.33]{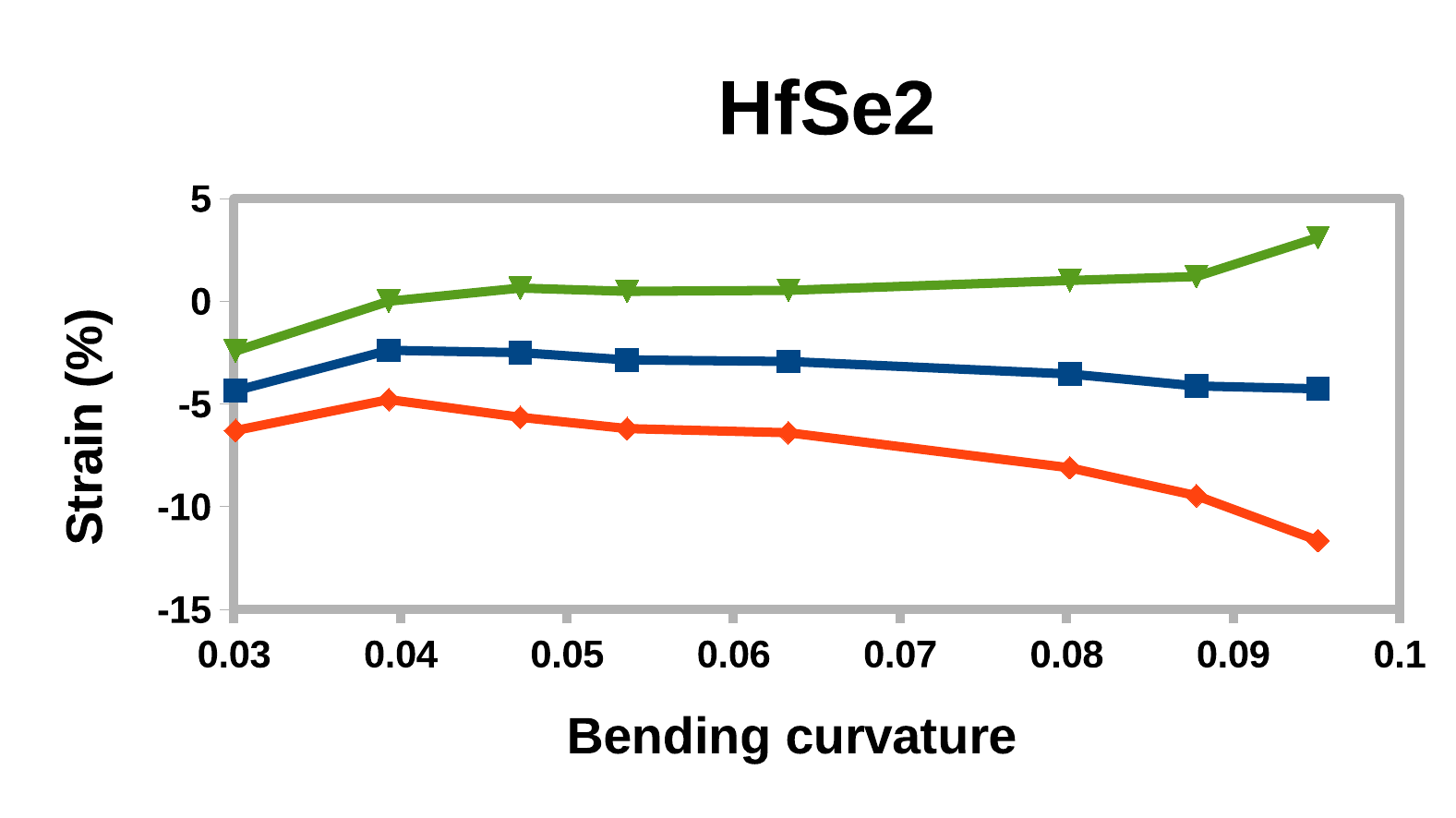}
 	\includegraphics[scale=0.33]{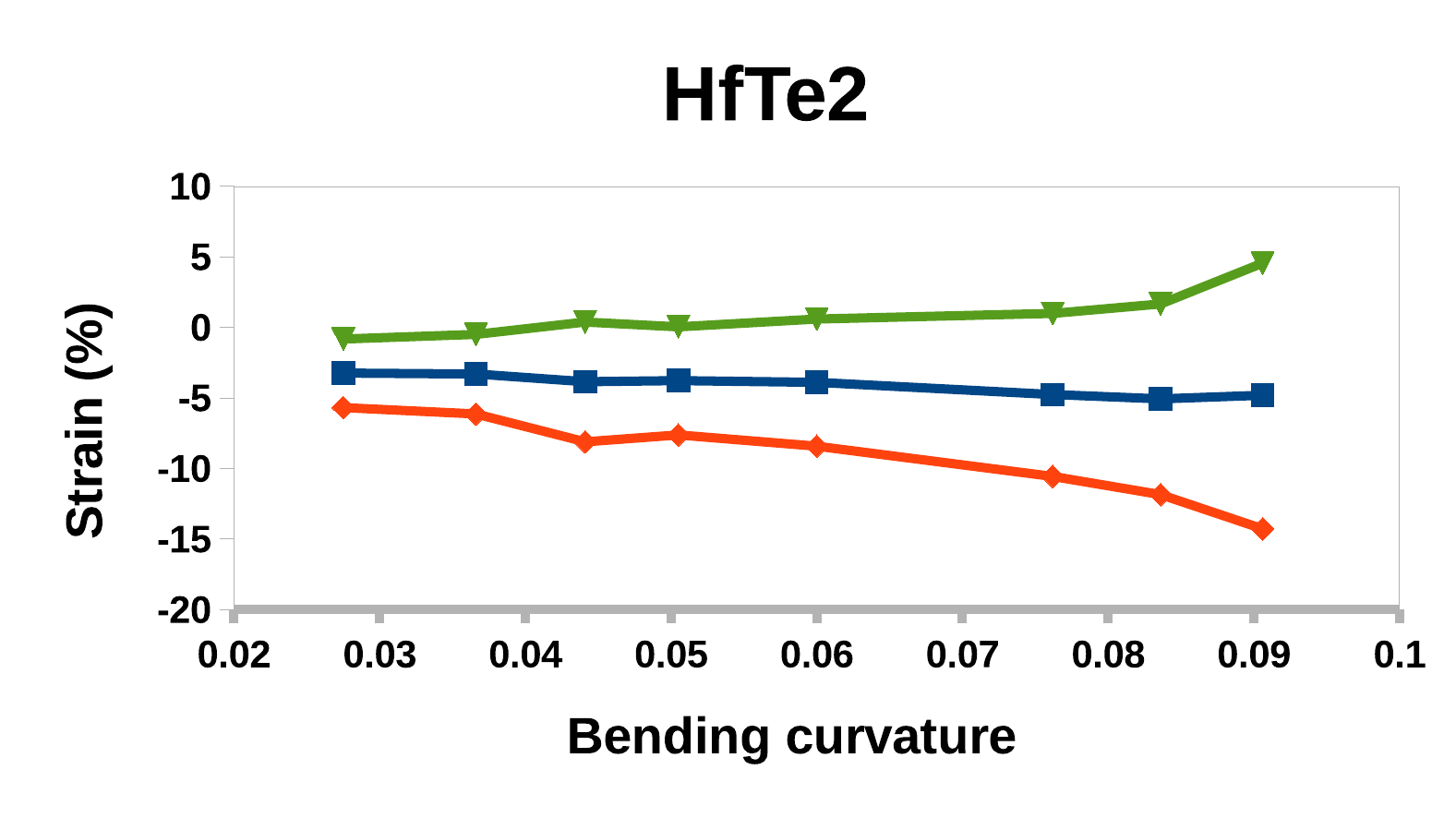}
 	\includegraphics[scale=0.33]{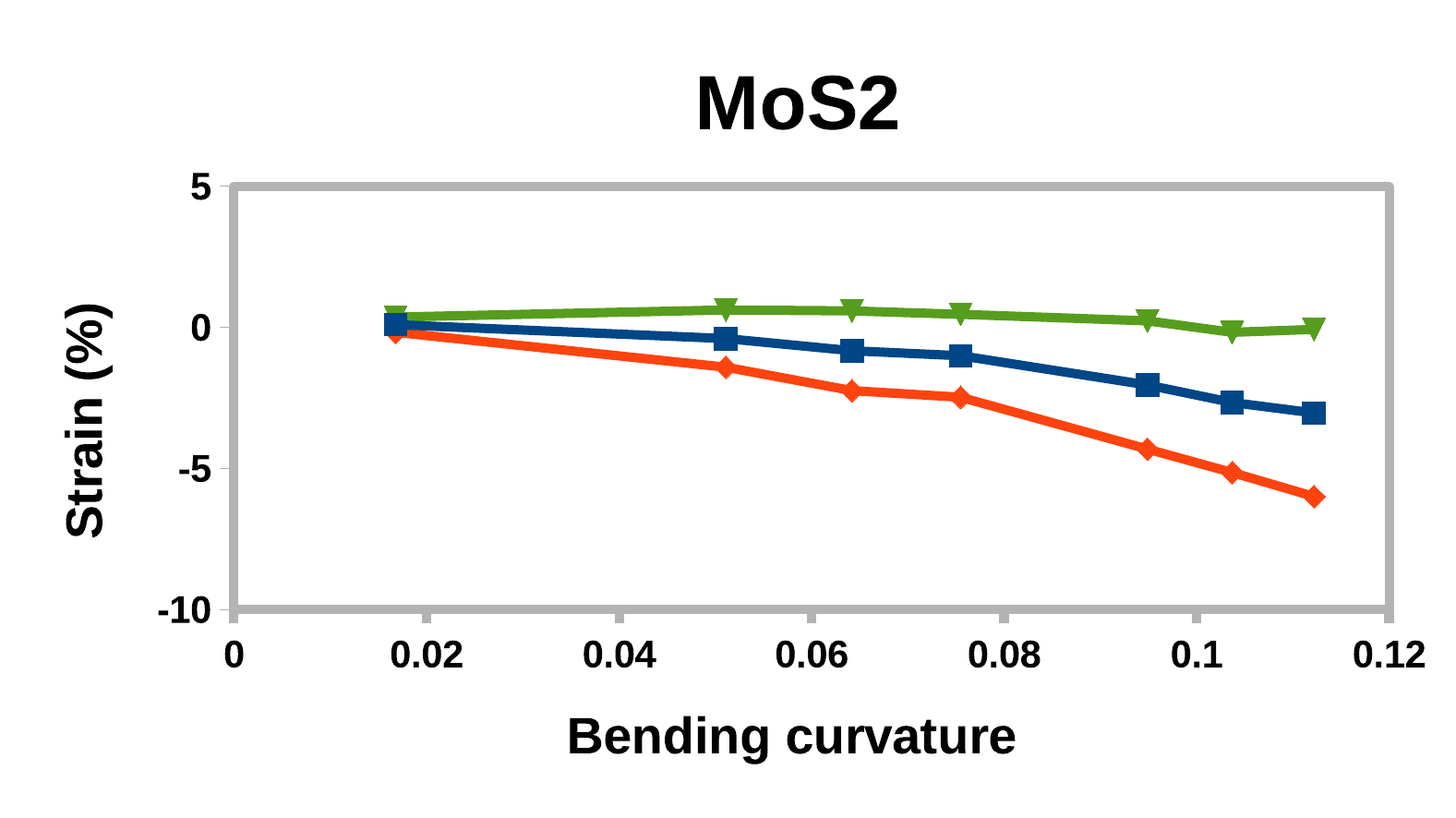}
 	\includegraphics[scale=0.33]{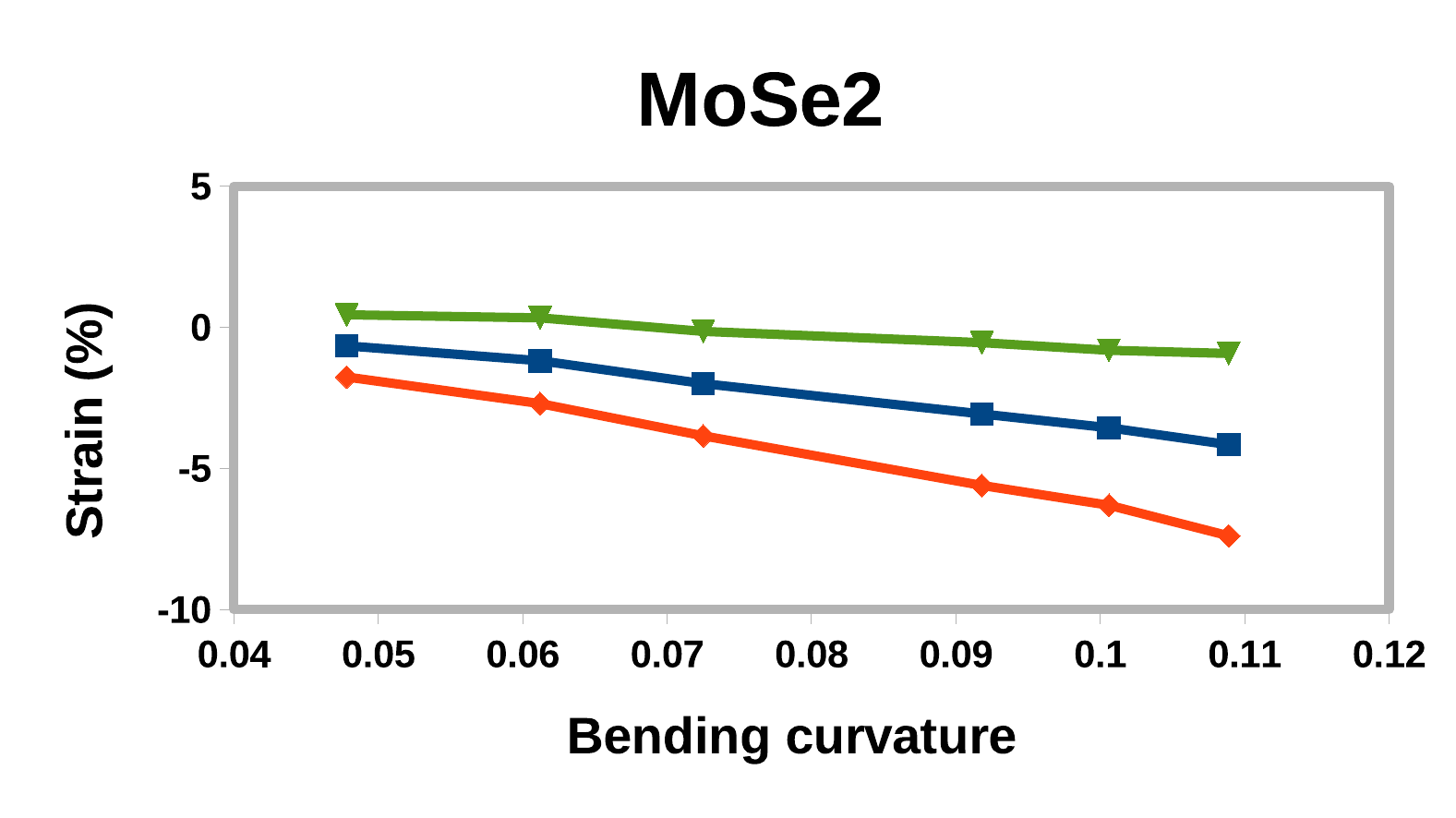}
 	\includegraphics[scale=0.33]{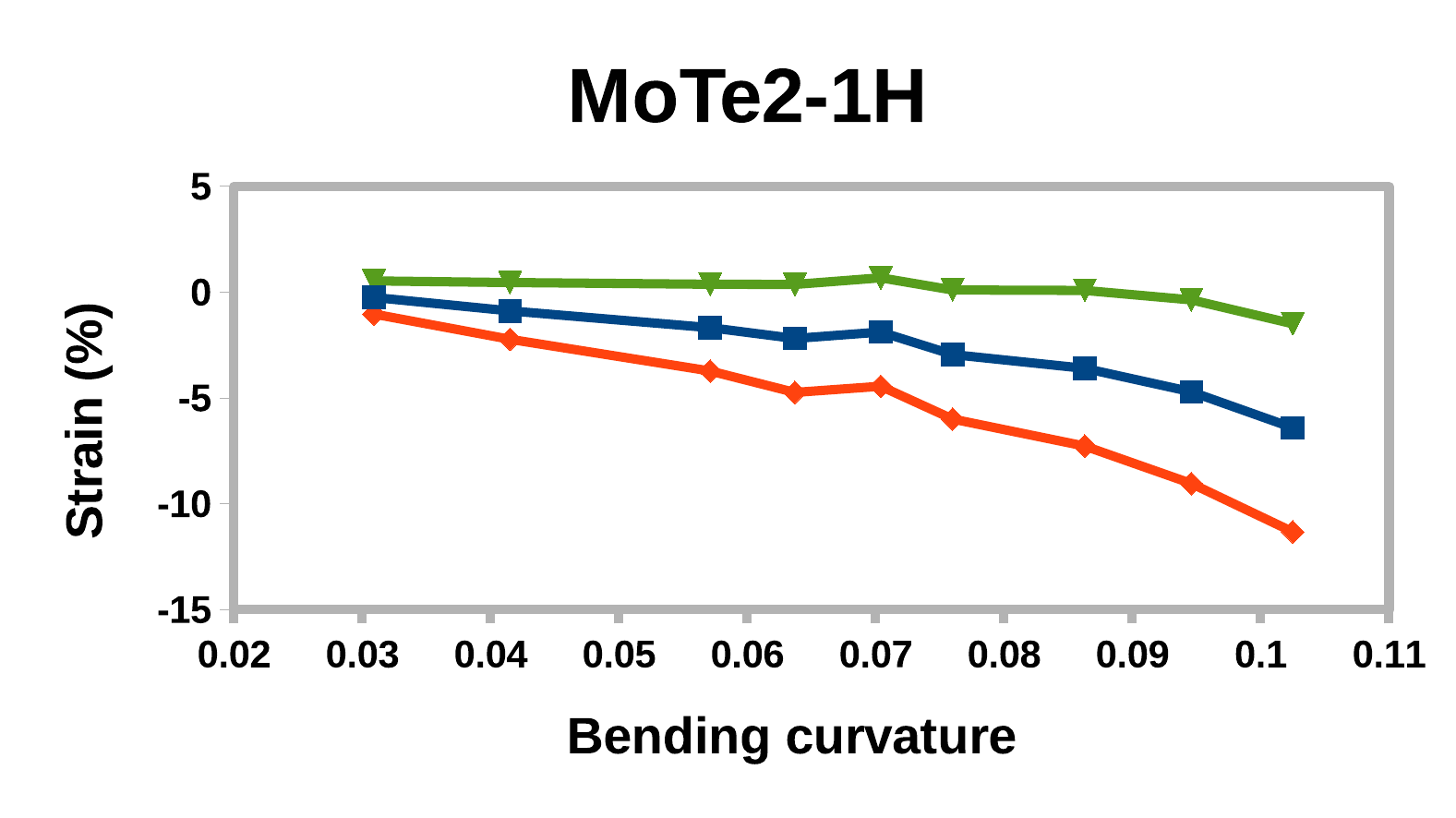}
 	\includegraphics[scale=0.33]{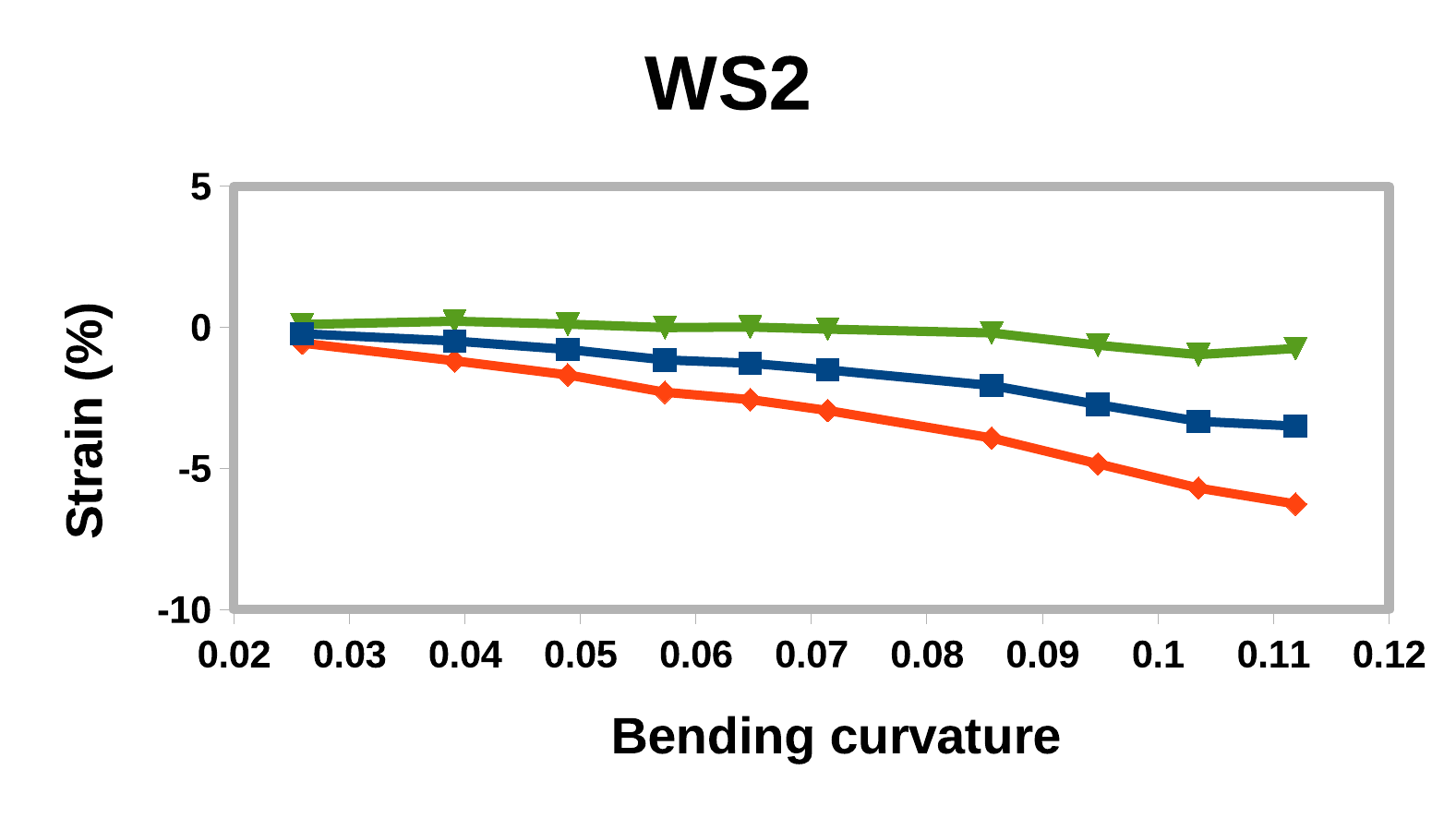}
 	\includegraphics[scale=0.33]{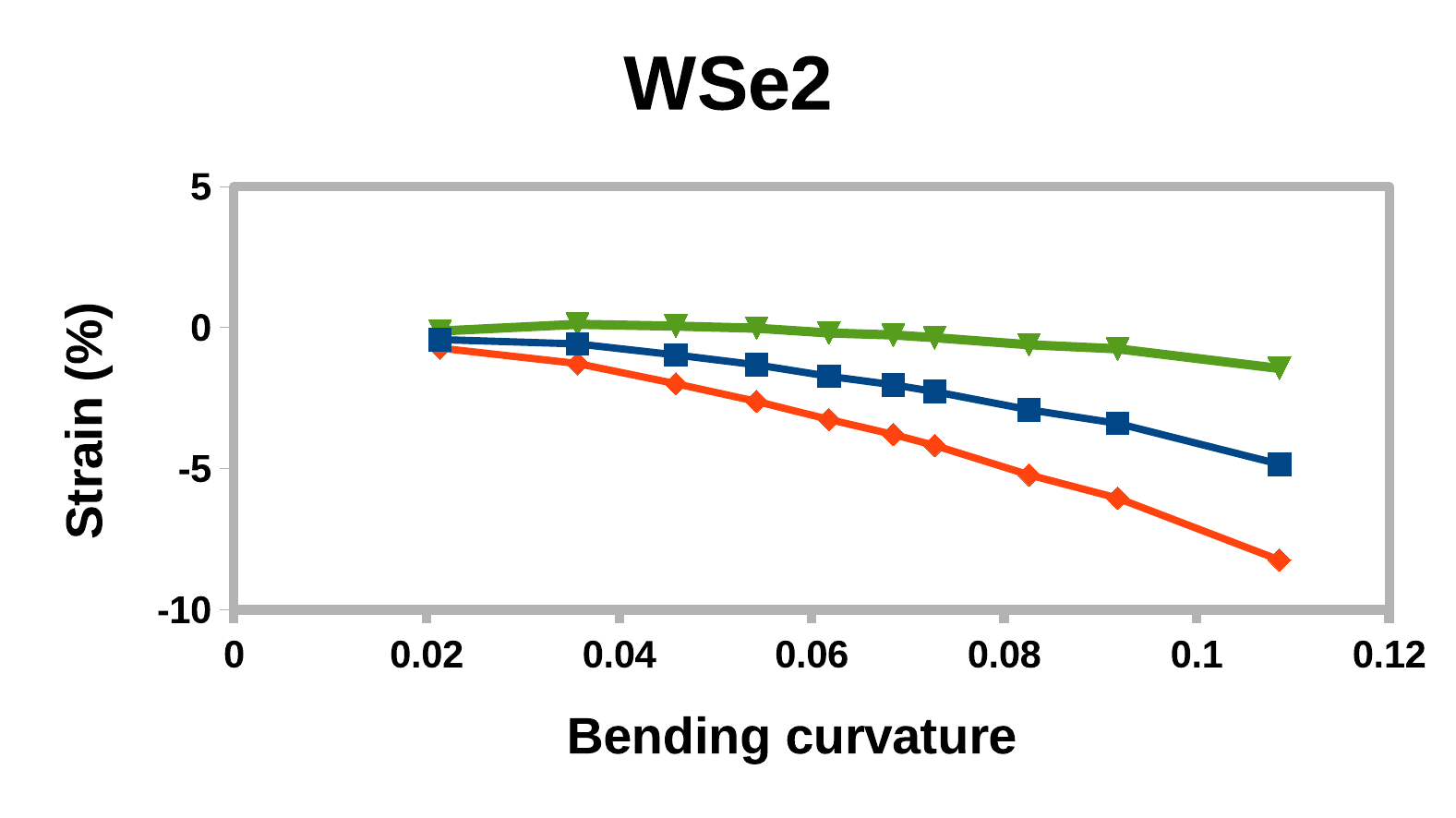}
 	\includegraphics[scale=0.33]{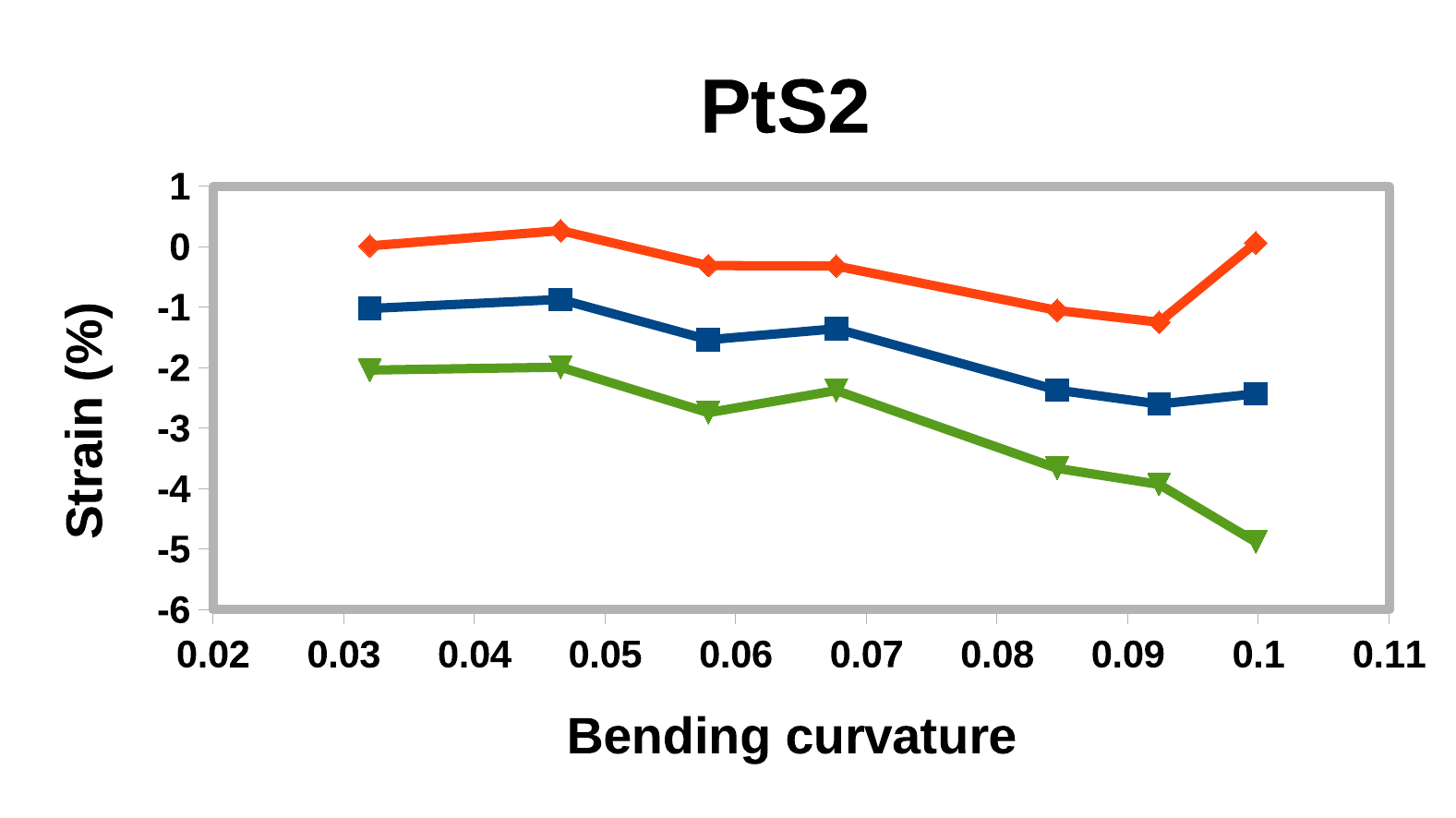}
 	\includegraphics[scale=0.33]{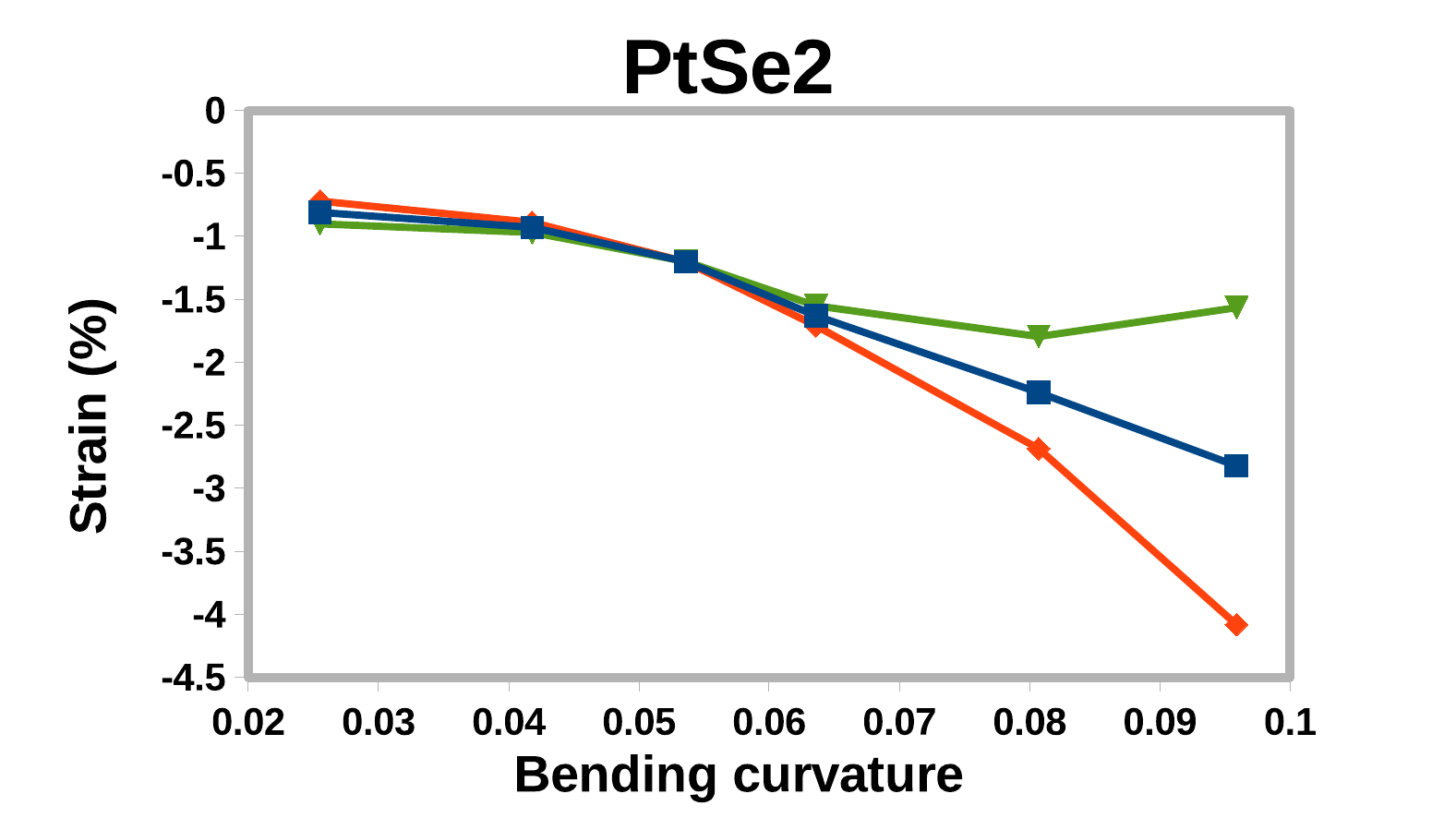}
 	\includegraphics[scale=0.33]{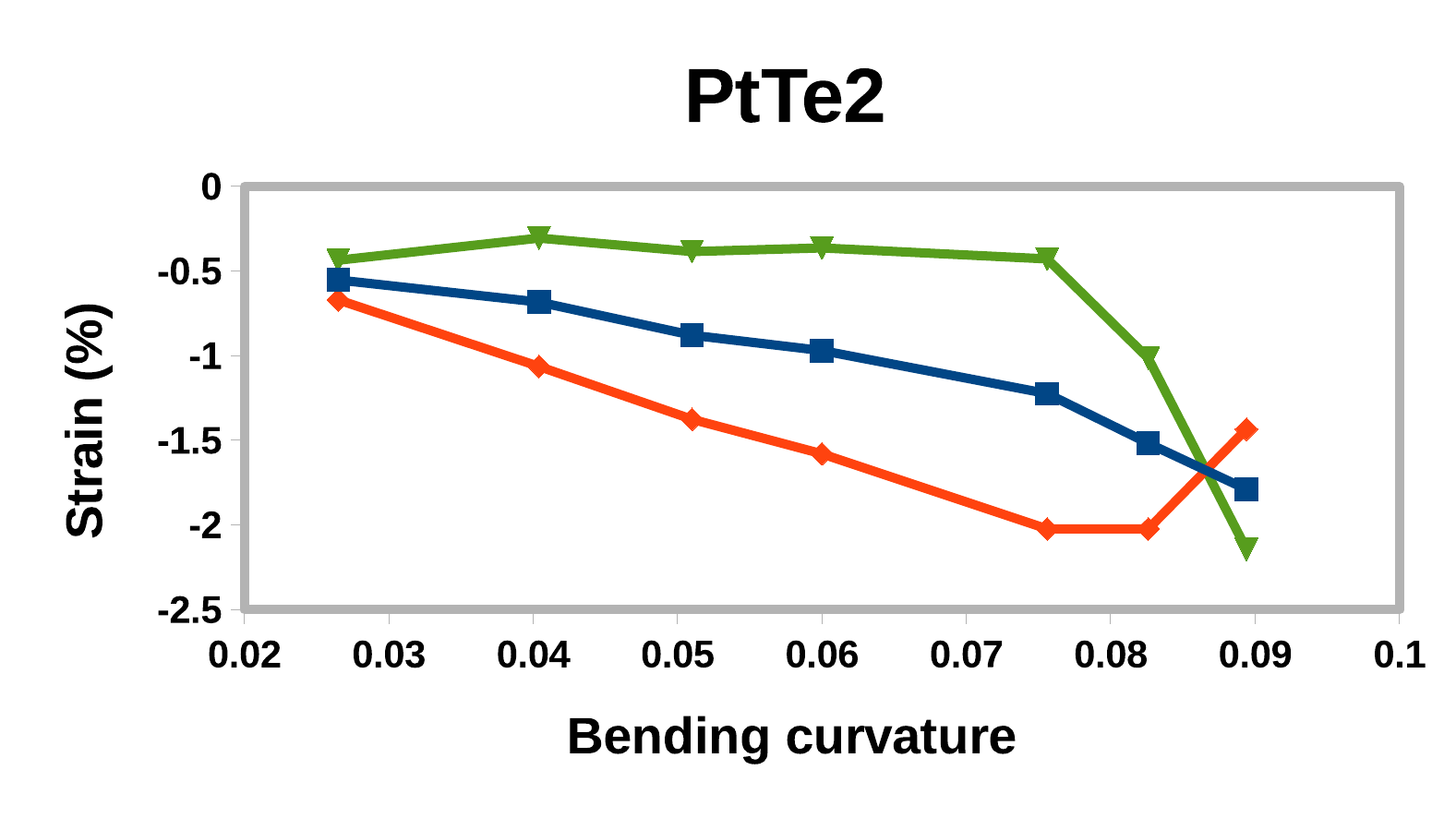}
 	\includegraphics[scale=0.33]{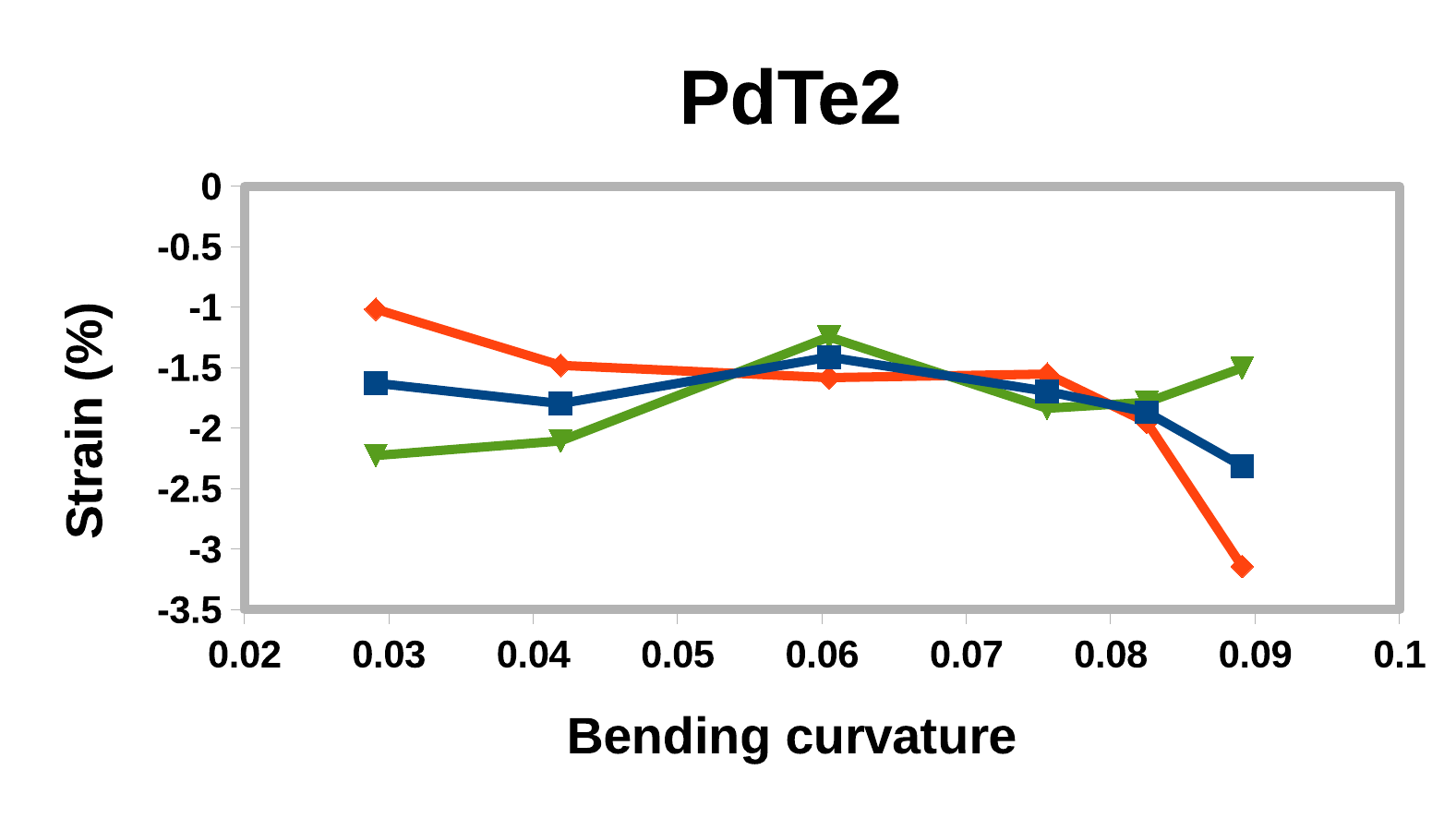}
 	\includegraphics[scale=0.33]{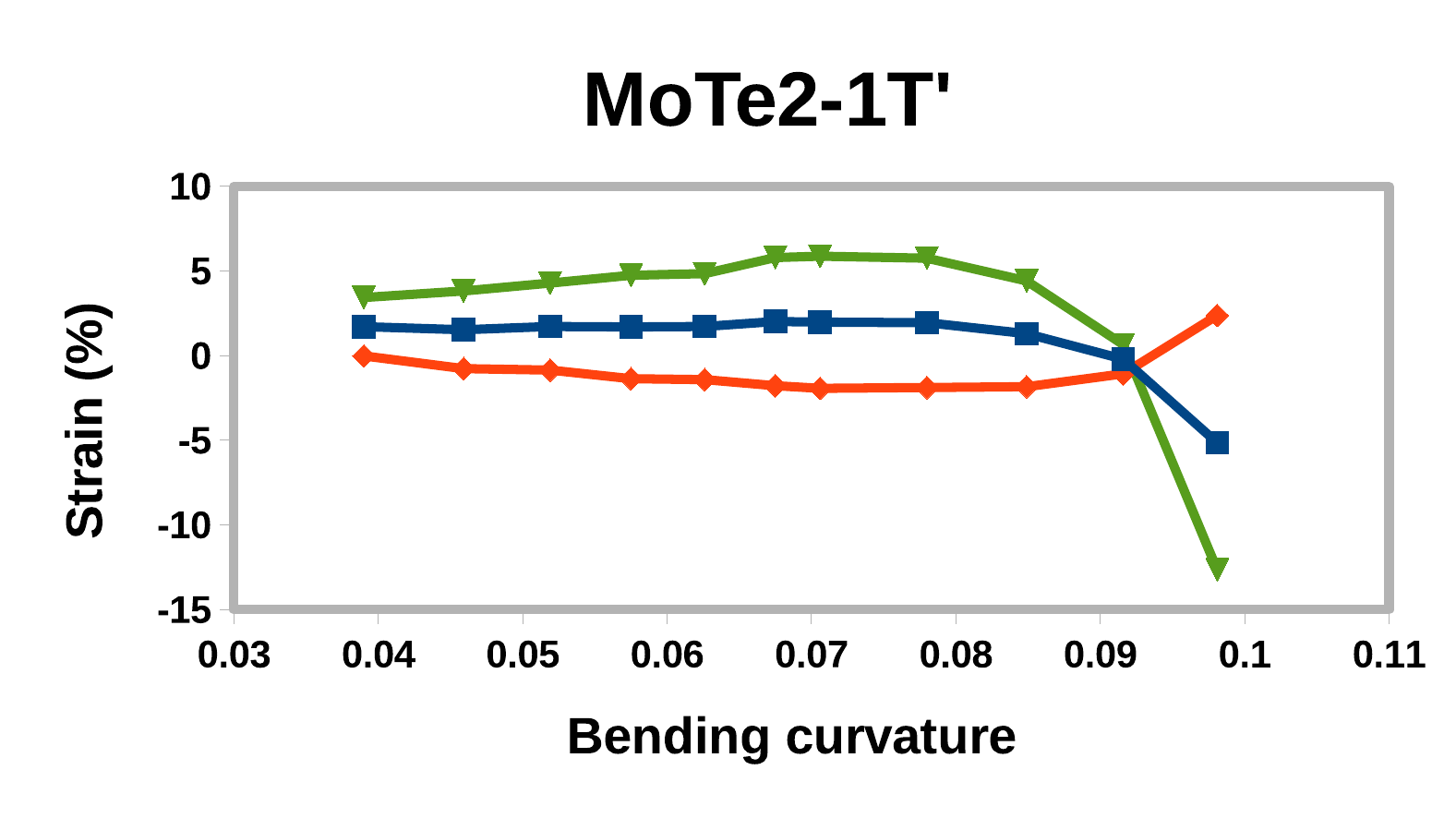}
 	\includegraphics[scale=0.33]{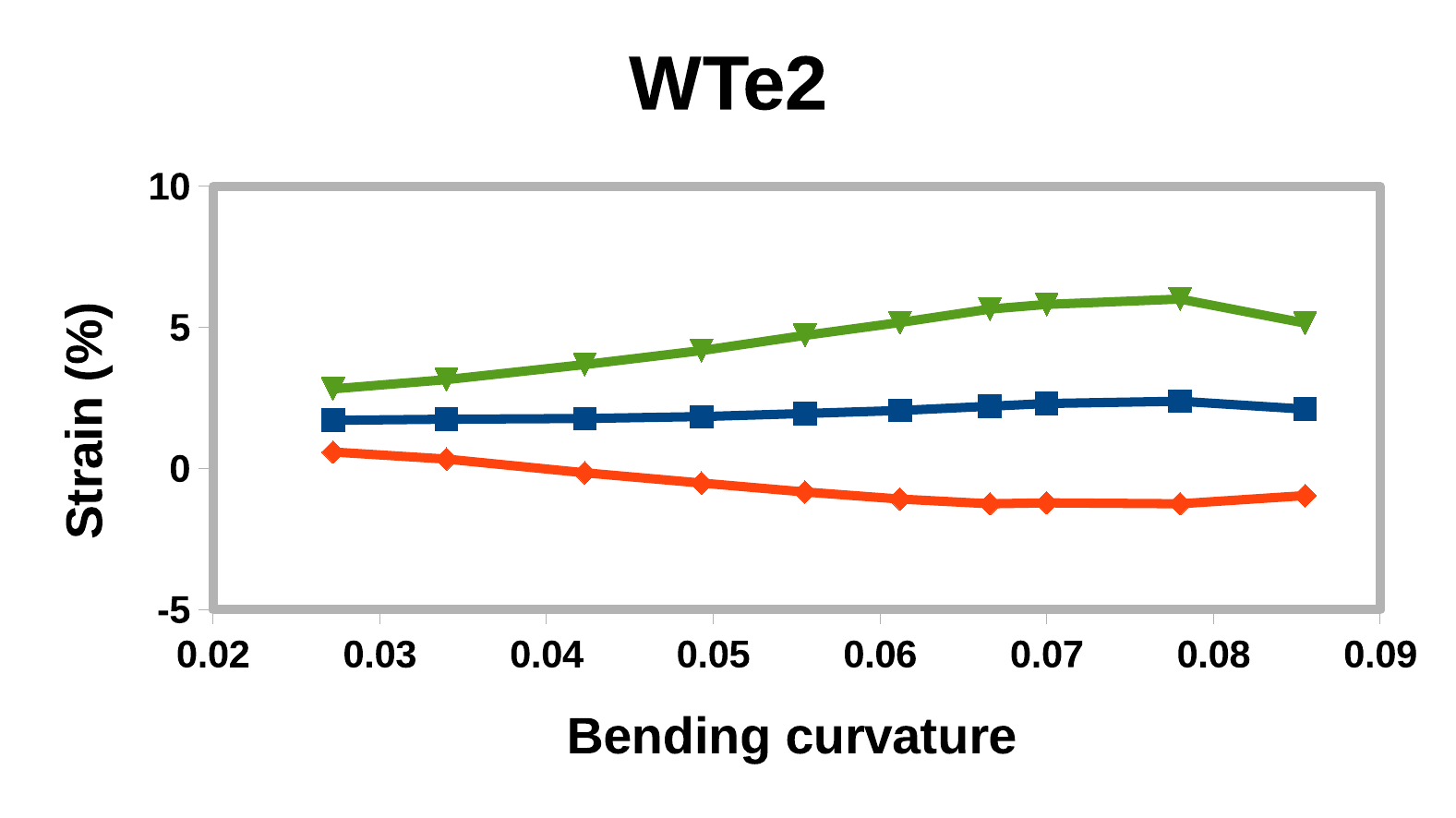}
 	\caption{The change in the physical thickness with respect to the flat nano-ribbon for the bending curvature around 0.09 $\AA^{-1}$. We utilized a 6$^{th}$ order polynomial fit to estimate the physical thickness. The {\color{blue} blue}, {\color{red} red}, and {\color{green} green} plots represent t$_{tot}$ (t$_{up}$ + t$_{dn}$), t$_{up}$, and t$_{dn}$ respectively (see figure 2 (III) in the main paper).}
 	\label{lab:thickness}
 	
 \end{figure}

 \begin{figure}[h!]
 	\renewcommand\thefigure{S3}
 	\includegraphics[scale=0.8]{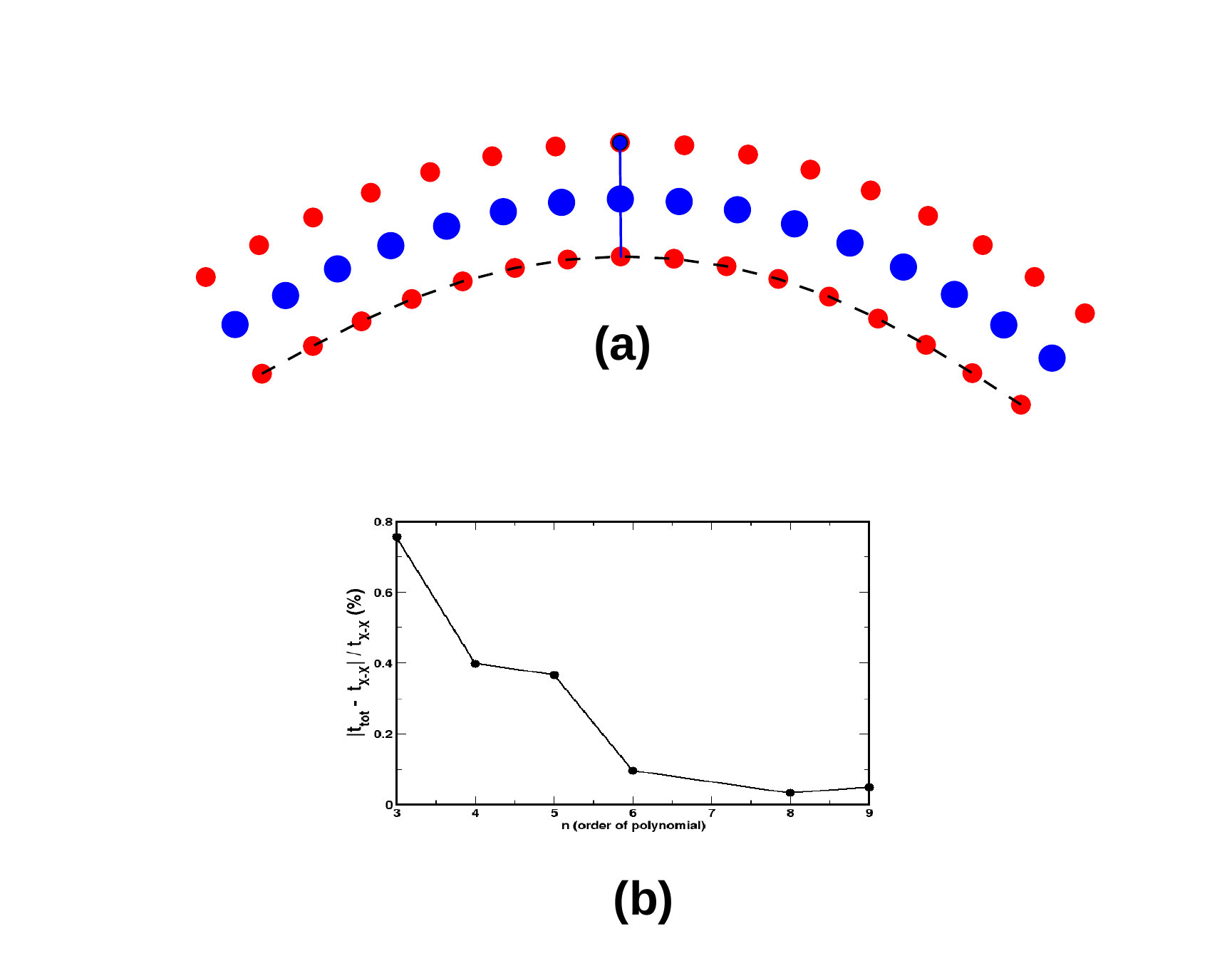}
 	\caption{Estimating the physical thickness at the curvature (vertex) region for 1H structure. (a) The layer is fitted with a n$^{th}$ order polynomial and the shortest distance from a point at the middle (A and B) to the fitted curve; t$_{tot}$ is the shortest distance from point A to the inner layer; t$_{up}$ is the shortest distance from point A to the middle layer; t$_{dn}$ is the shortest distance from point B to the inner layer.(b) an absolute error with respect to polynomial order; t$_{X-X}$ is the distance between point A and C. A sixth order polynomial is sufficient to estimate the physical thickness for 1H, 1T, and 1T$^\prime$ structures.}
 	\label{fig:1T-BG}
 \end{figure}

 \begin{figure}[h!]
 	\renewcommand\thefigure{S4}
 	\text{I}
 	\includegraphics[scale=1.2]{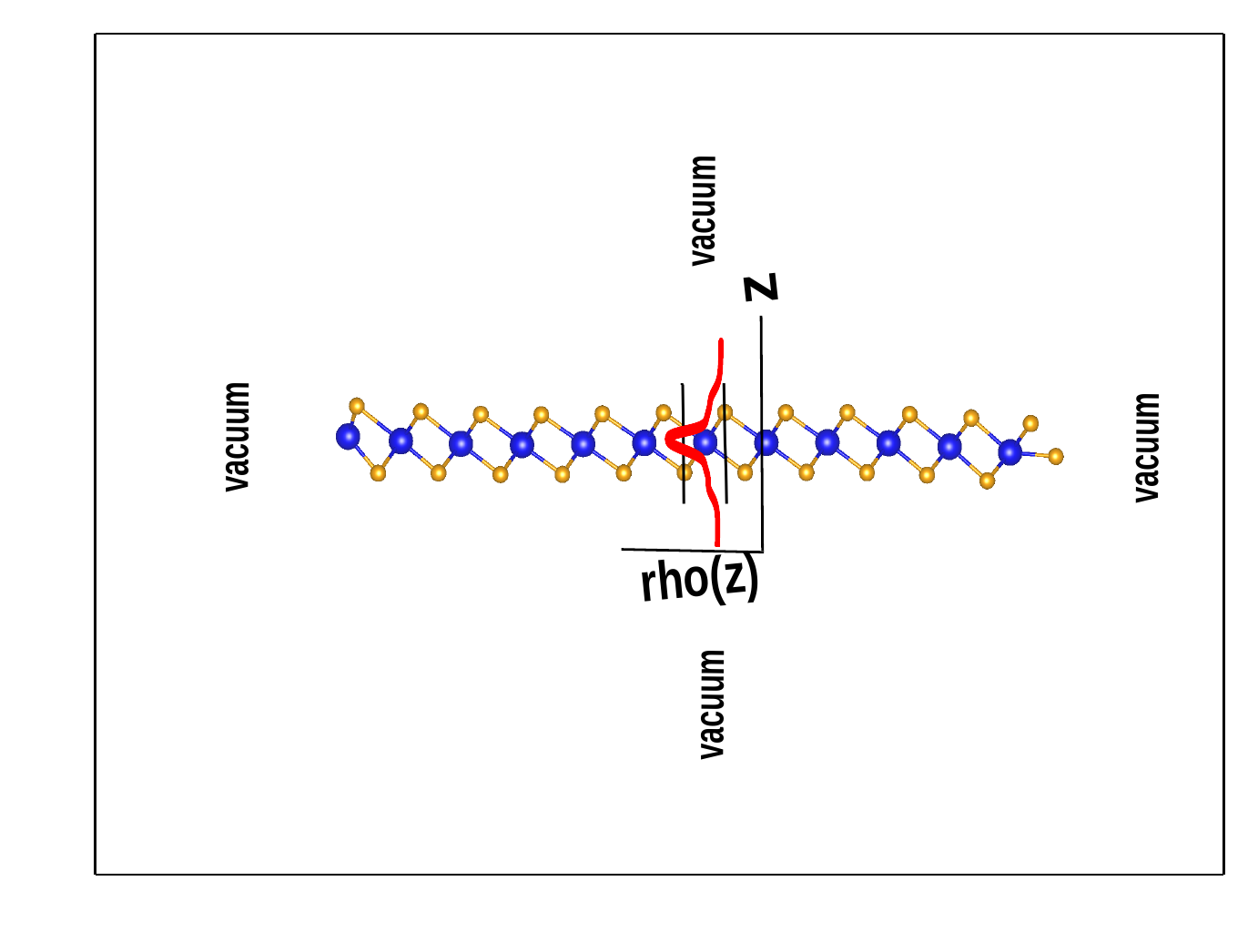}
 	
 	\includegraphics[scale=0.33]{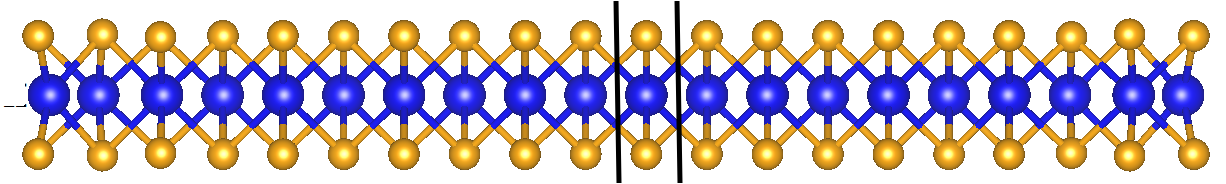}
 	\includegraphics[scale=0.5]{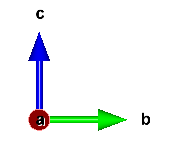}
 	\includegraphics[scale=0.25]{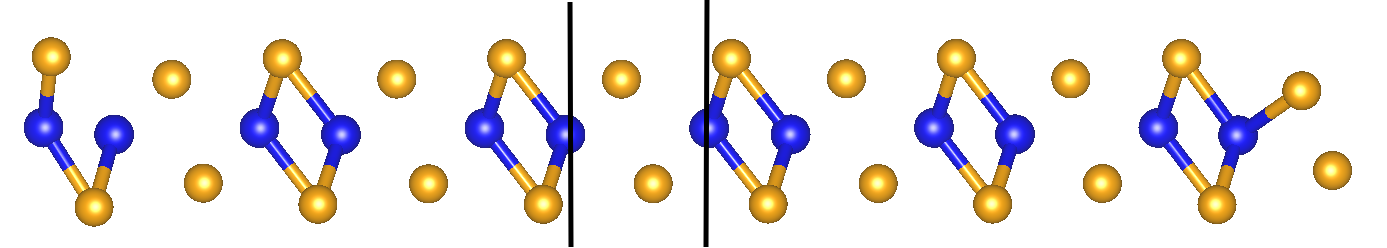}

 	\caption{A narrow window is taken at the middle of the nano-ribbon. The local electronic charge density is calculated along the out-of-plane direction (c- axis) within the narrow window. The first, second, and the third figure correspond to 1T, 1H, and 1T$^\prime$ respectively.}
 	\label{fig:window}	
 \end{figure}

 \begin{figure}[h!]
 	\renewcommand\thefigure{S5}
 	\caption{Variation of the local electronic charge distribution with respect to bending curvature. The distance between 2 vertical red lines represent the d$_{X-X}$ distance of the monolayer bulk.}
 	\includegraphics[scale=0.35]{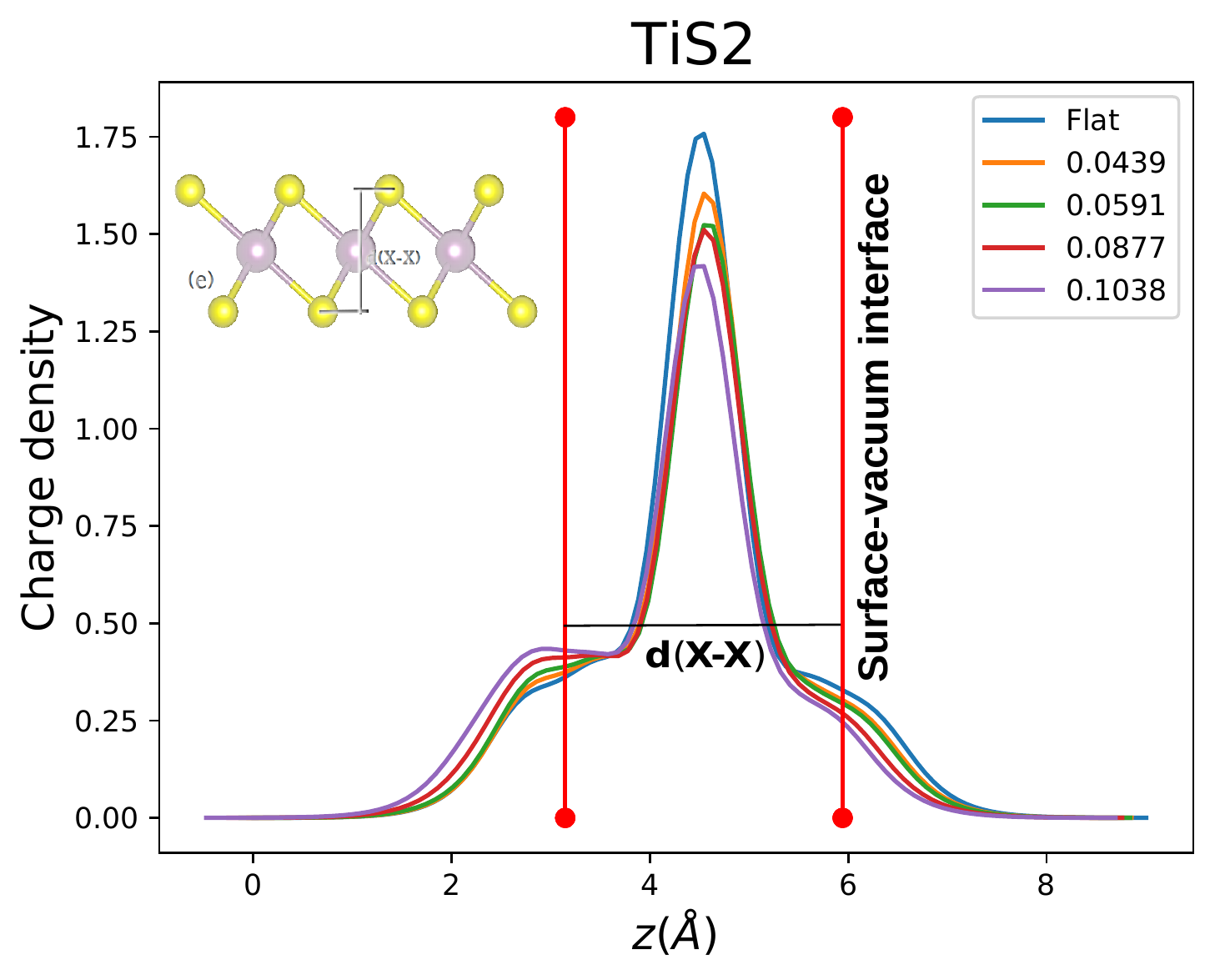}
 	\includegraphics[scale=0.35]{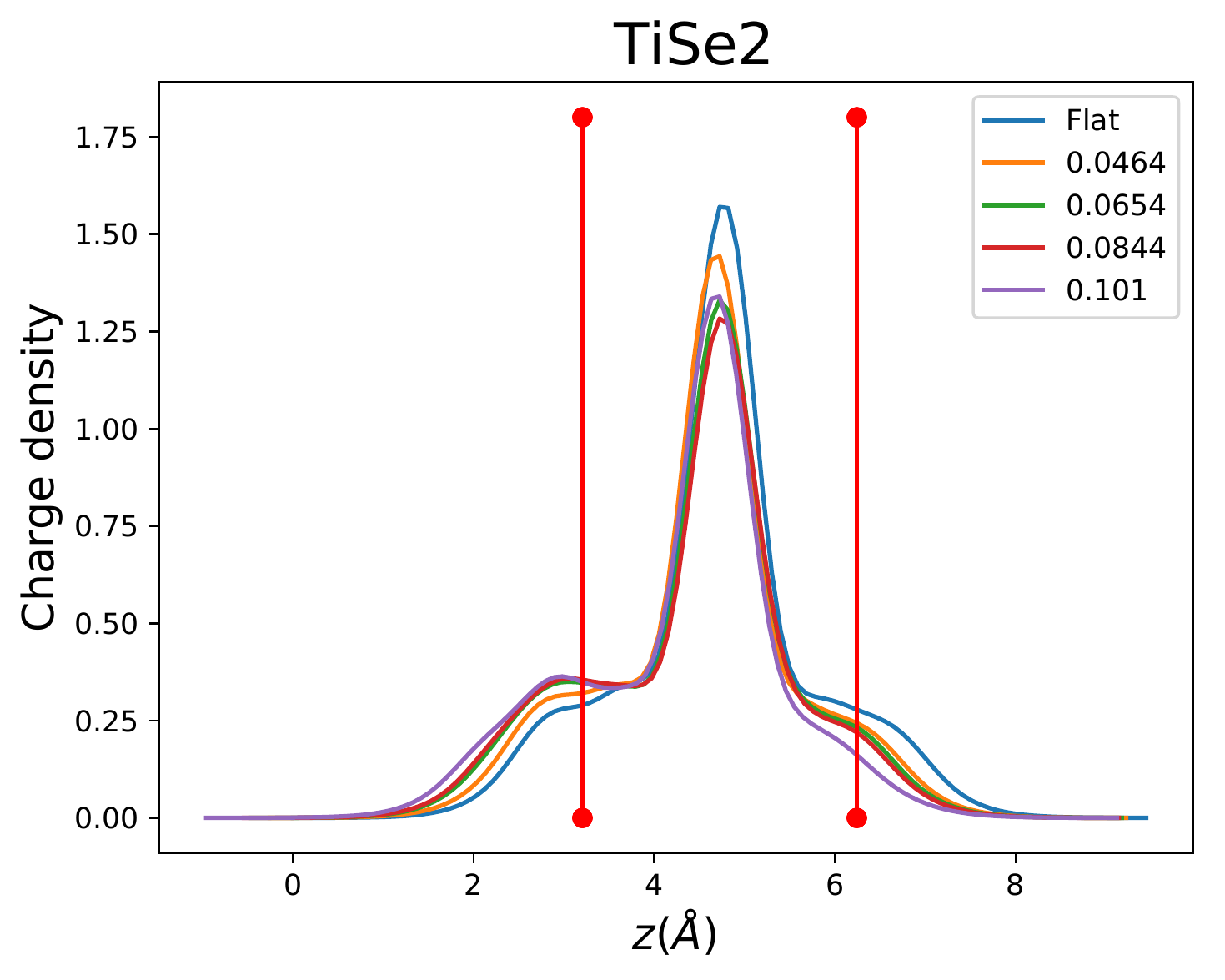}
 	\includegraphics[scale=0.35]{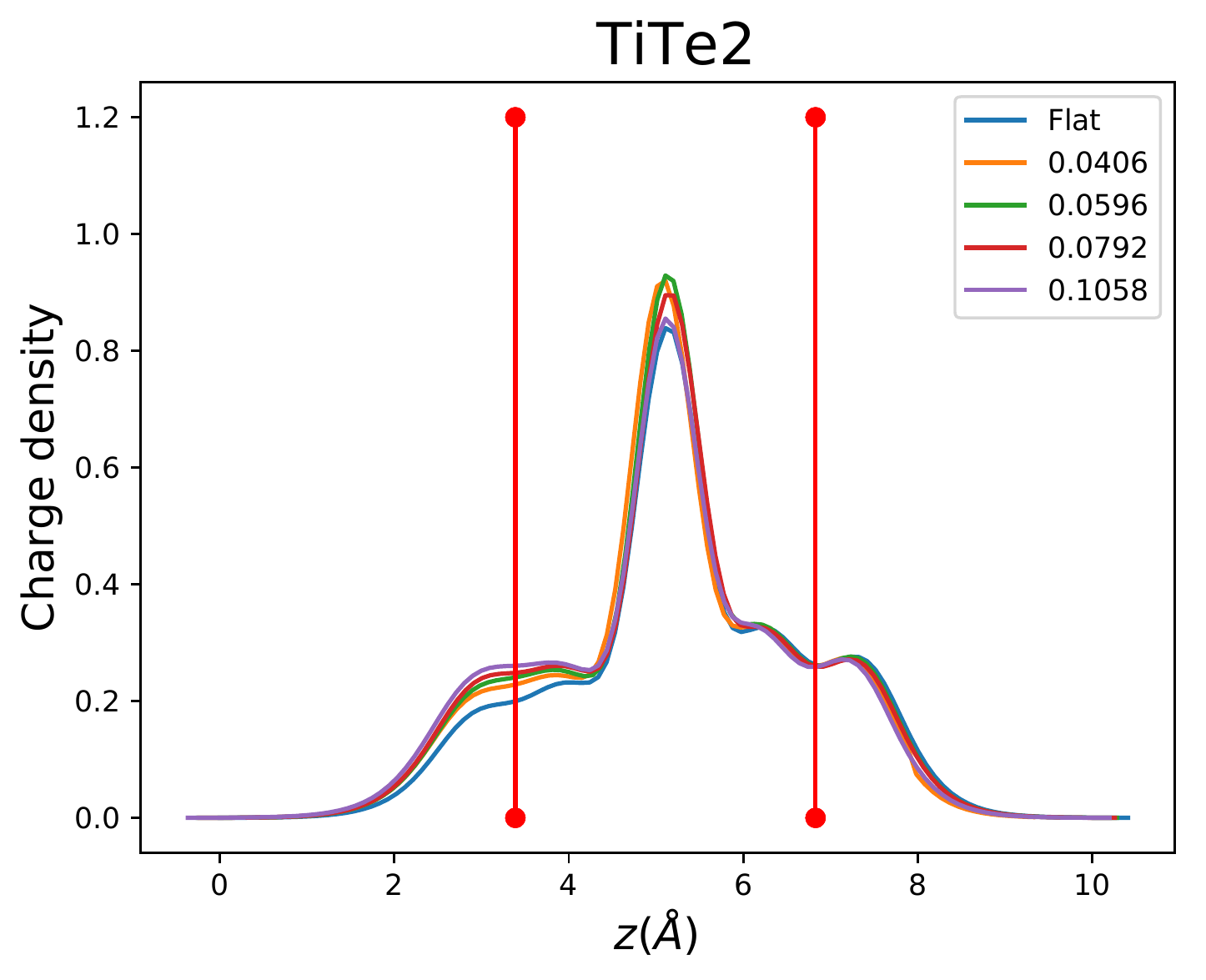}
 	\includegraphics[scale=0.35]{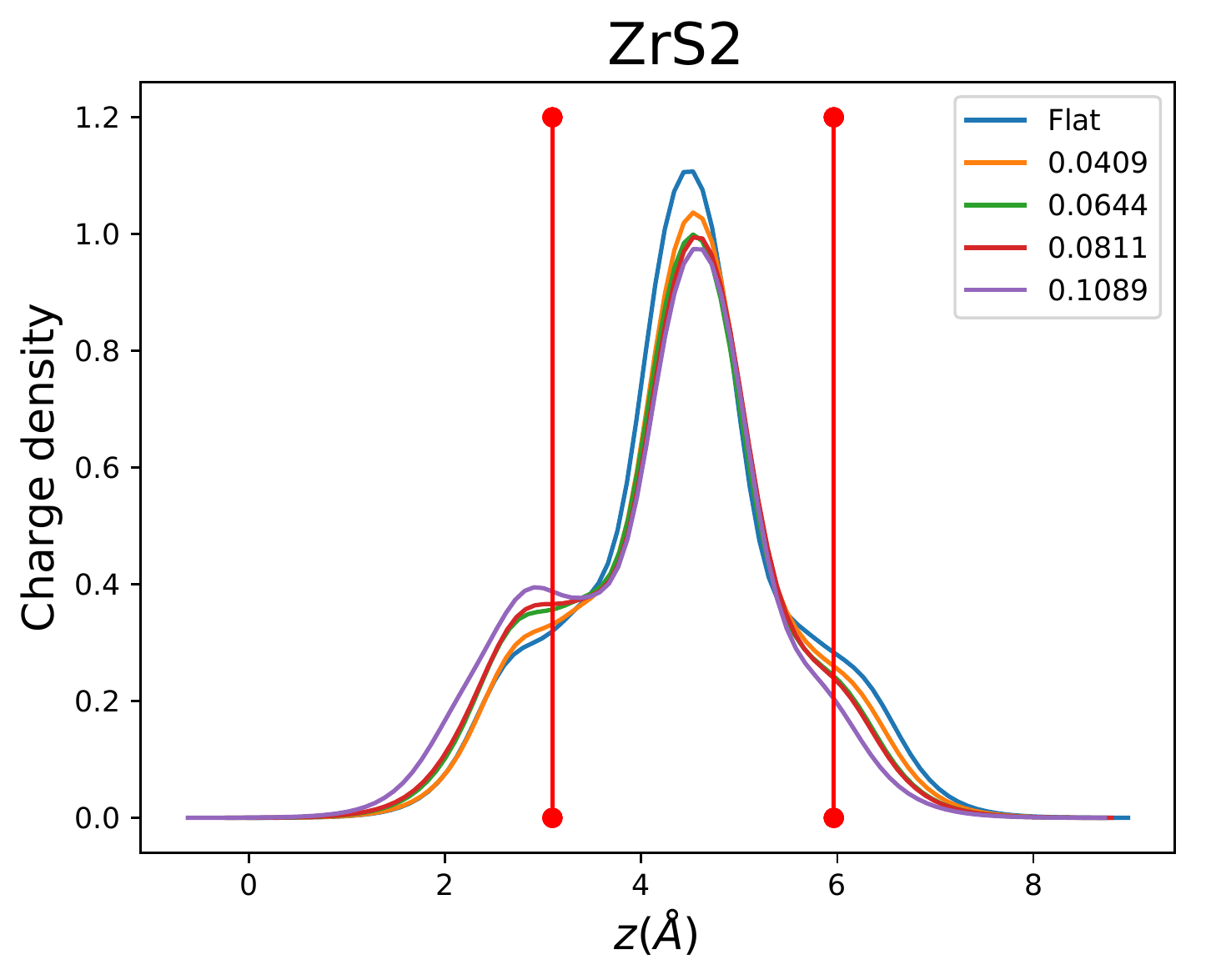}
 	\includegraphics[scale=0.35]{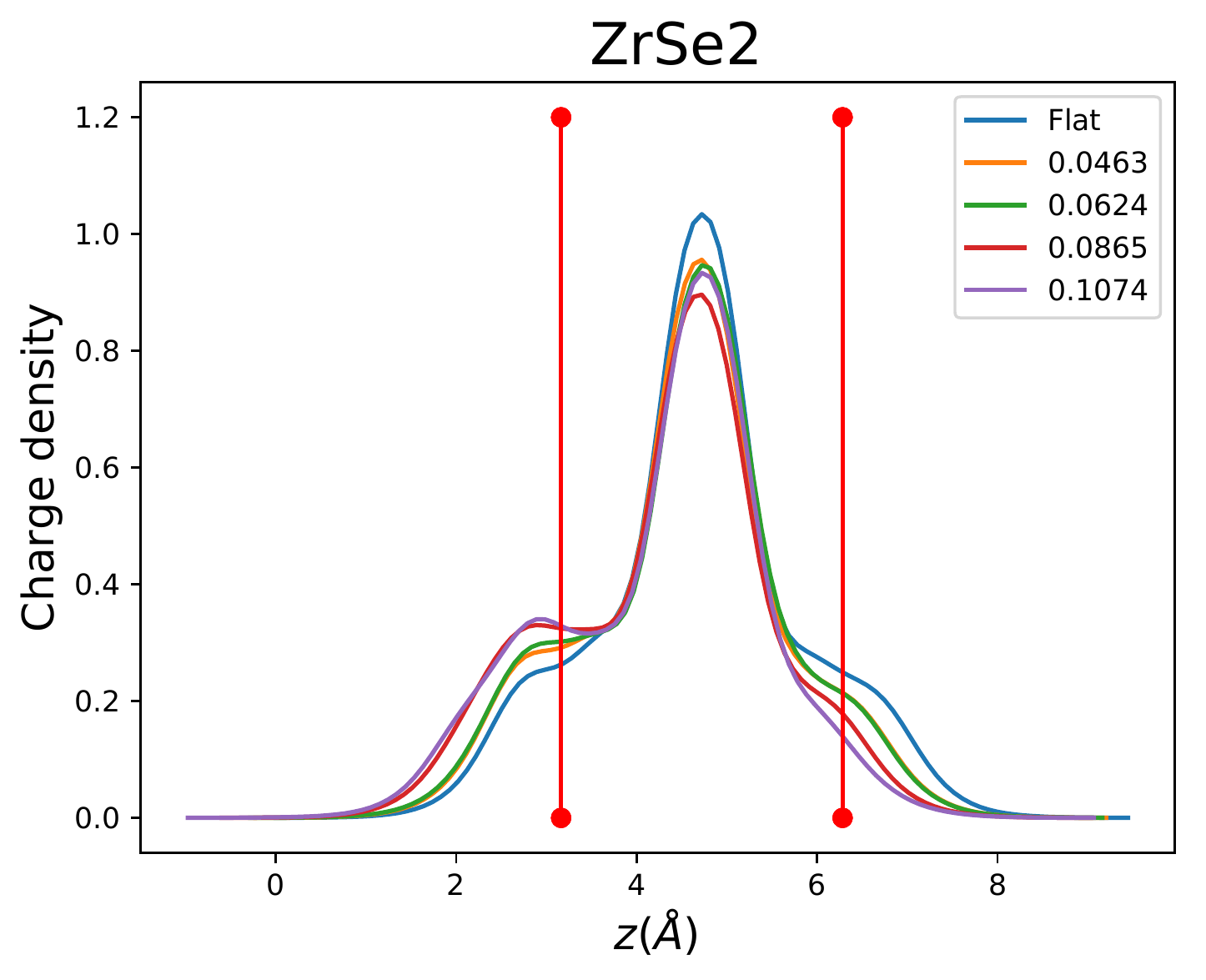}
 	\includegraphics[scale=0.35]{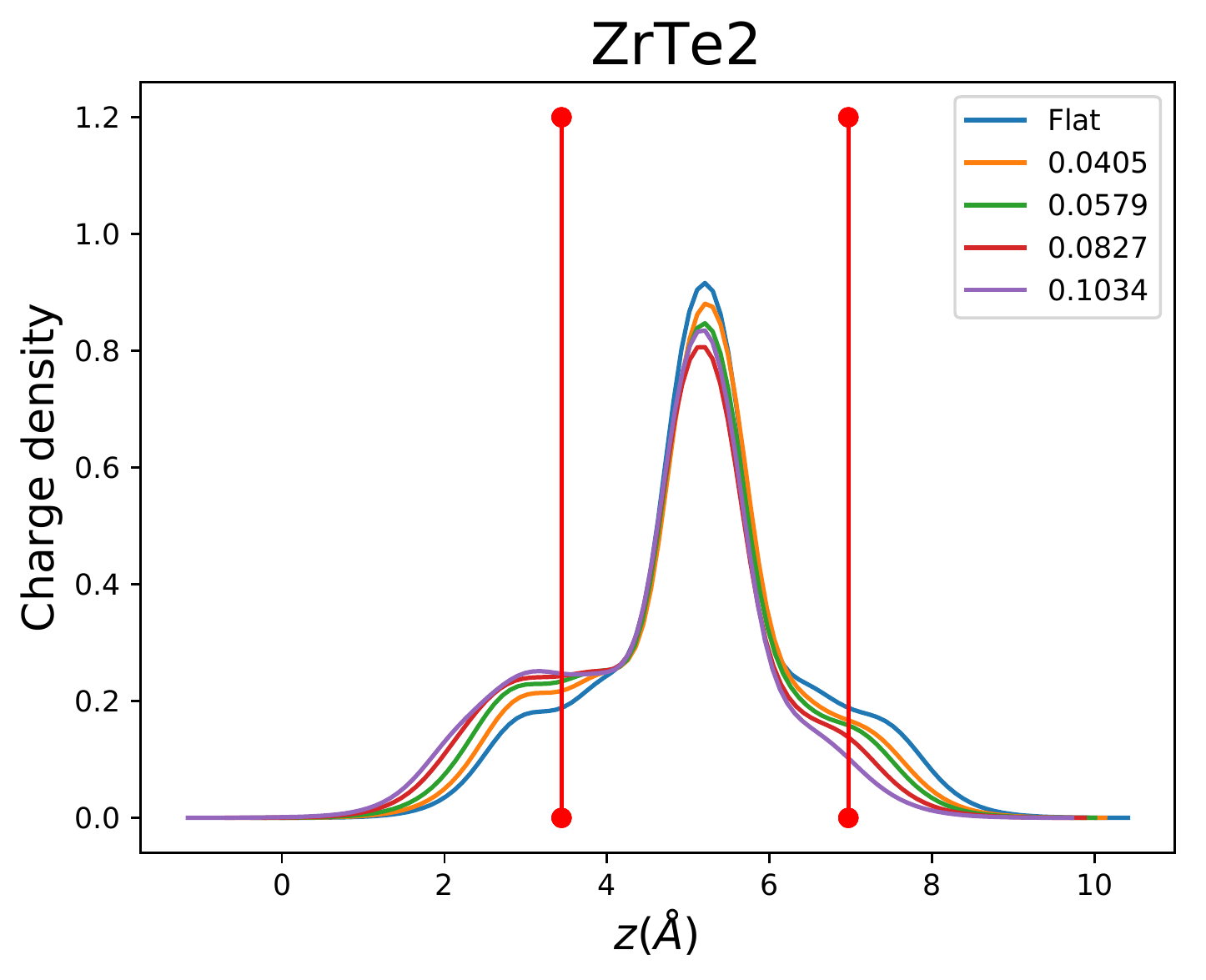}
 	\includegraphics[scale=0.35]{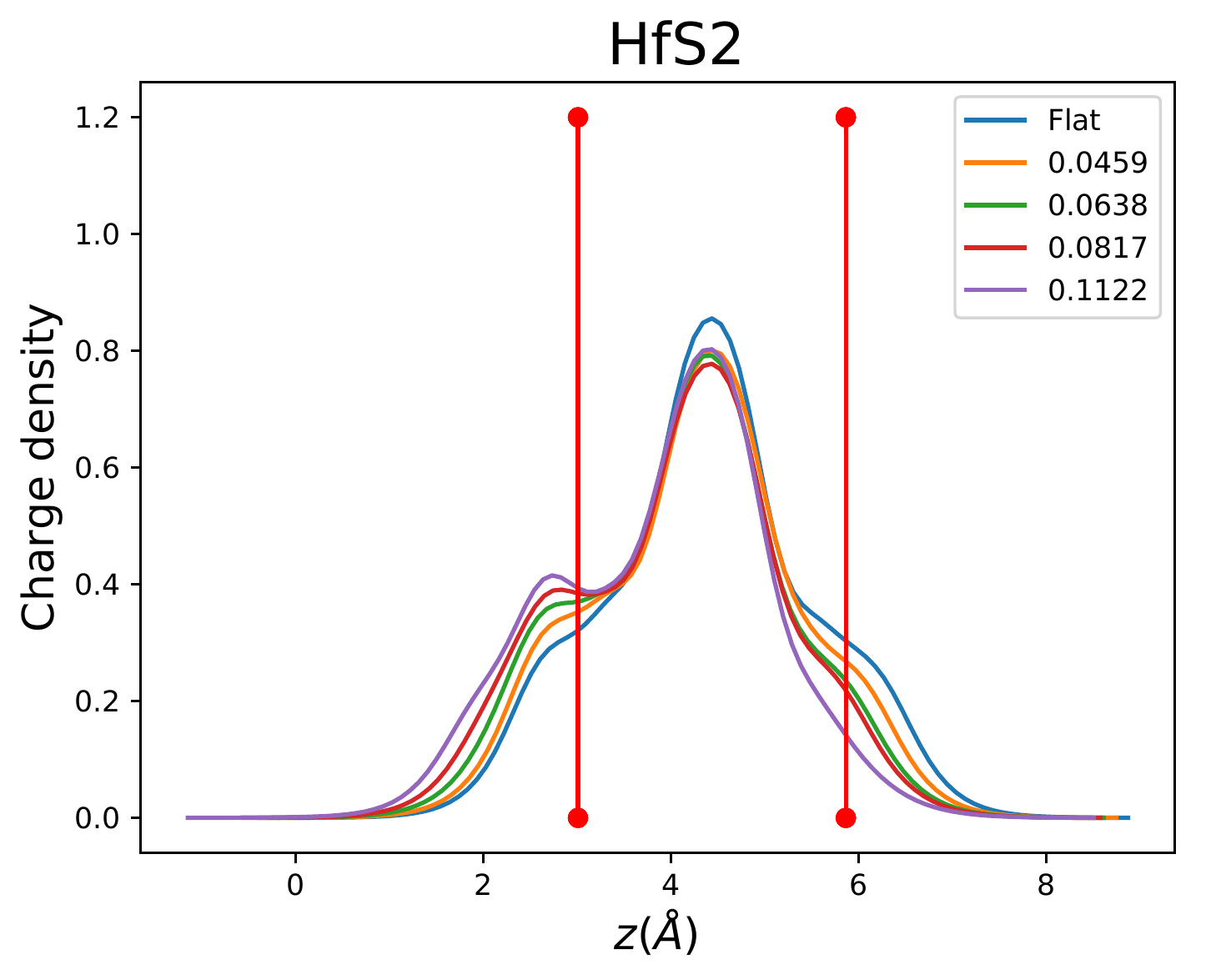}
 	\includegraphics[scale=0.35]{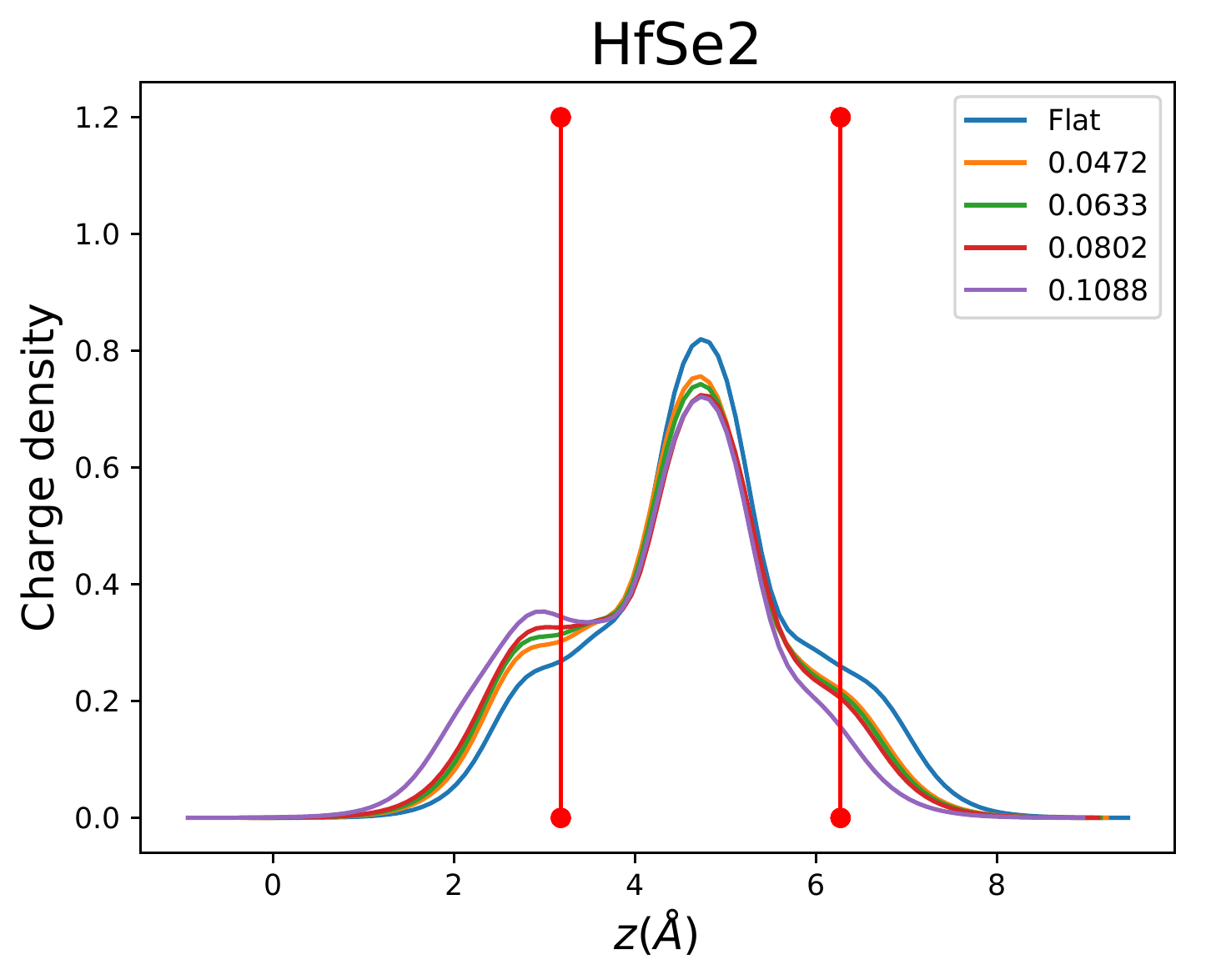}
 	\includegraphics[scale=0.35]{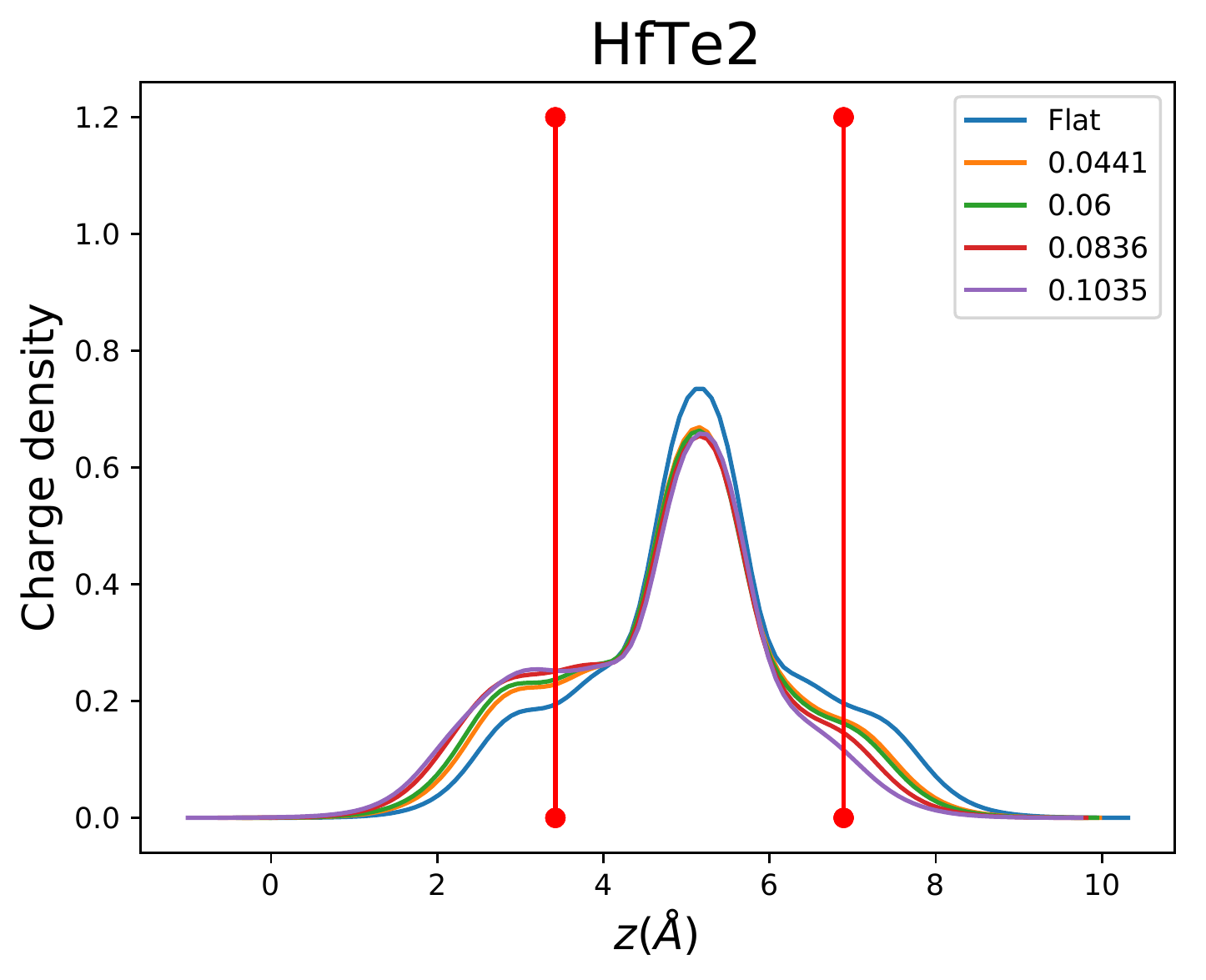}
 	\includegraphics[scale=0.35]{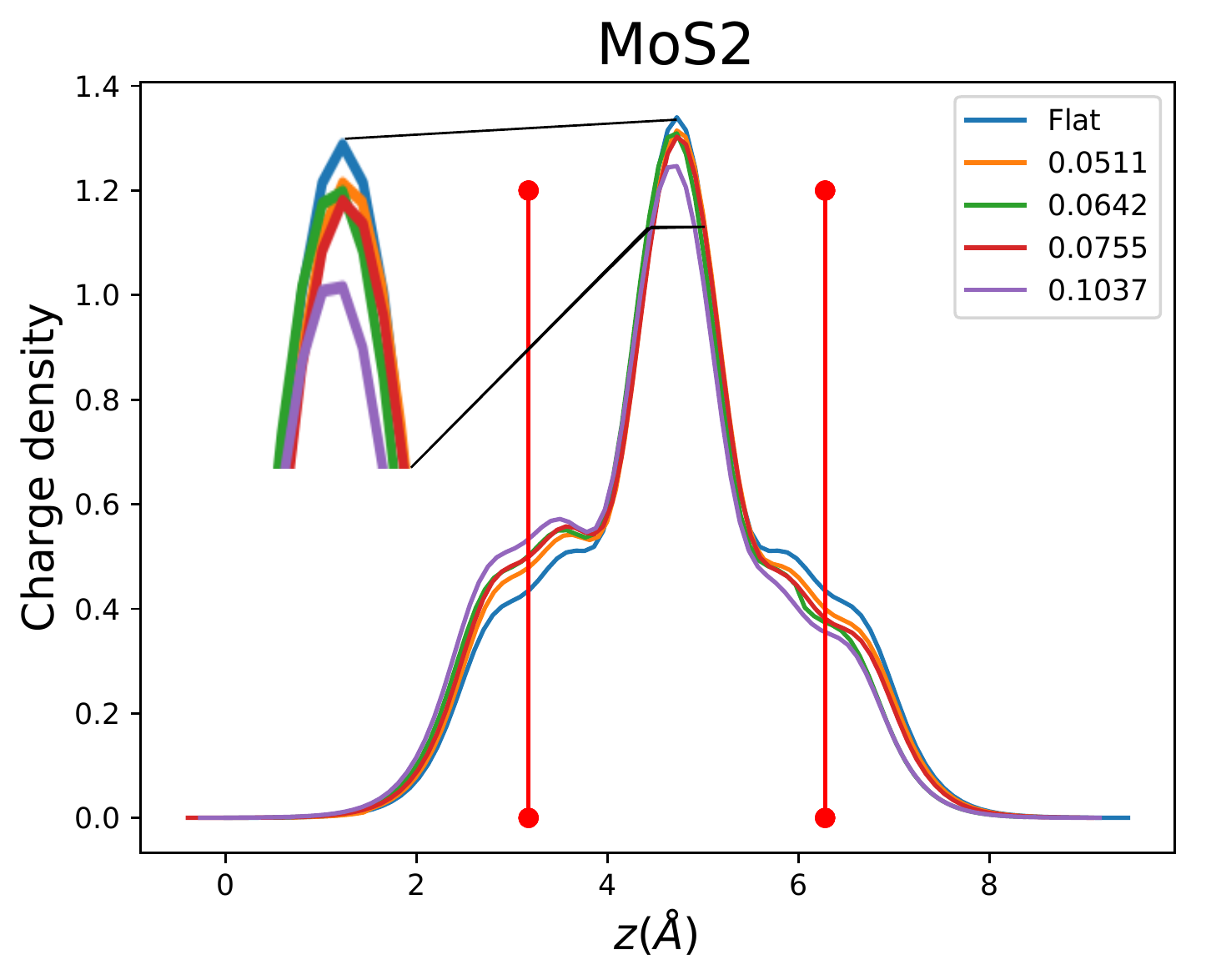}
 	\includegraphics[scale=0.35]{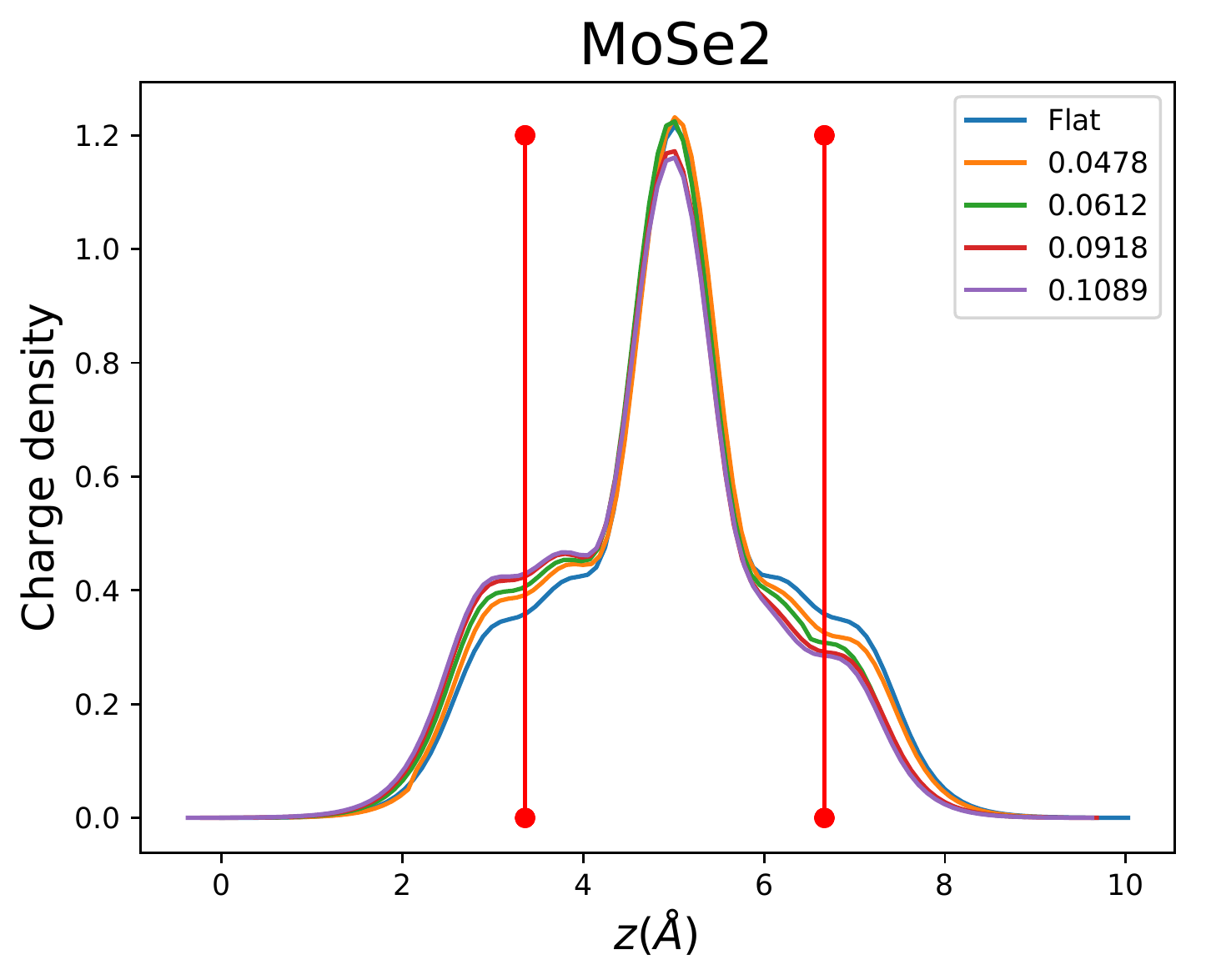}
 	\includegraphics[scale=0.35]{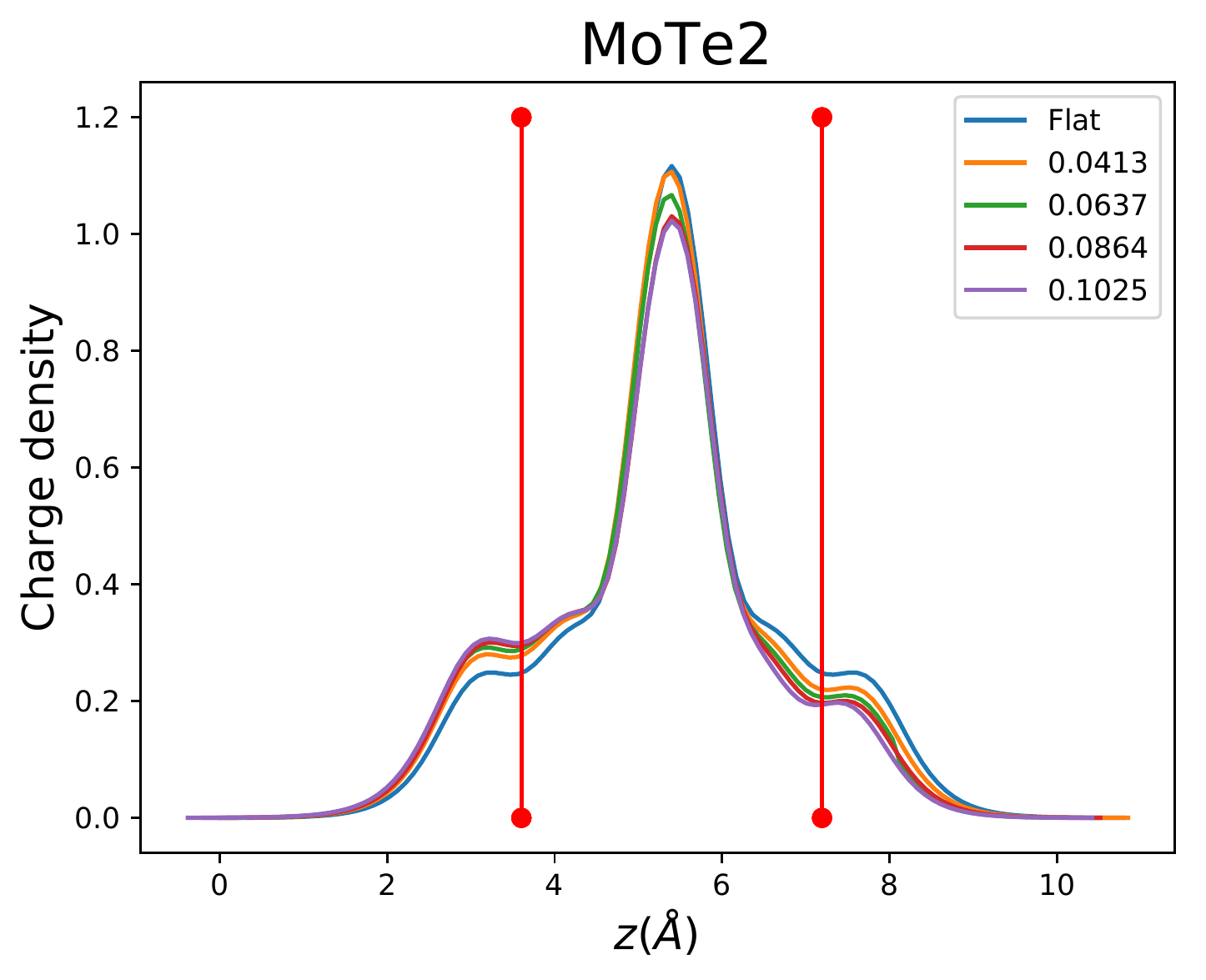}
 	\includegraphics[scale=0.35]{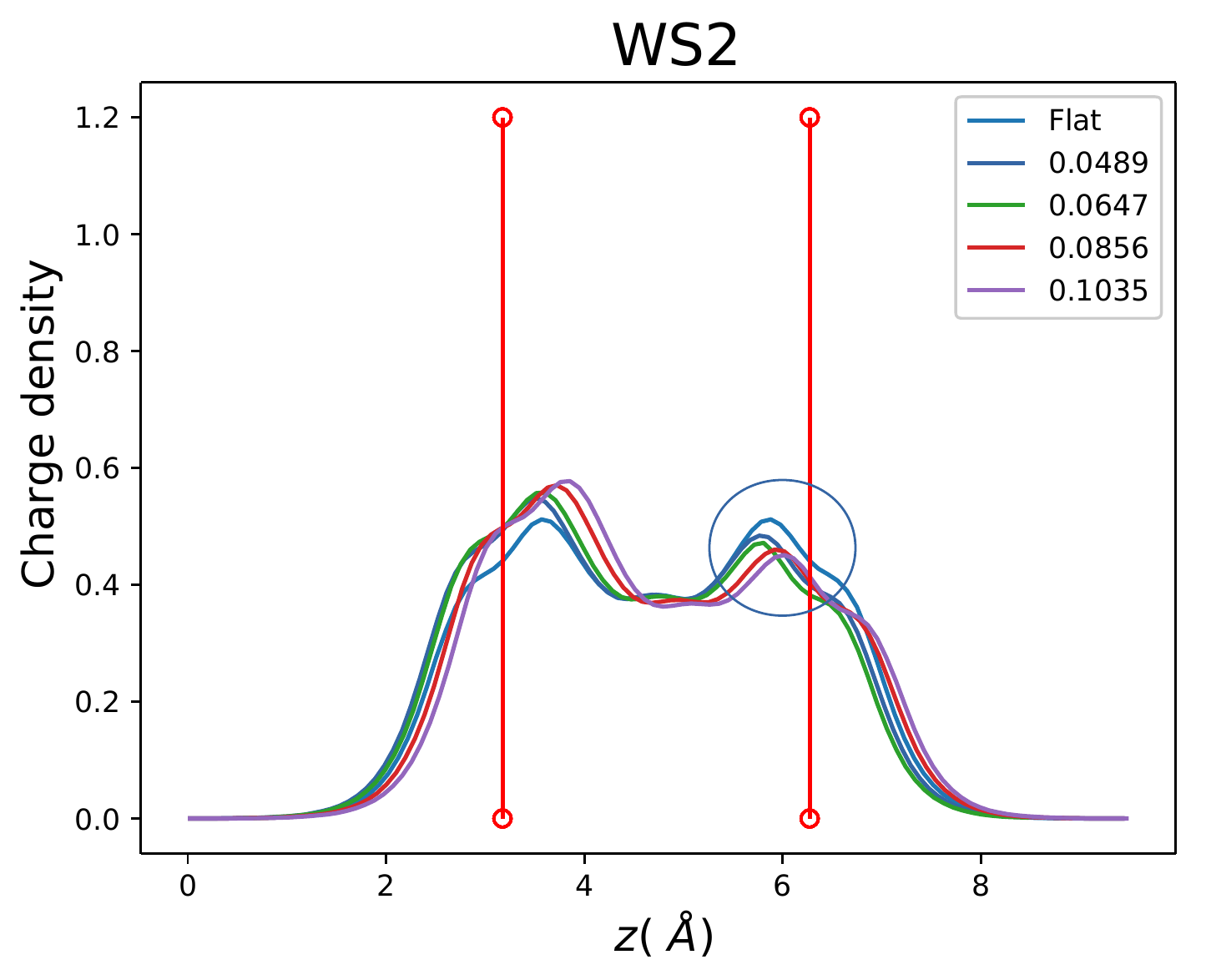}
 	\includegraphics[scale=0.35]{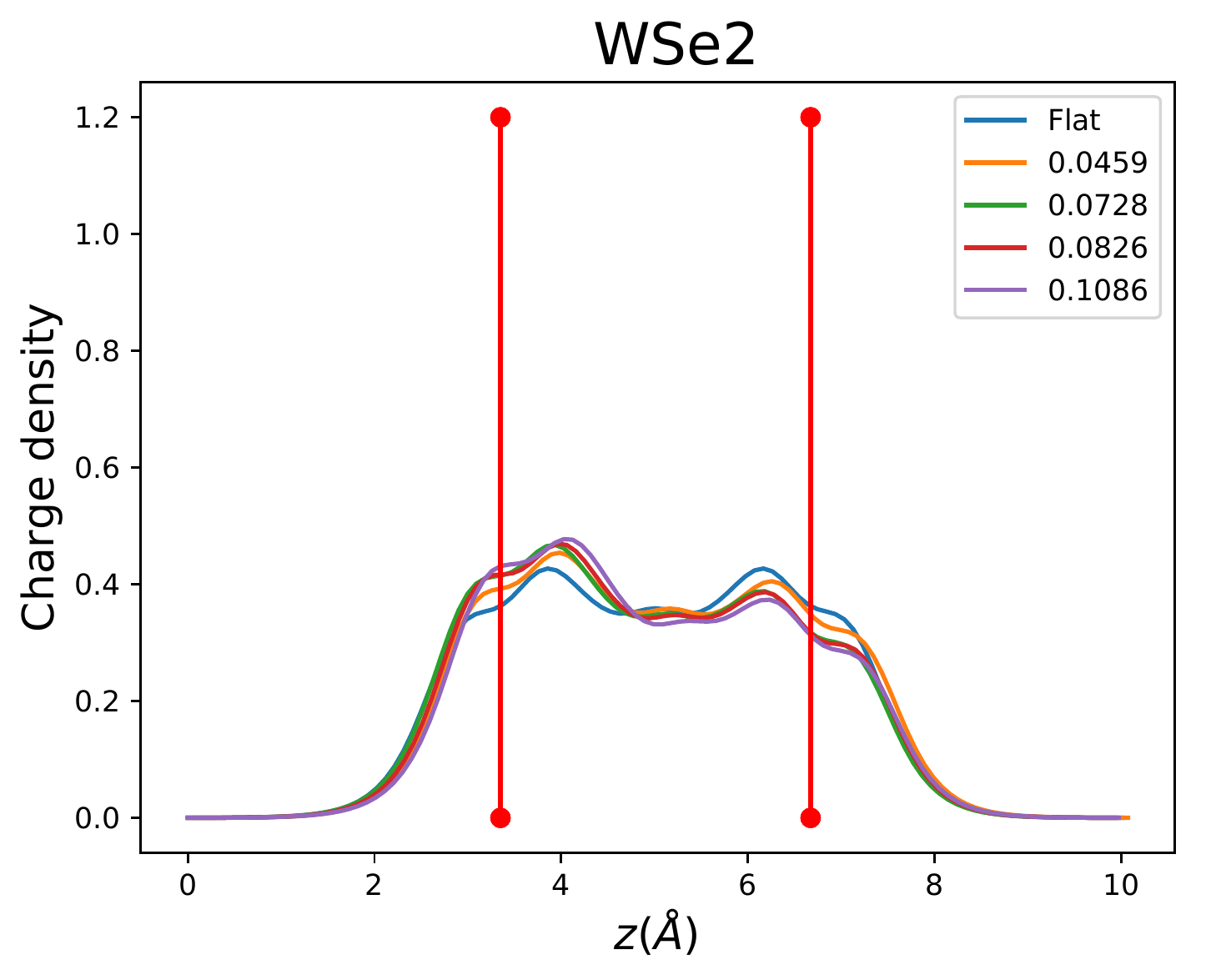}
 	\includegraphics[scale=0.35]{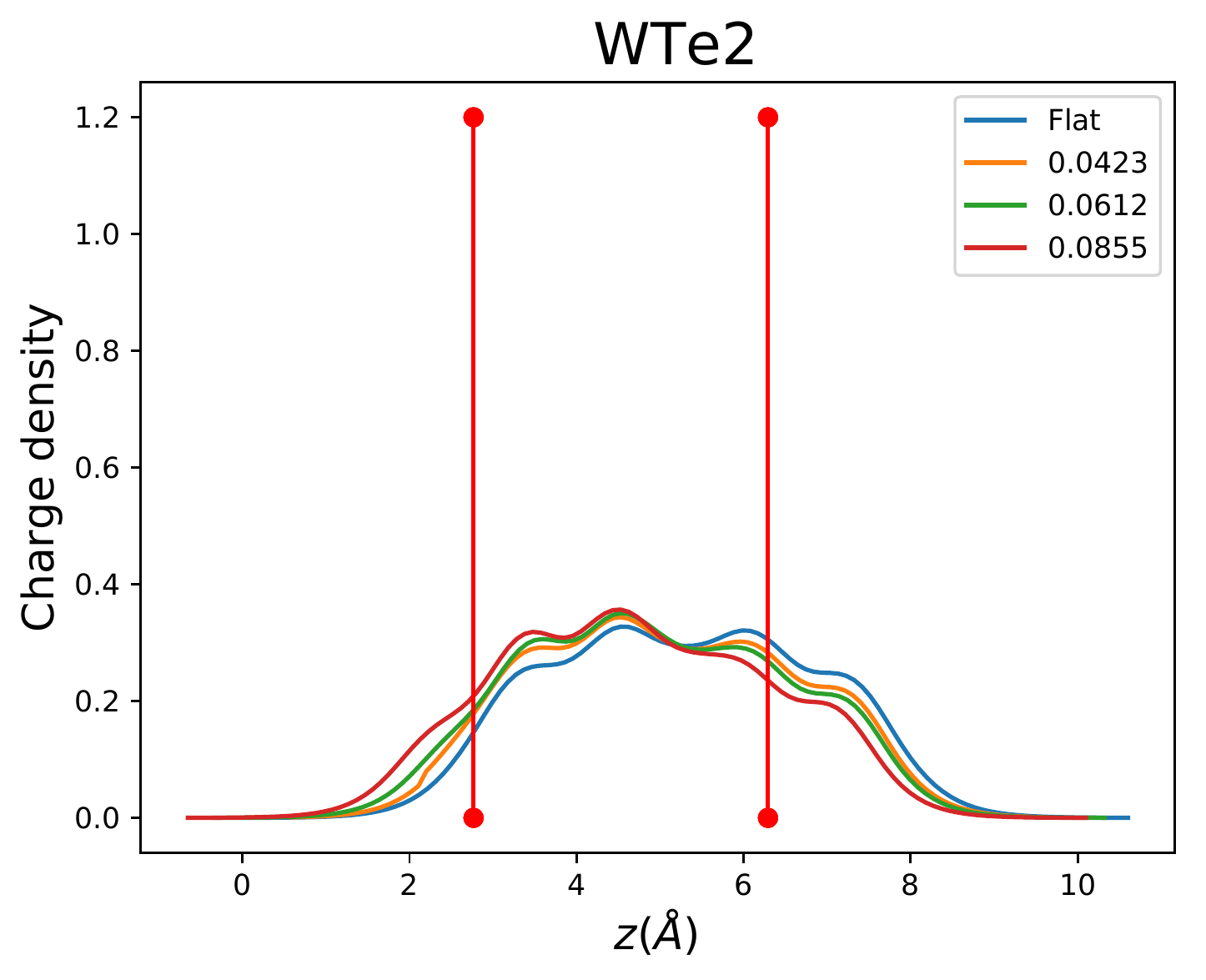}
 	\label{fig:log-charge}
 	
 \end{figure}
 \begin{figure}
 	\includegraphics[scale=0.35]{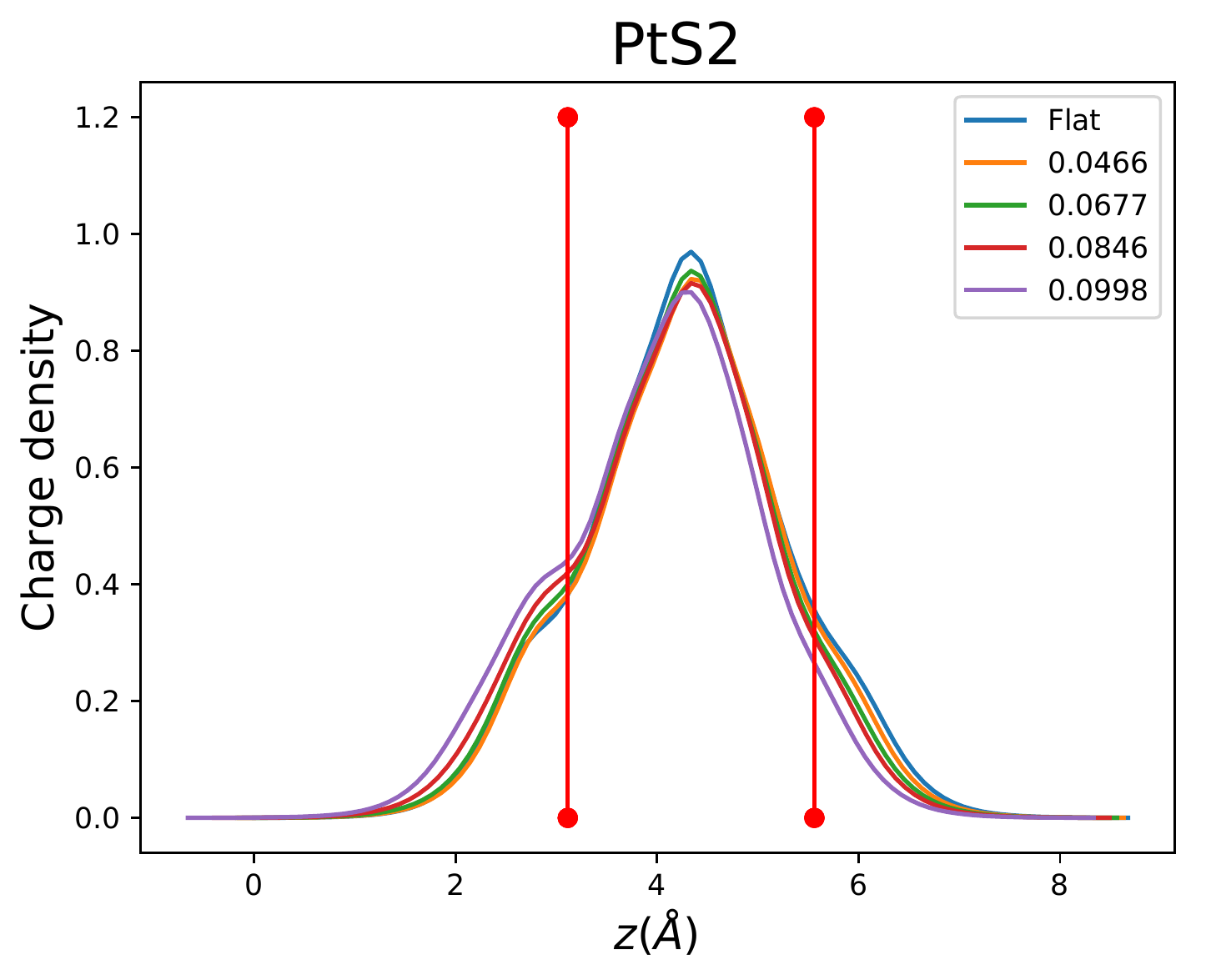}
 	\includegraphics[scale=0.35]{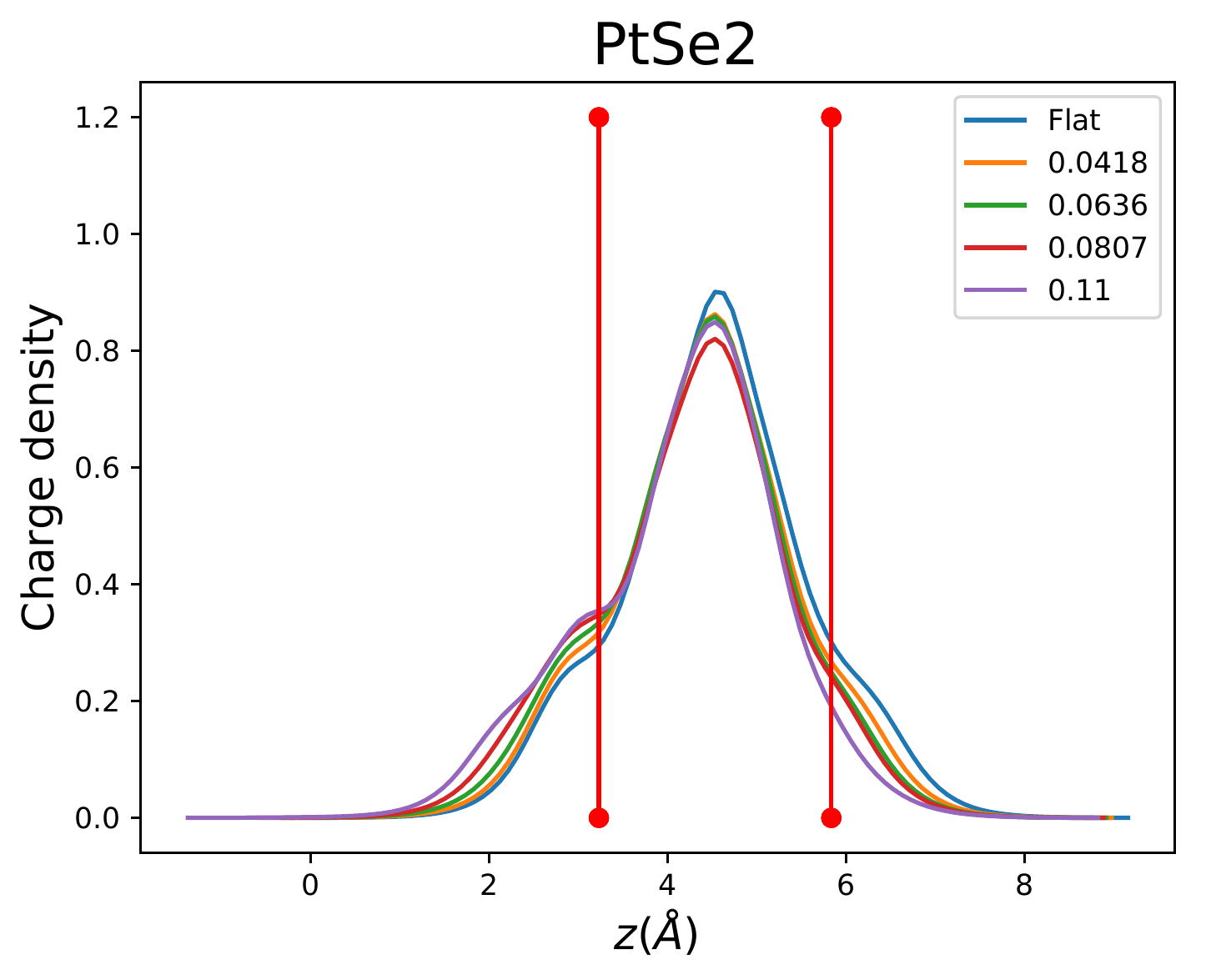}
 	\includegraphics[scale=0.35]{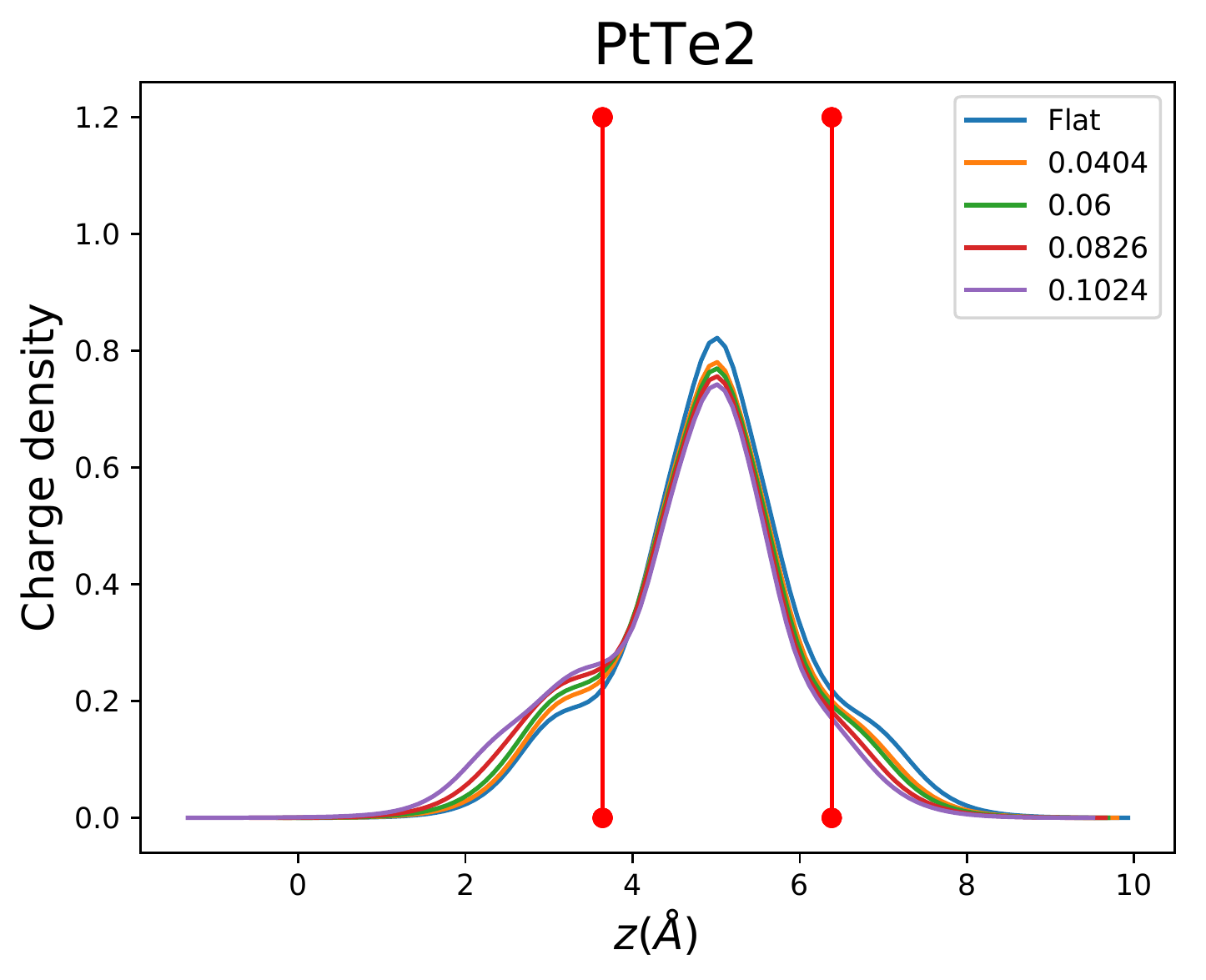}
 	\includegraphics[scale=0.35]{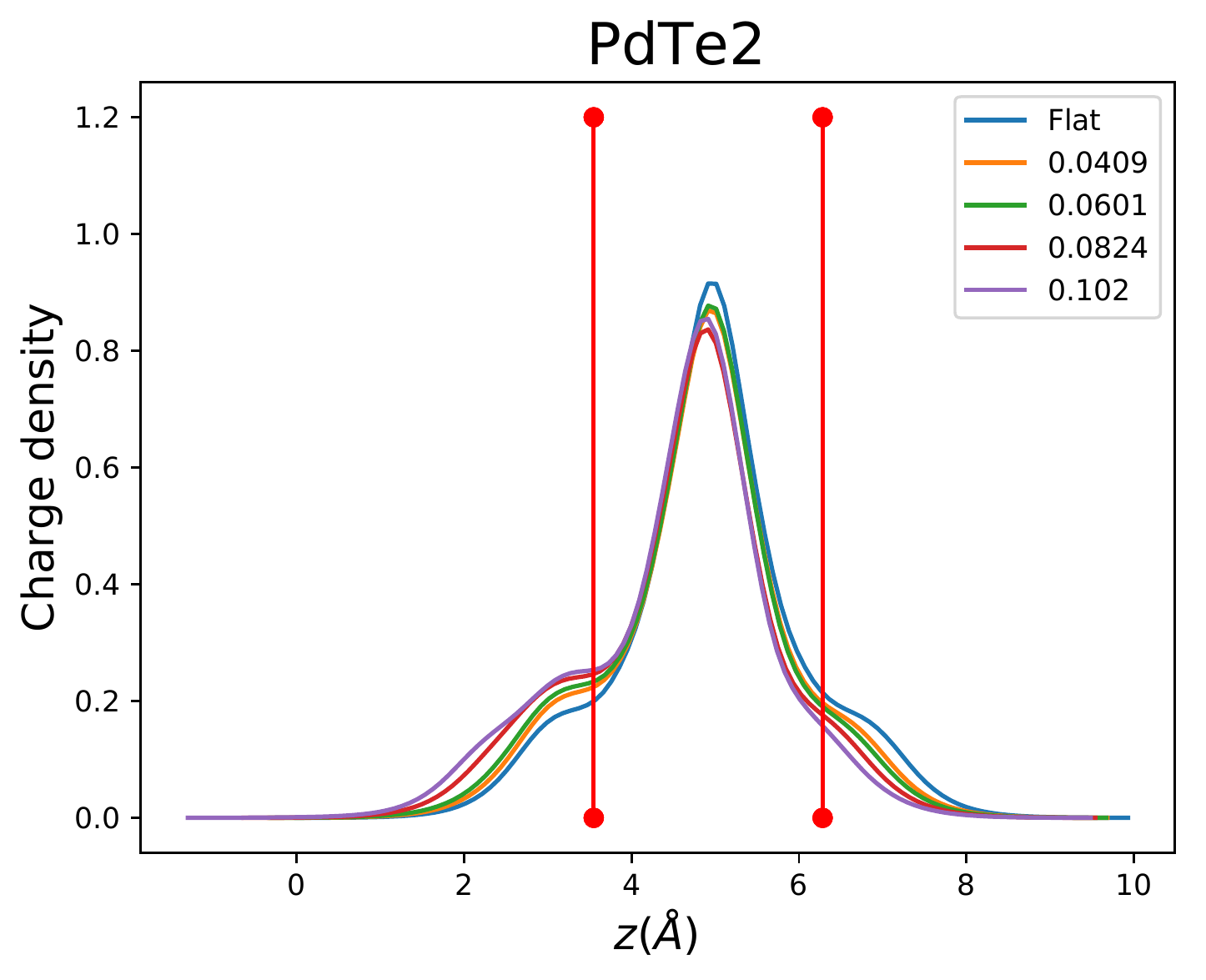}
 	\includegraphics[scale=0.35]{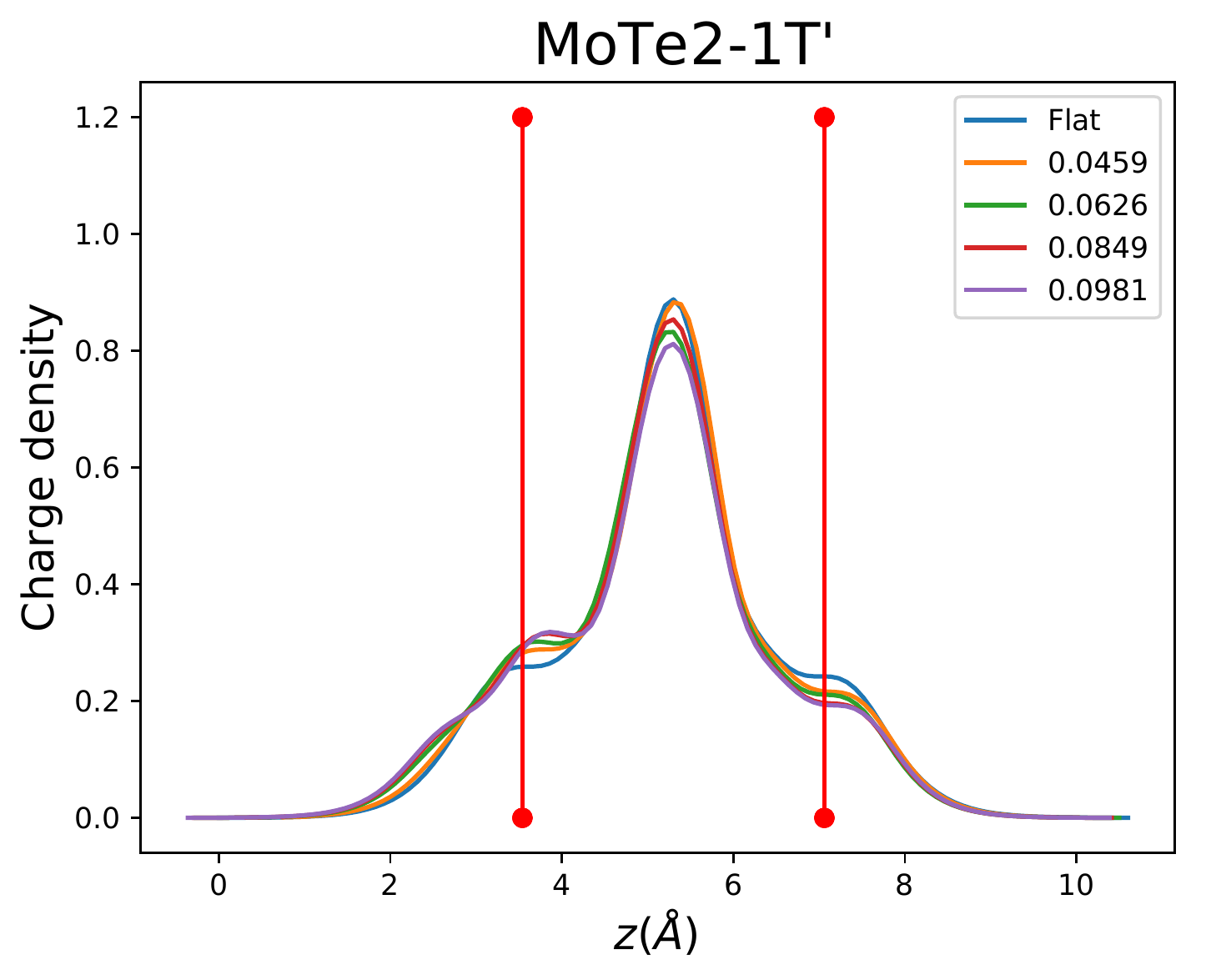}
 \end{figure}

 \begin{figure}[h!]
 	\renewcommand\thefigure{S6}
 	\includegraphics[height=1in, width=1.5in]{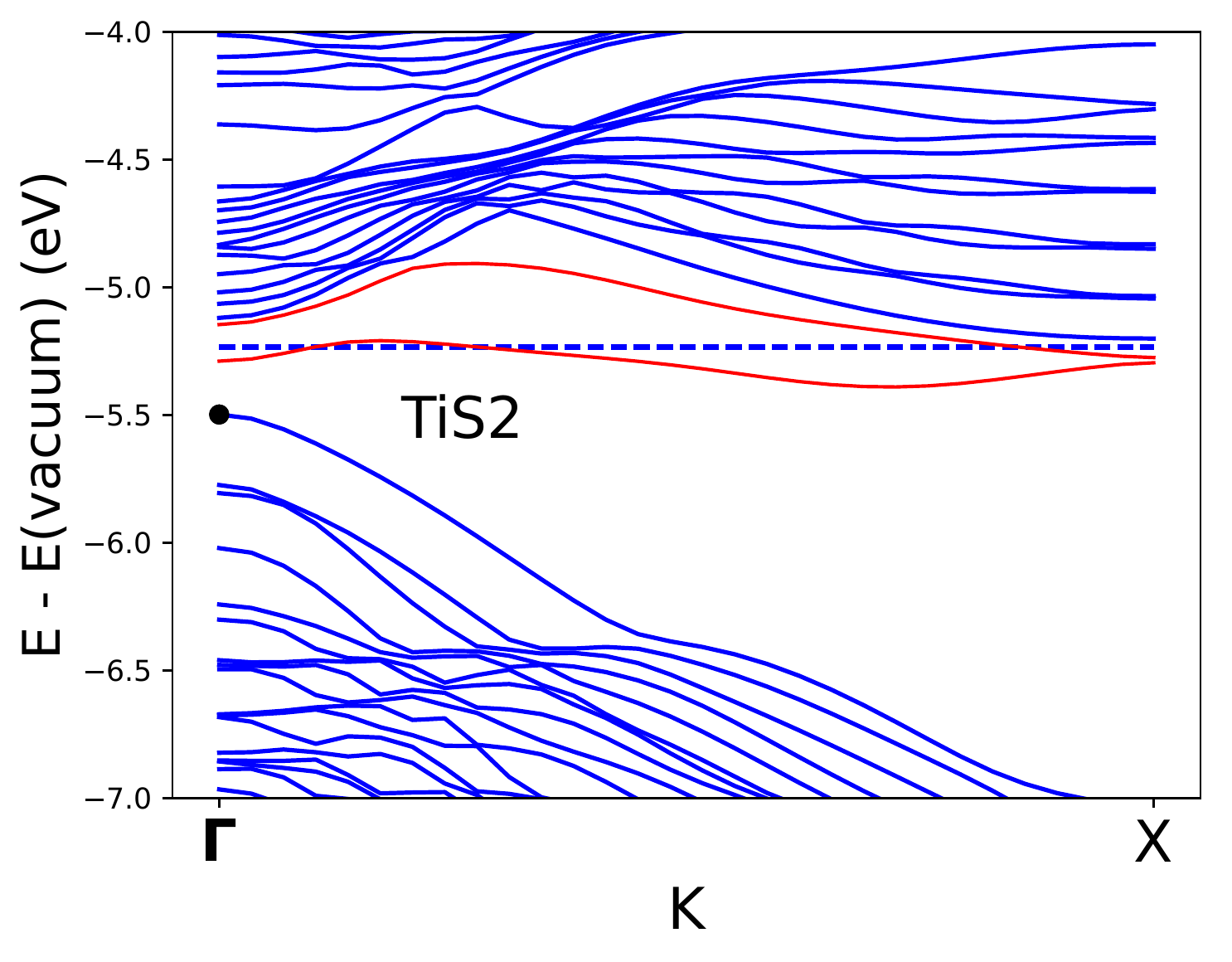}
 	\includegraphics[height=0.9in, width=1.5in]{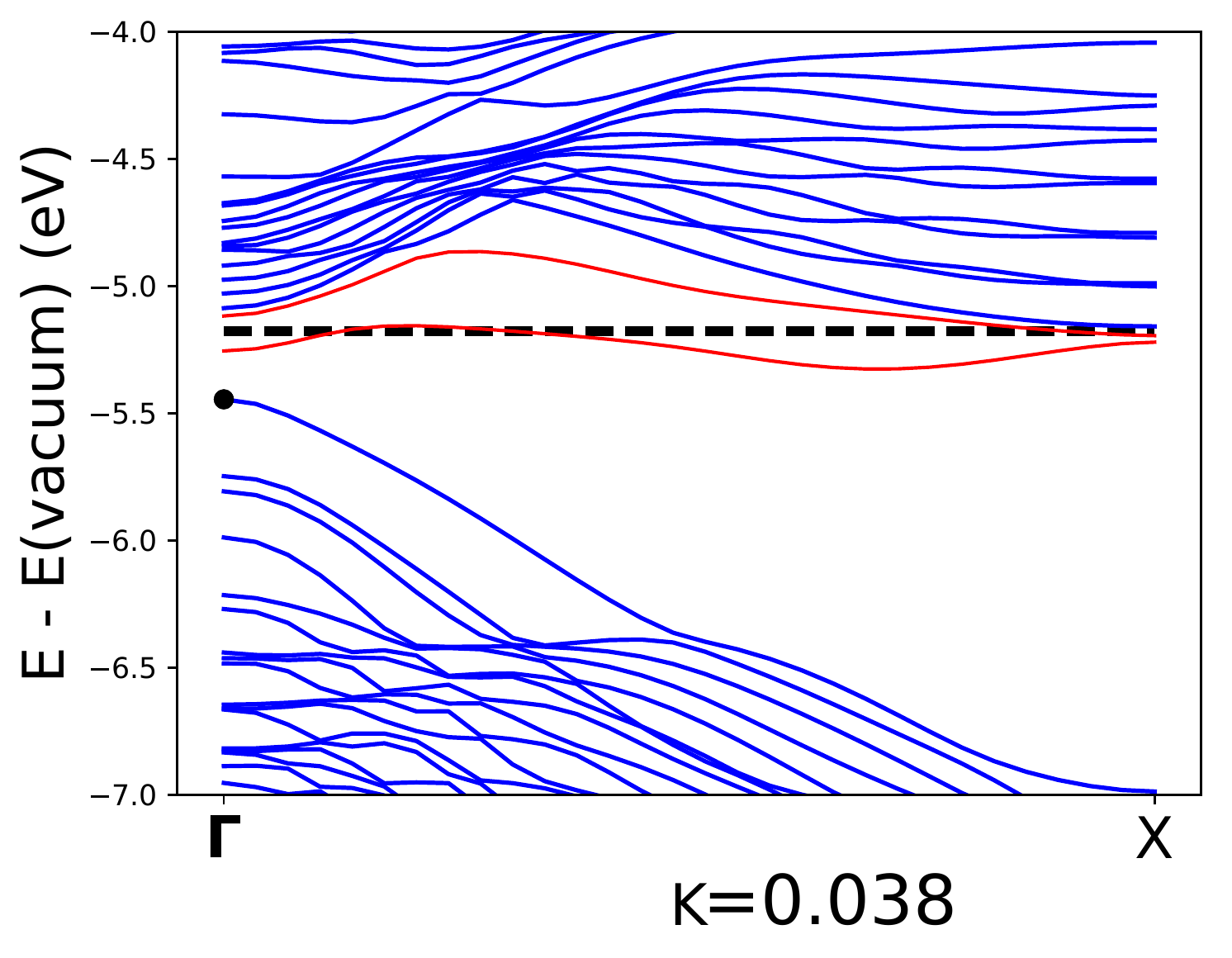}
 	\includegraphics[height=0.9in, width=1.5in]{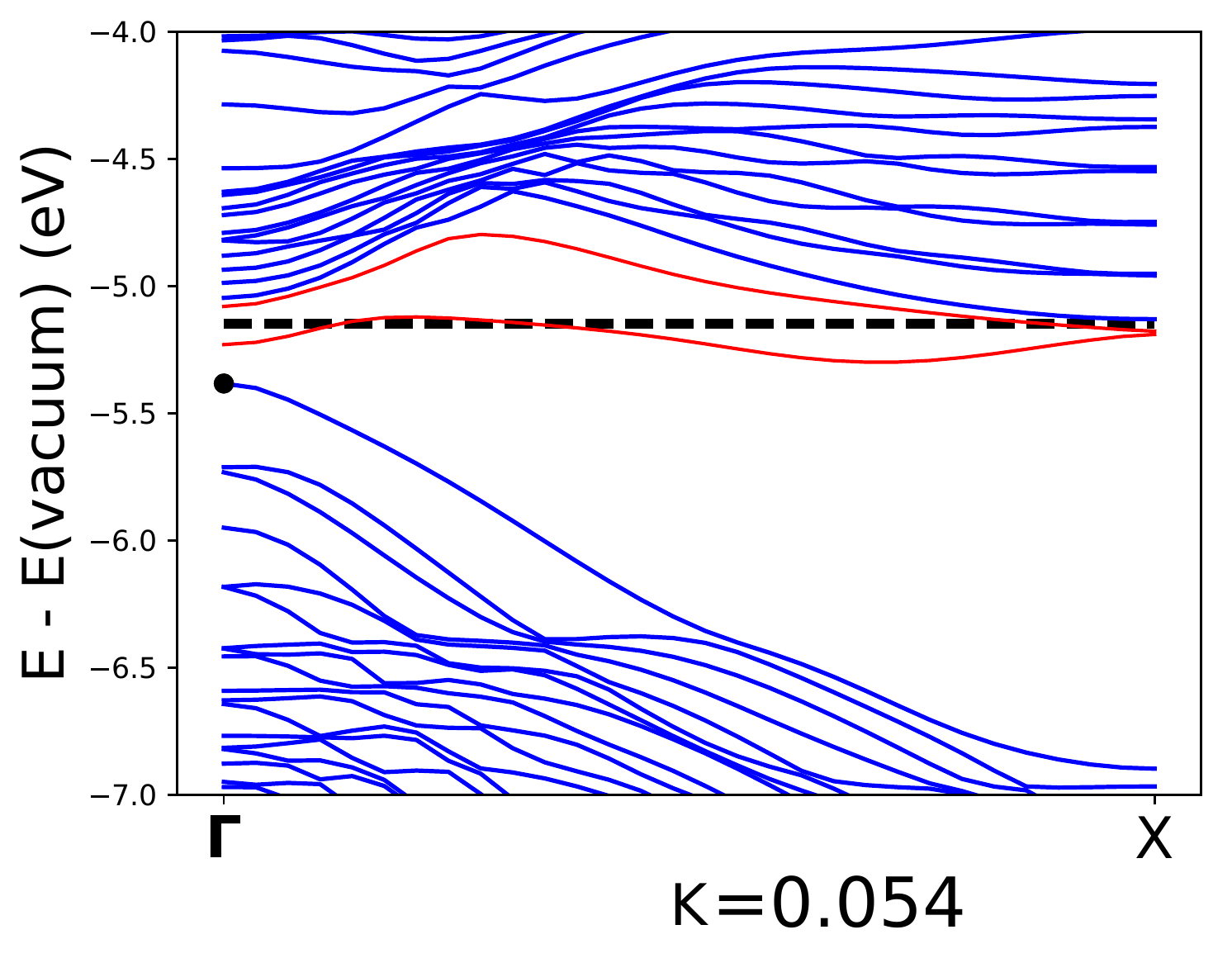}
 	\includegraphics[height=0.9in, width=1.5in]{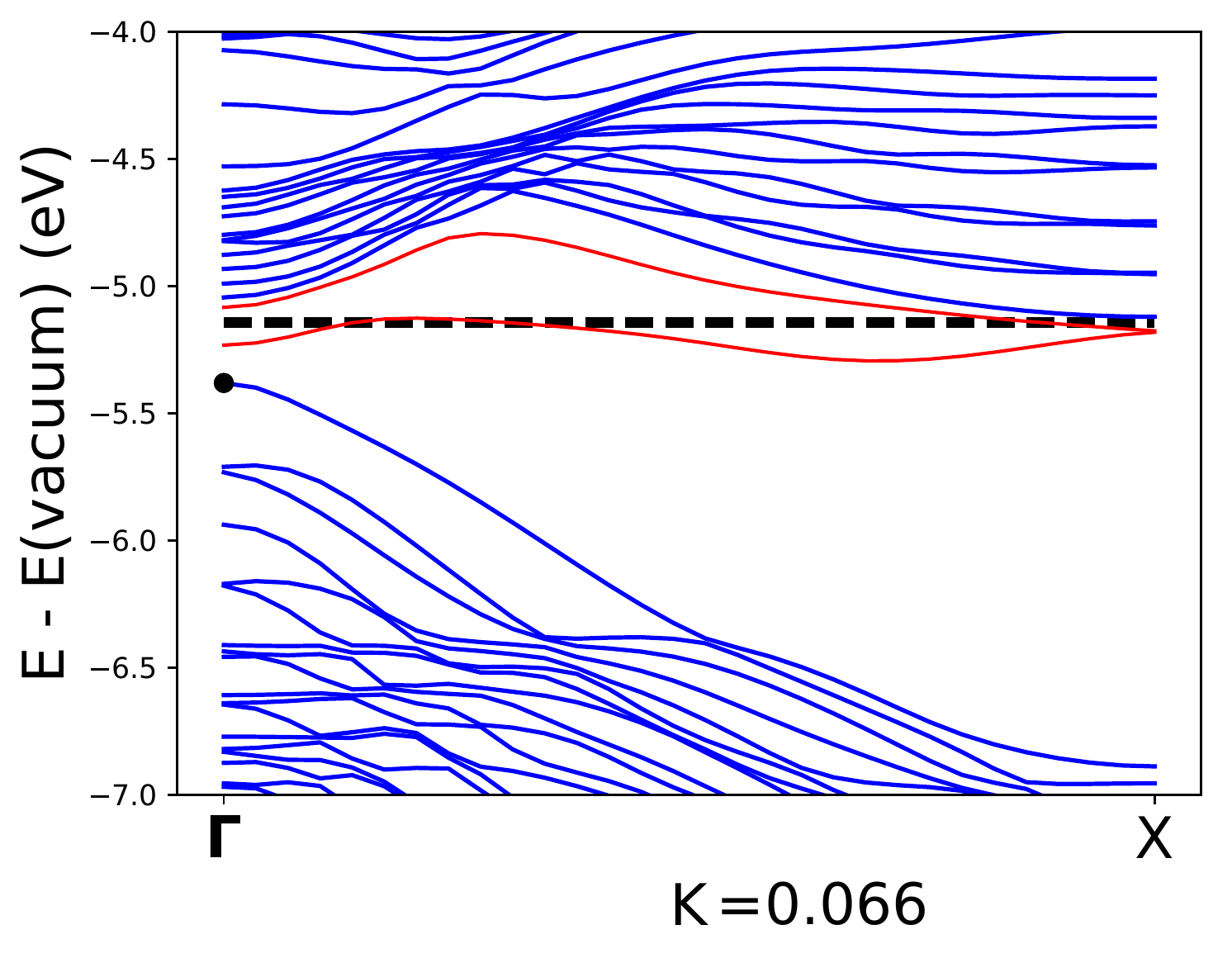}
 	\includegraphics[height=0.9in, width=1.5in]{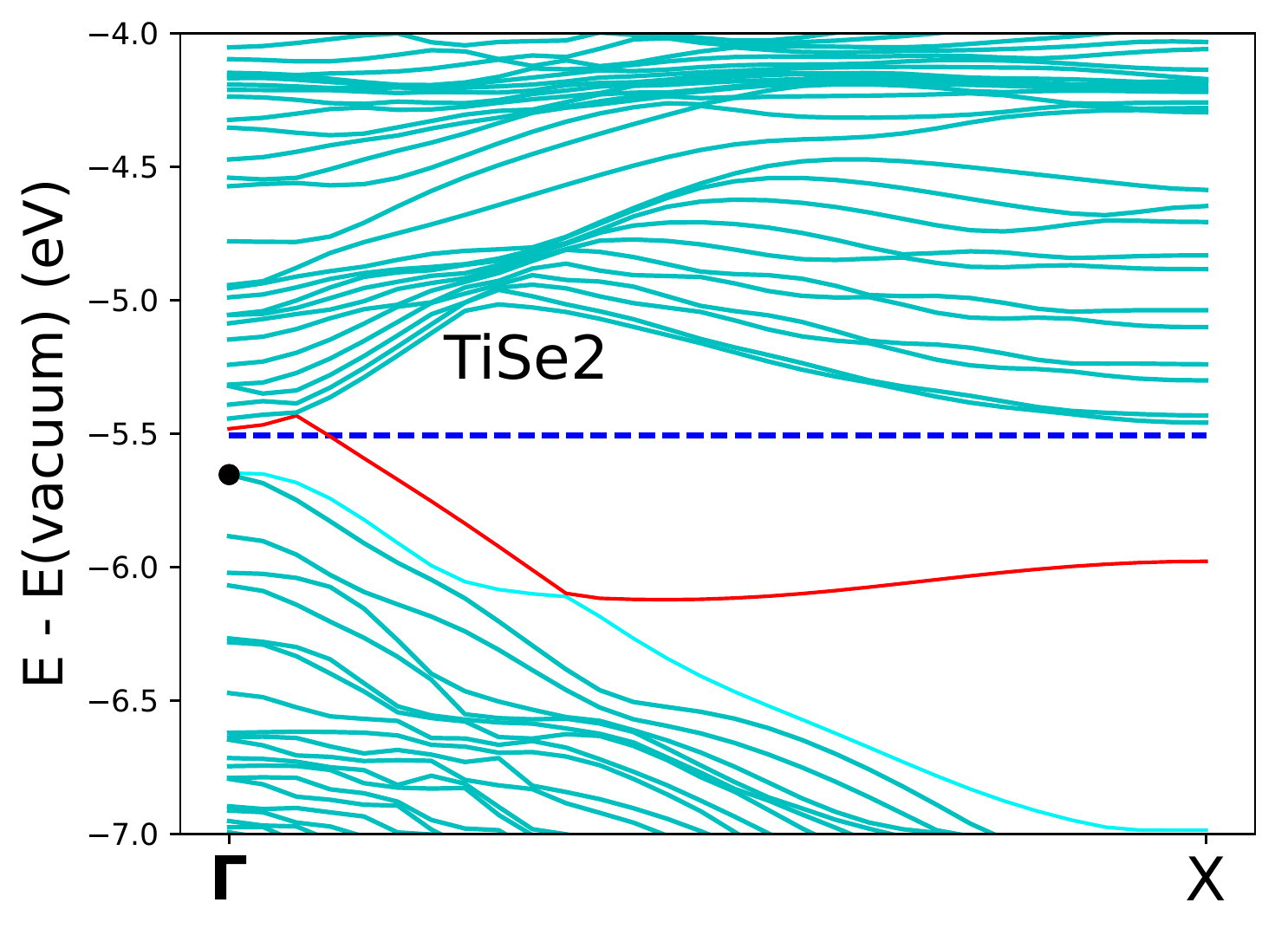}
 	\includegraphics[height=0.9in, width=1.5in]{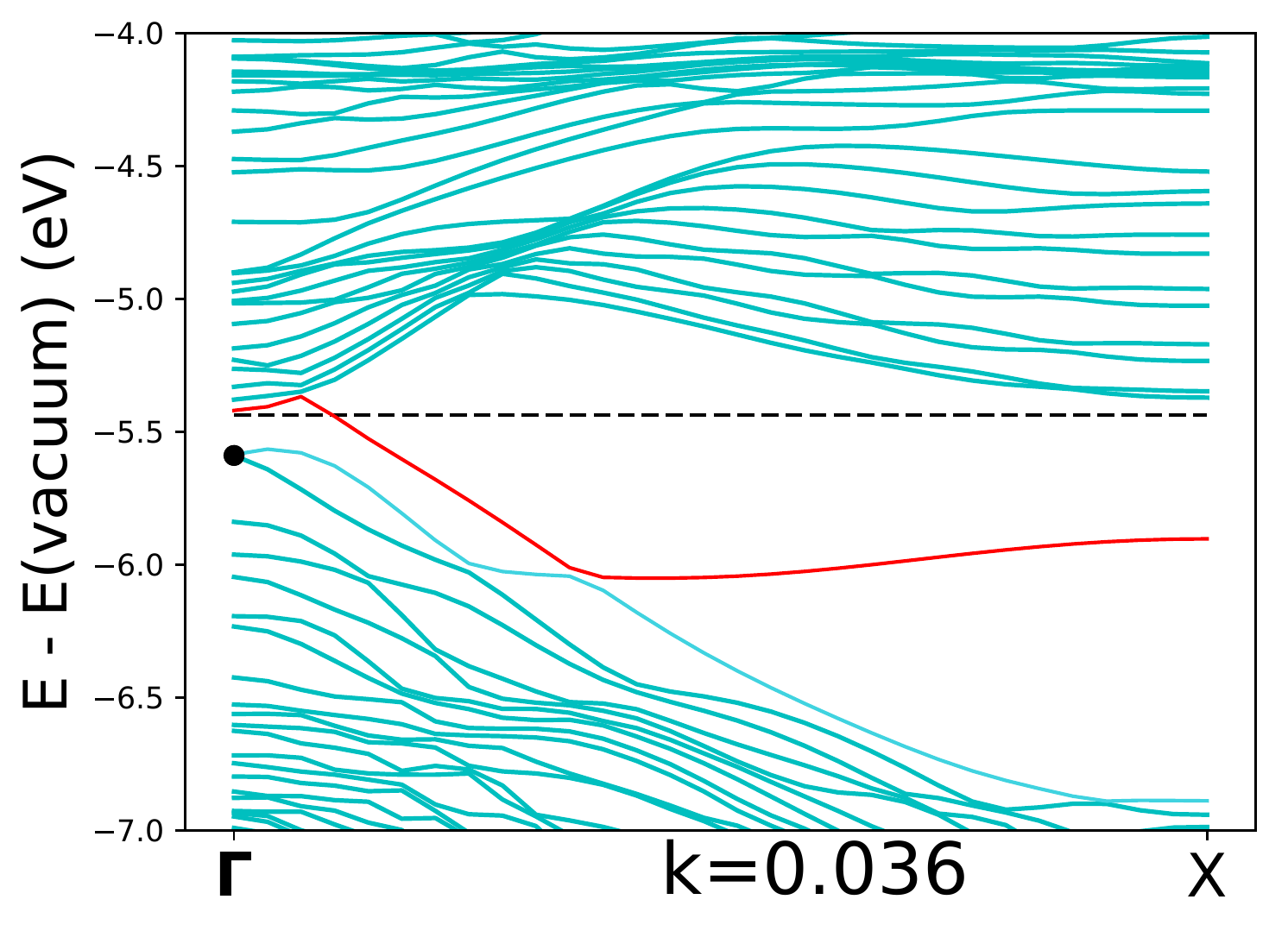}
 	\includegraphics[height=0.9in, width=1.5in]{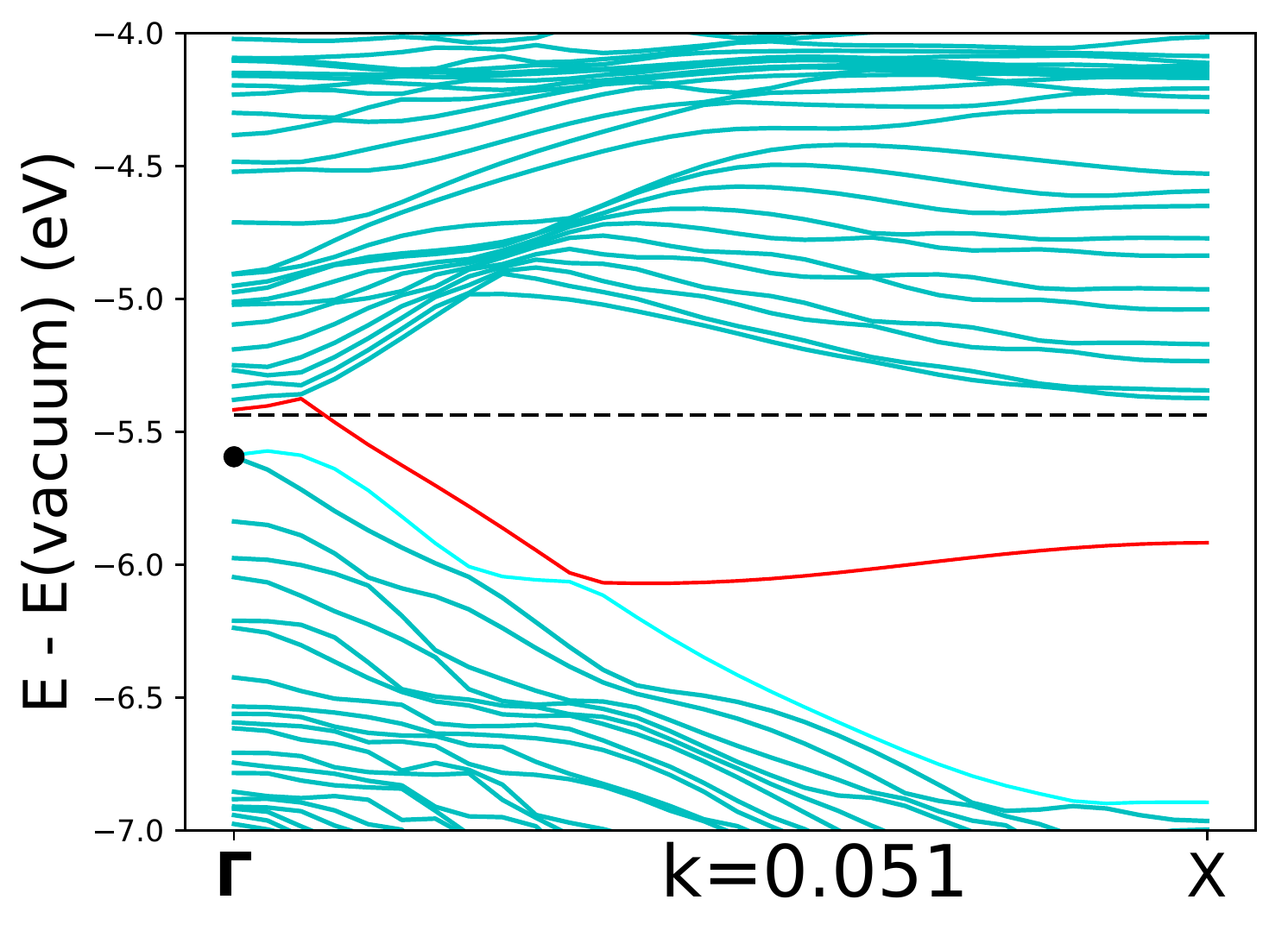}
 	\includegraphics[height=0.9in, width=1.5in]{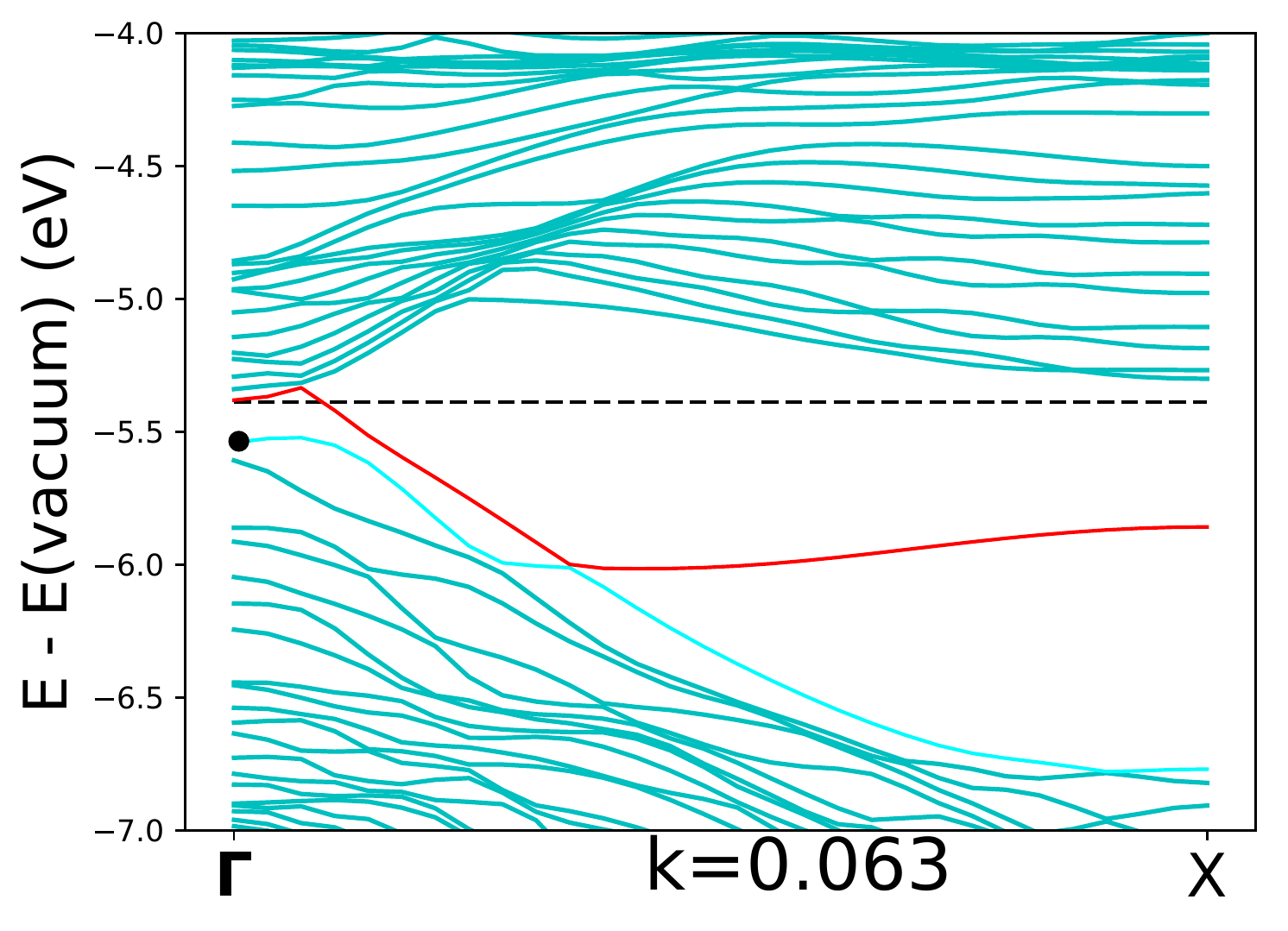}
 	\includegraphics[height=0.9in, width=1.5in]{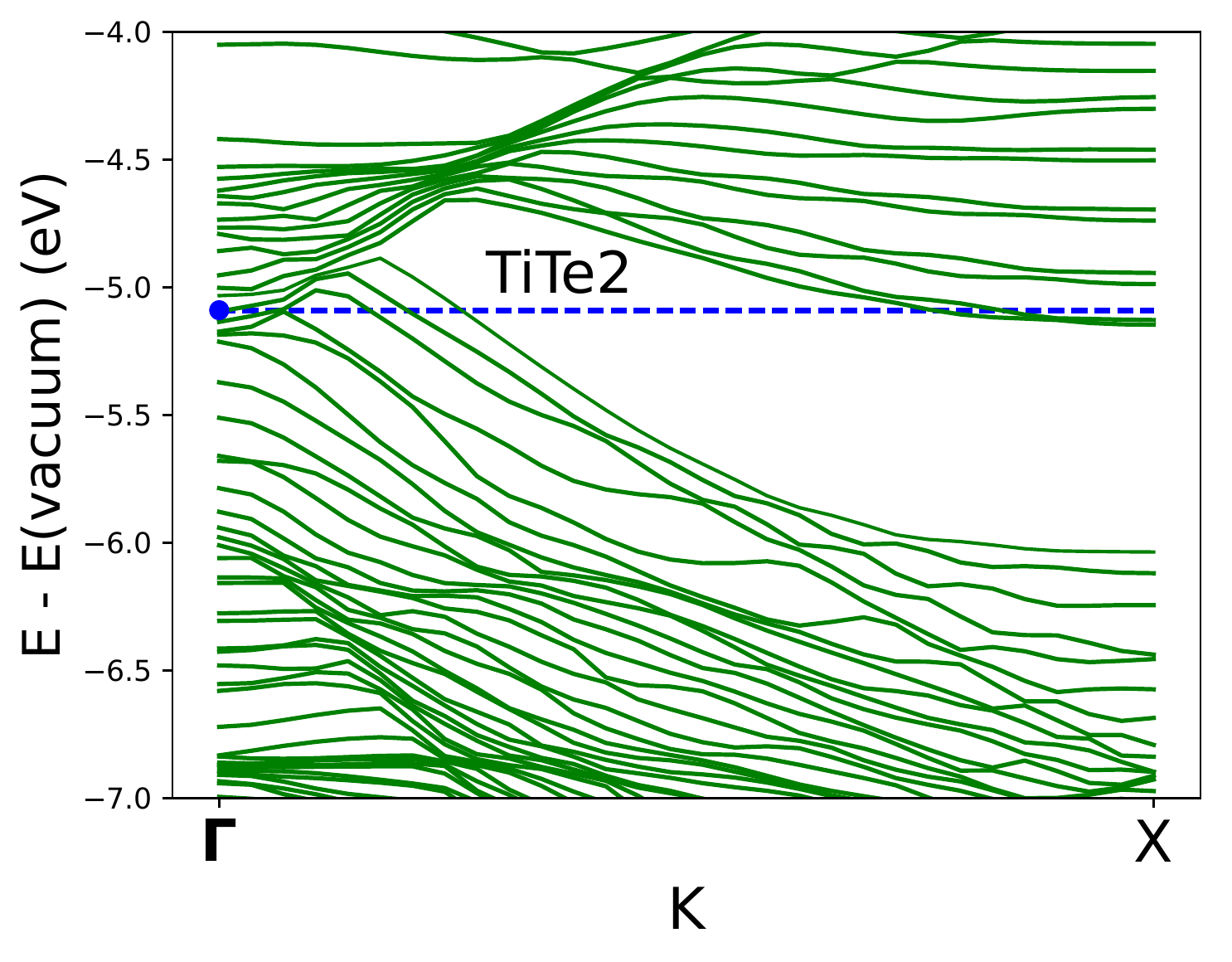}
 	\includegraphics[height=0.9in, width=1.5in]{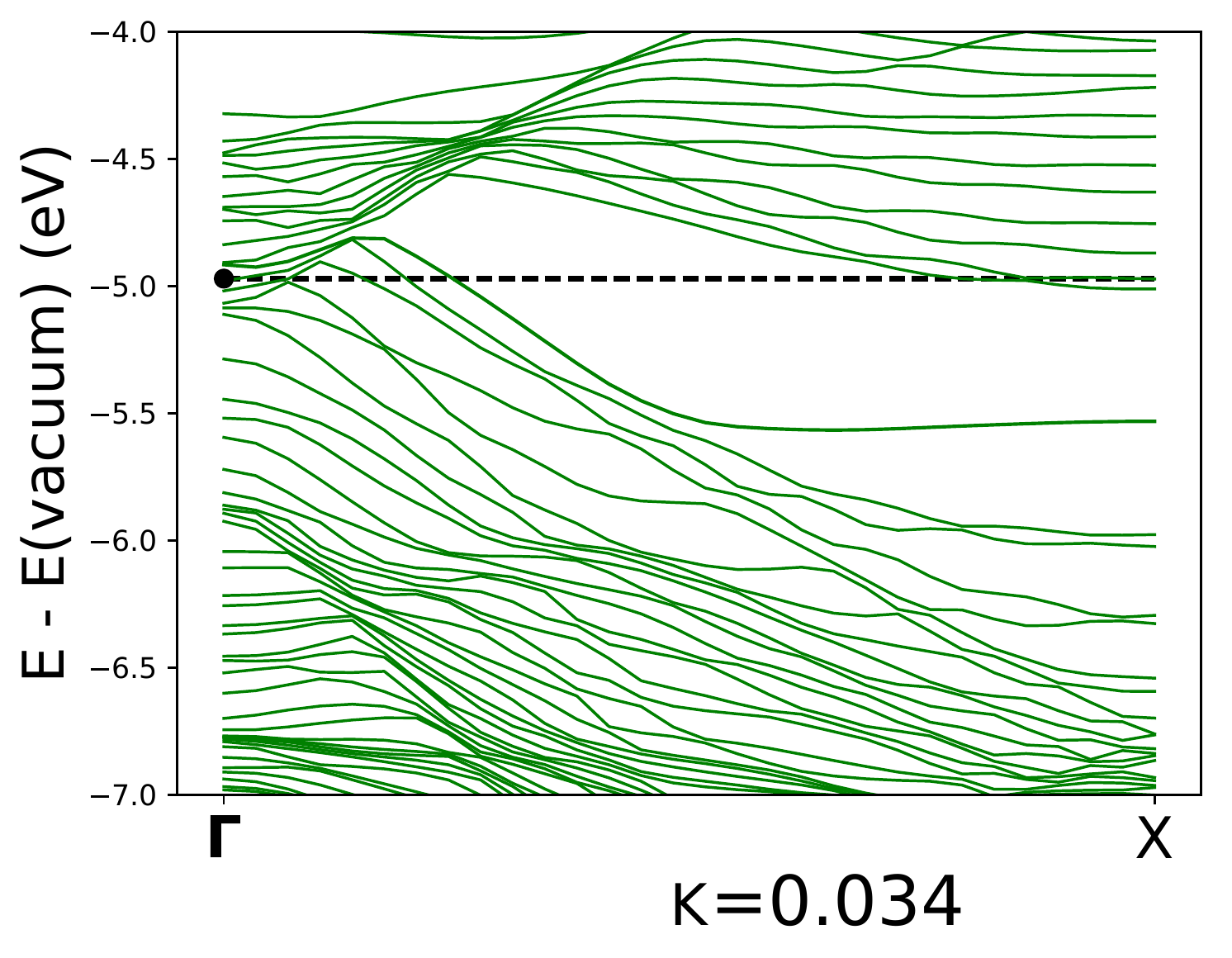}
 	\includegraphics[height=0.9in, width=1.5in]{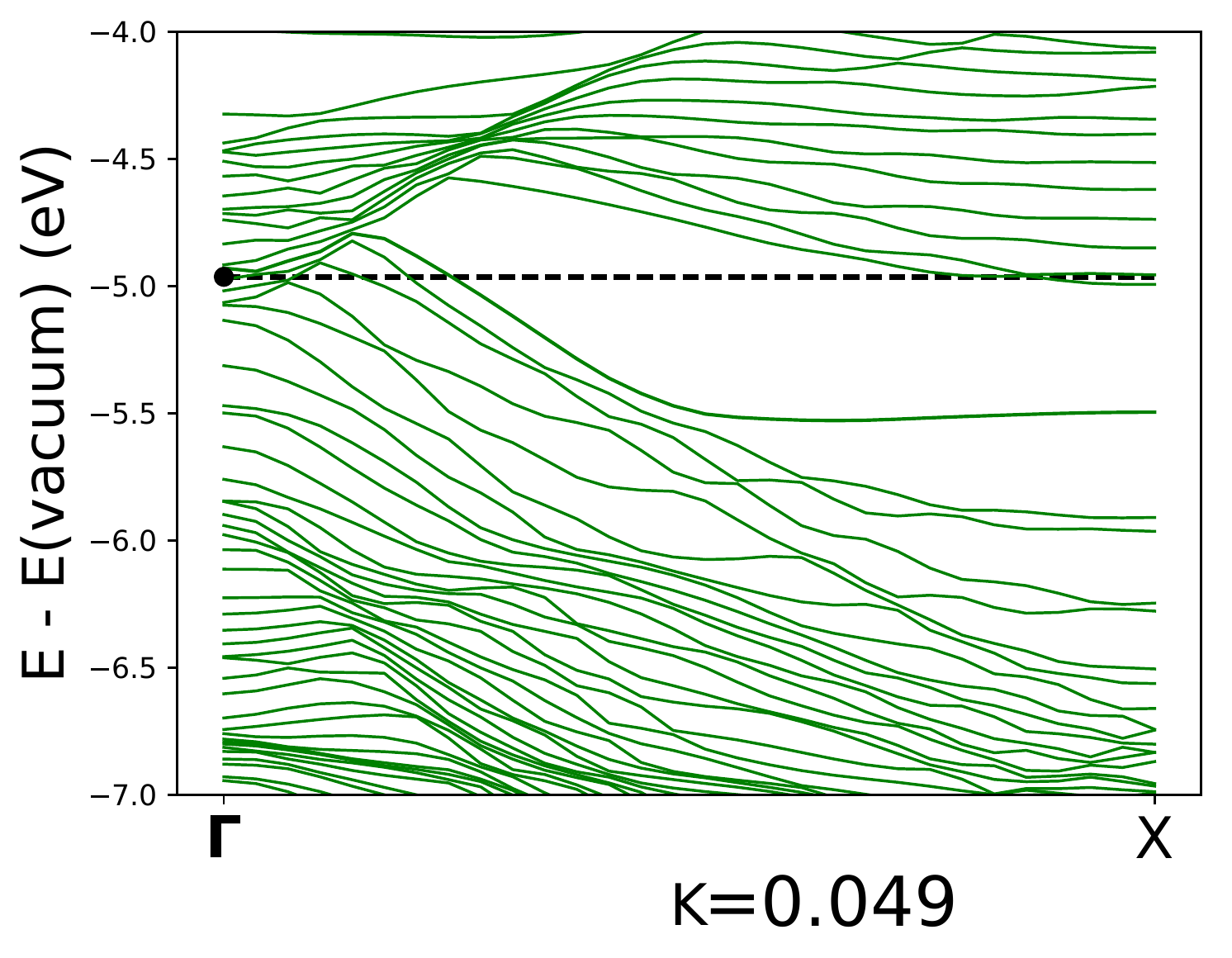}
 	\includegraphics[height=0.9in, width=1.5in]{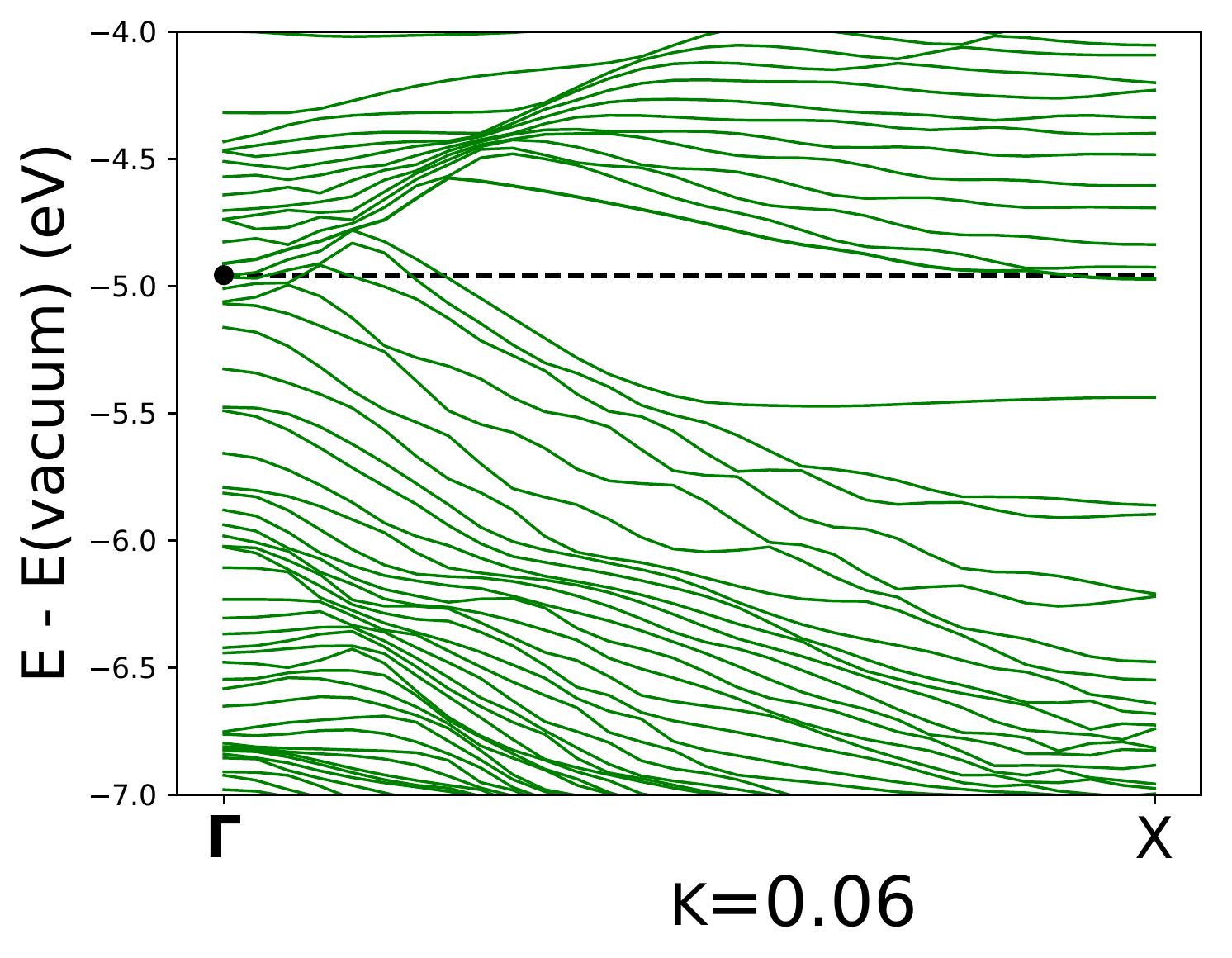}
 	\includegraphics[height=0.9in, width=1.5in]{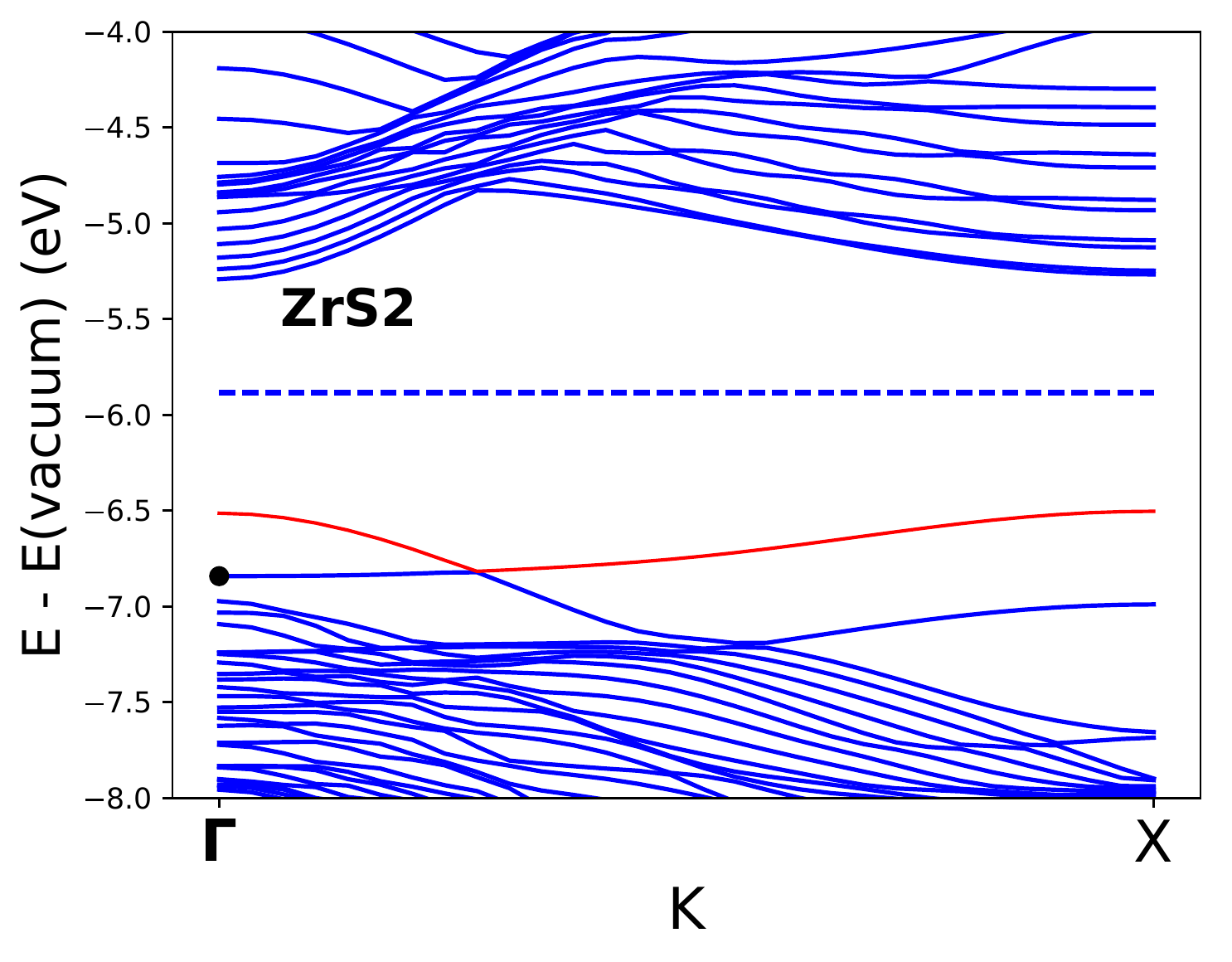}
 	\includegraphics[height=0.9in, width=1.5in]{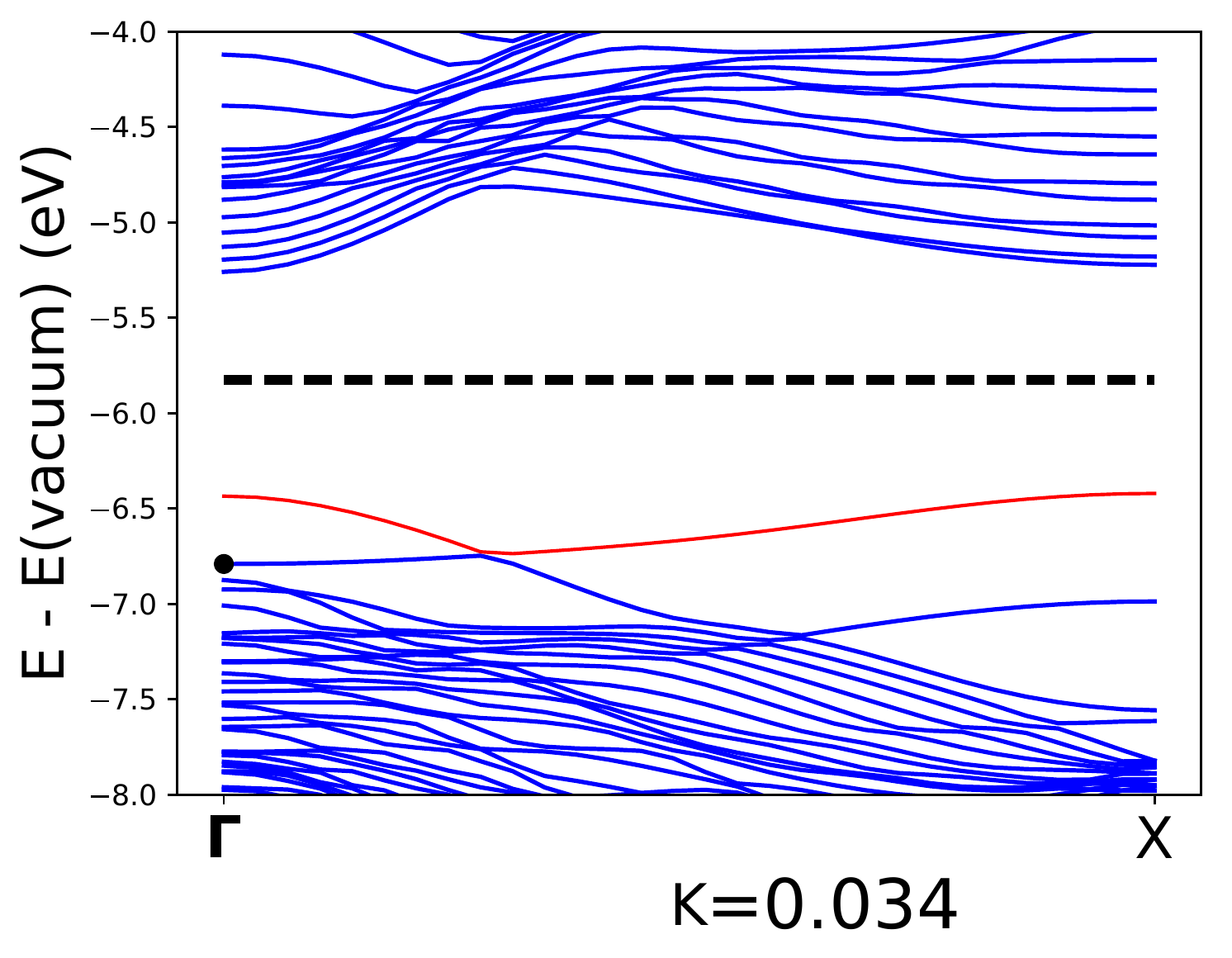}
 	\includegraphics[height=0.9in, width=1.5in]{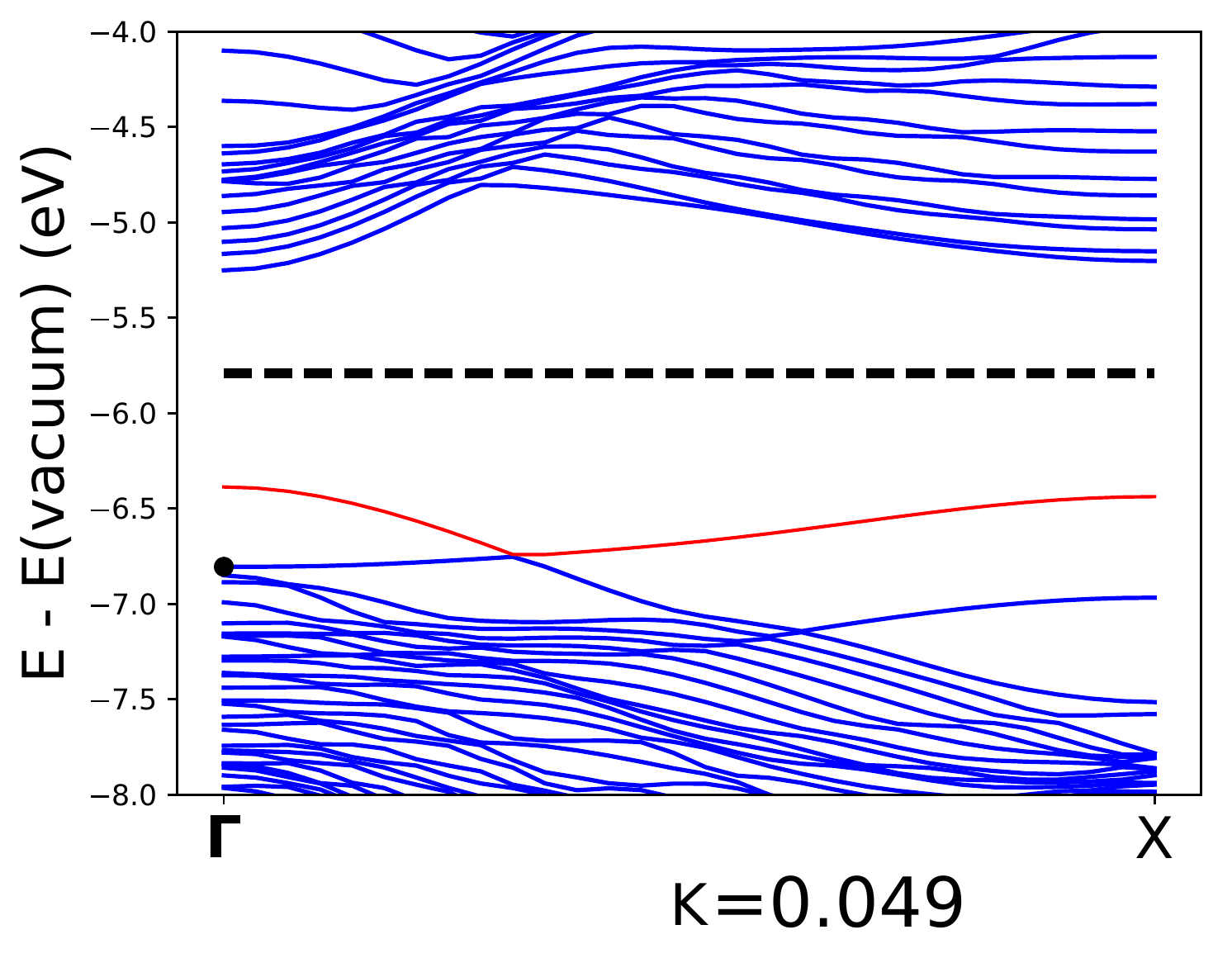}
 	\includegraphics[height=0.9in, width=1.5in]{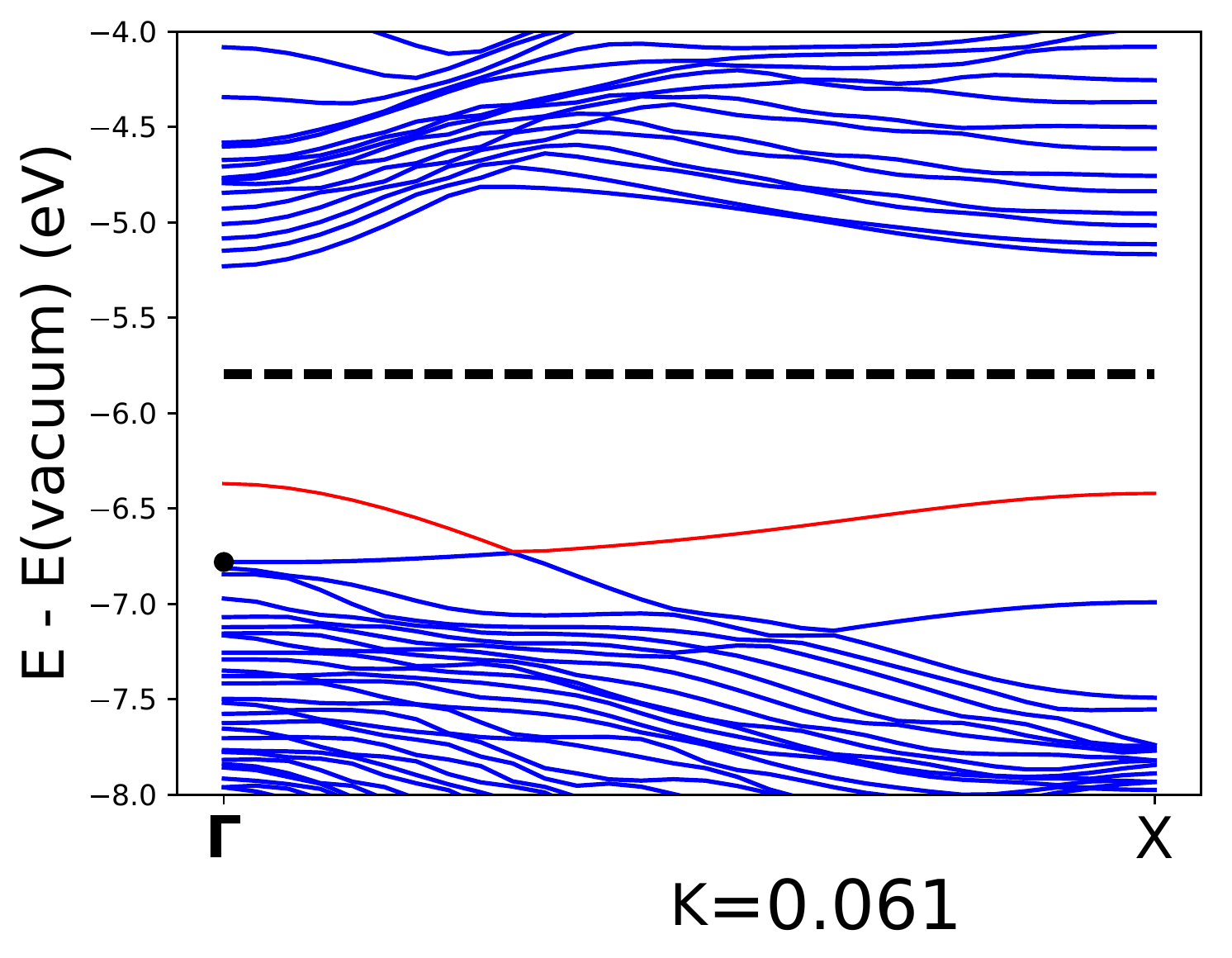}
 	\includegraphics[height=0.9in, width=1.5in]{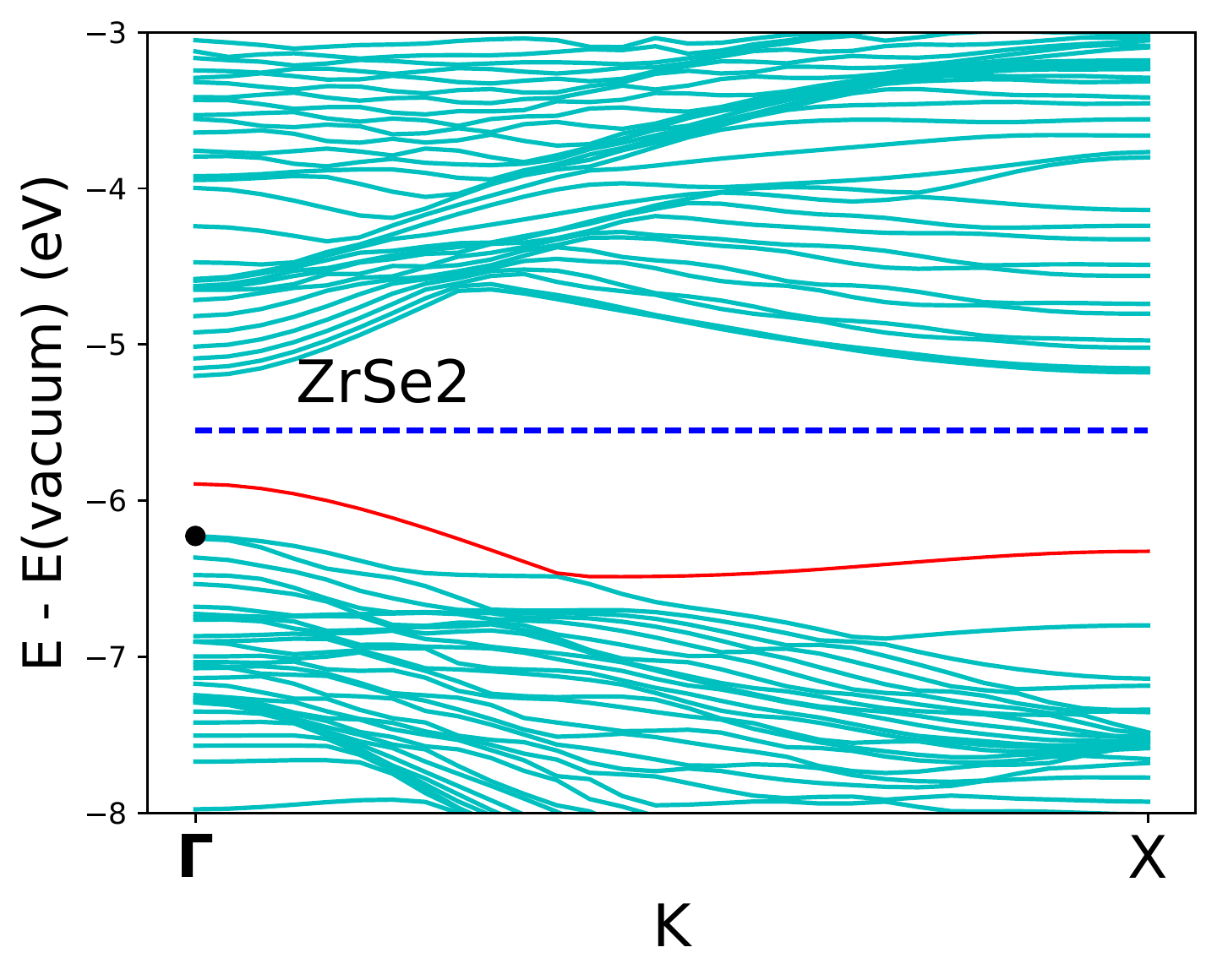}
 	\includegraphics[height=0.9in, width=1.5in]{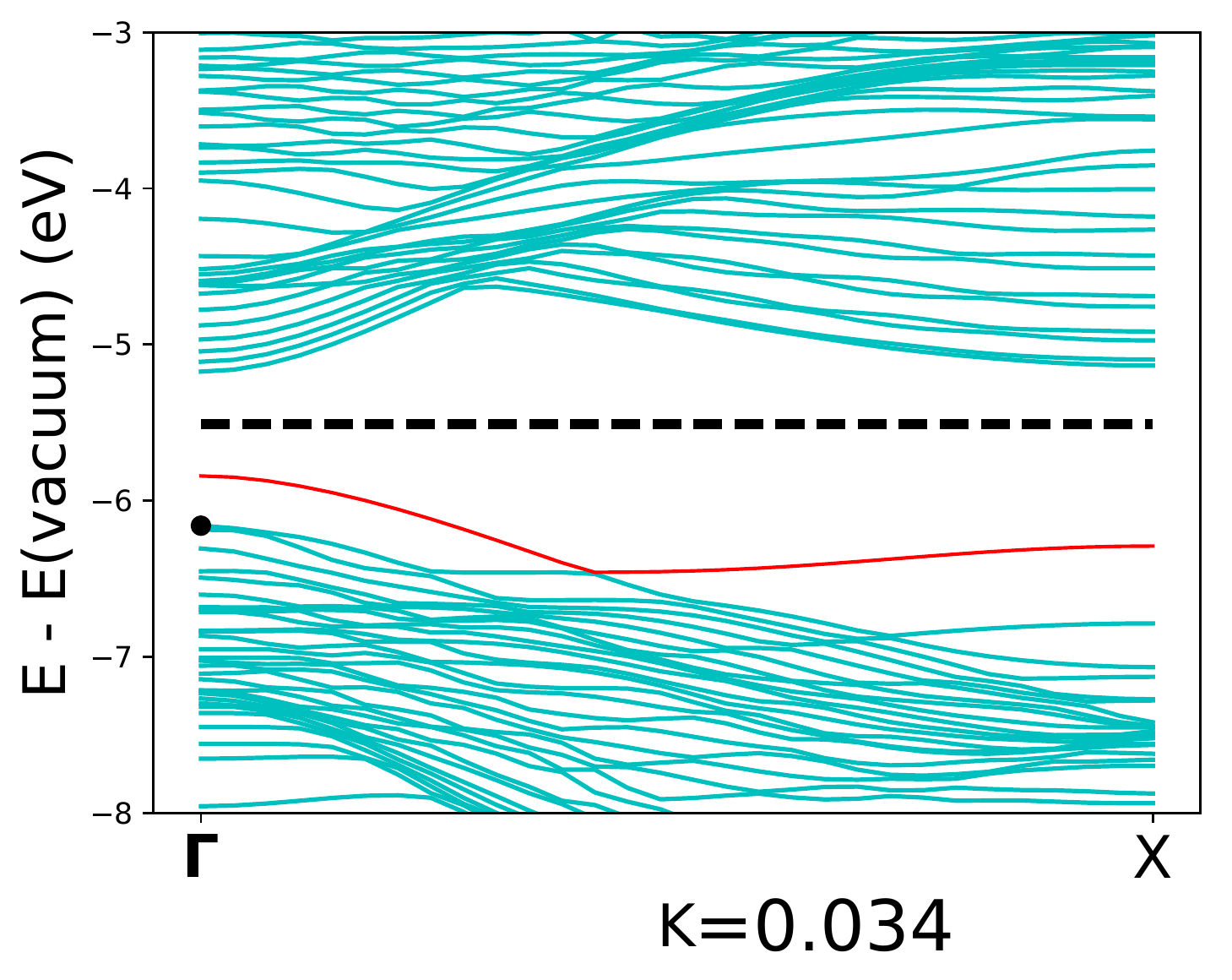}
 	\includegraphics[height=0.9in, width=1.5in]{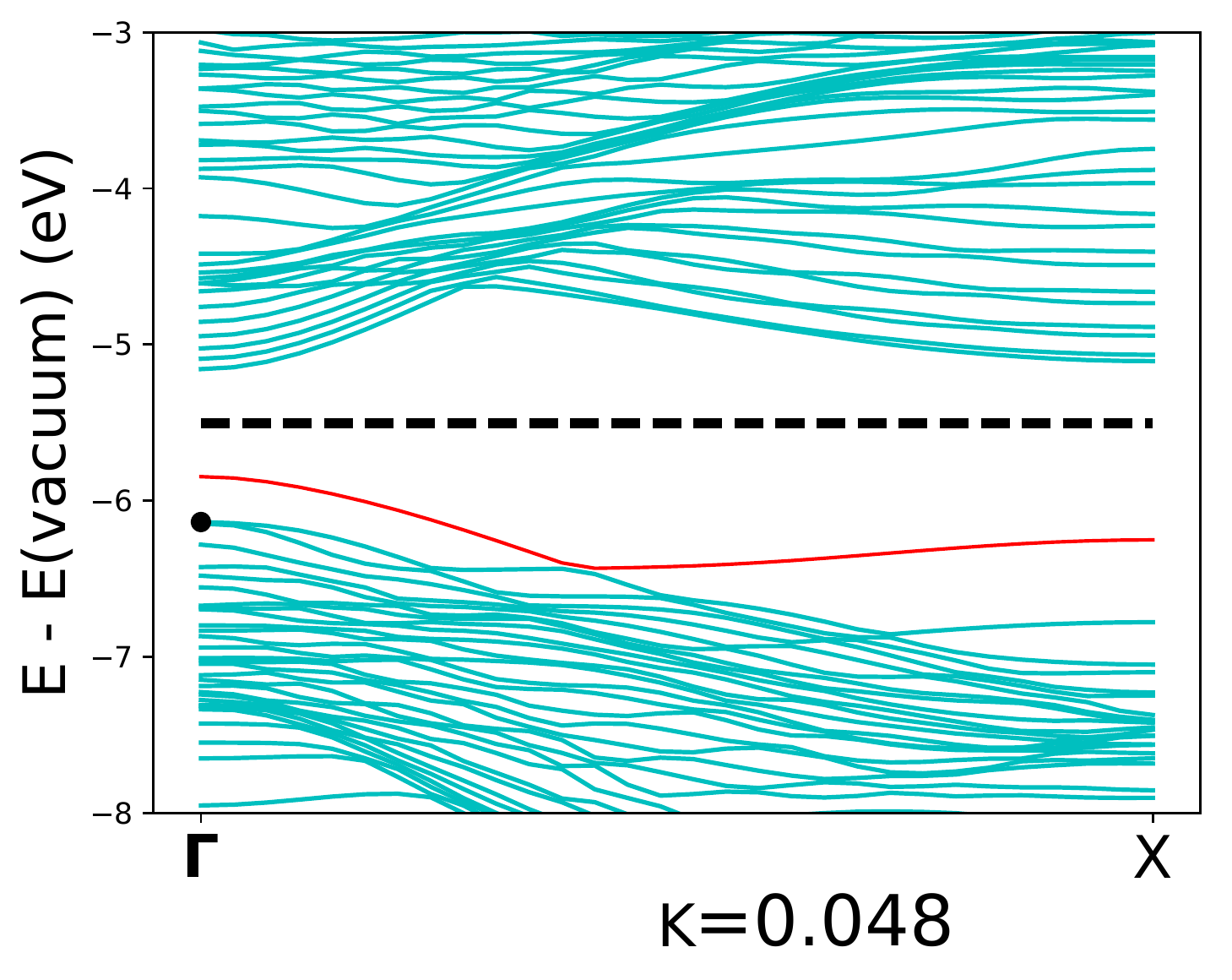}
 	\includegraphics[height=0.9in, width=1.5in]{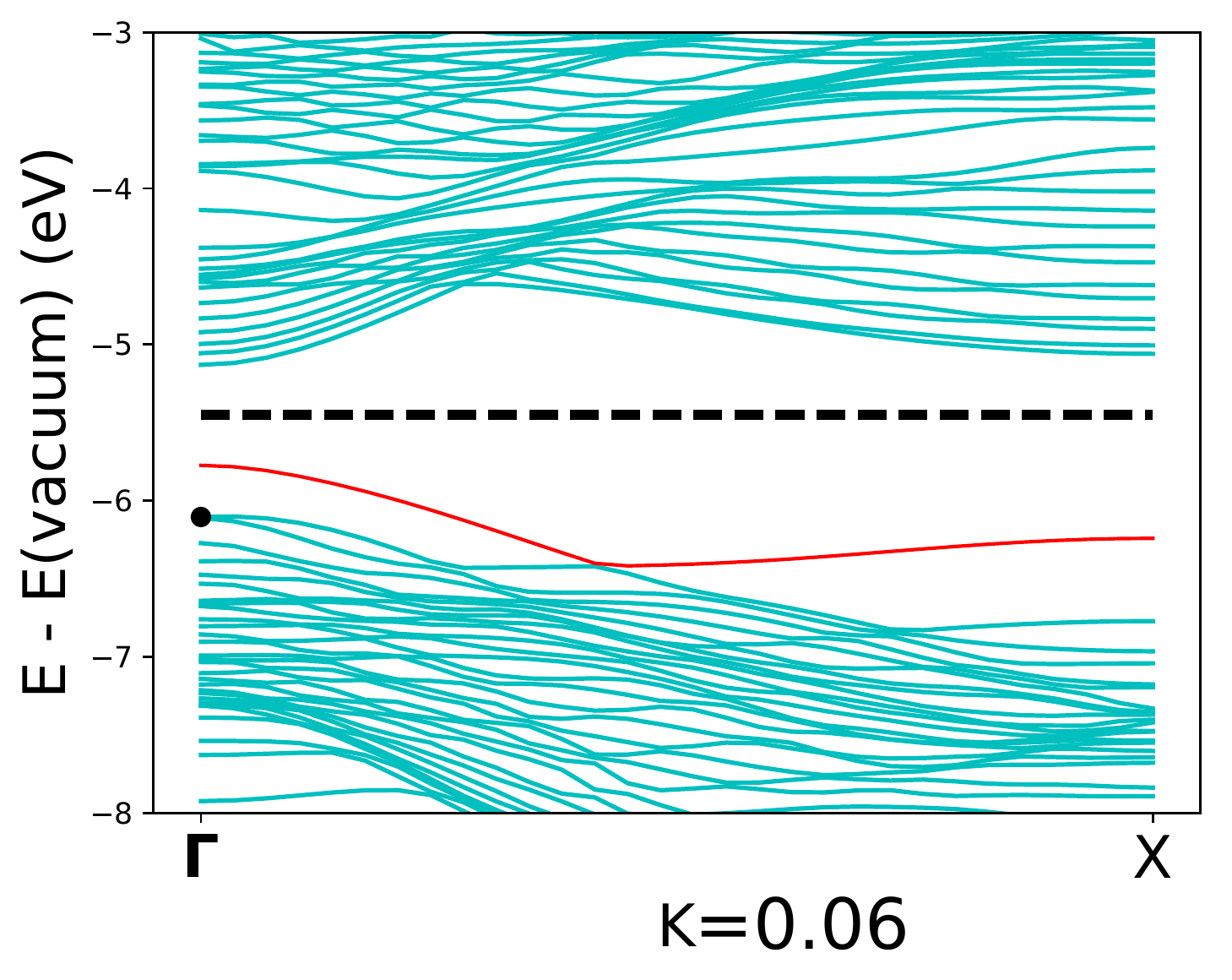}
 	\includegraphics[height=0.9in, width=1.5in]{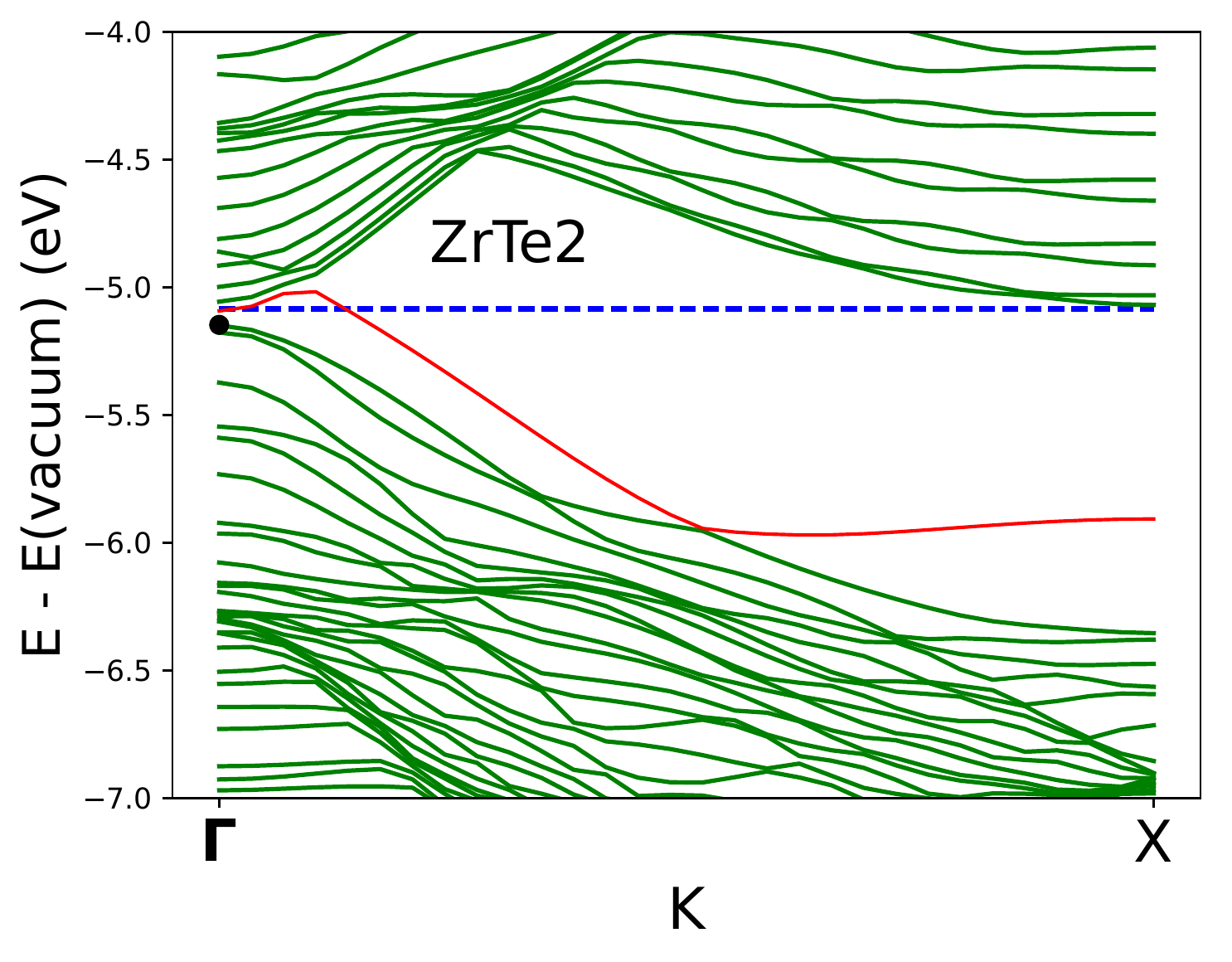}
 	\includegraphics[height=0.9in, width=1.5in]{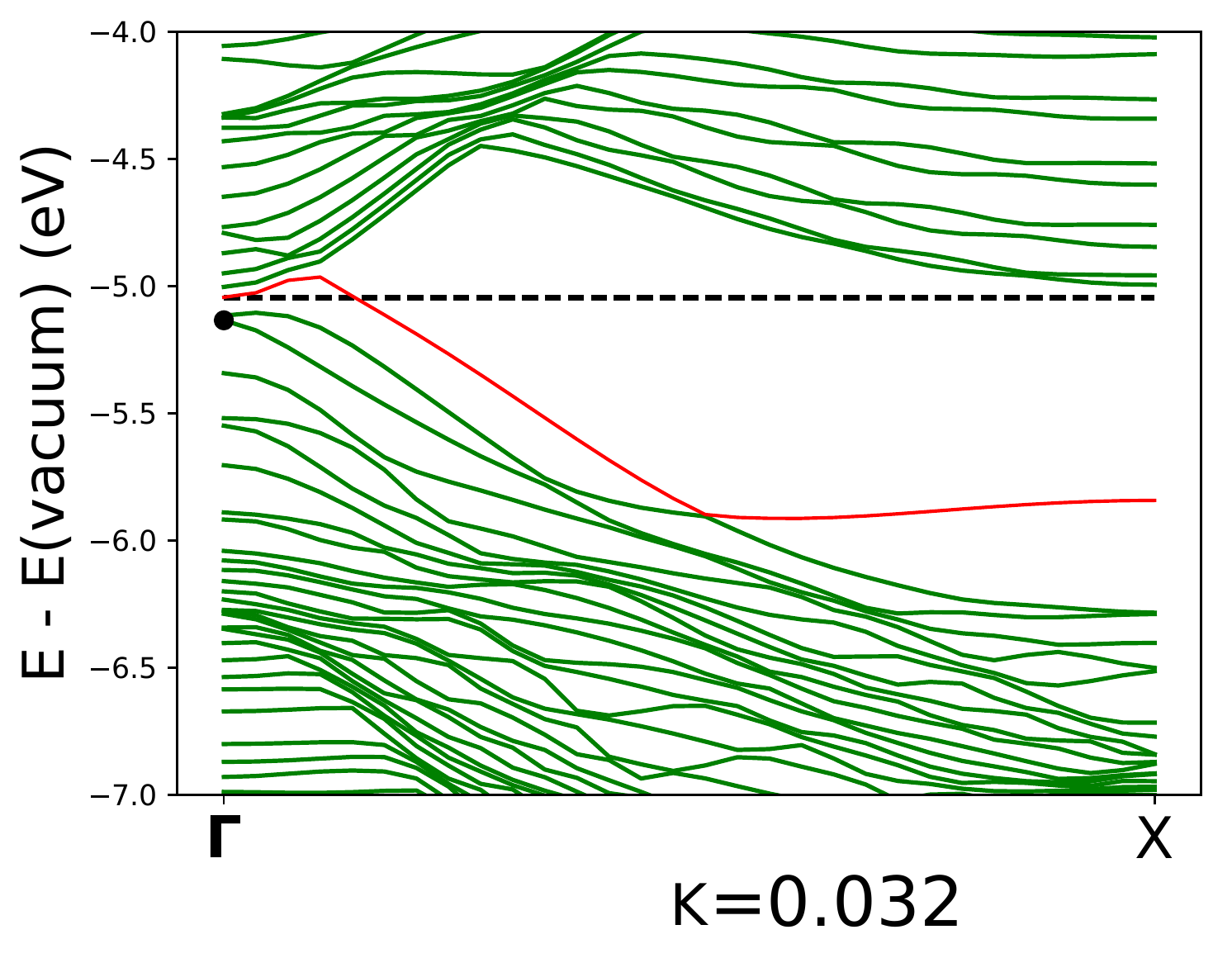}
 	\includegraphics[height=0.9in, width=1.5in]{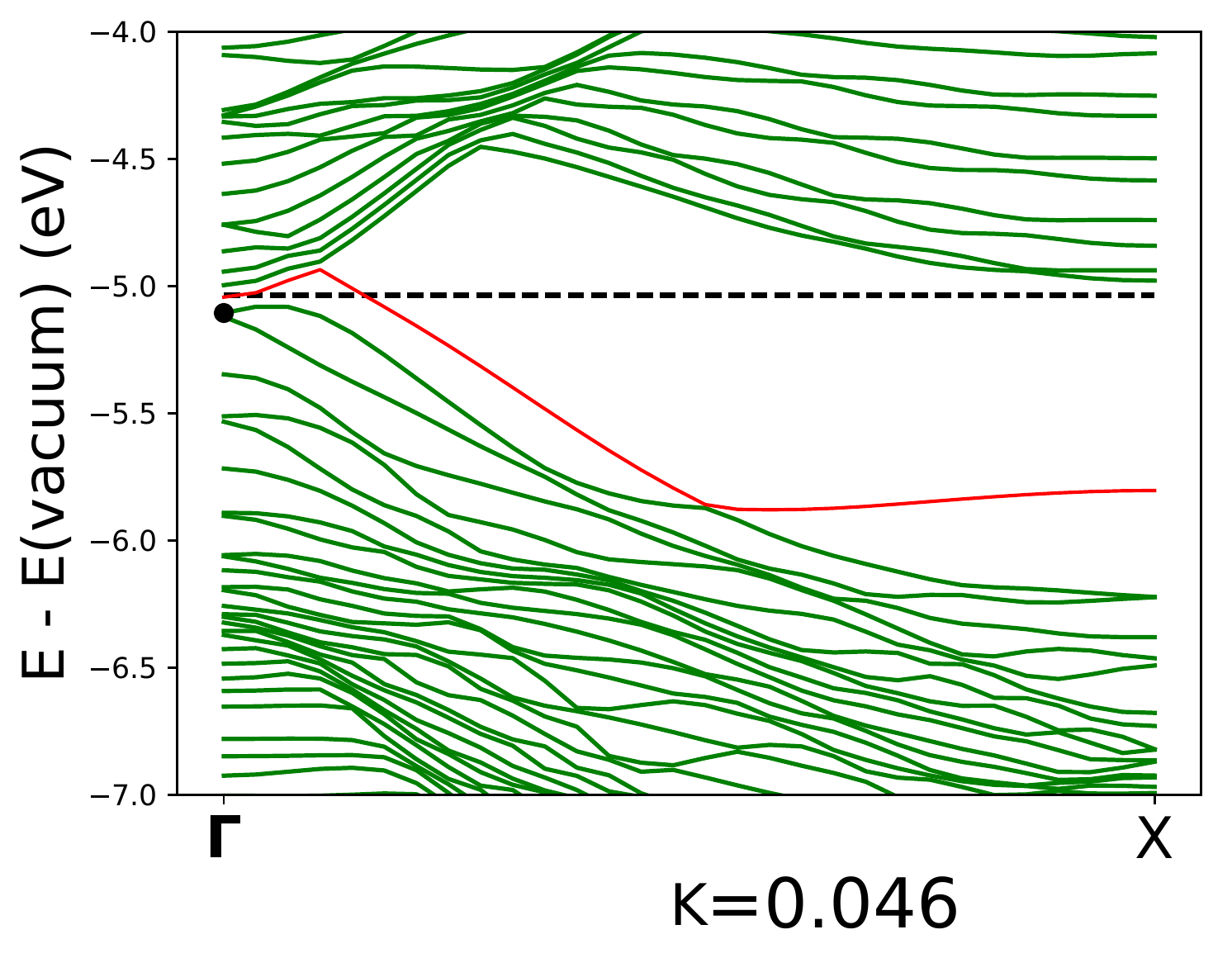}
 	\includegraphics[height=0.9in, width=1.5in]{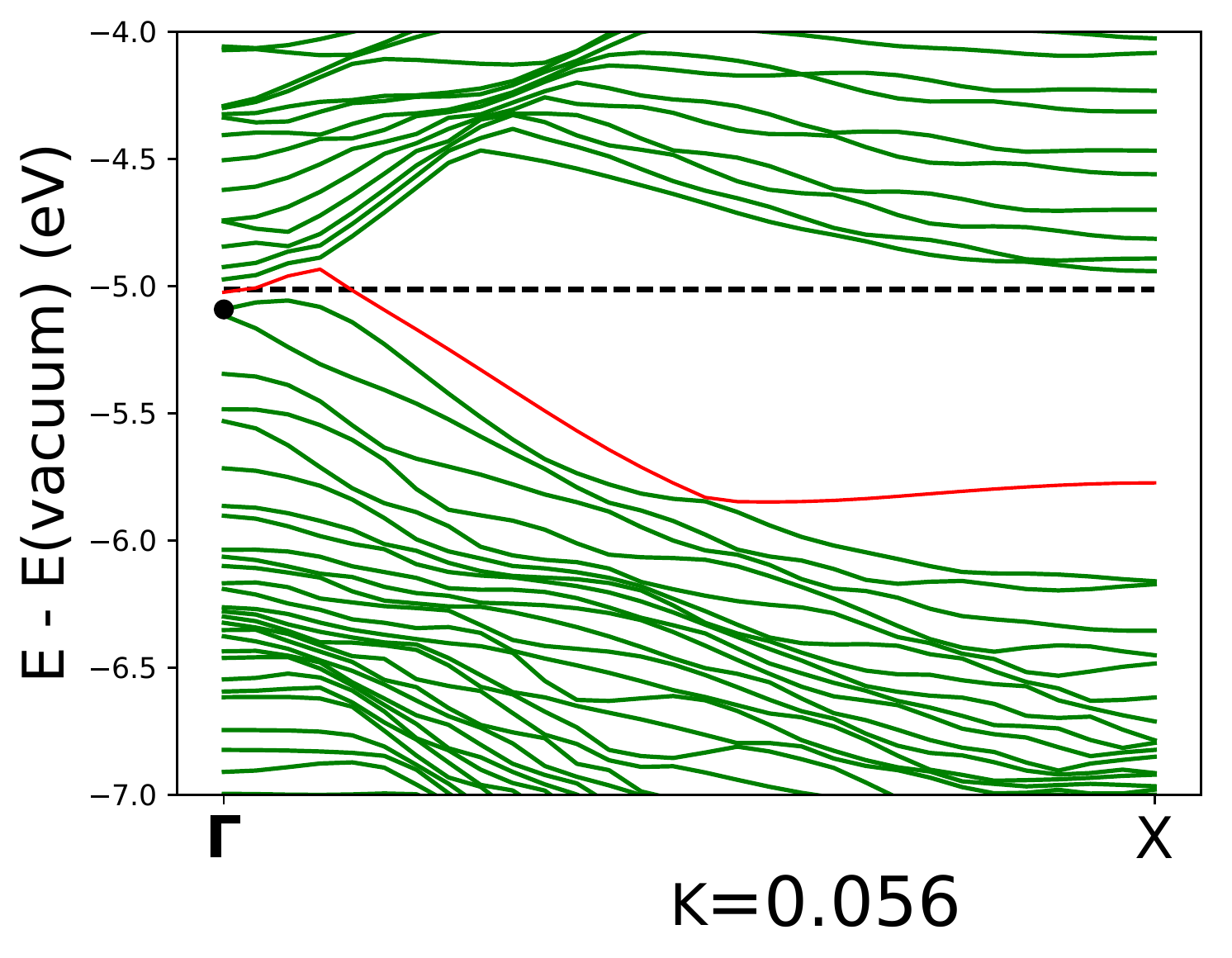}
 	\includegraphics[height=0.9in, width=1.5in]{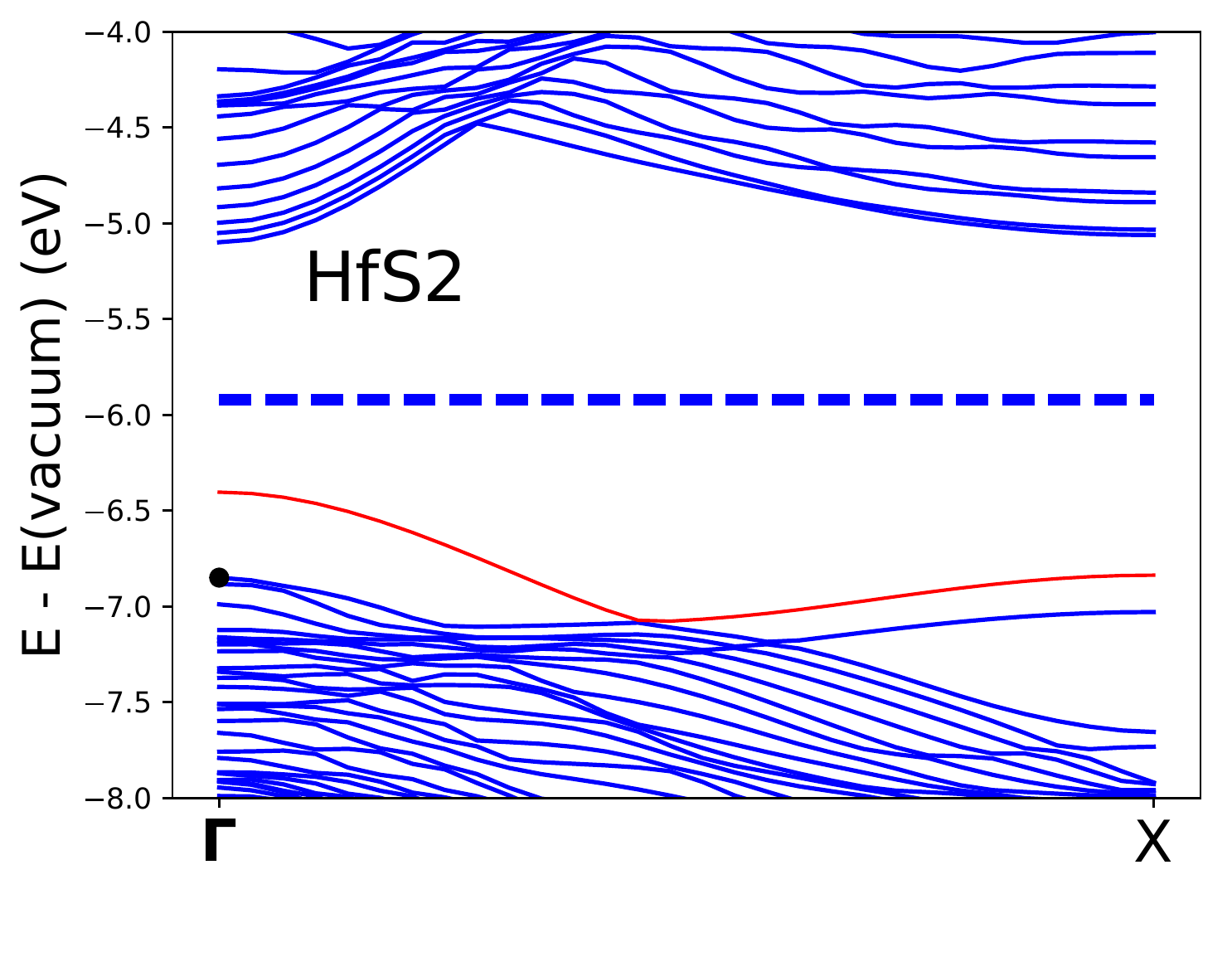}
 	\includegraphics[height=0.9in, width=1.5in]{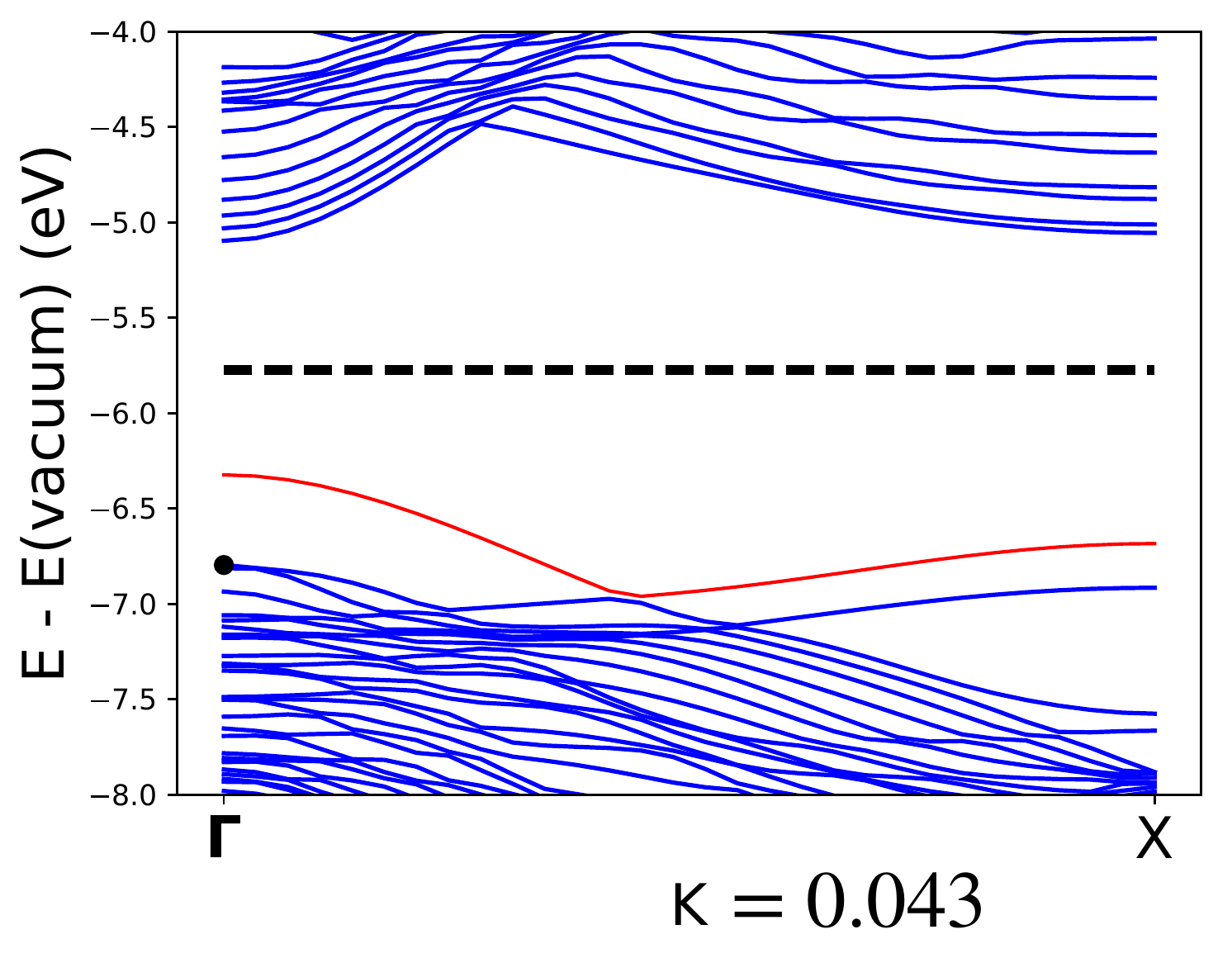}
 	\includegraphics[height=0.9in, width=1.5in]{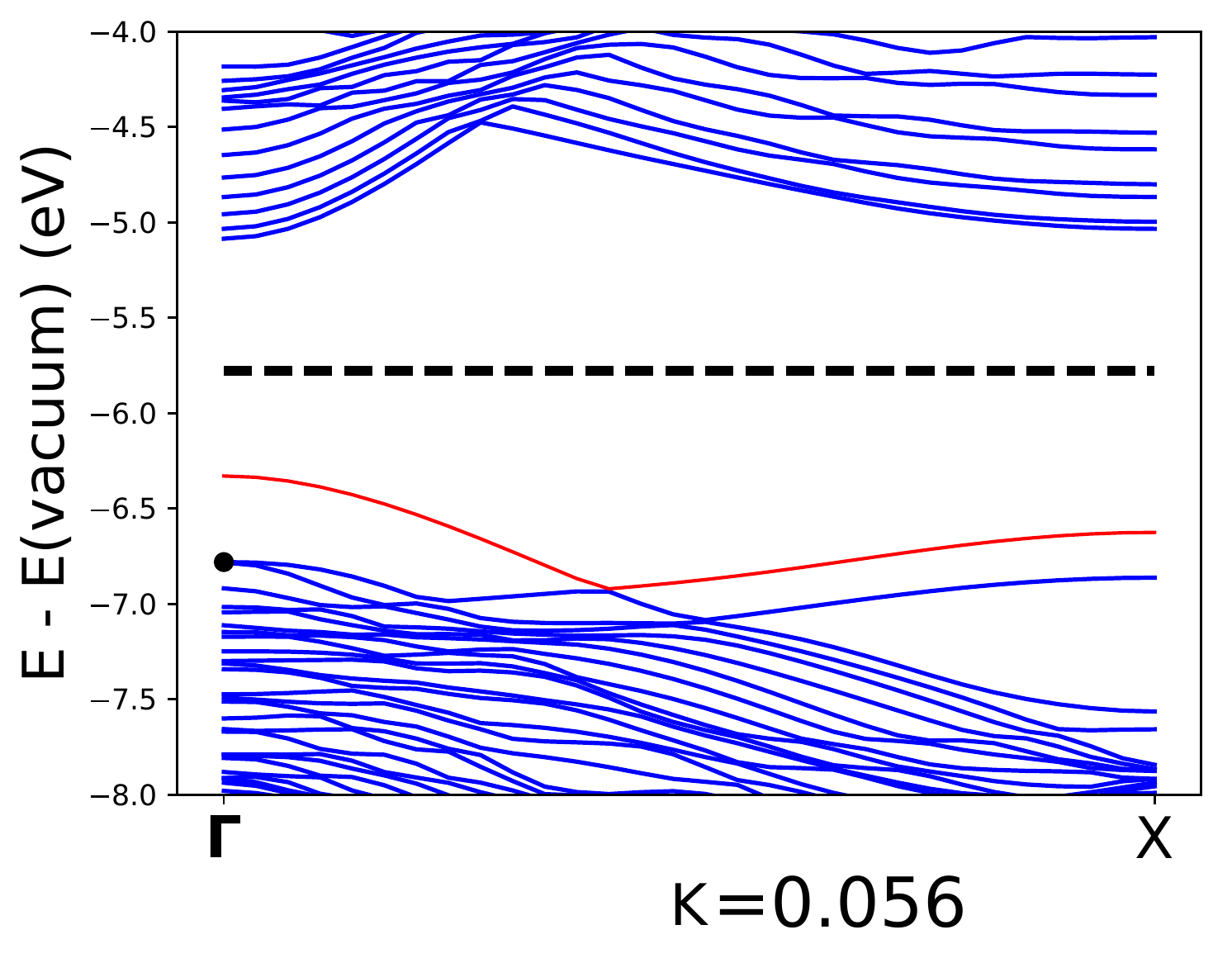}
 	\includegraphics[height=0.9in, width=1.5in]{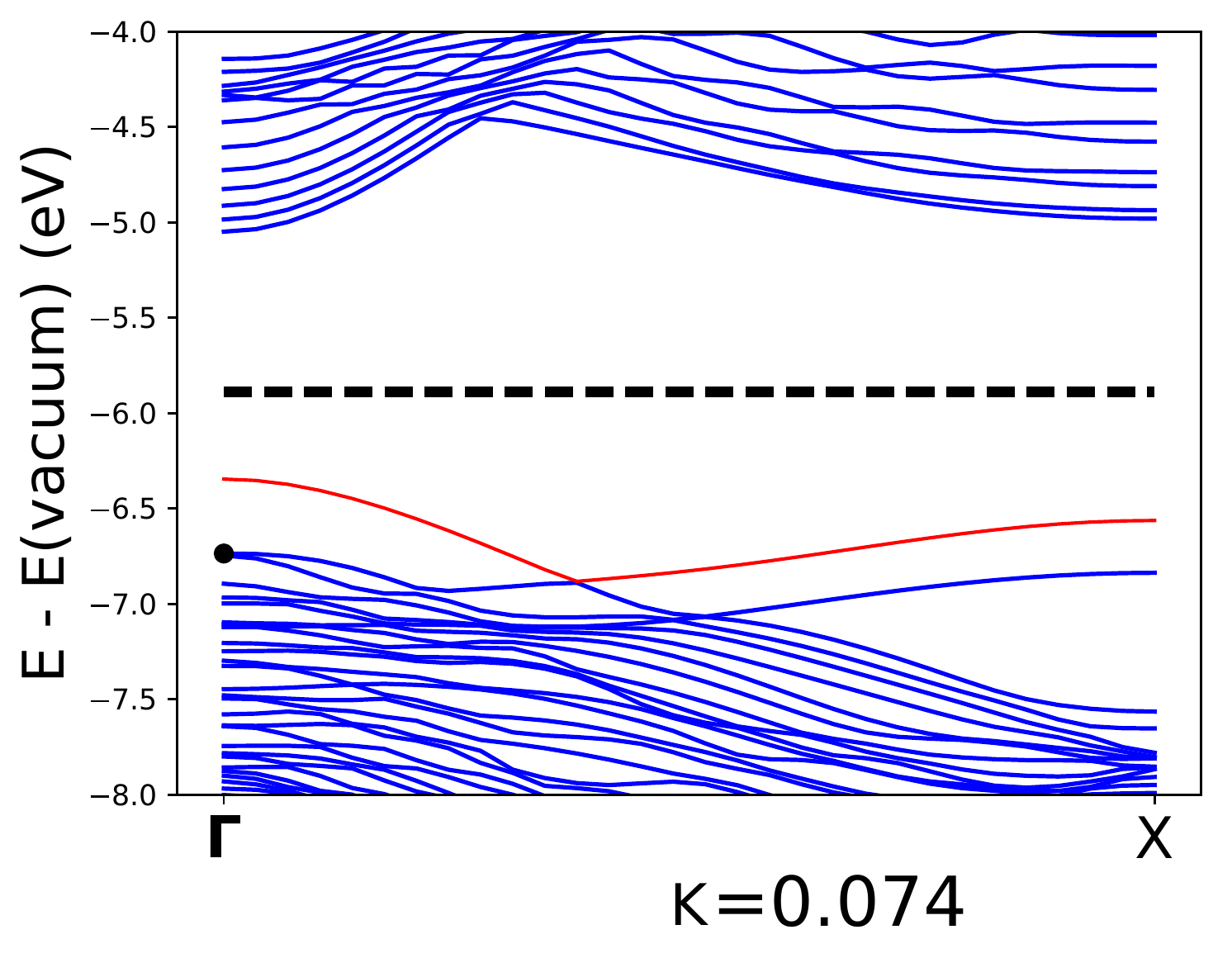}
 	\includegraphics[height=0.9in, width=1.5in]{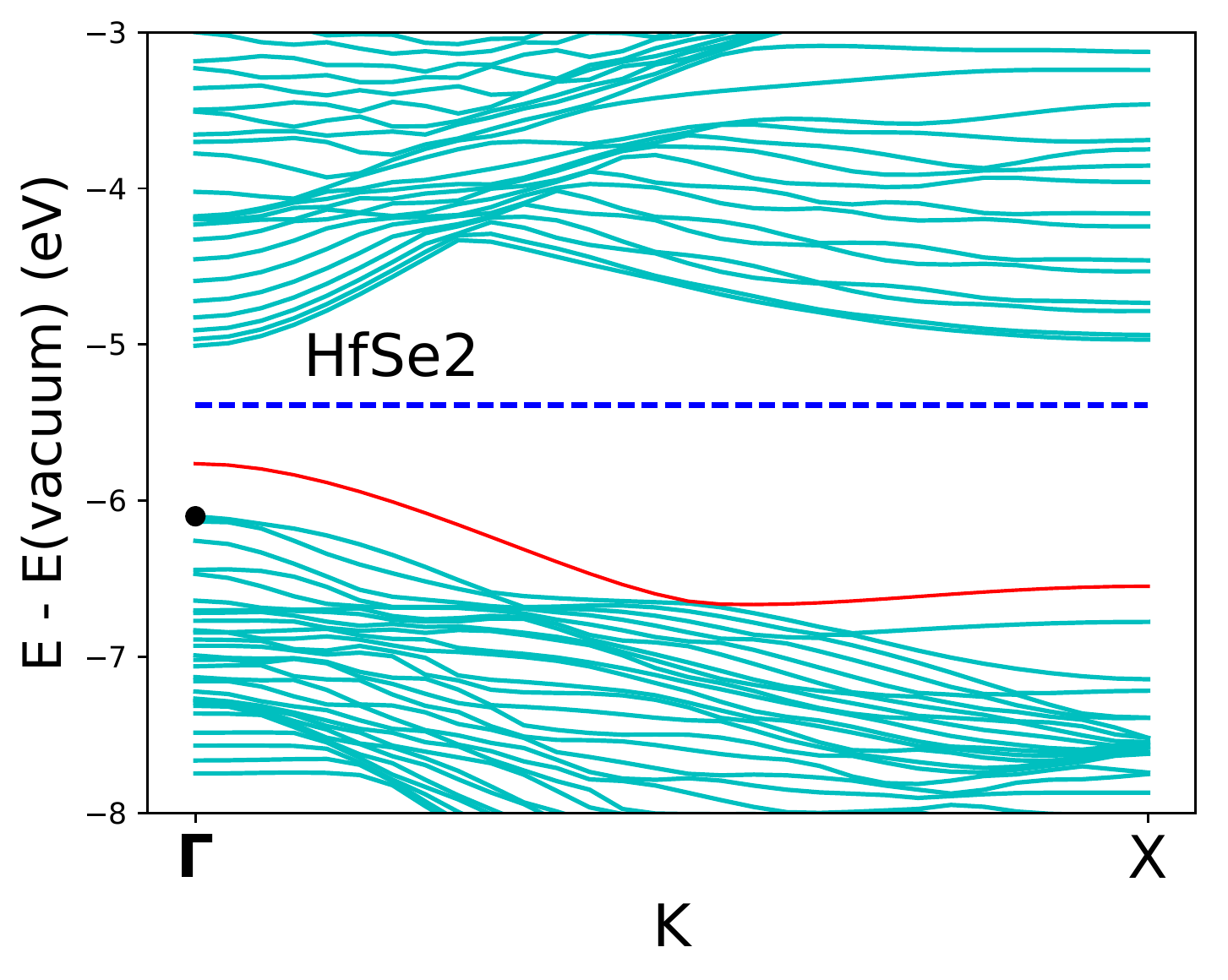}
 	\includegraphics[height=0.9in, width=1.5in]{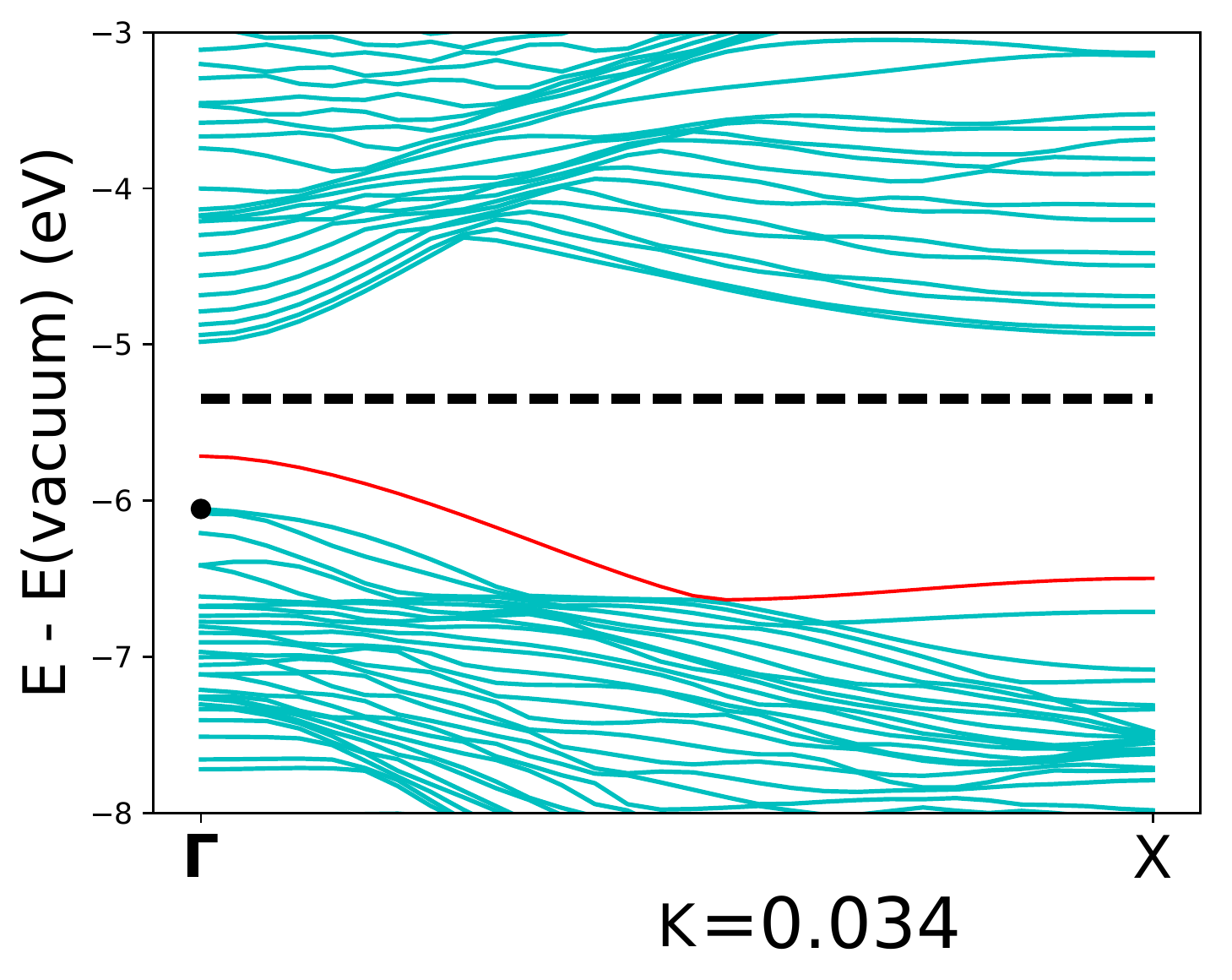}
 	\includegraphics[height=0.9in, width=1.5in]{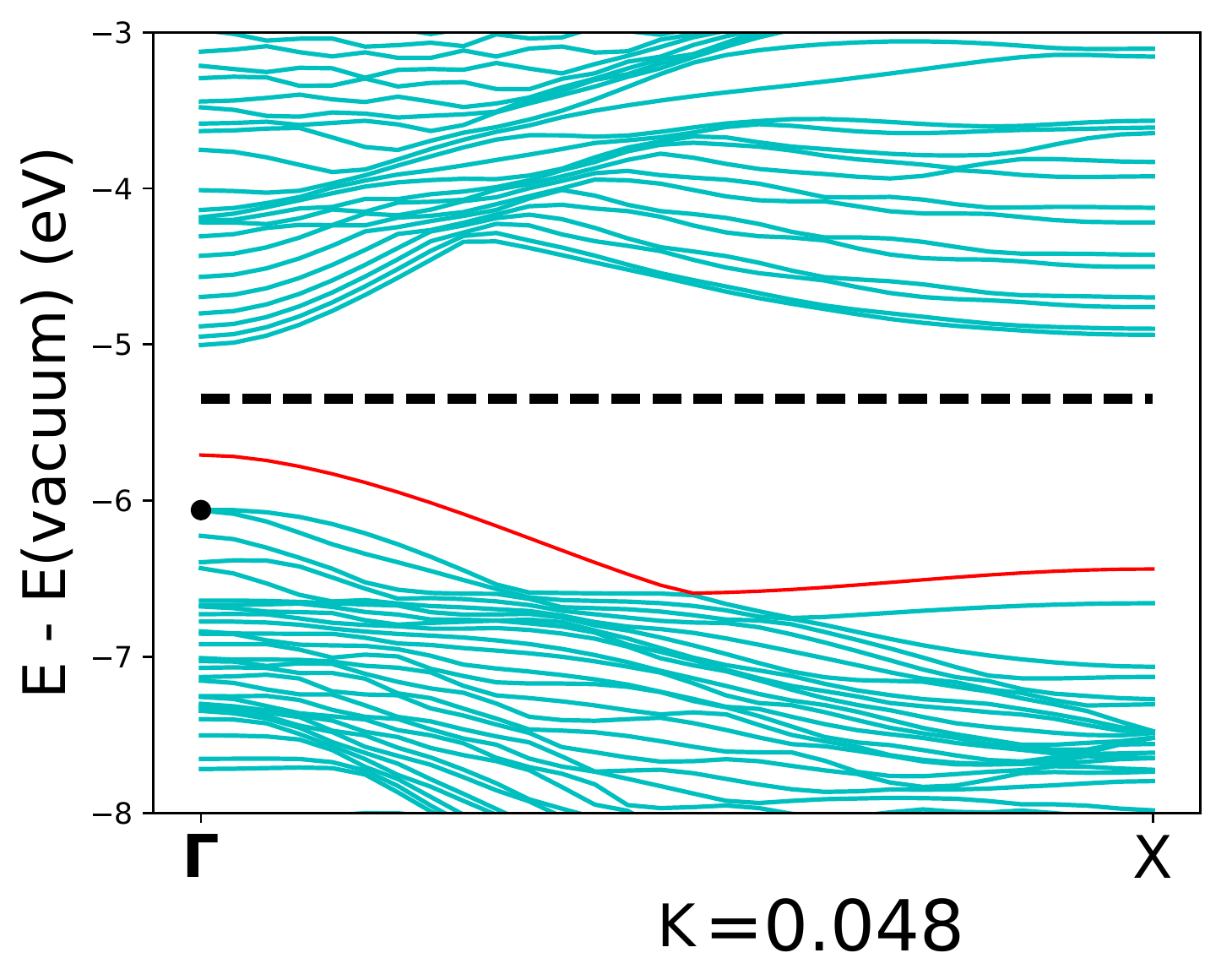}
 	\includegraphics[height=0.9in, width=1.5in]{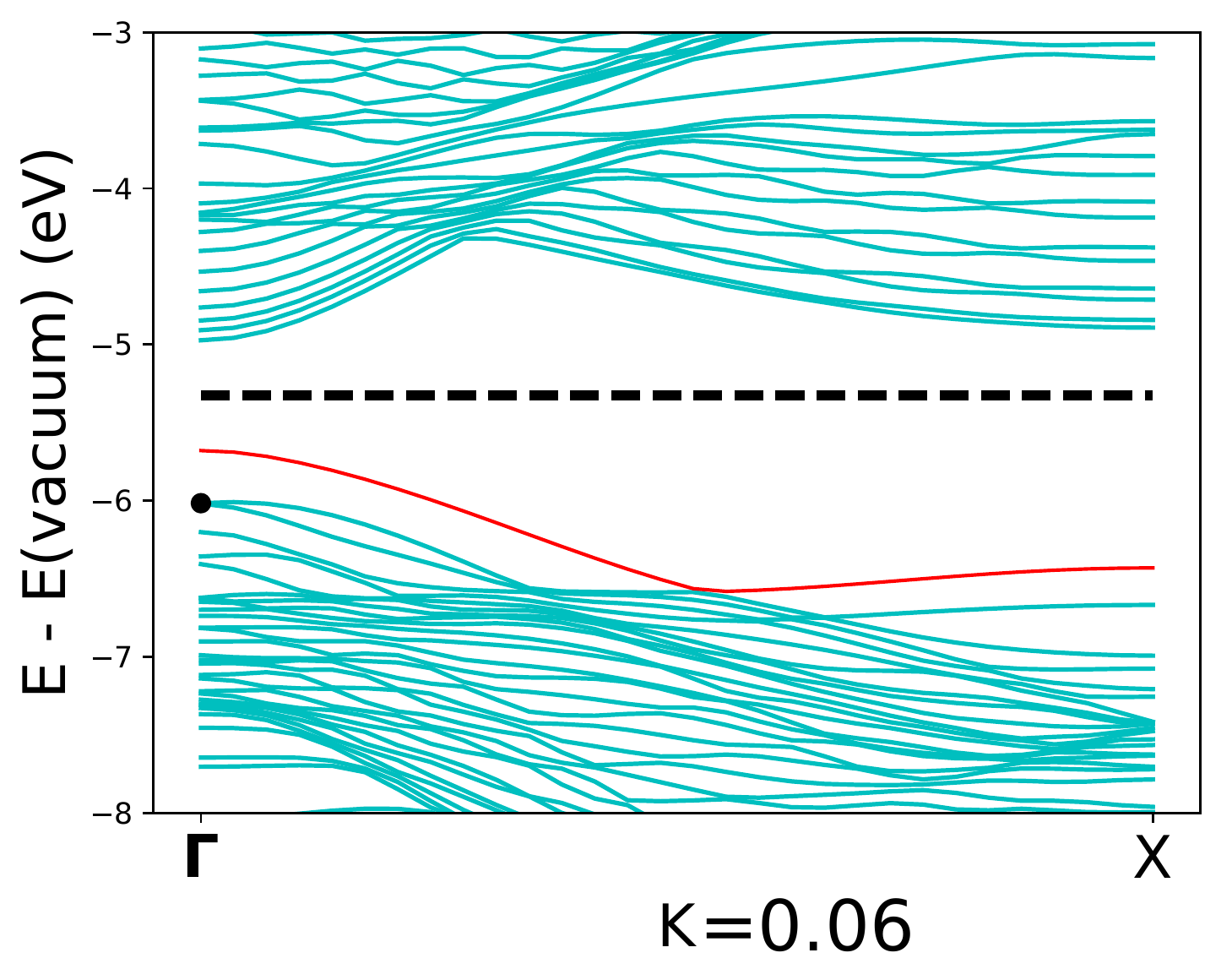}
 	\includegraphics[height=0.9in, width=1.5in]{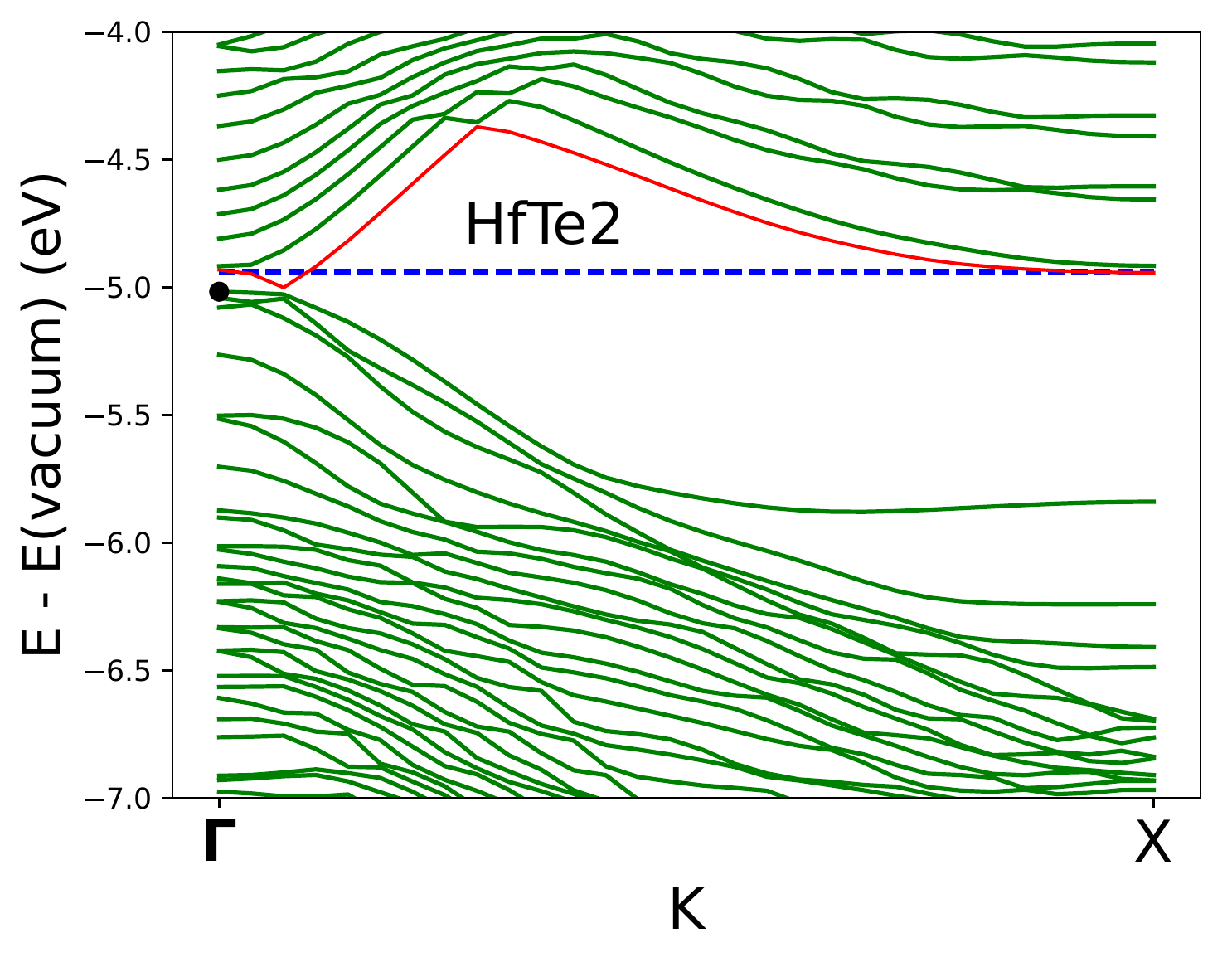}
 	\includegraphics[height=0.9in, width=1.5in]{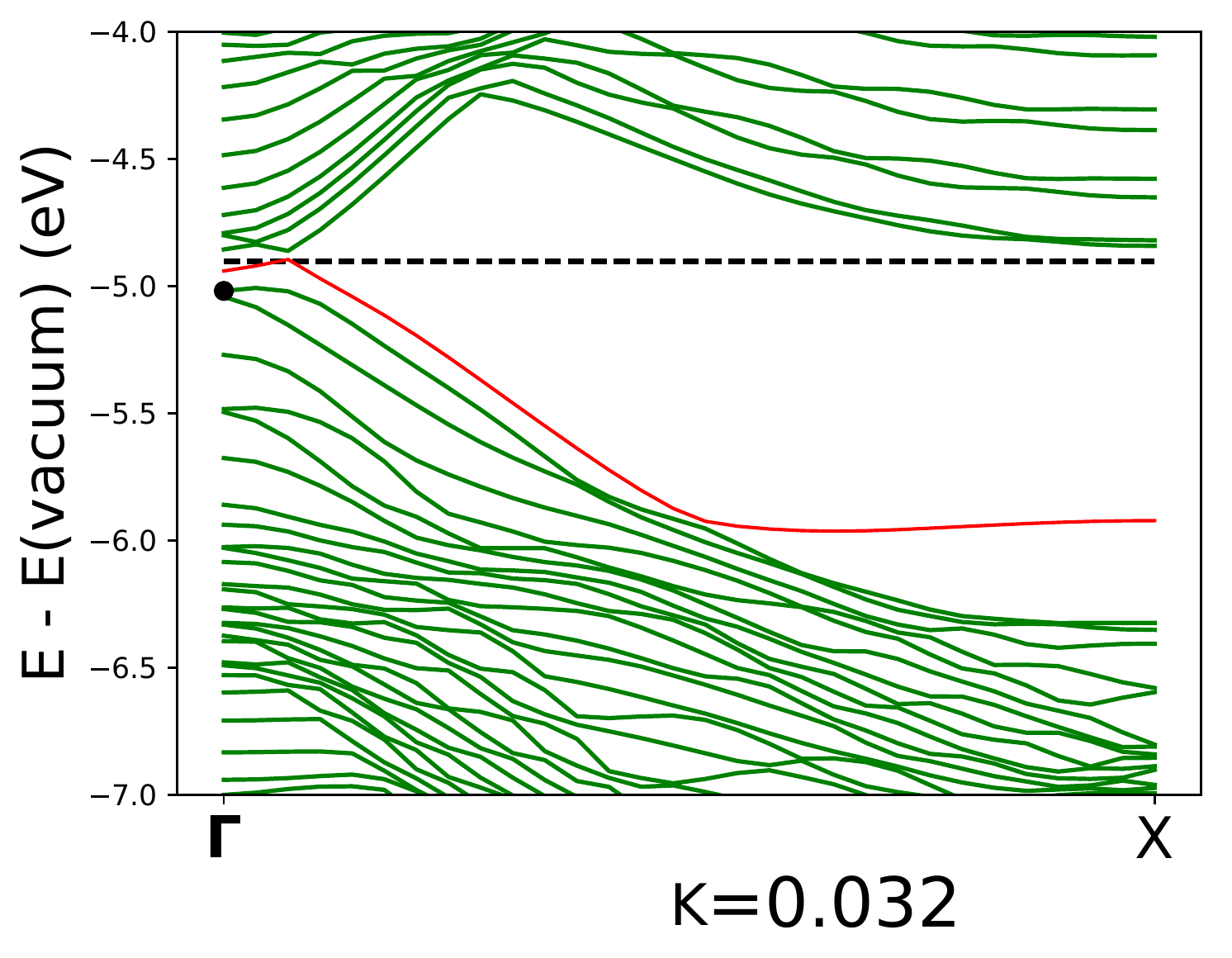}
 	\includegraphics[height=0.9in, width=1.5in]{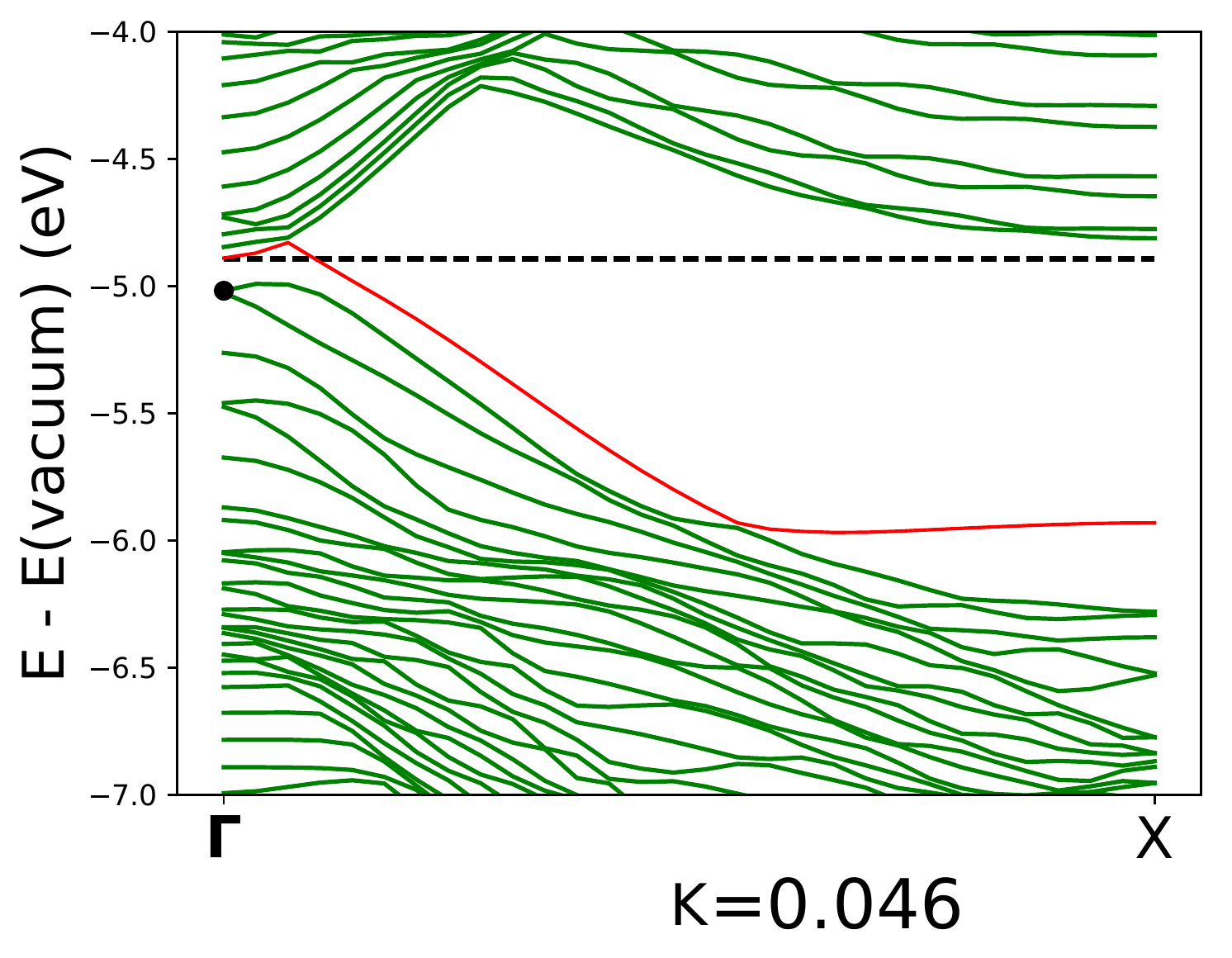}
 	\includegraphics[height=0.9in, width=1.5in]{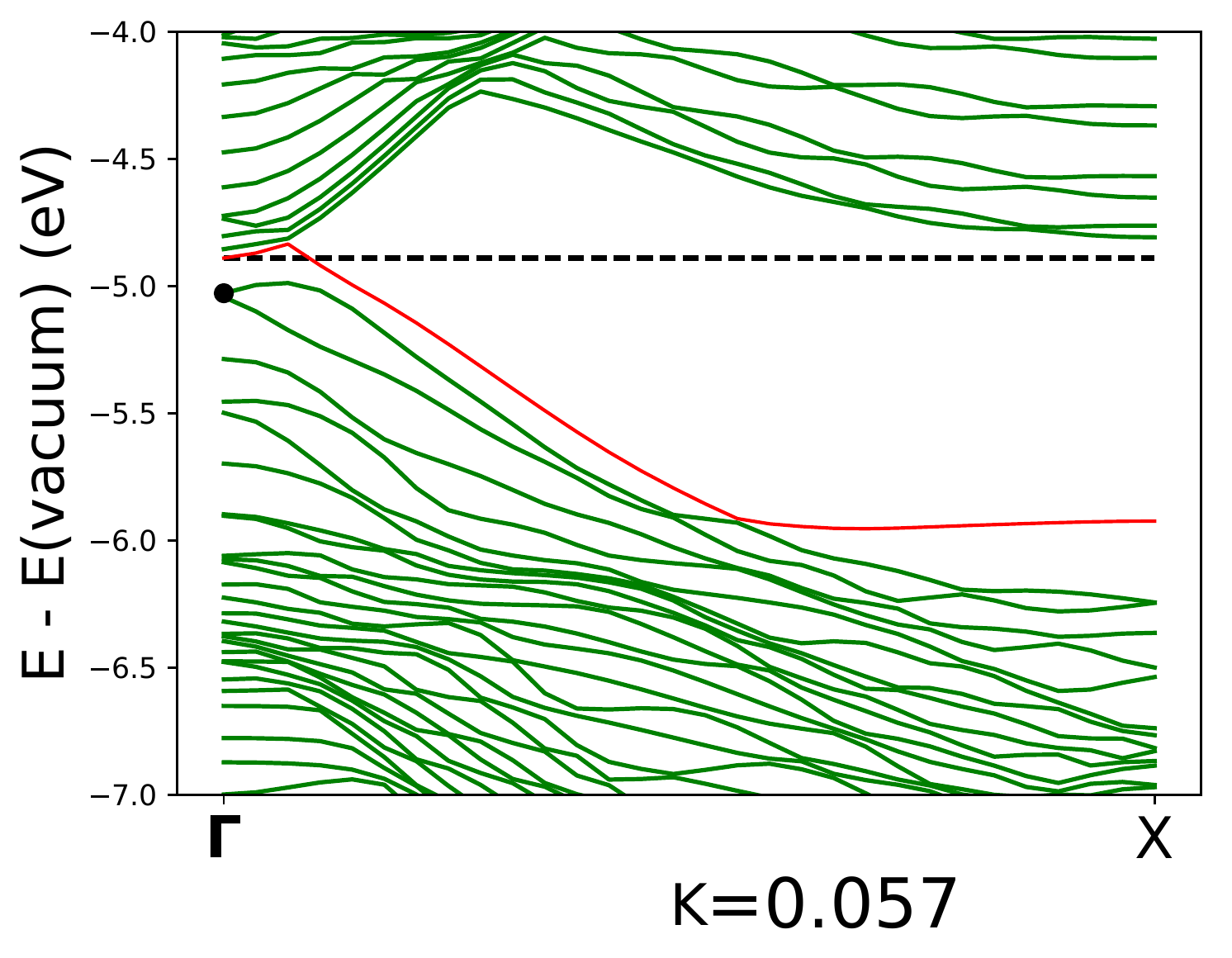}
 	\caption{Band structures with respect to vacuum corresponding to group IV TMDs; \textbf{TiS$_2$}, \textbf{TiSe$_2$}, \textbf{TiTe$_2$}, \textbf{ZrS$_2$}, \textbf{ZrSe$_2$}, \textbf{ZrTe$_2$}, \textbf{HfS$_2$}, \textbf{HfSe$_2$}, and \textbf{HfTe$_2$}. $\kappa$ is the bending curvature ($\AA^{-1}$).}
 	\label{fig:band-IV}
 \end{figure}
 
 \begin{figure}[h!]
 	\renewcommand\thefigure{S7}
 	\includegraphics[height=1in, width=1.5in]{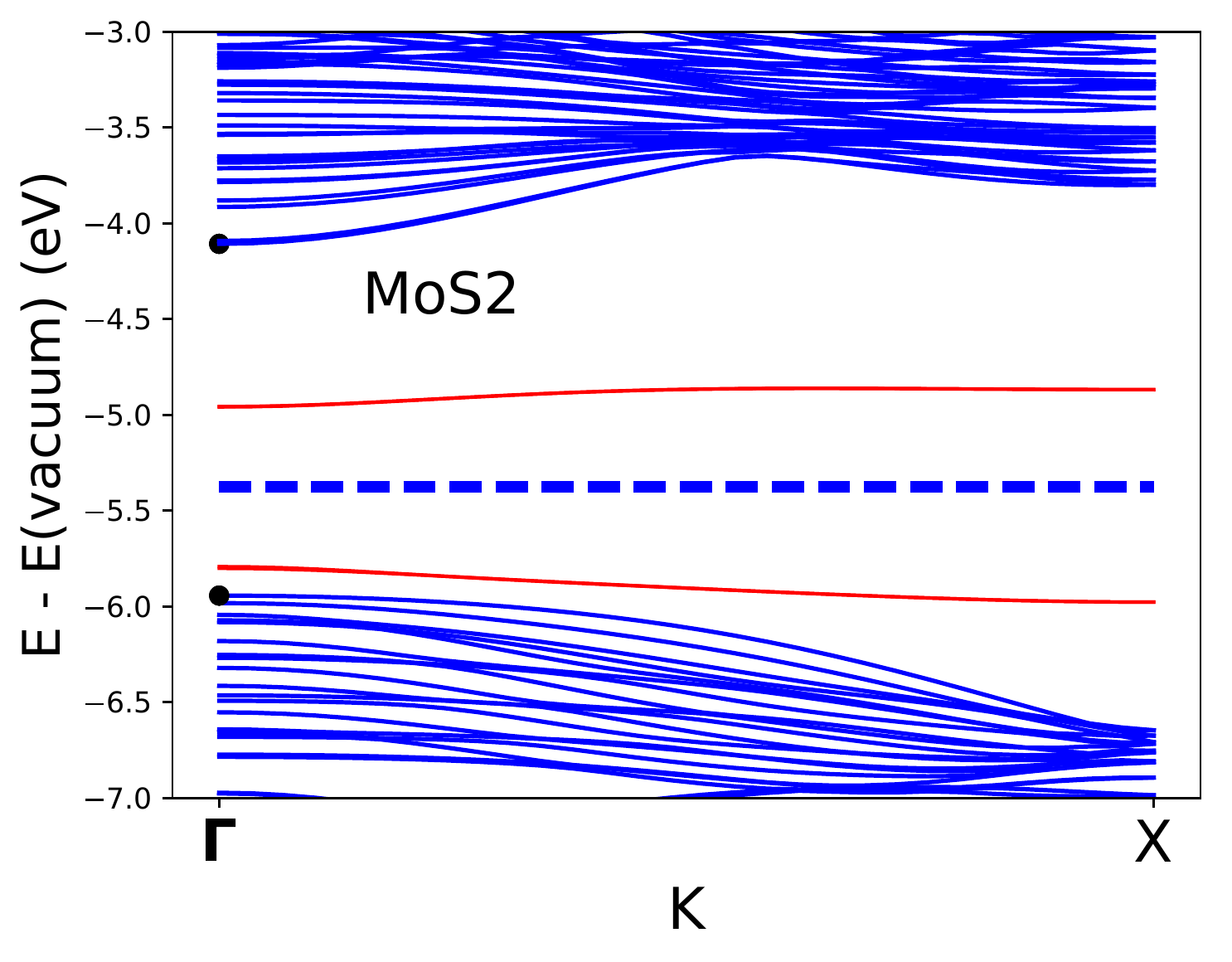}
 	\includegraphics[height=1in, width=1.5in]{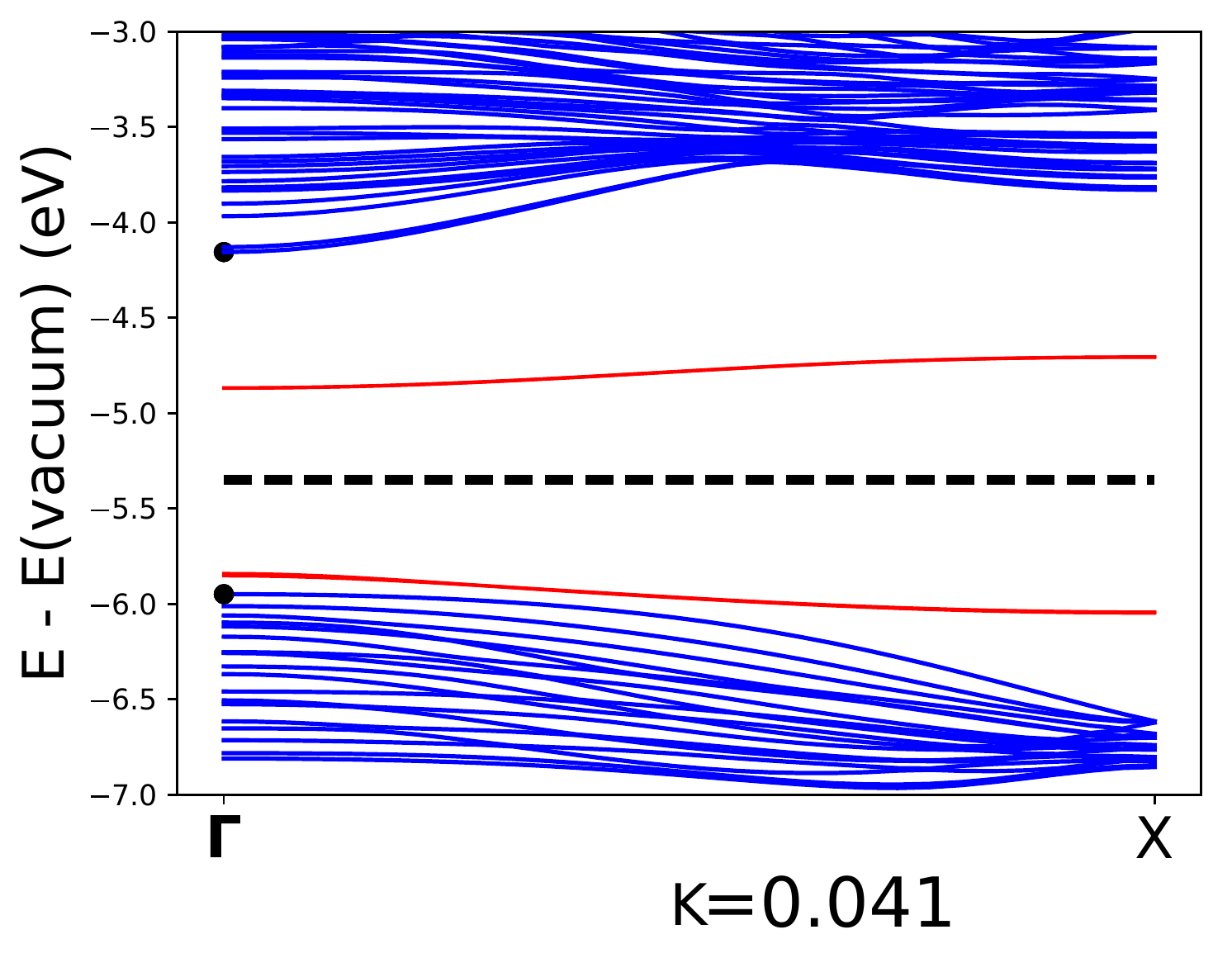}
 	\includegraphics[height=1in, width=1.5in]{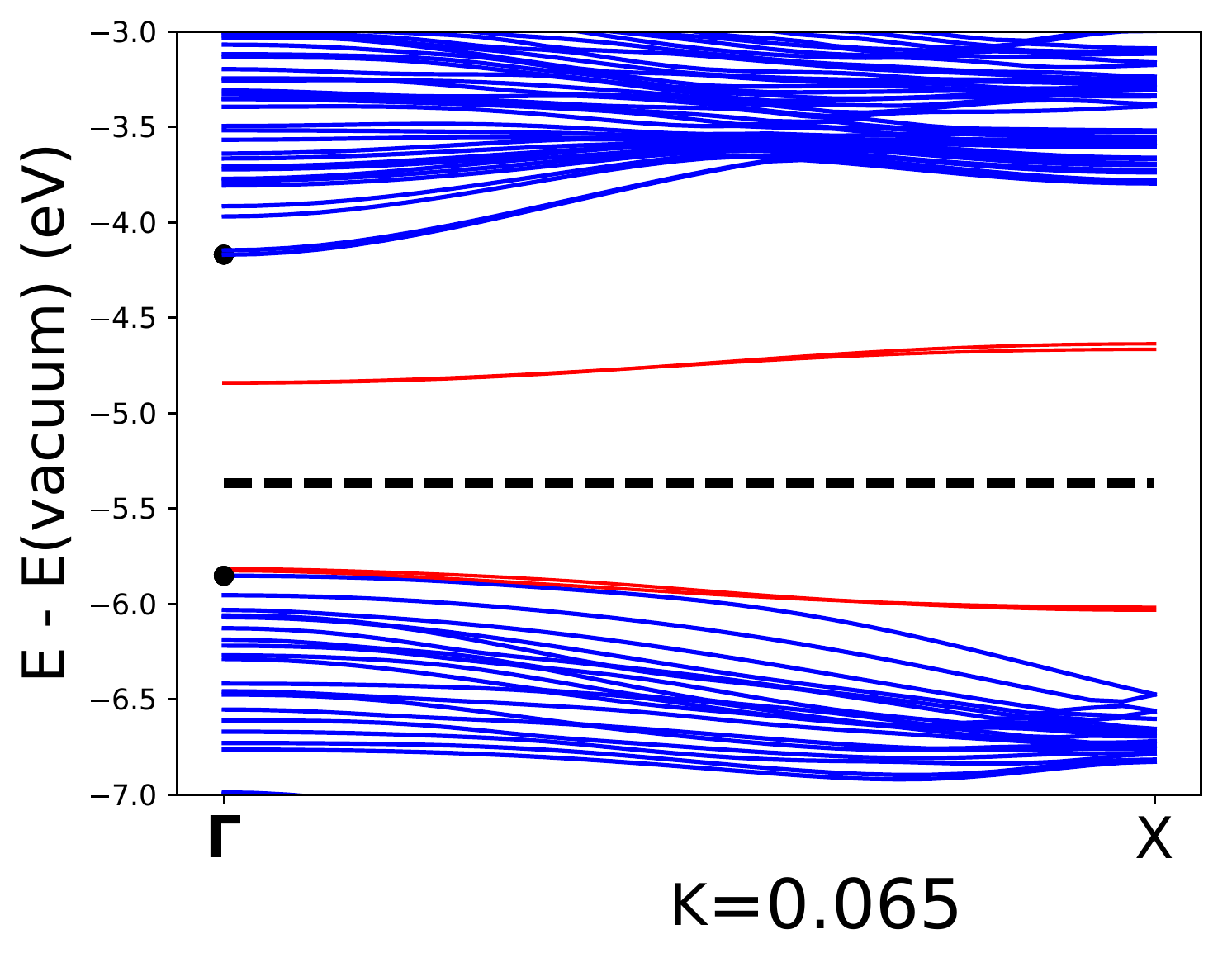}
 	\includegraphics[height=1in, width=1.5in]{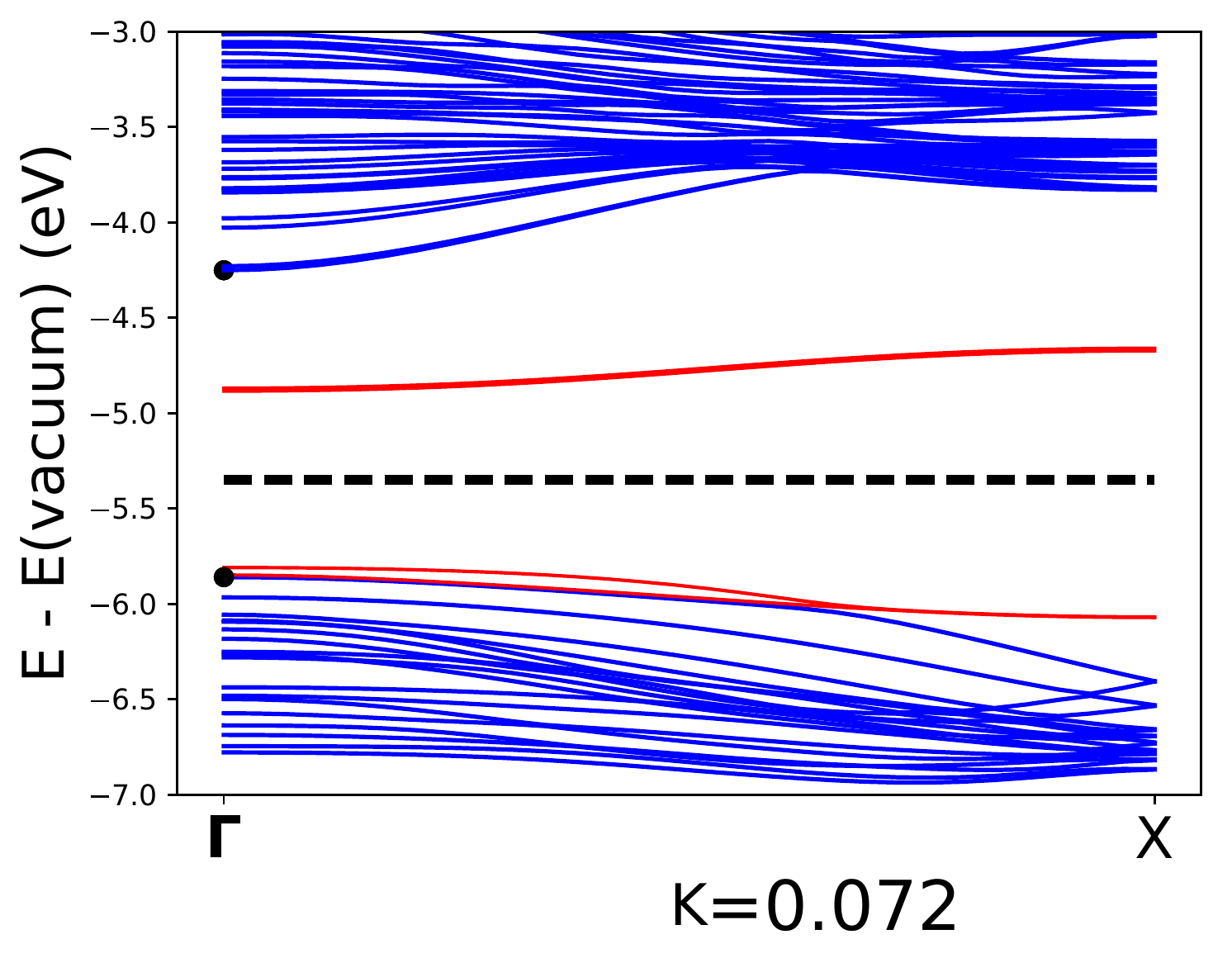}
 	\includegraphics[height=1in, width=1.5in]{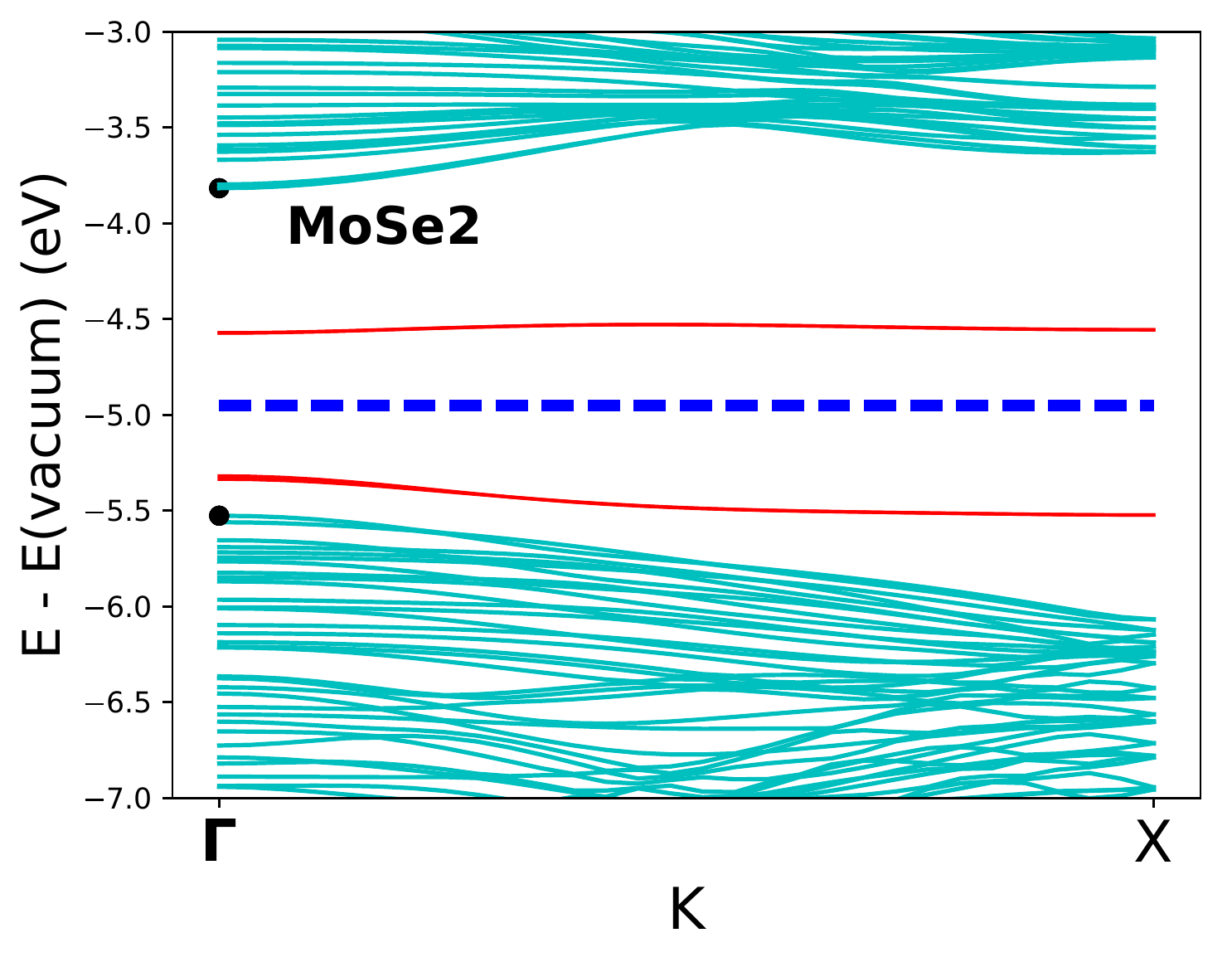}
 	\includegraphics[height=1in, width=1.5in]{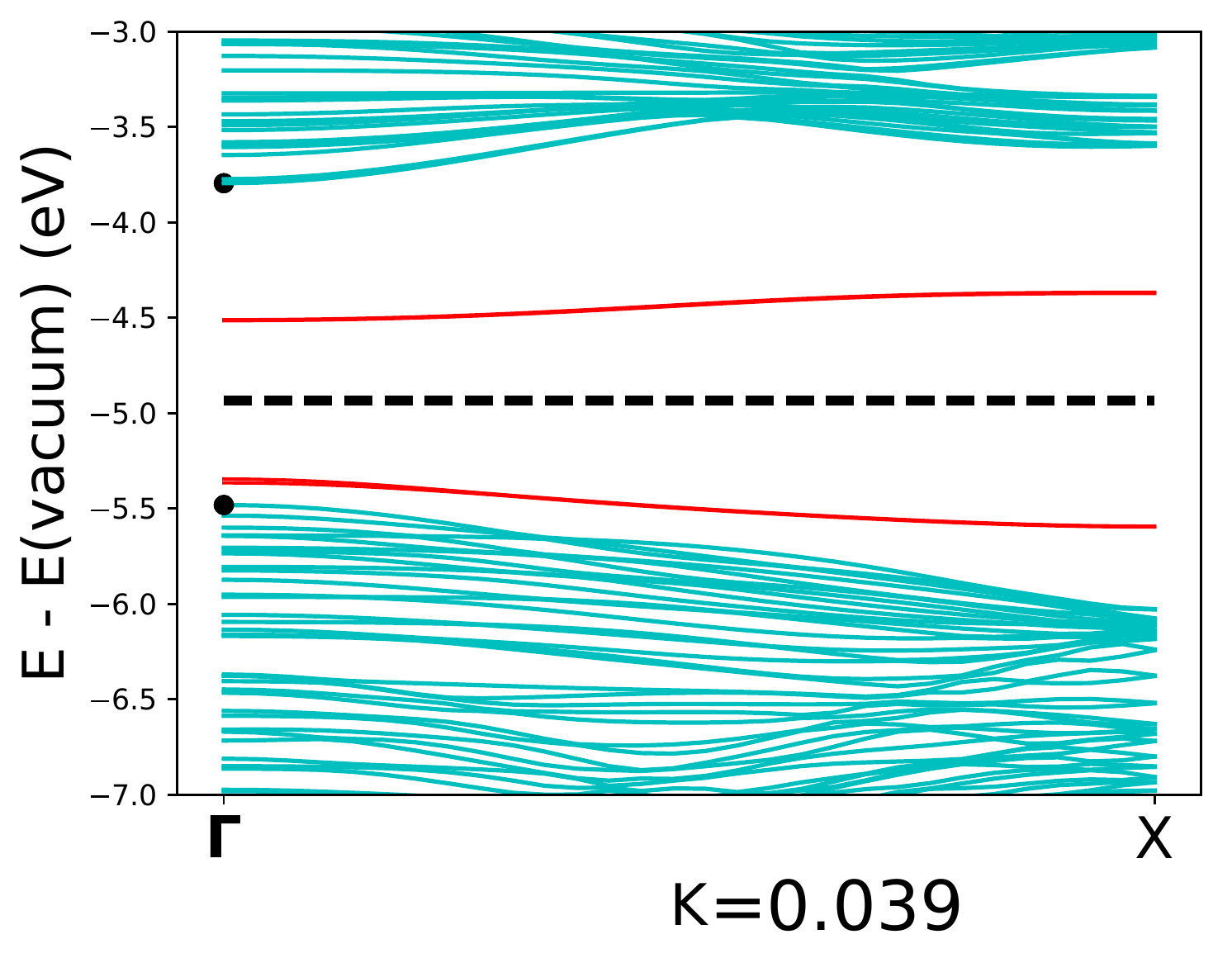}
 	\includegraphics[height=1in, width=1.5in]{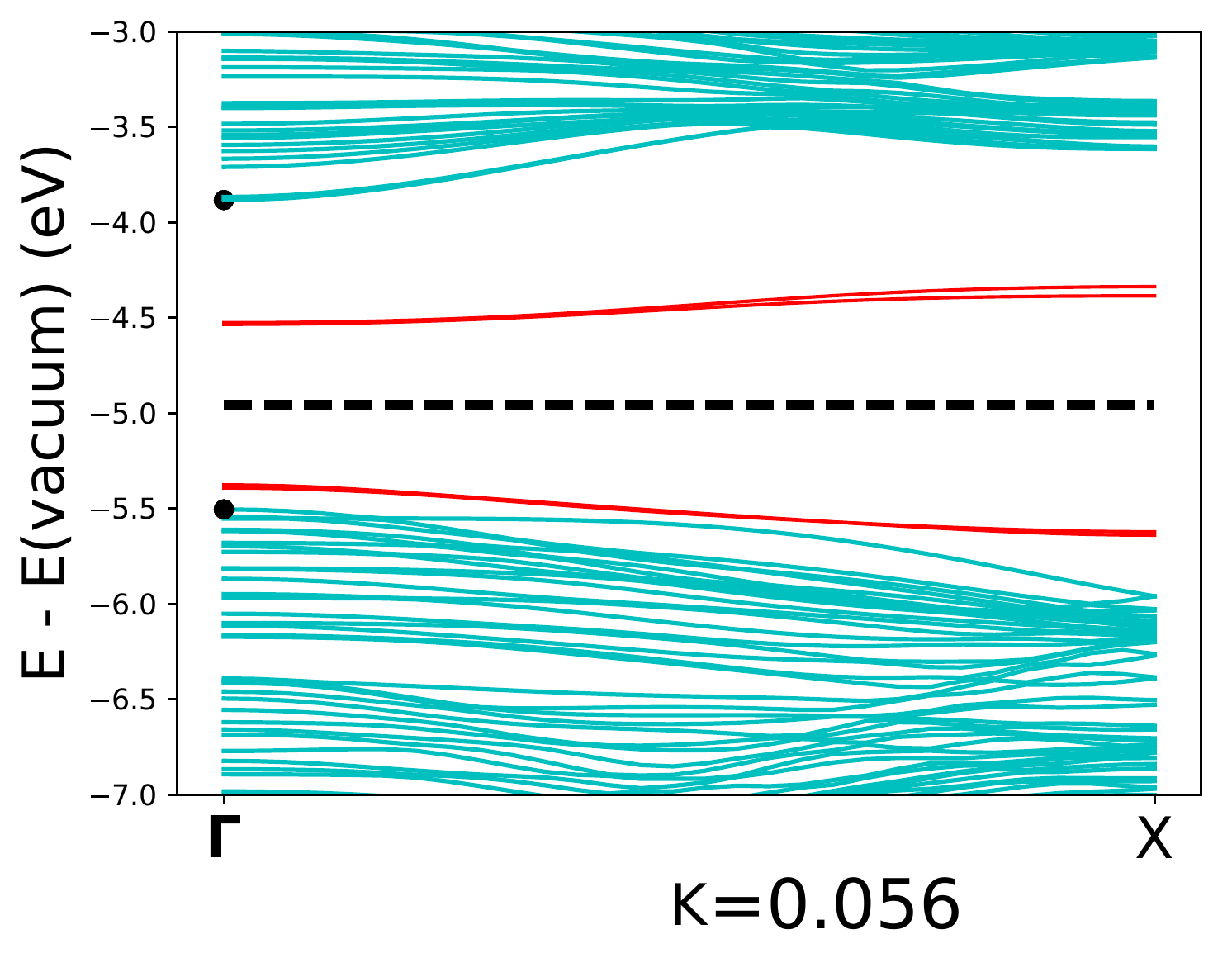}
 	\includegraphics[height=1in, width=1.5in]{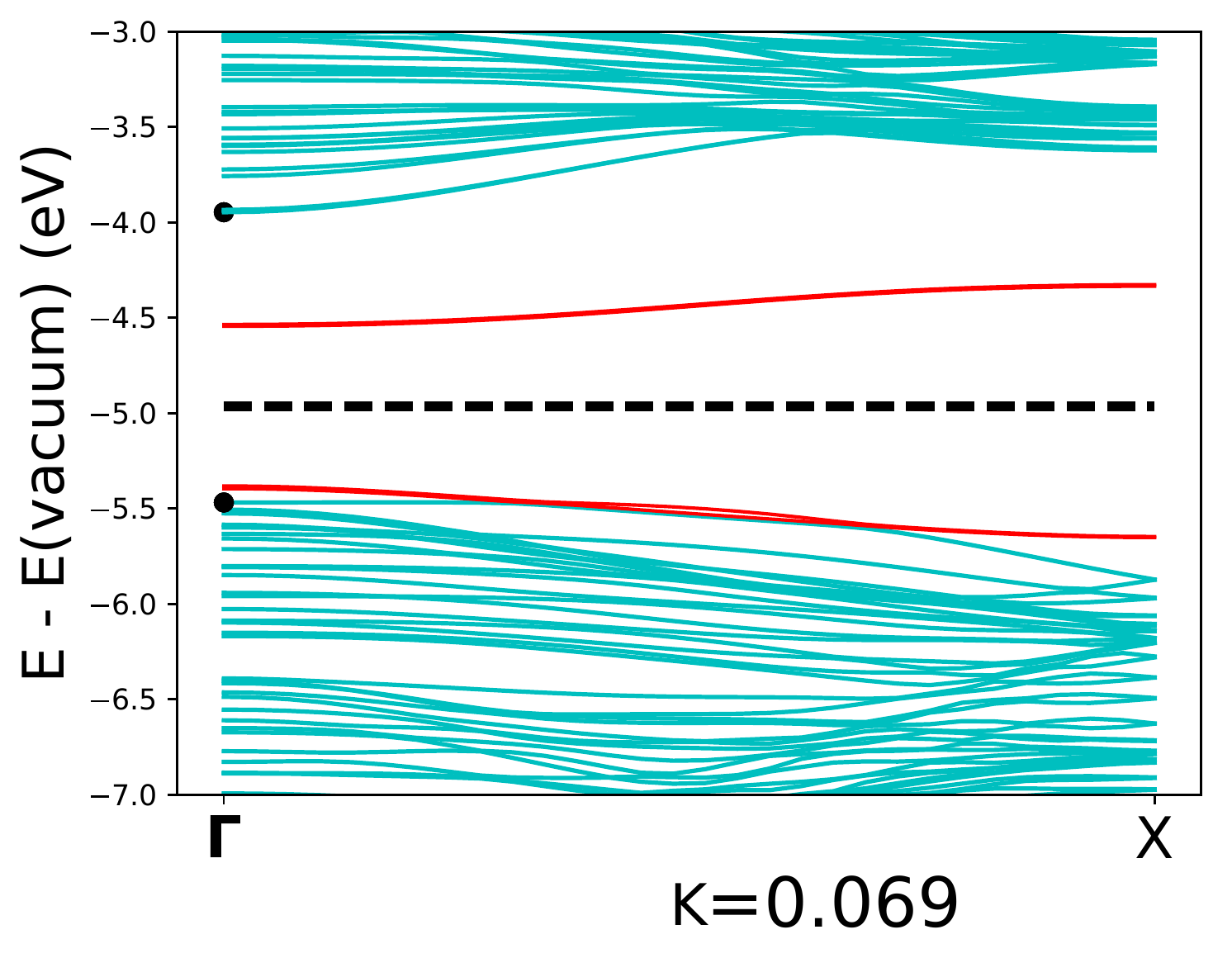}
 	\includegraphics[height=1in, width=1.5in]{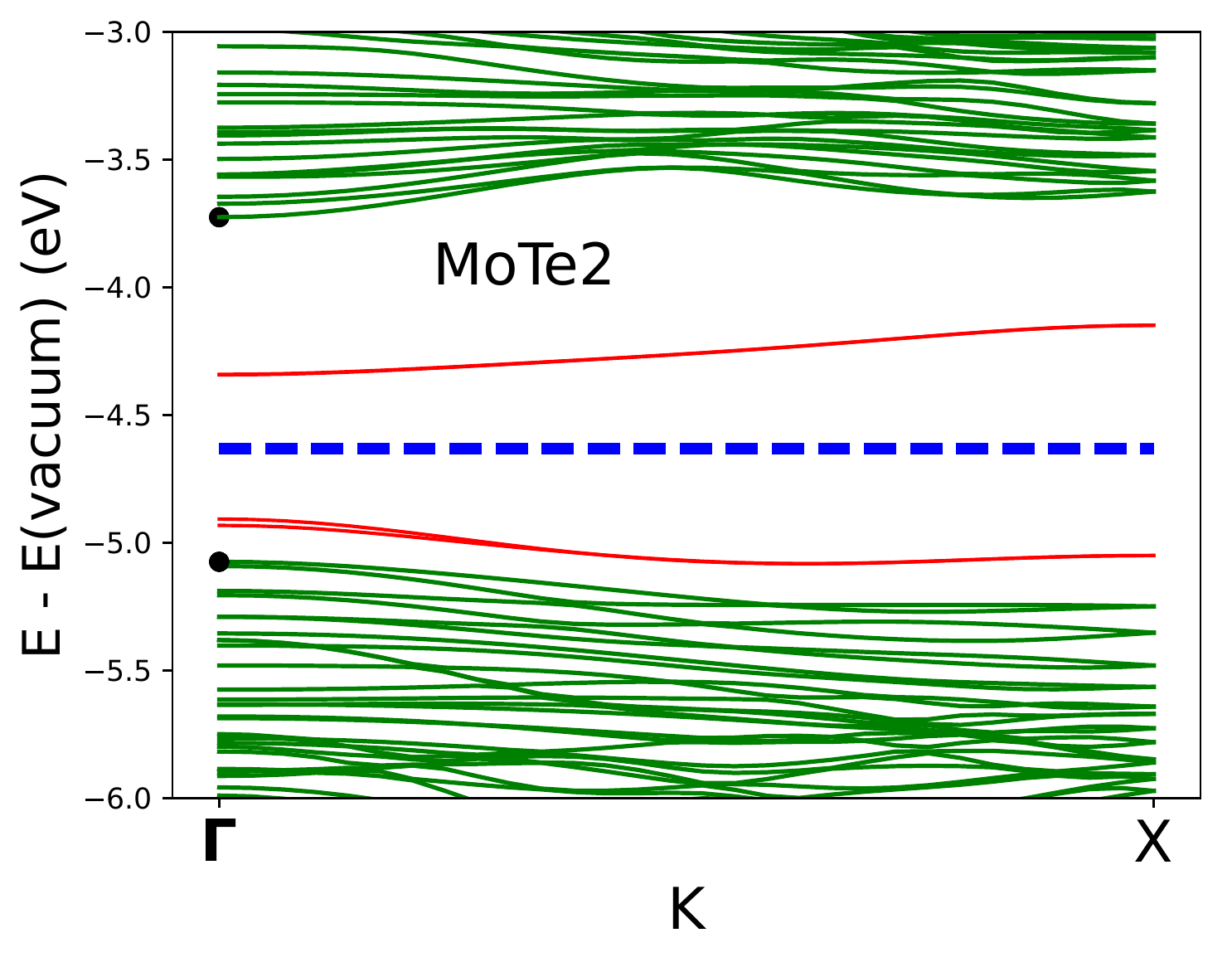}
 	\includegraphics[height=1in, width=1.5in]{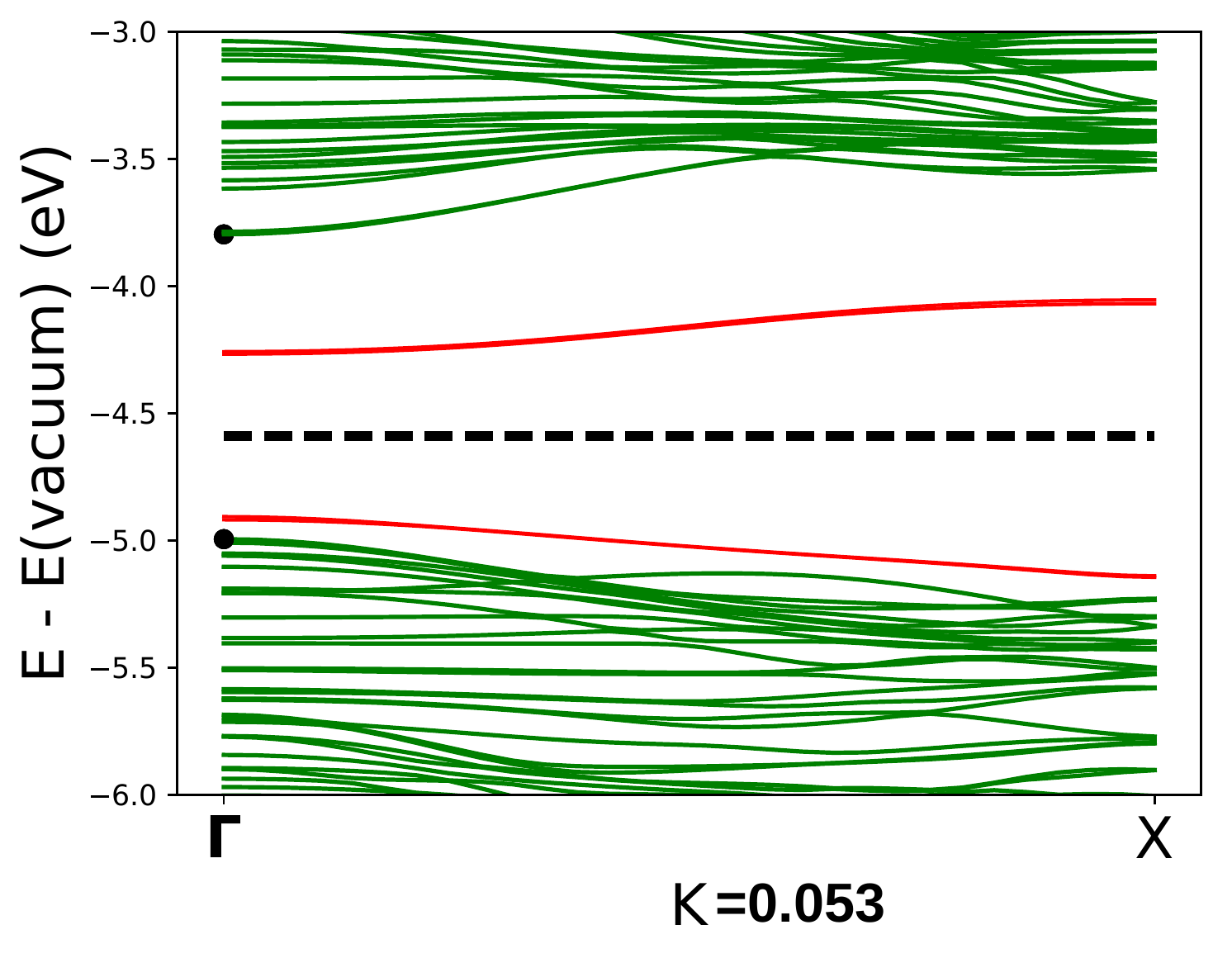}
 	\includegraphics[height=1in, width=1.5in]{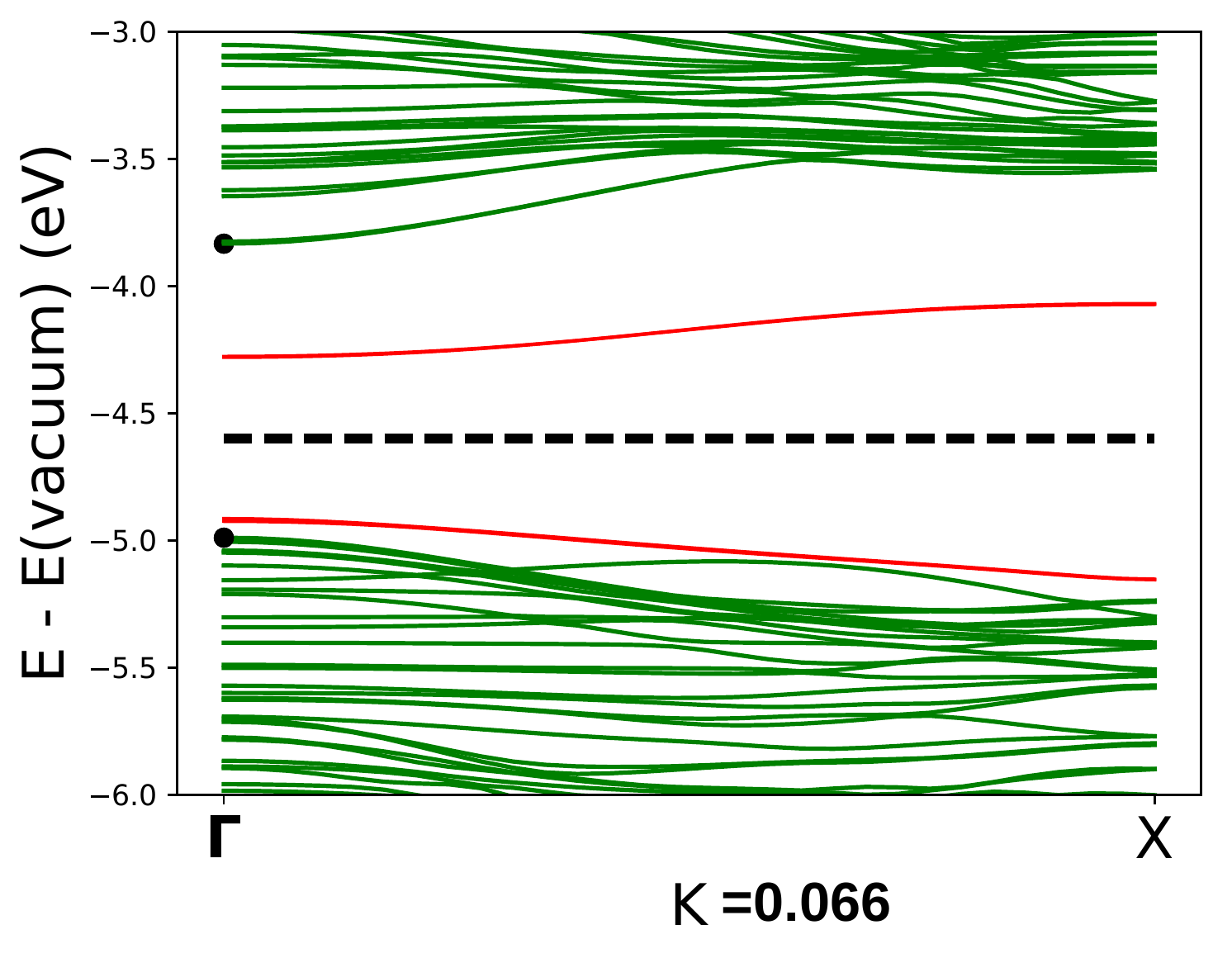}
 	\includegraphics[height=1in, width=1.5in]{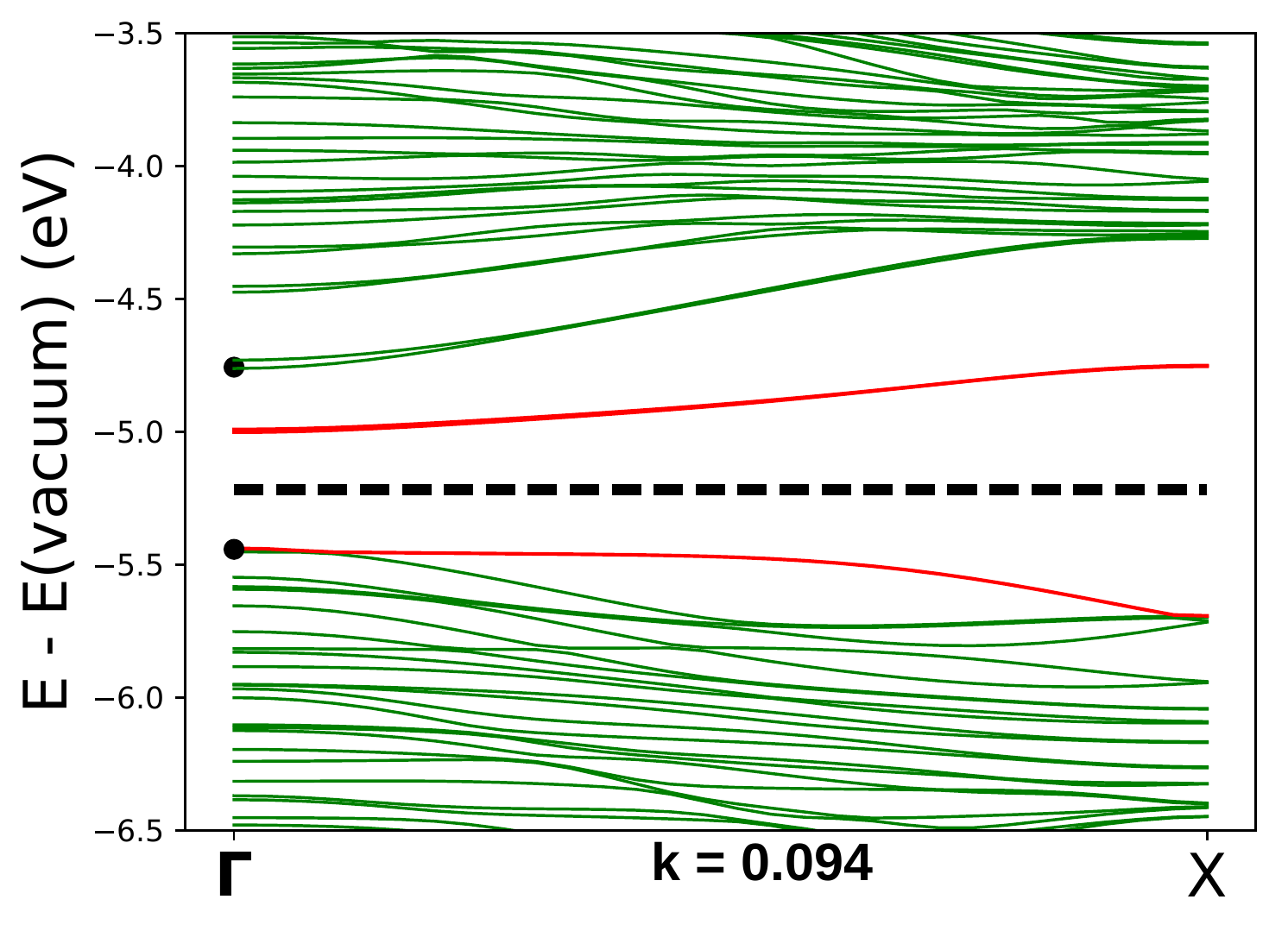}
 	\includegraphics[height=1in, width=1.5in]{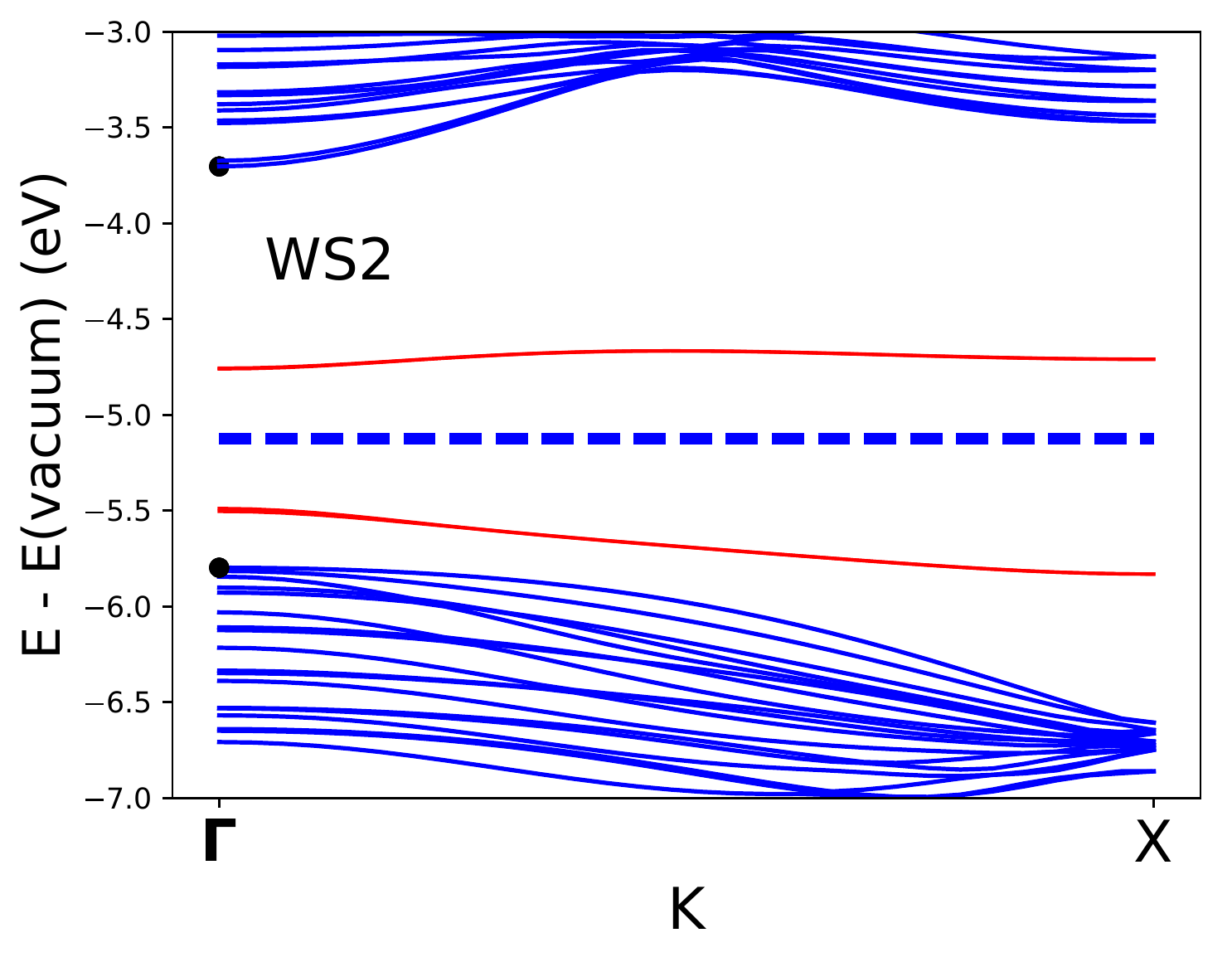}
 	\includegraphics[height=1in, width=1.5in]{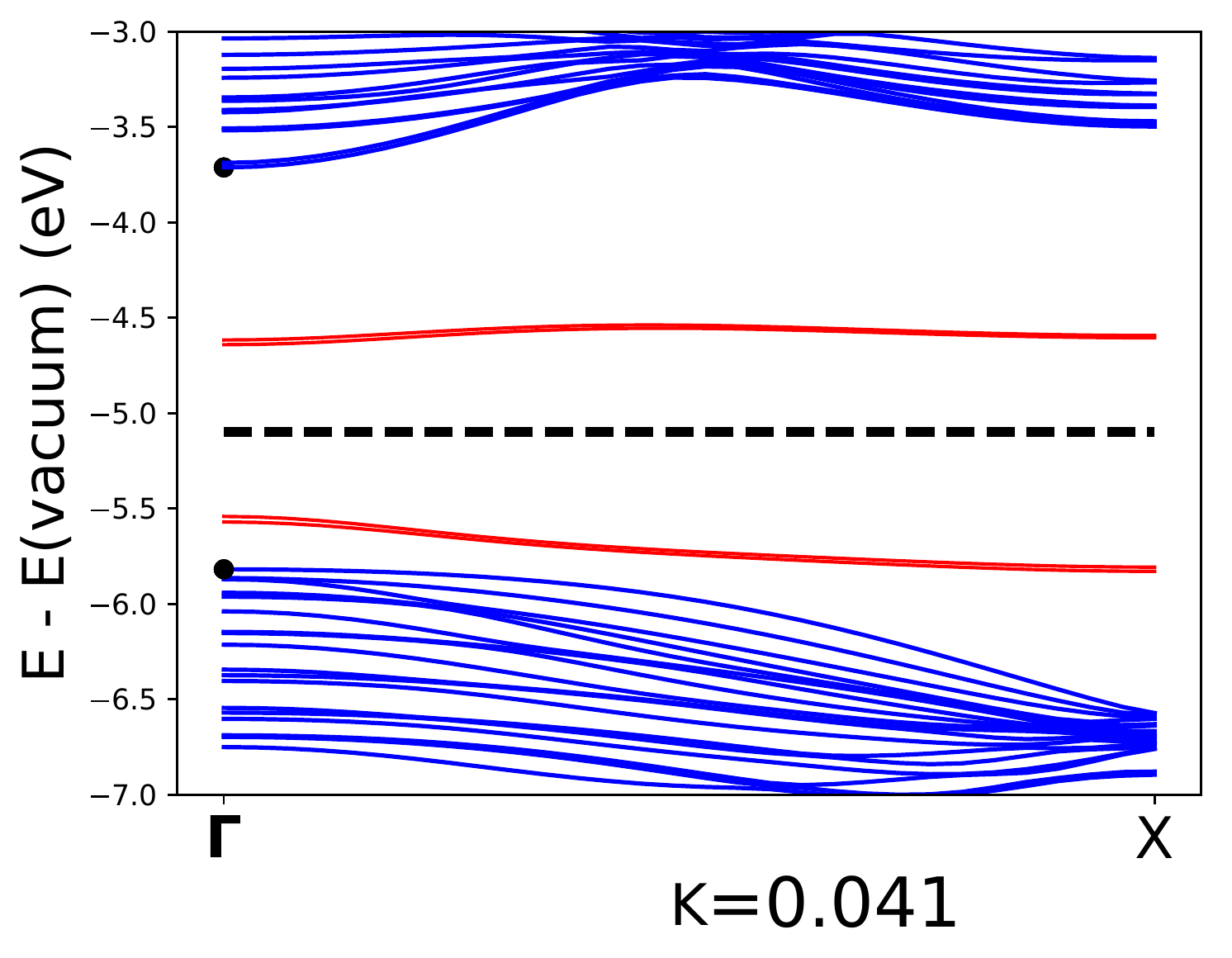}
 	\includegraphics[height=1in, width=1.5in]{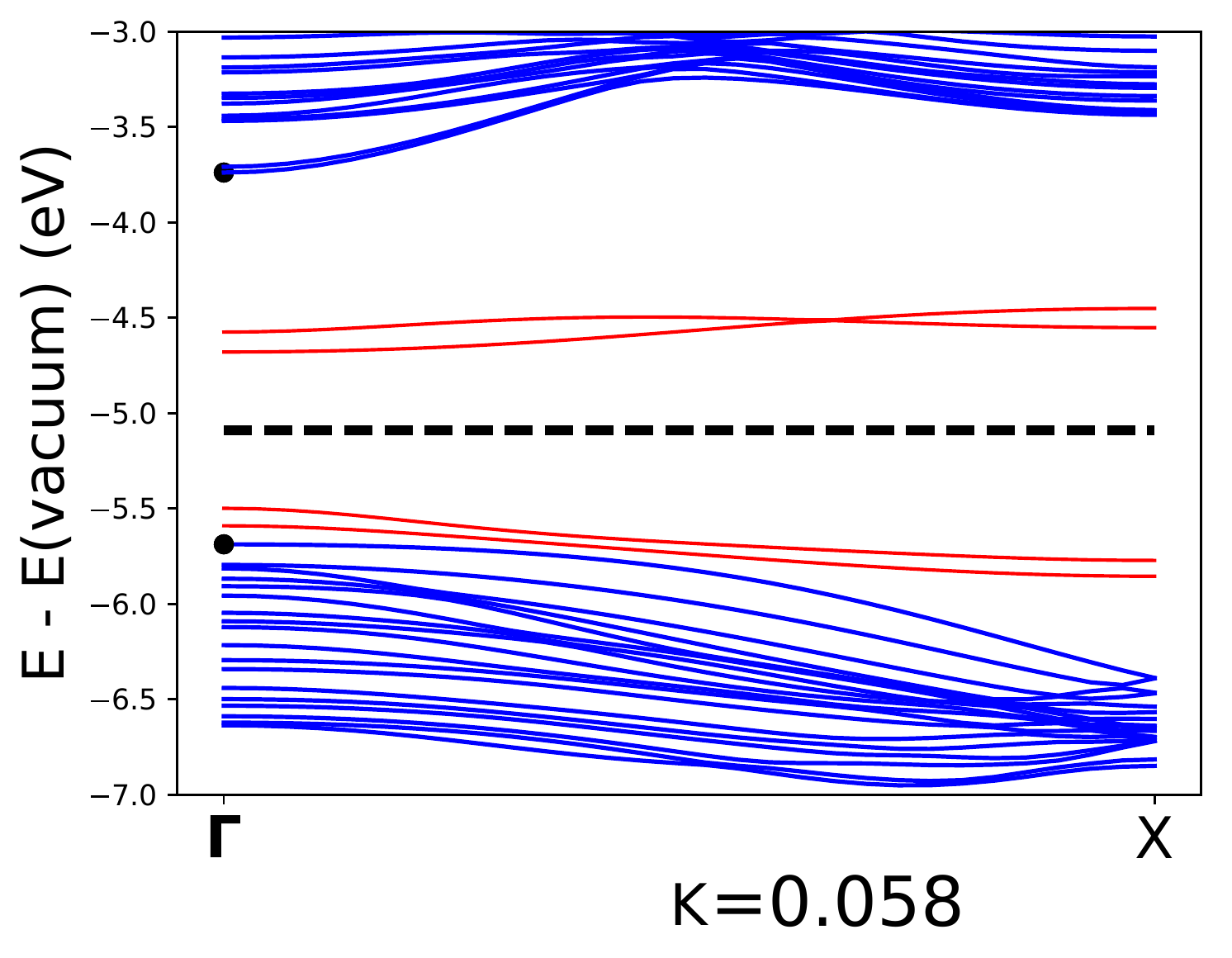}
 	\includegraphics[height=1in, width=1.5in]{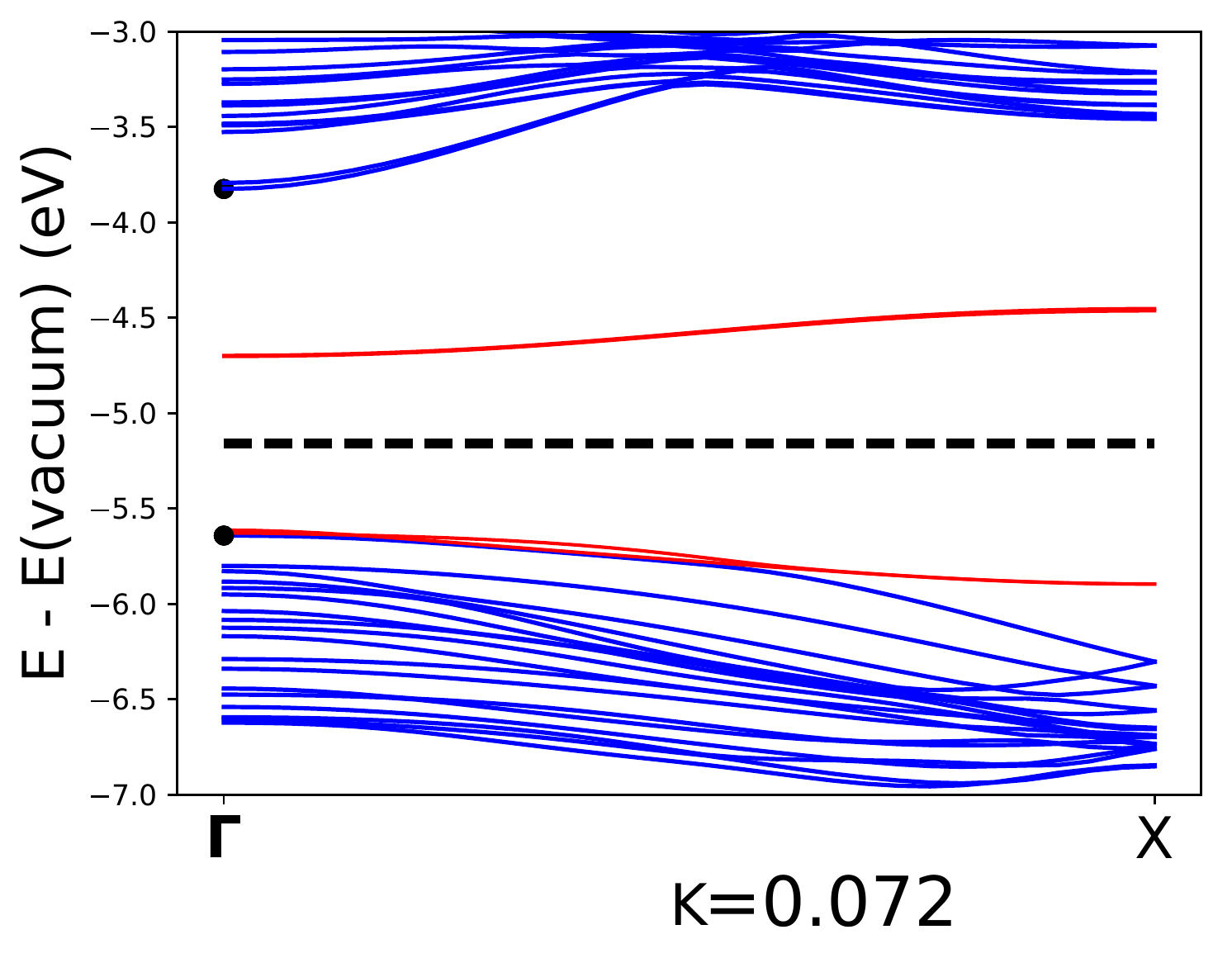}
 	\includegraphics[height=1in, width=1.5in]{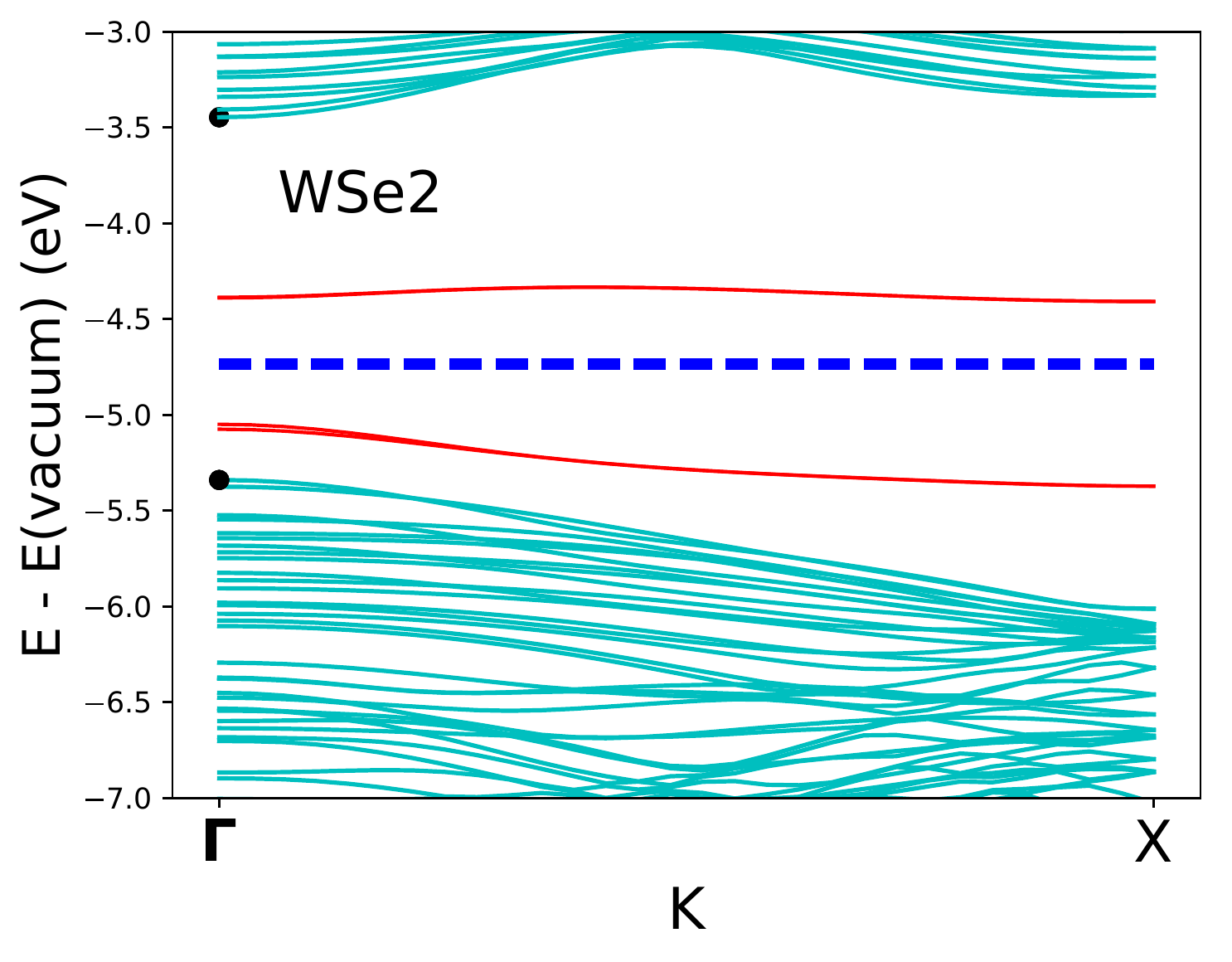}
 	\includegraphics[height=1in, width=1.5in]{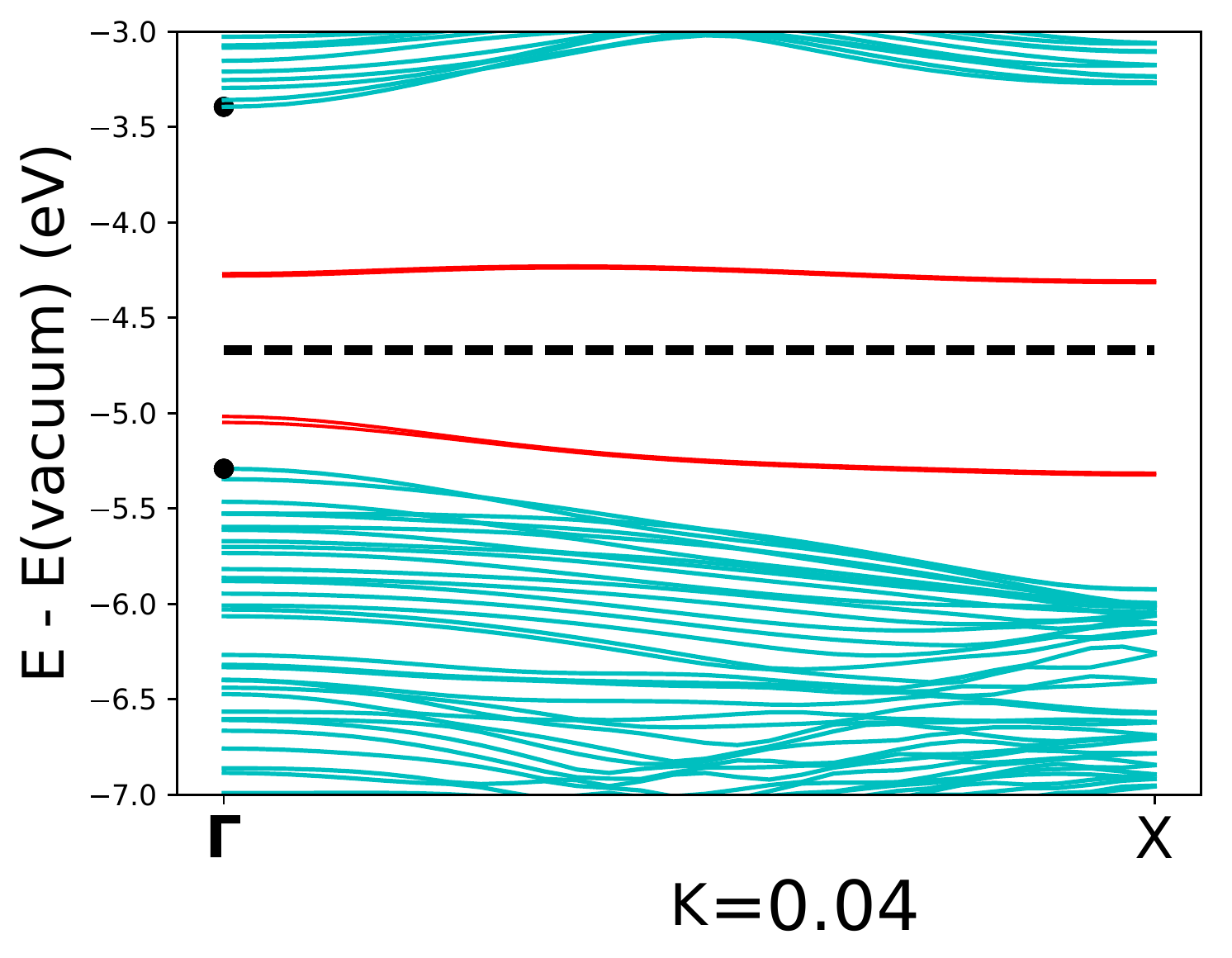}
 	\includegraphics[height=1in, width=1.5in]{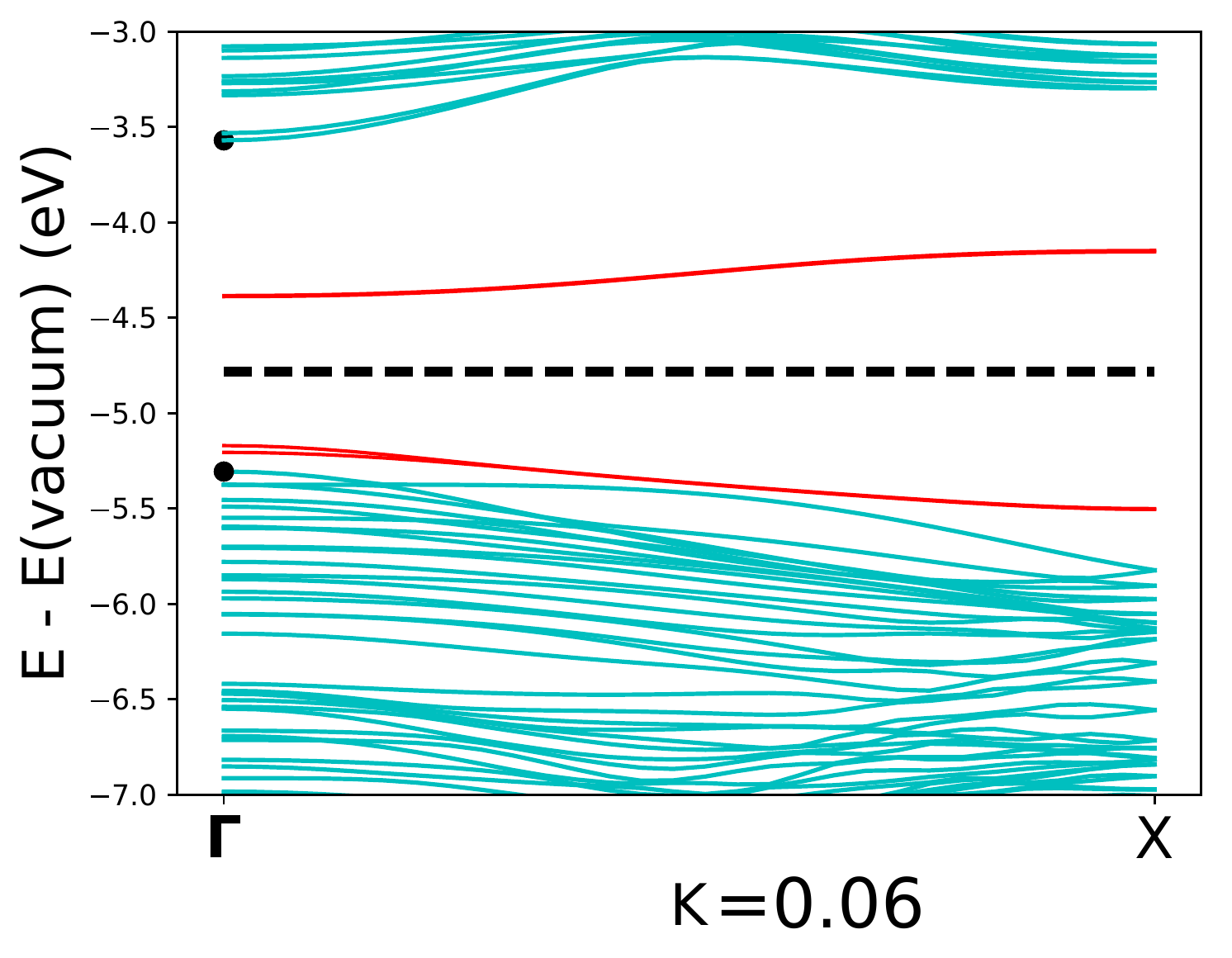}
 	\includegraphics[height=1in, width=1.5in]{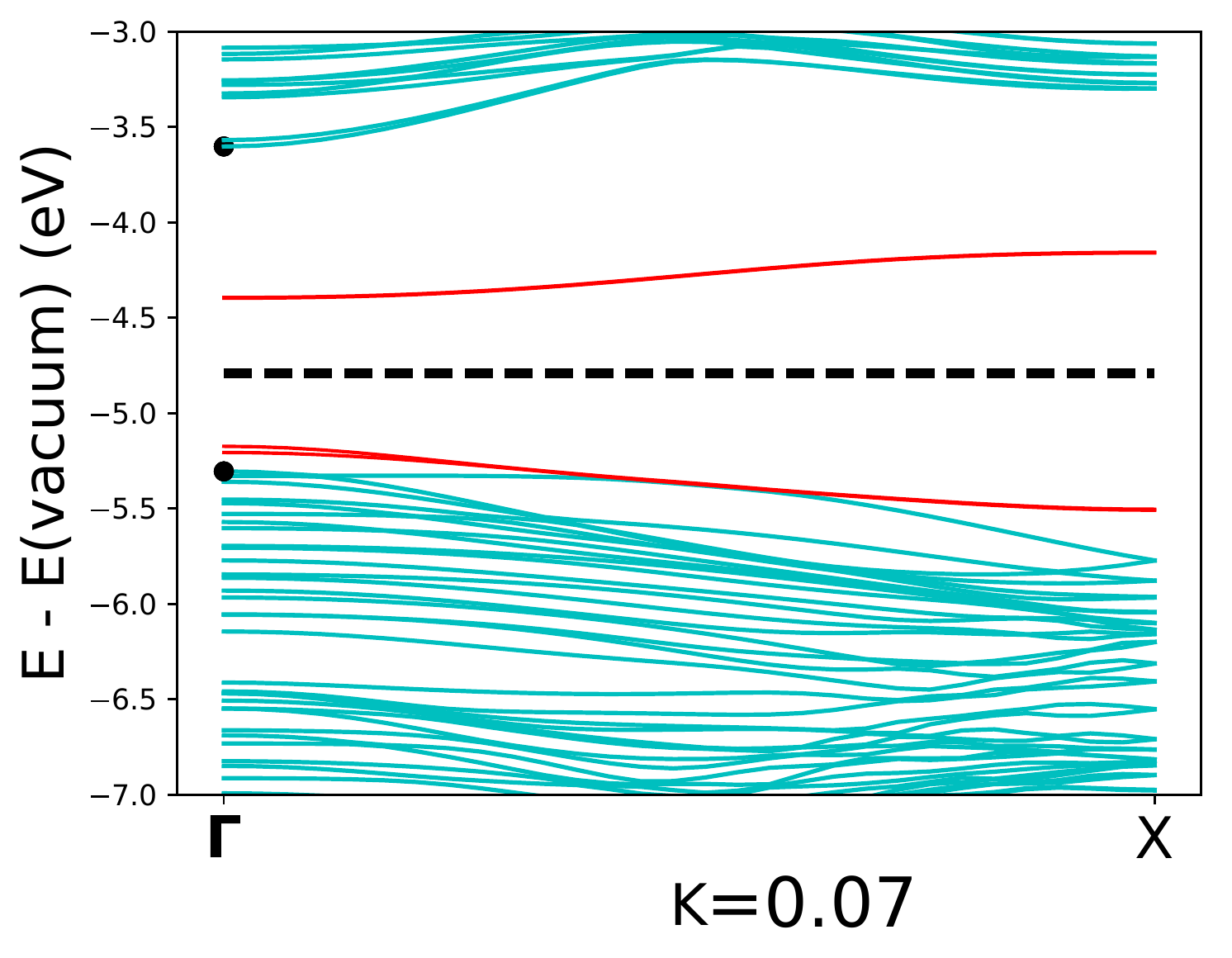}
 	\includegraphics[height=1in, width=1.5in]{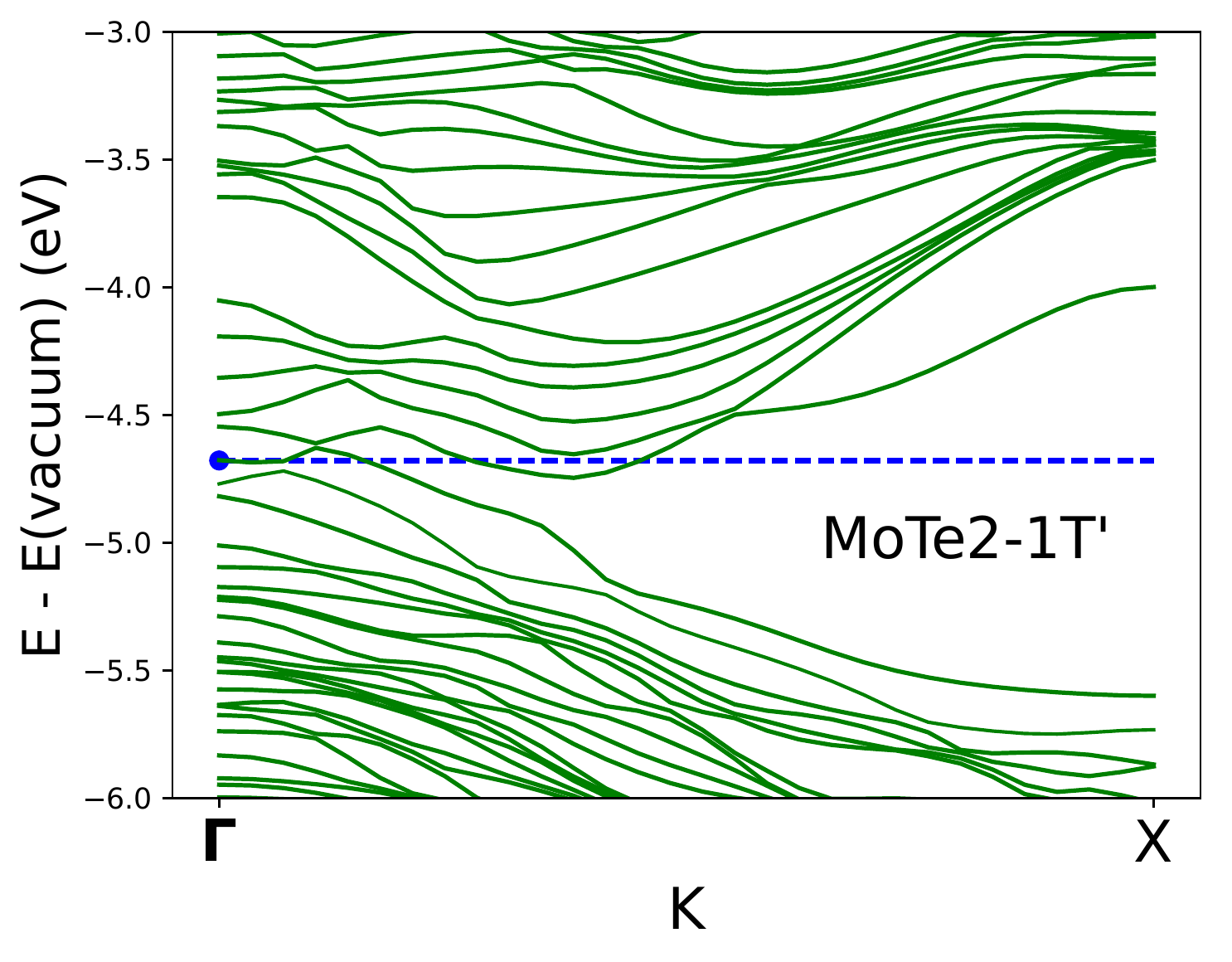}
 	\includegraphics[height=1in, width=1.5in]{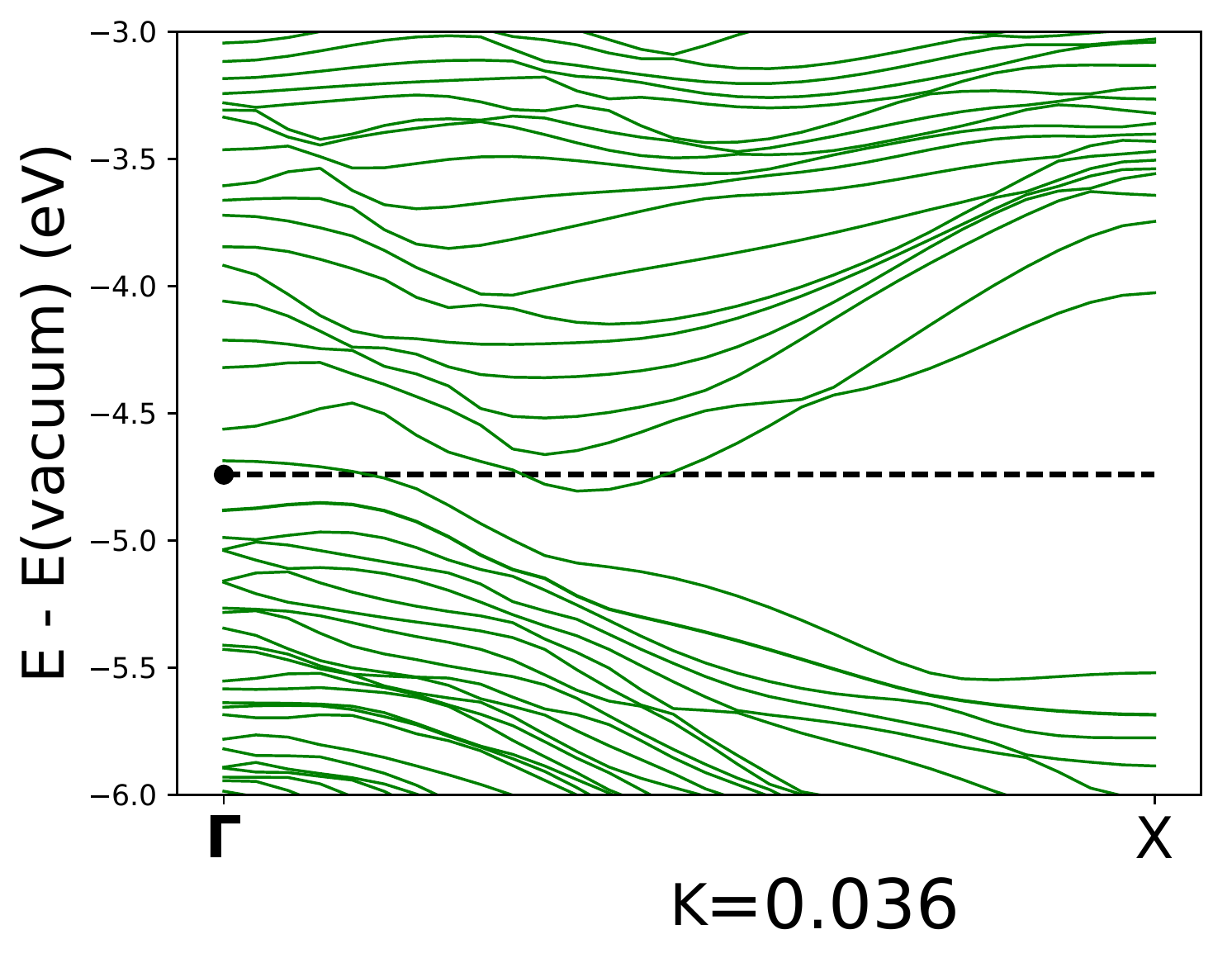}
 	\includegraphics[height=1in, width=1.5in]{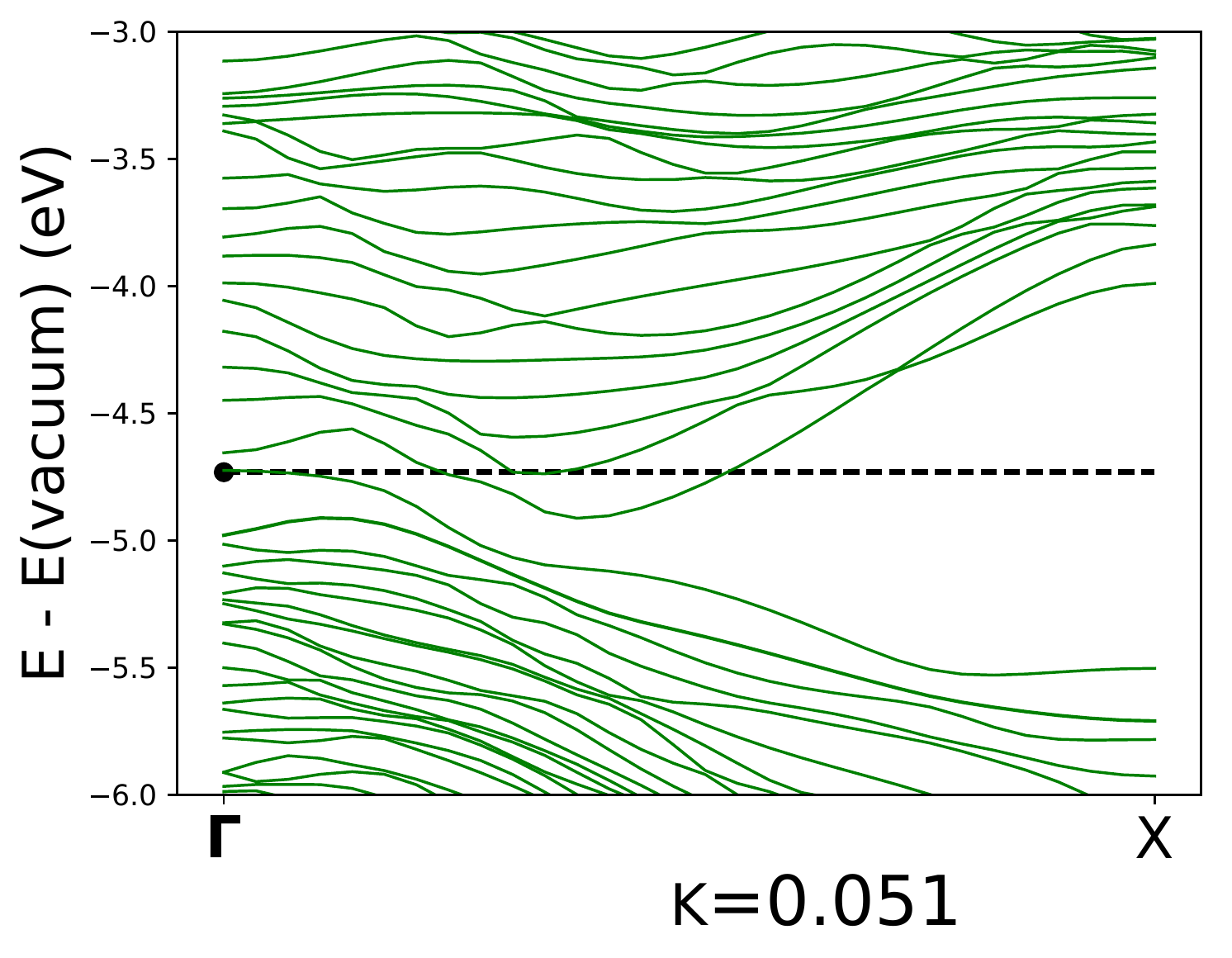}
 	\includegraphics[height=1in, width=1.5in]{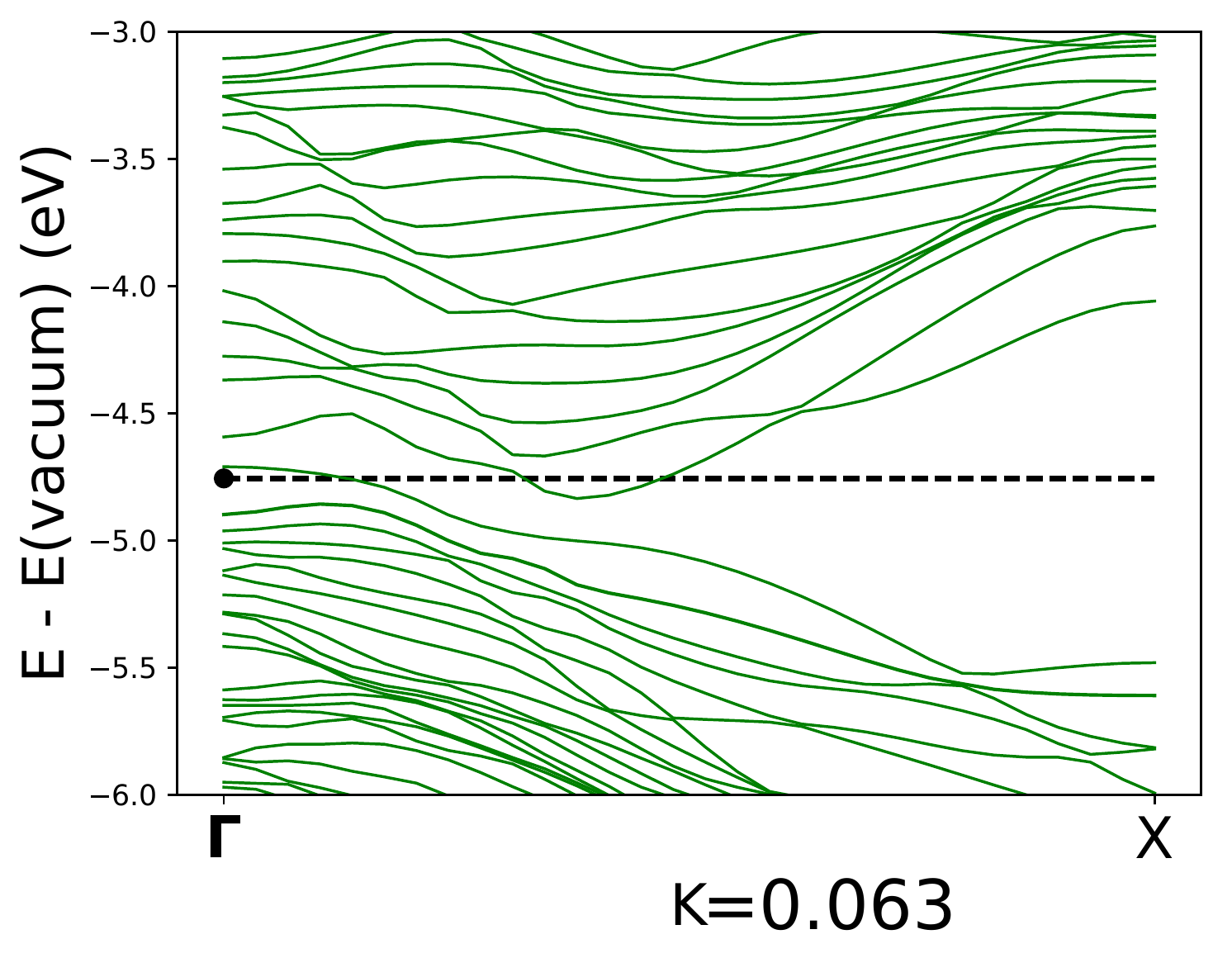}
 	\includegraphics[height=1in, width=1.5in]{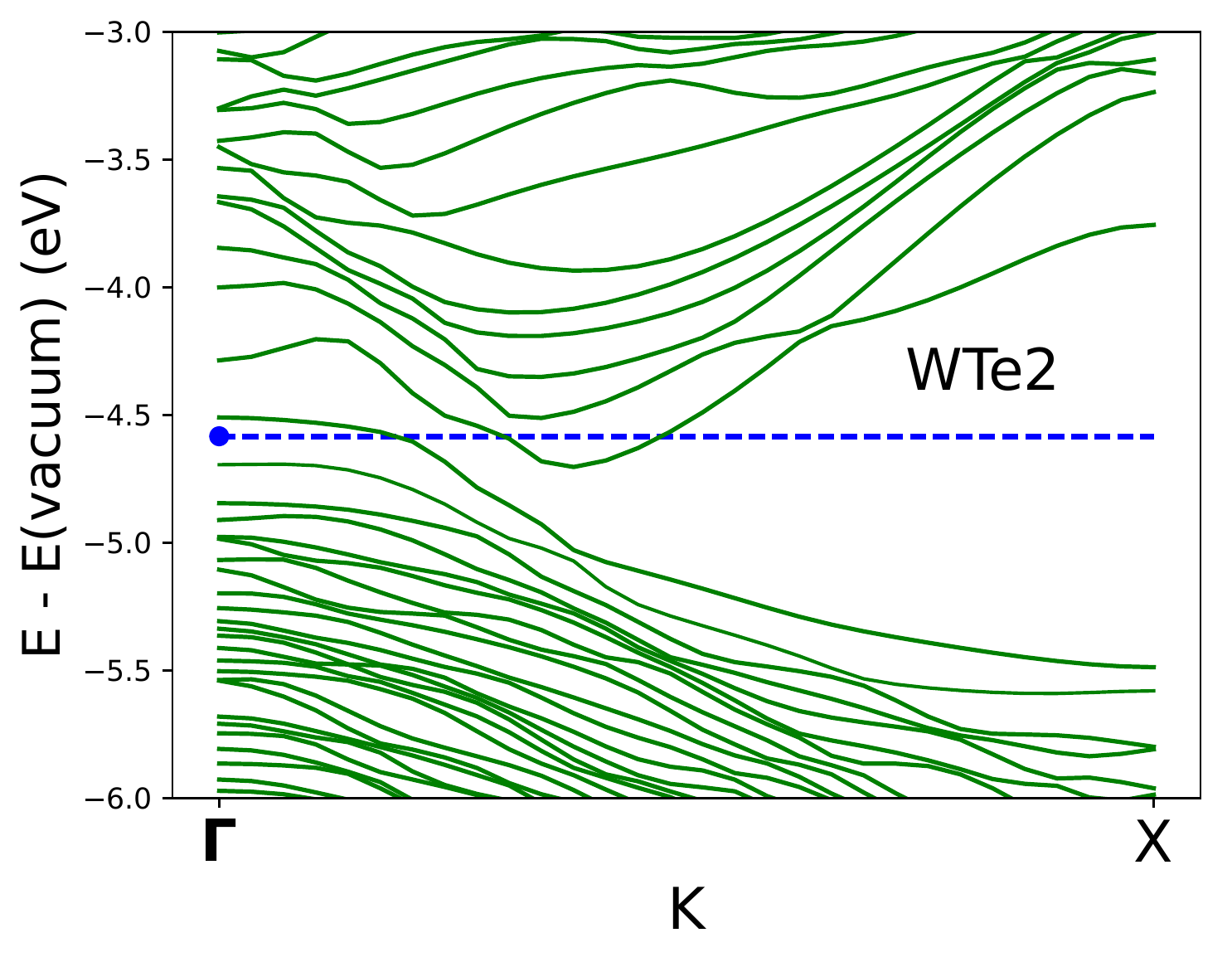}
 	\includegraphics[height=1in, width=1.5in]{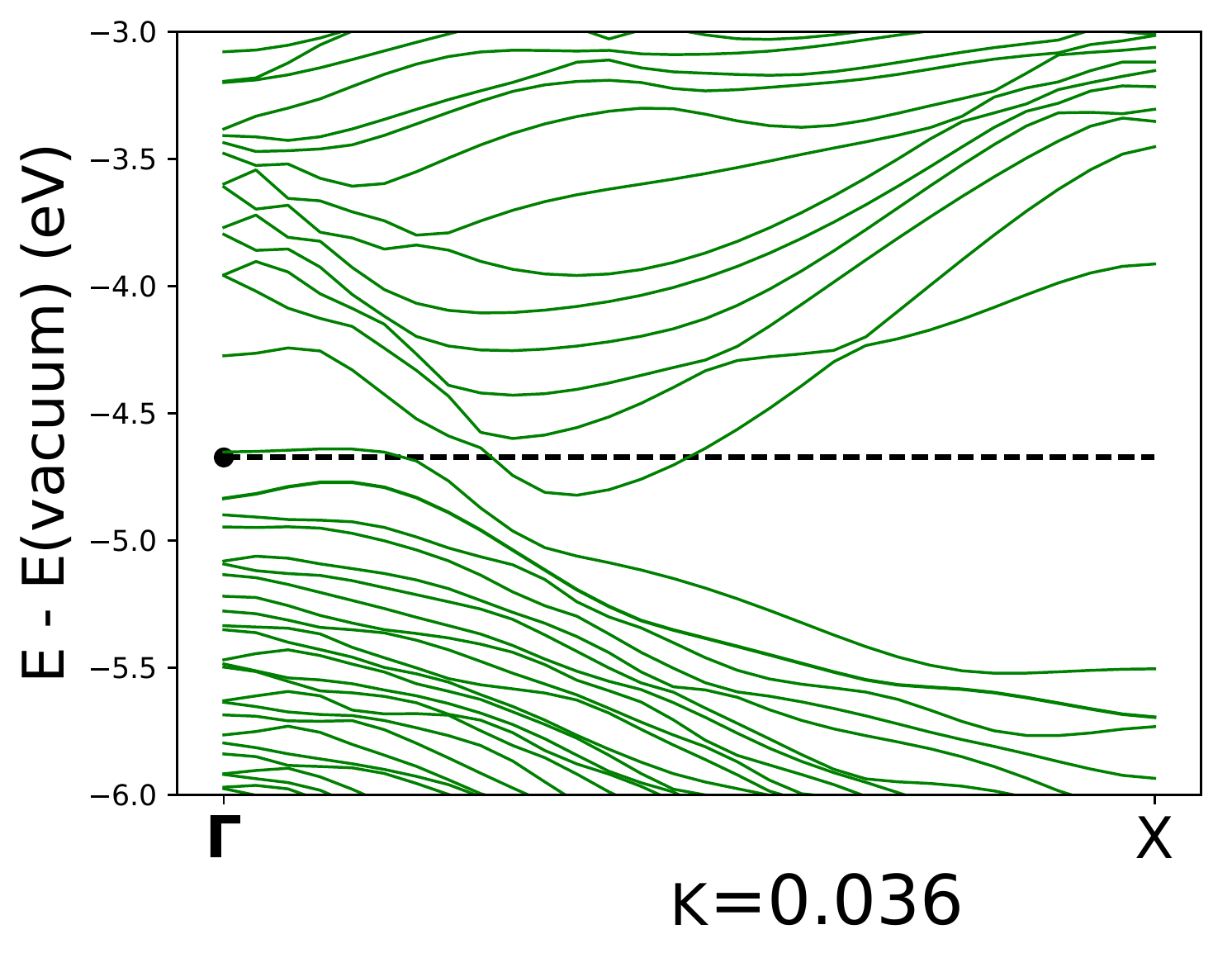}
 	\includegraphics[height=1in, width=1.5in]{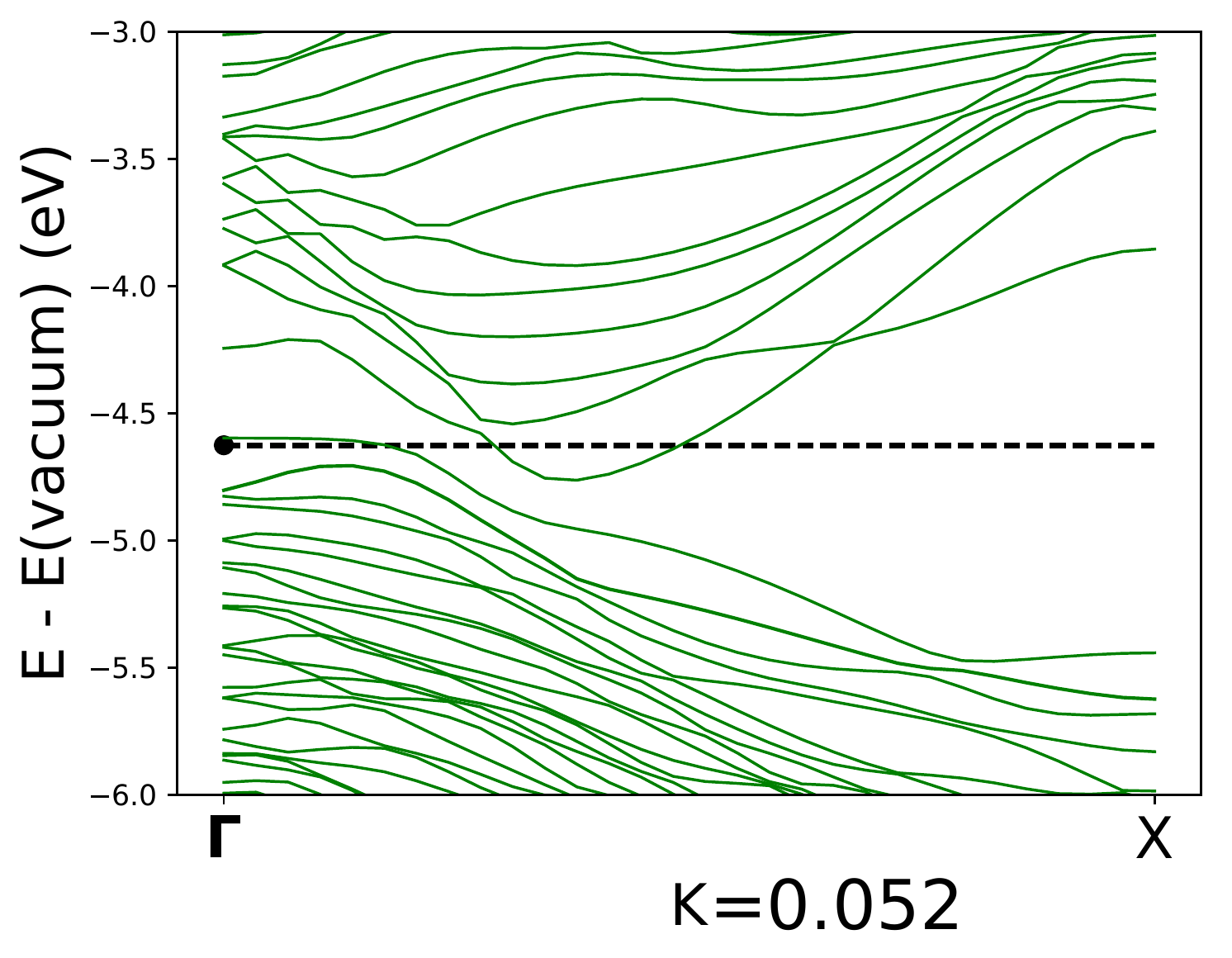}
 	\includegraphics[height=1in, width=1.5in]{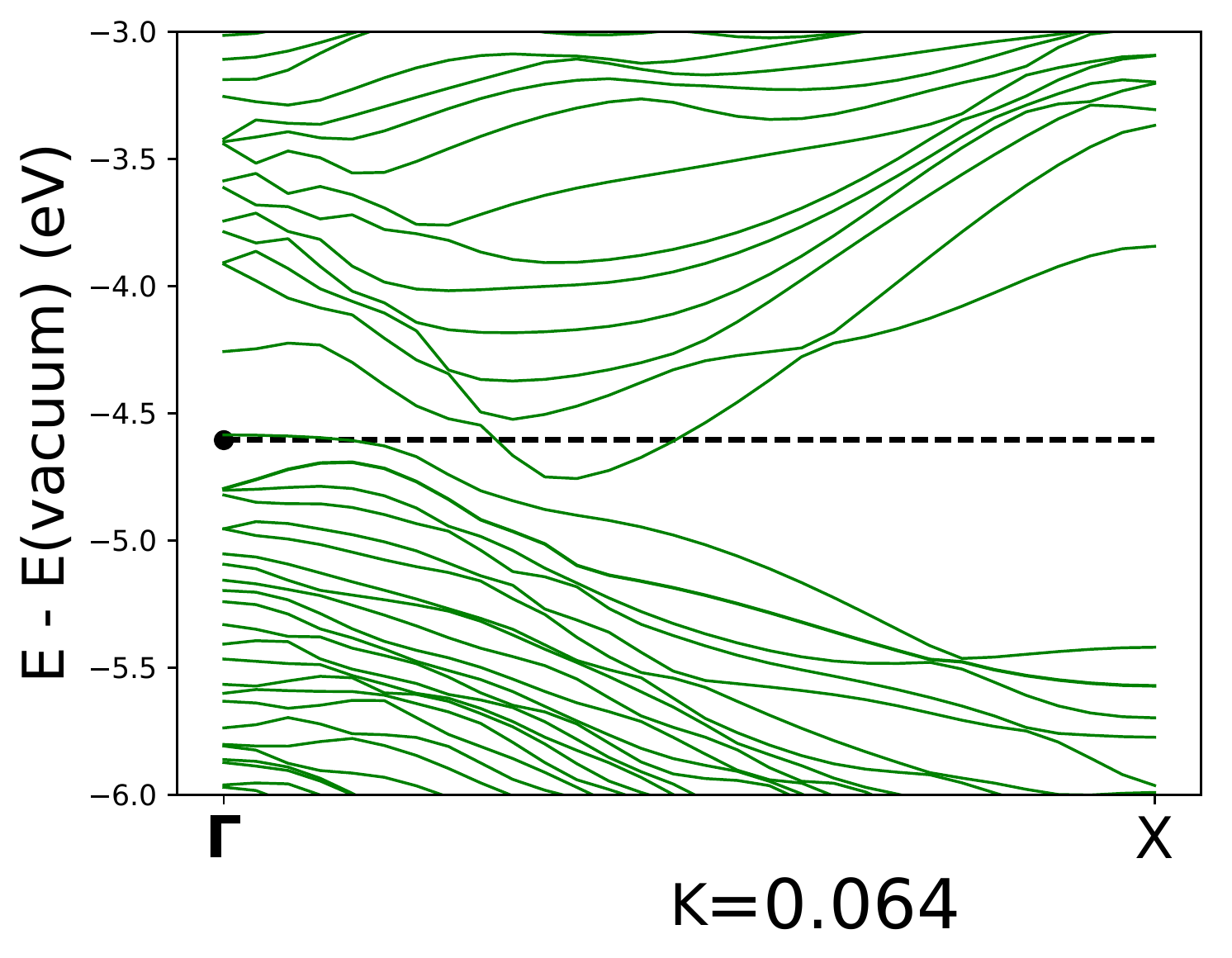}
 	\caption{Band structures with respect to vacuum for group VI TMDs; \textbf{MoS$_2$}, \textbf{MoSe$_2$}, \textbf{MoTe$_2$}, \textbf{MoTe$_2$-1T$^\prime$}, \textbf{WS$_2$}, \textbf{WSe$_2$}, and \textbf{WTe$_2$}. $\kappa$ is the bending curvature ($\AA^{-1}$).}
 	\label{fig:band-VI}
 \end{figure}

 \begin{figure}[h!]
 	\renewcommand\thefigure{S8}
 	\includegraphics[height=1in, width=1.5in]{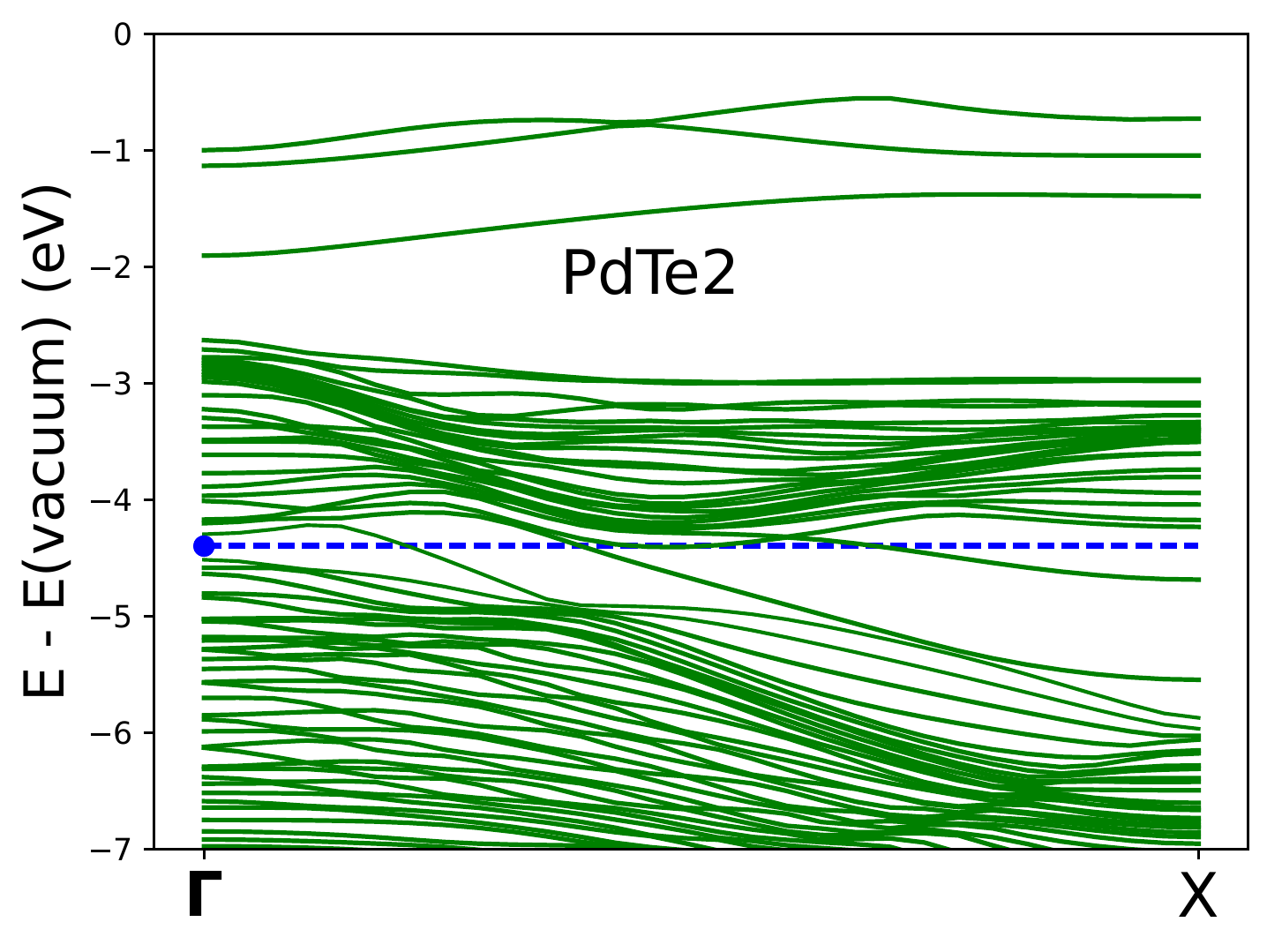}
 	\includegraphics[height=1in, width=1.5in]{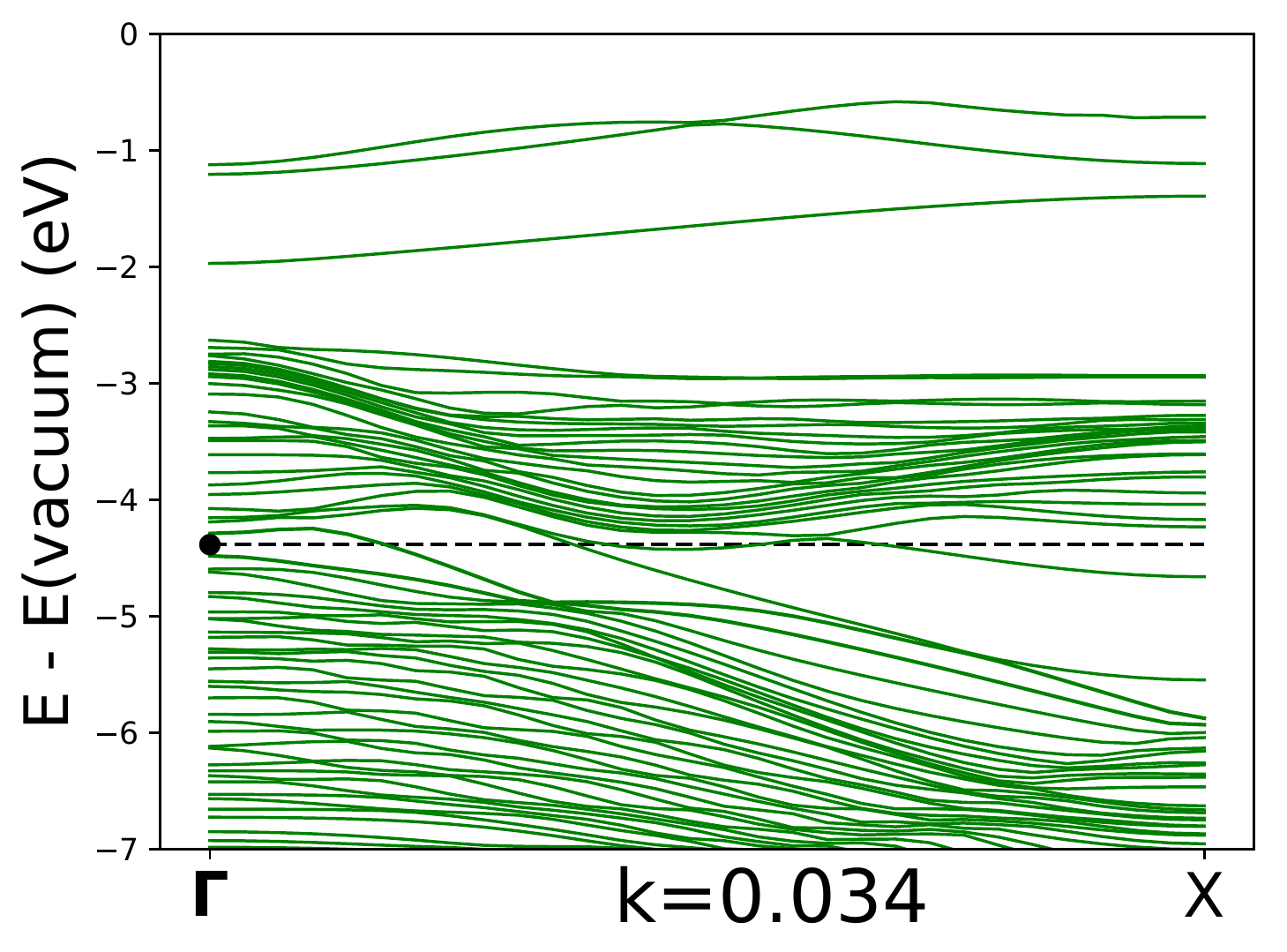}
 	\includegraphics[height=1in, width=1.5in]{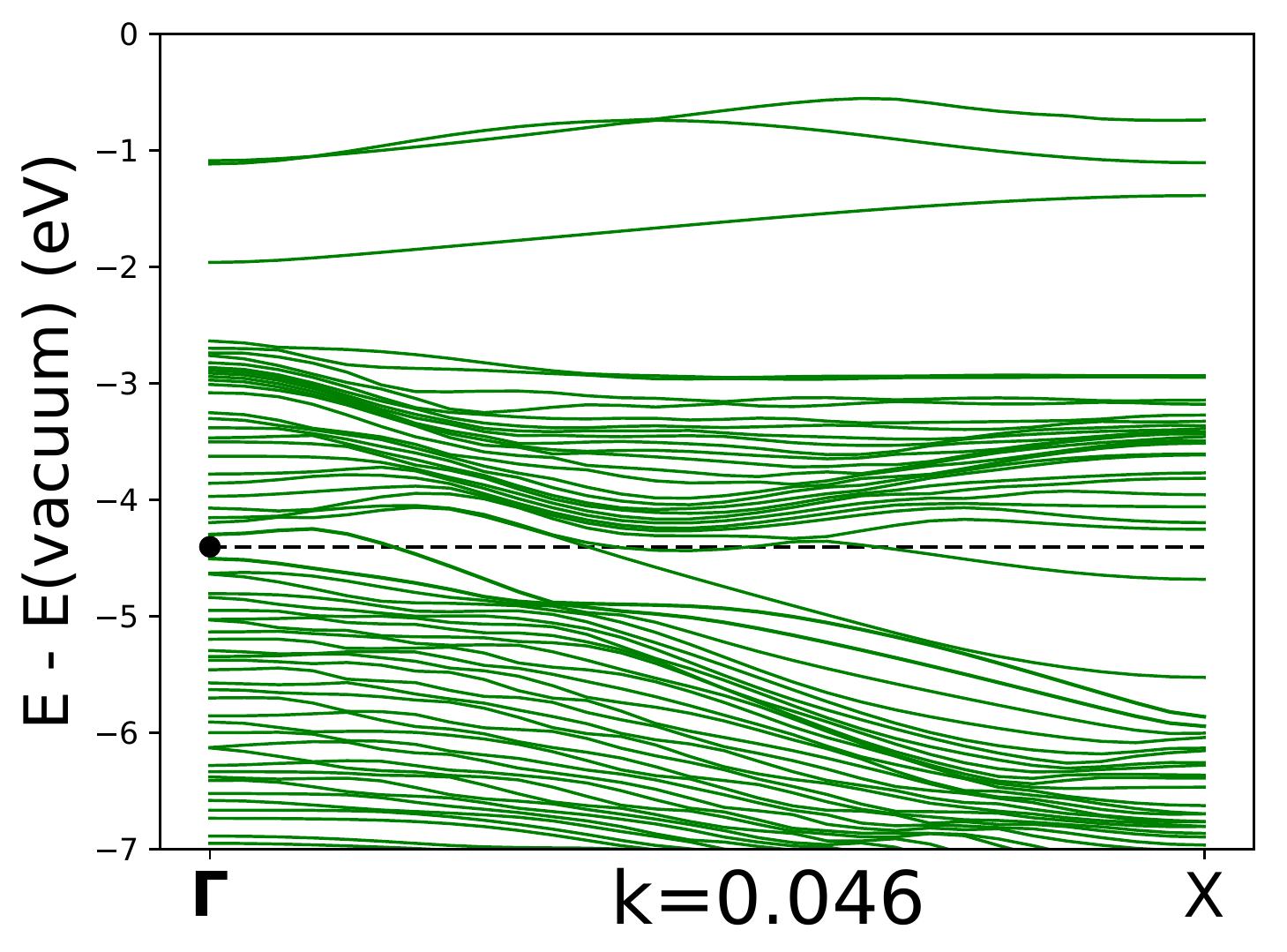}
 	\includegraphics[height=1in, width=1.5in]{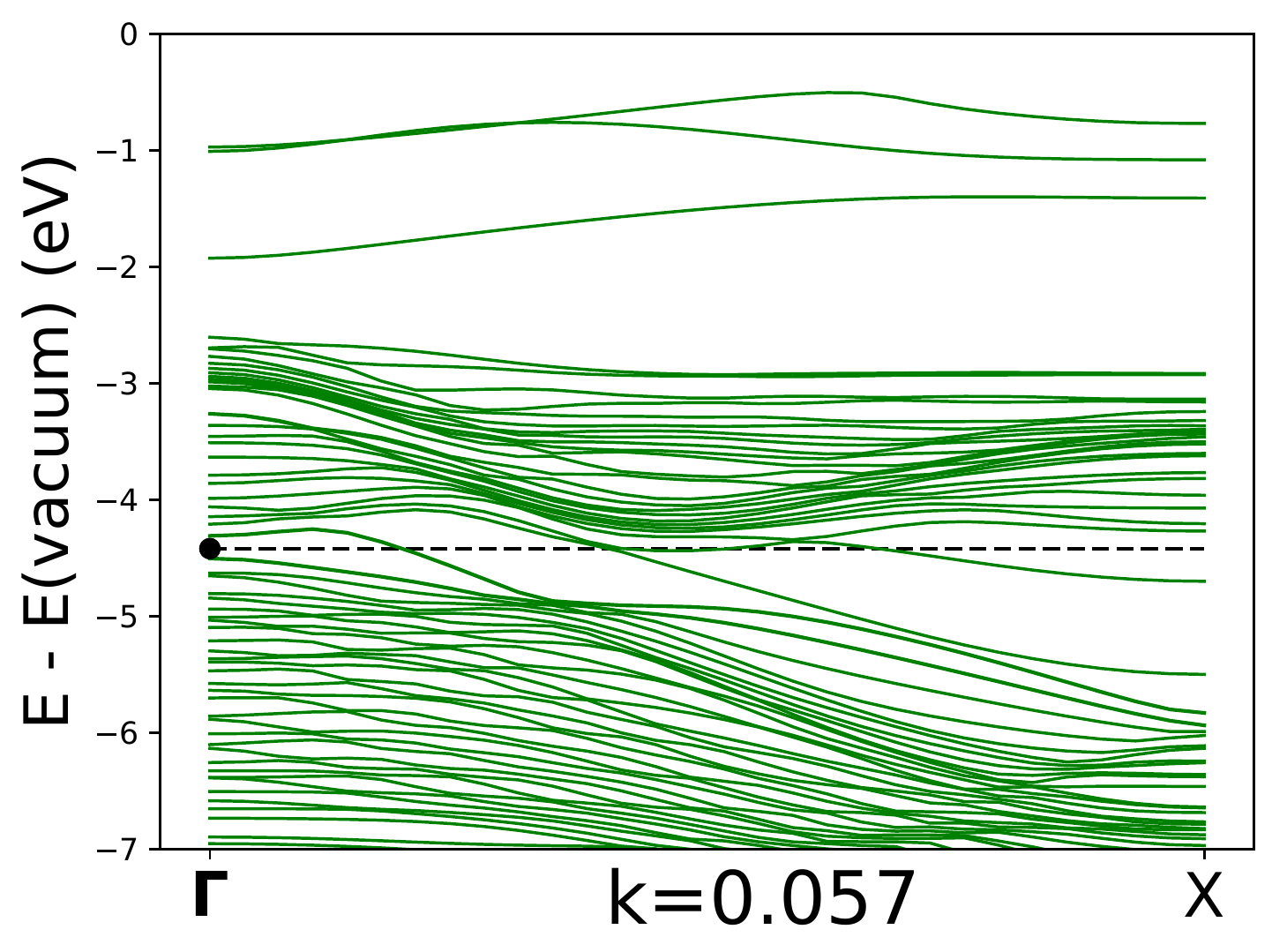}
 	\includegraphics[height=1in, width=1.5in]{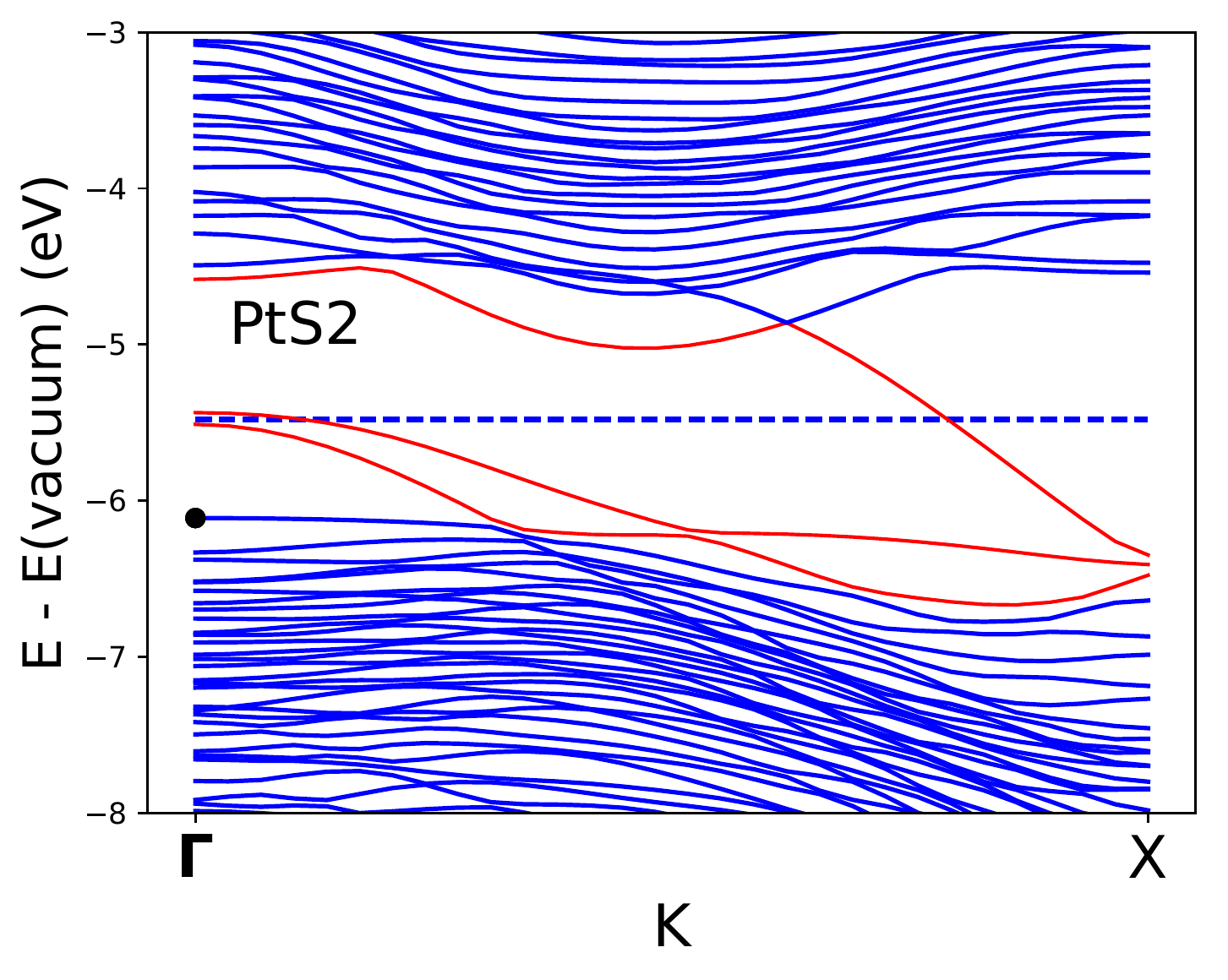}
 	\includegraphics[height=1in, width=1.5in]{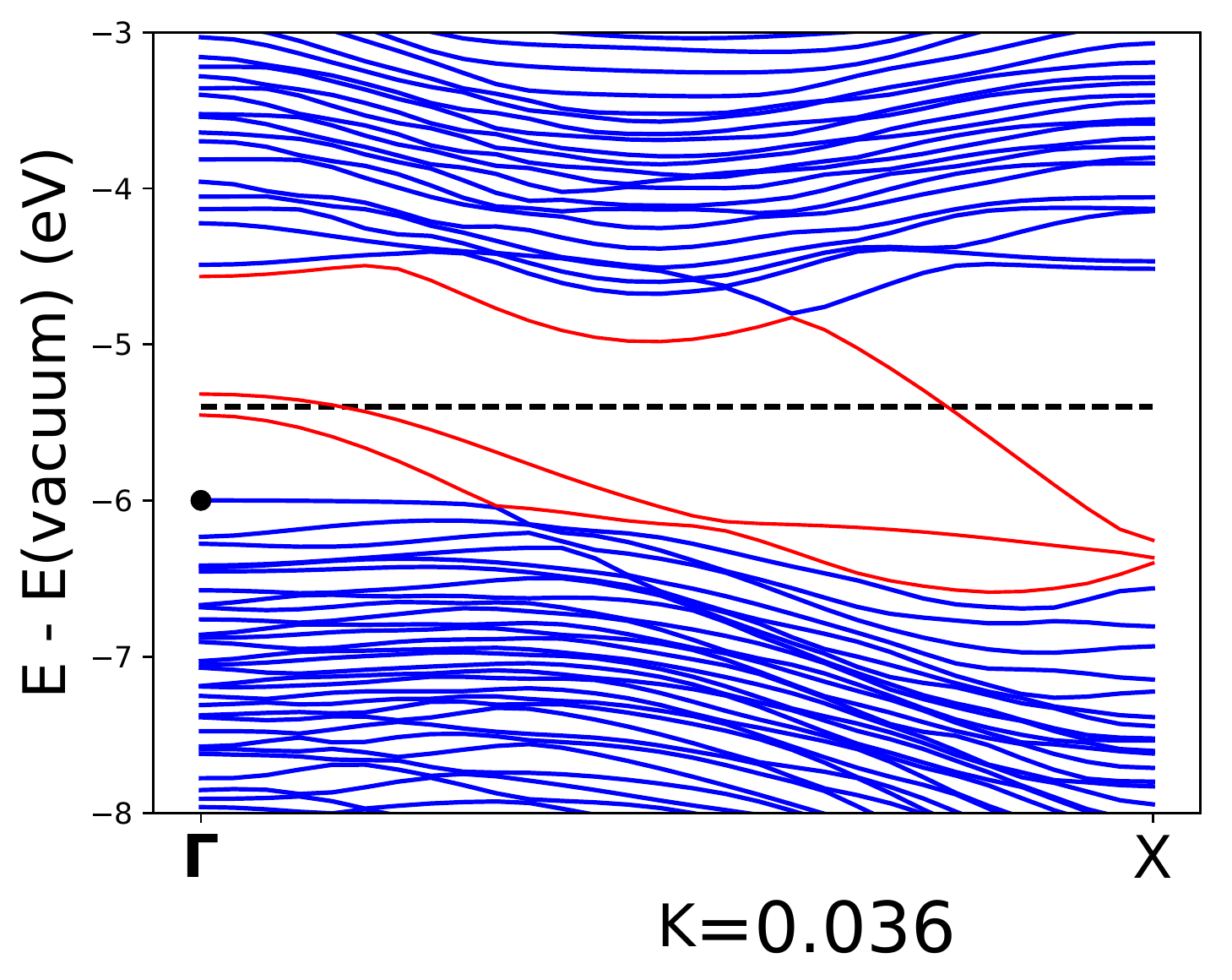}
 	\includegraphics[height=1in, width=1.5in]{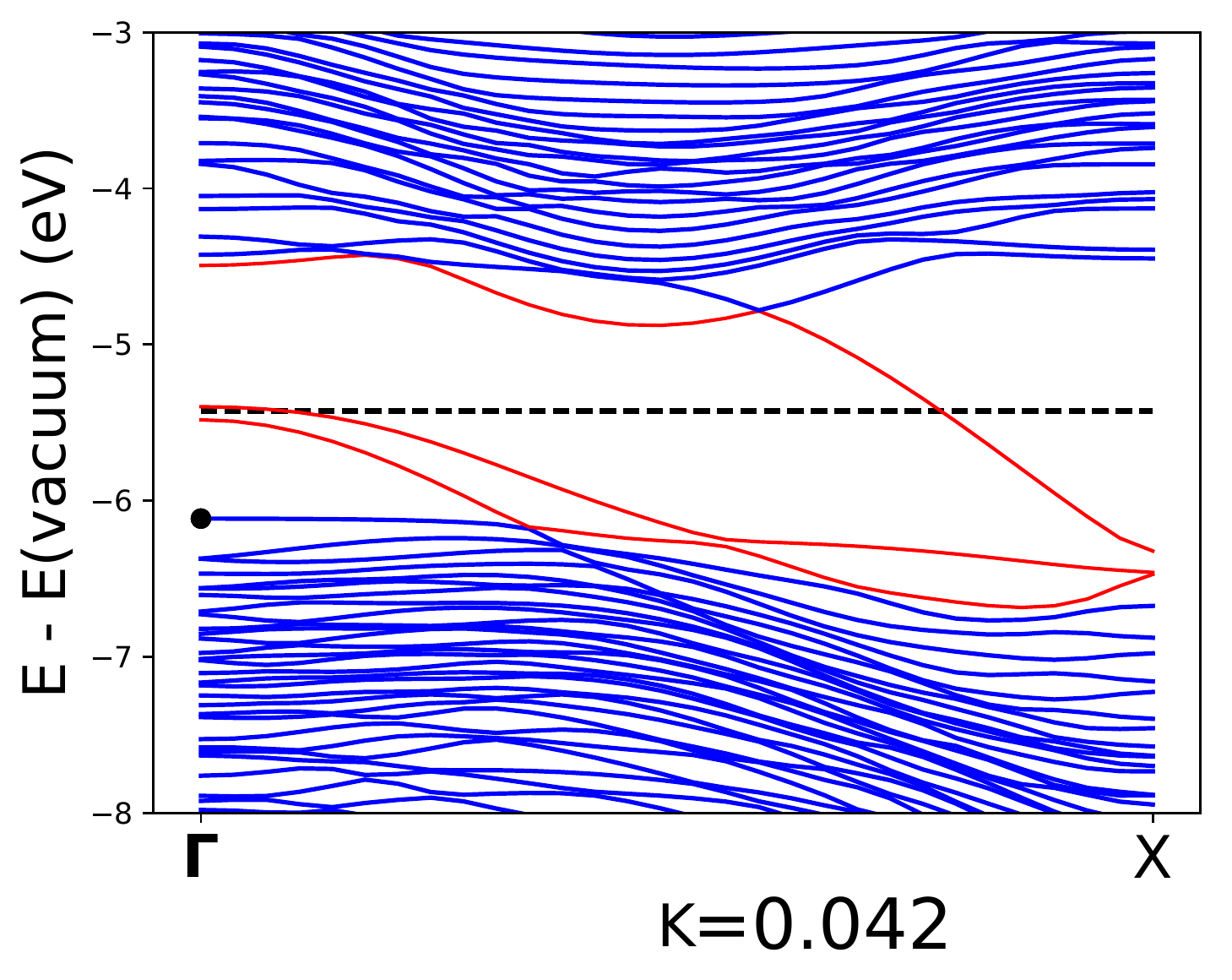}
 	\includegraphics[height=1in, width=1.5in]{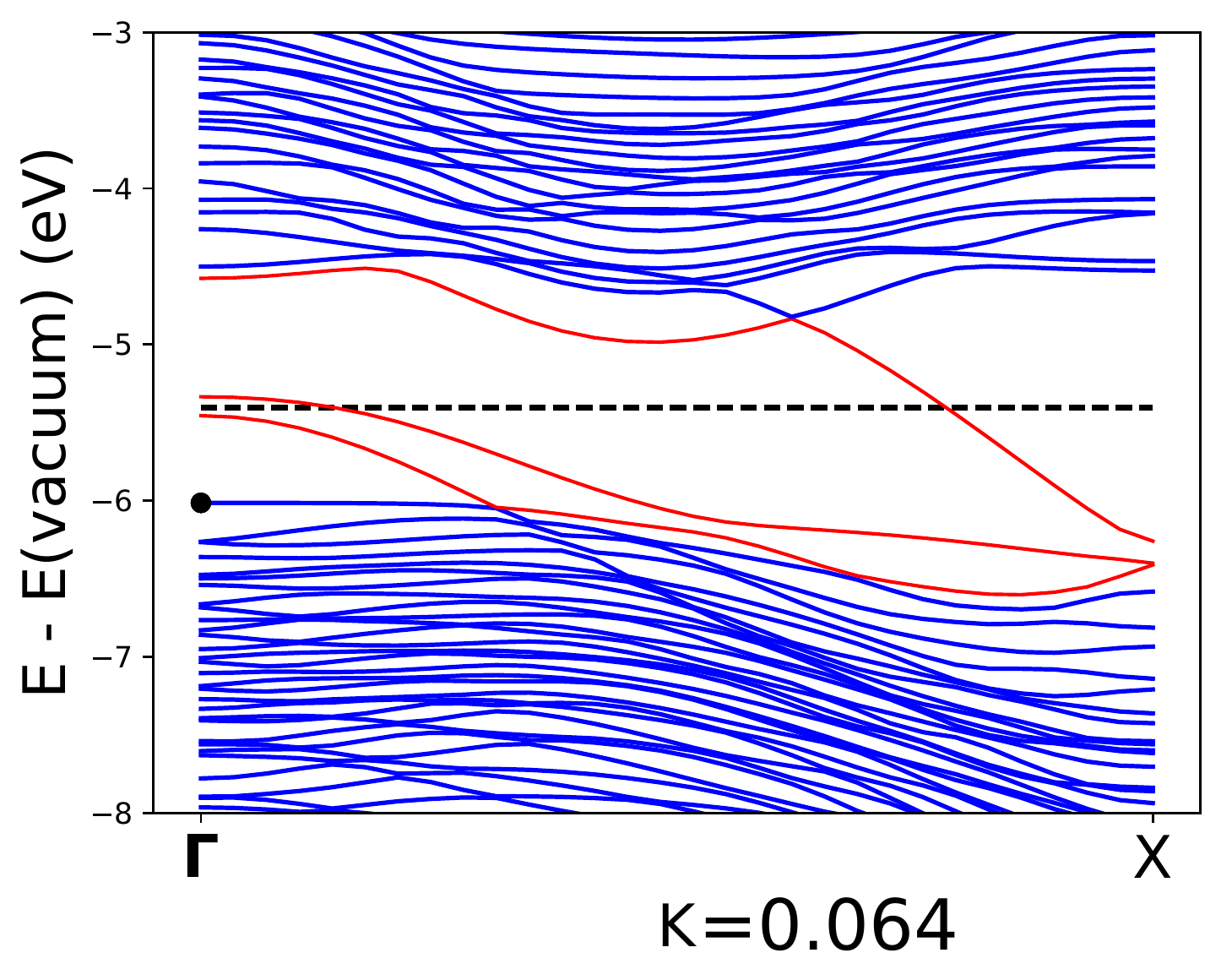}
 	\includegraphics[height=1in, width=1.5in]{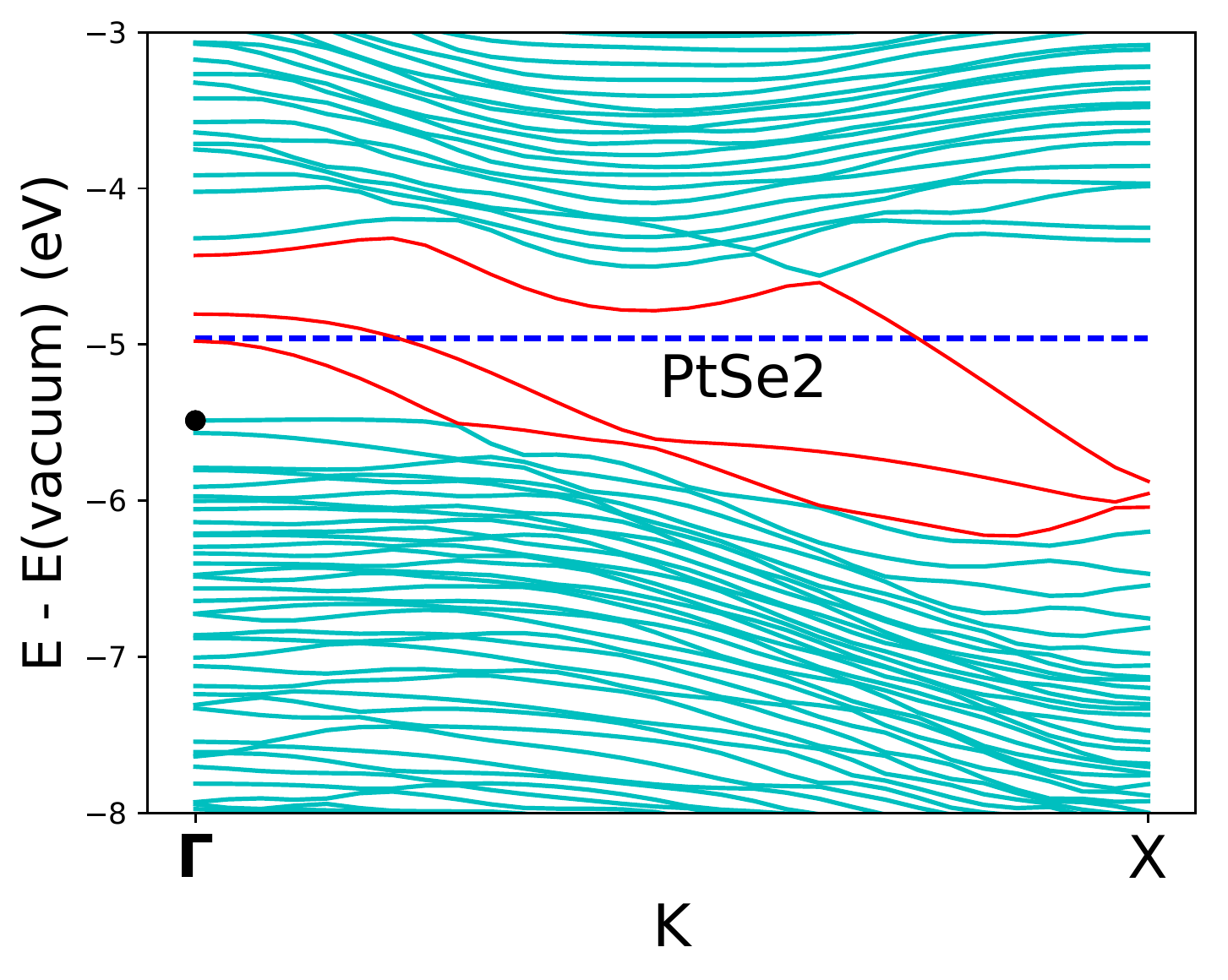}
 	\includegraphics[height=1in, width=1.5in]{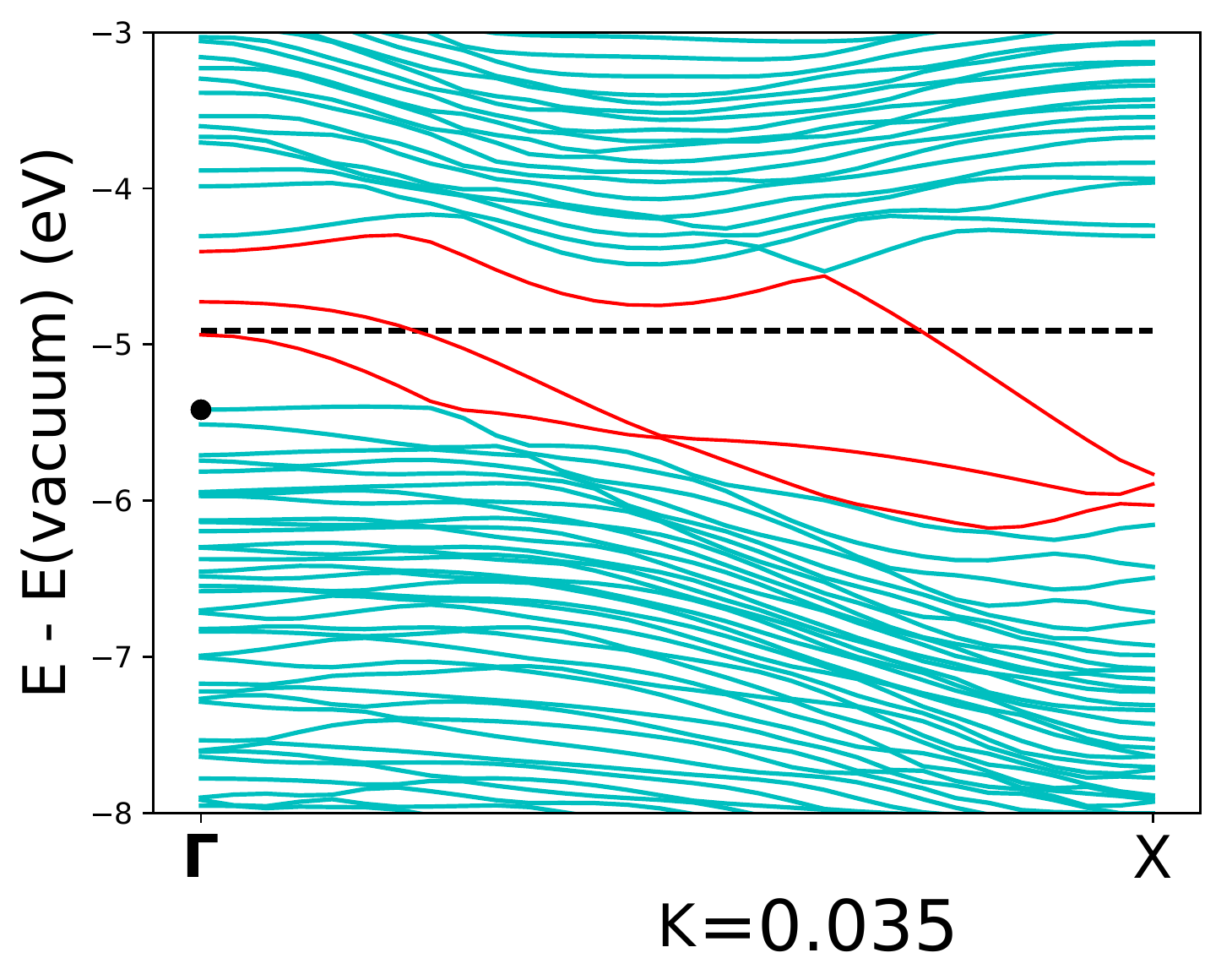}
 	\includegraphics[height=1in, width=1.5in]{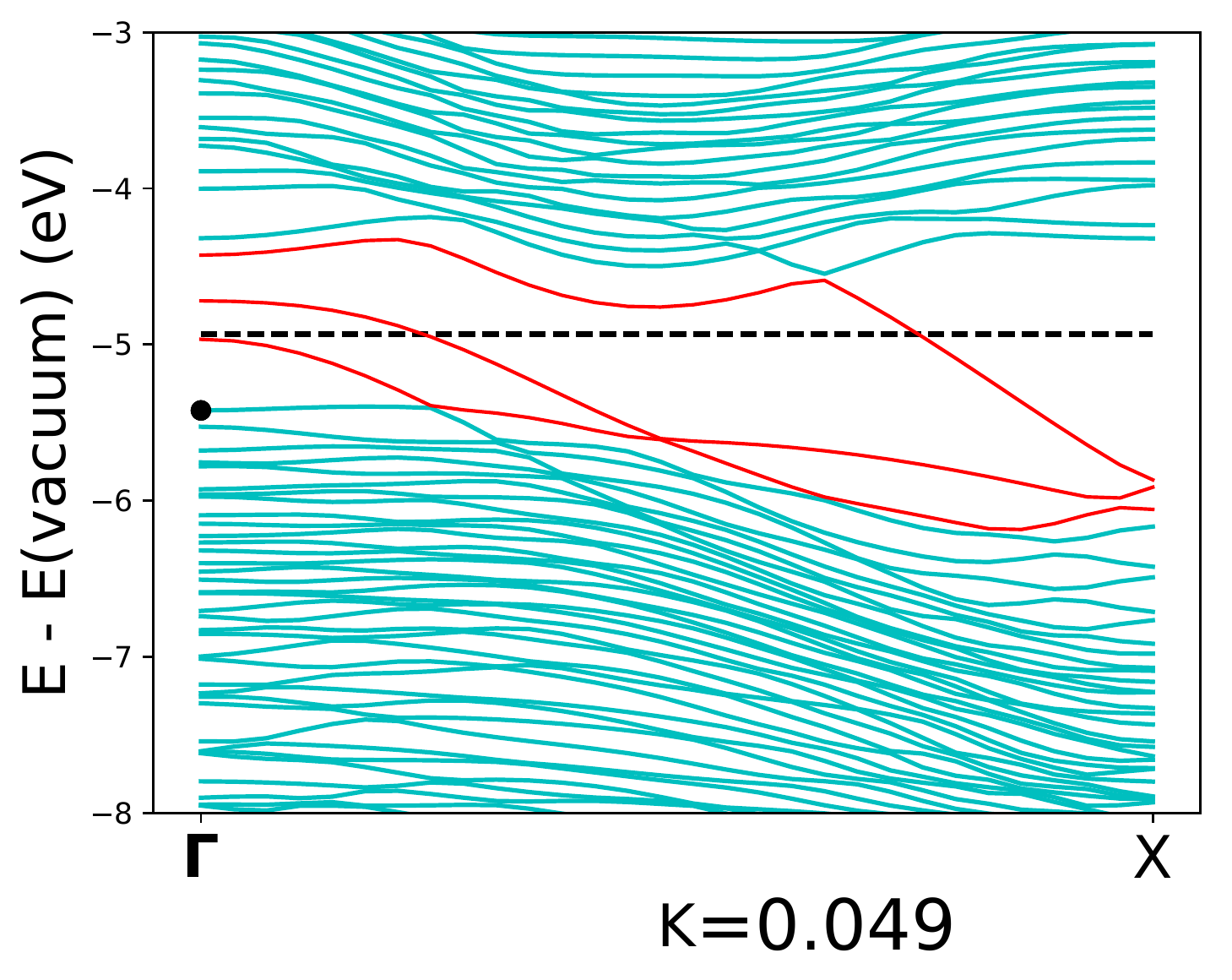}
 	\includegraphics[height=1in, width=1.5in]{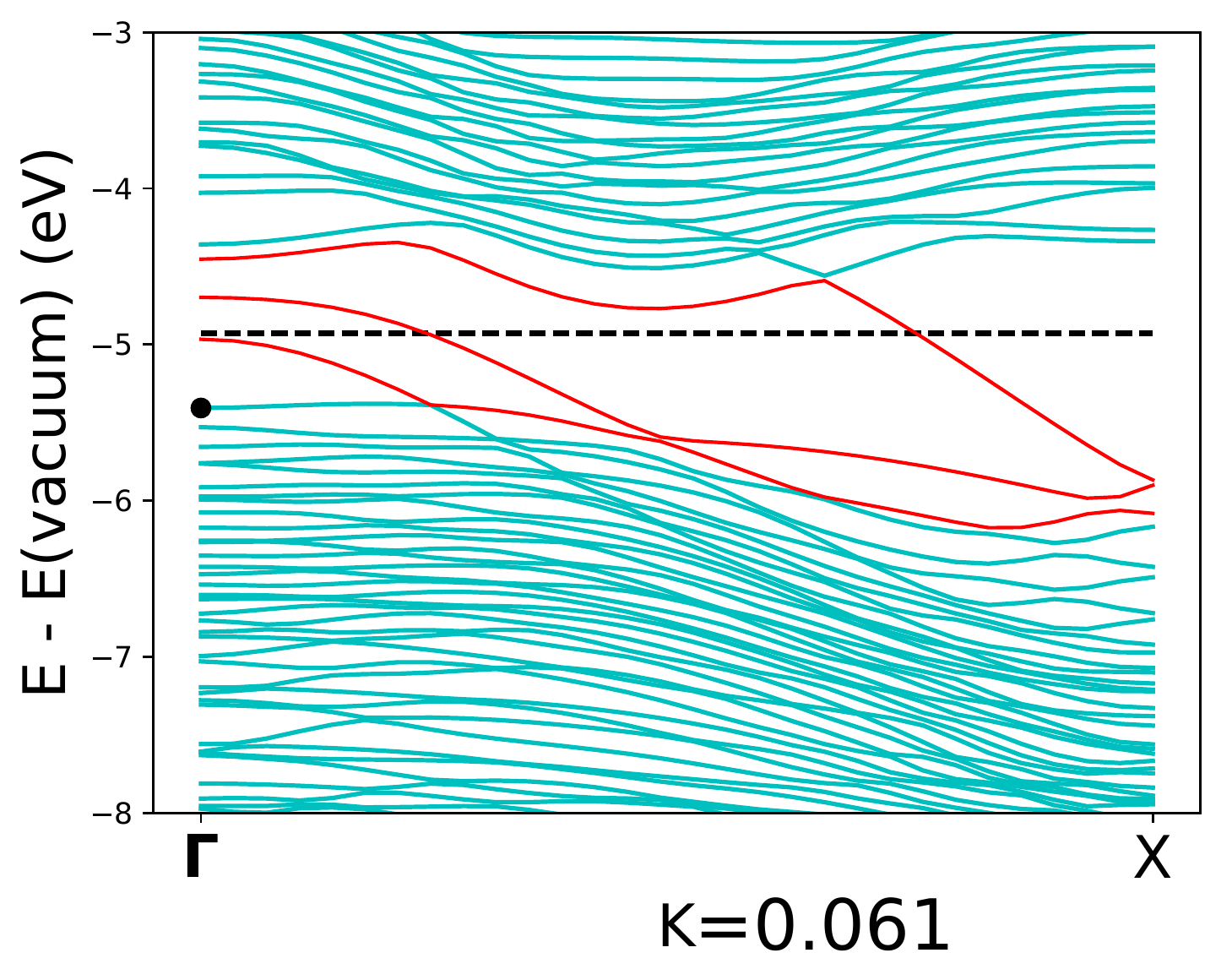}
 	\includegraphics[height=1in, width=1.5in]{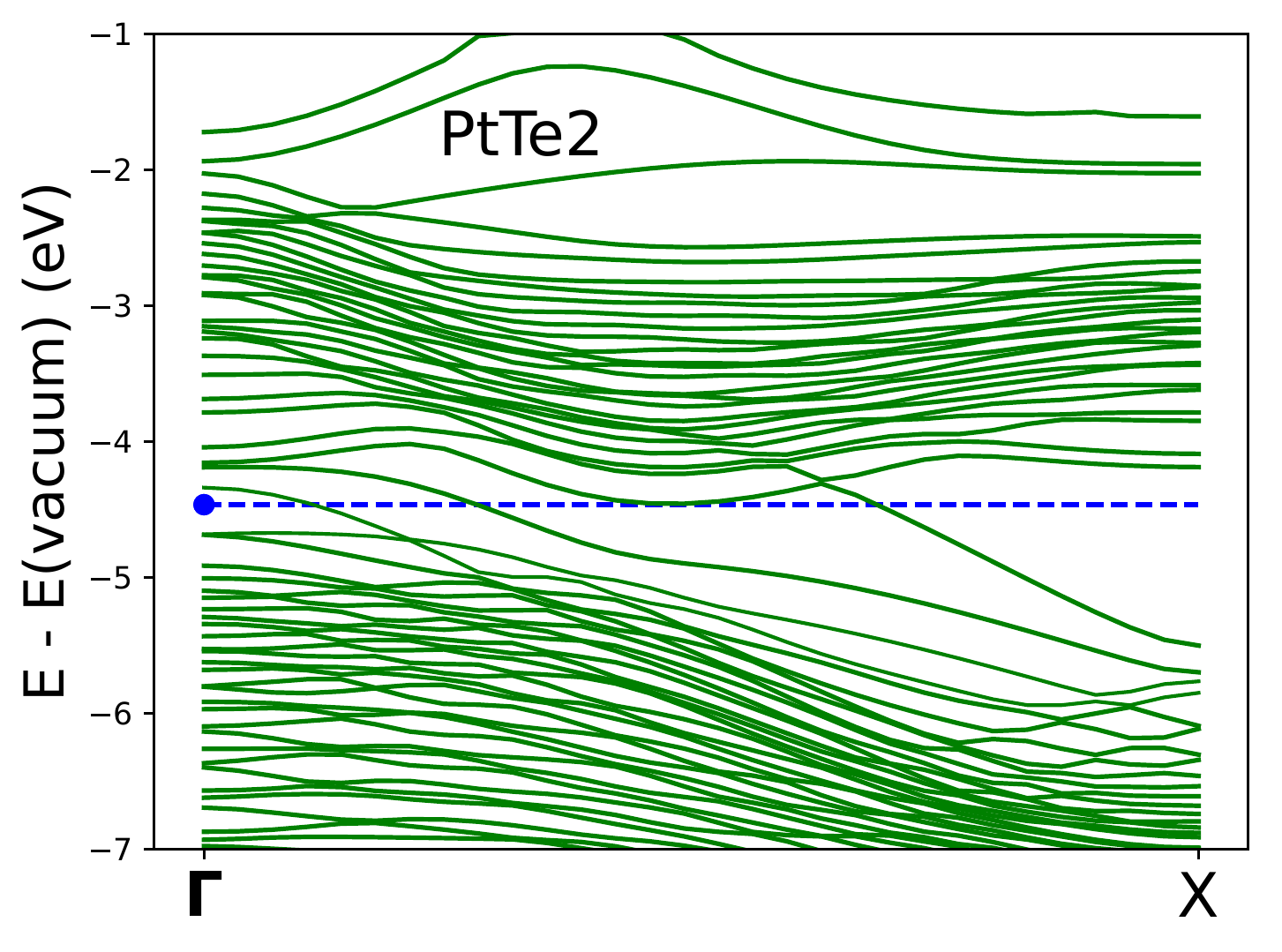}
 	\includegraphics[height=1in, width=1.5in]{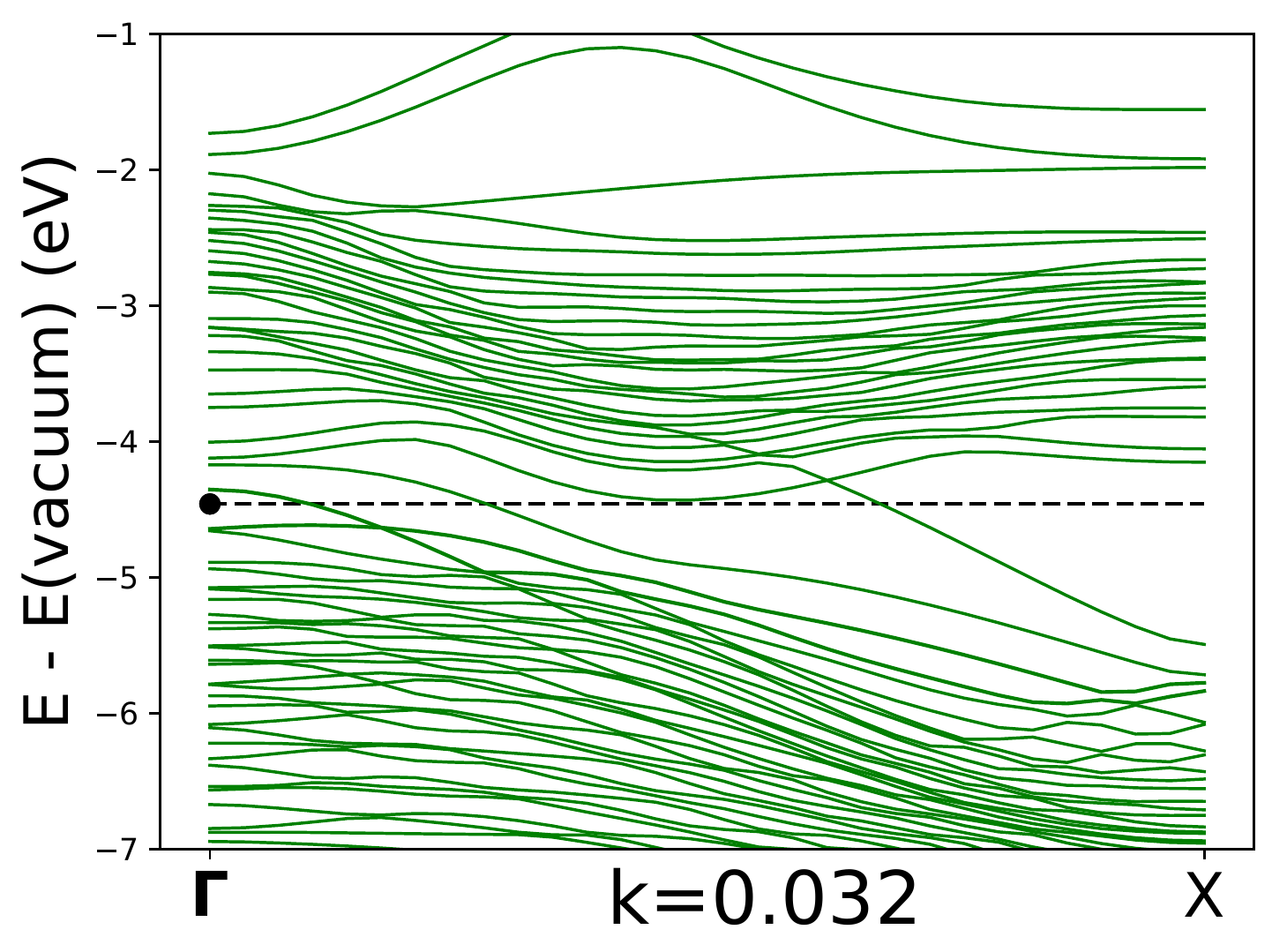}
 	\includegraphics[height=1in, width=1.5in]{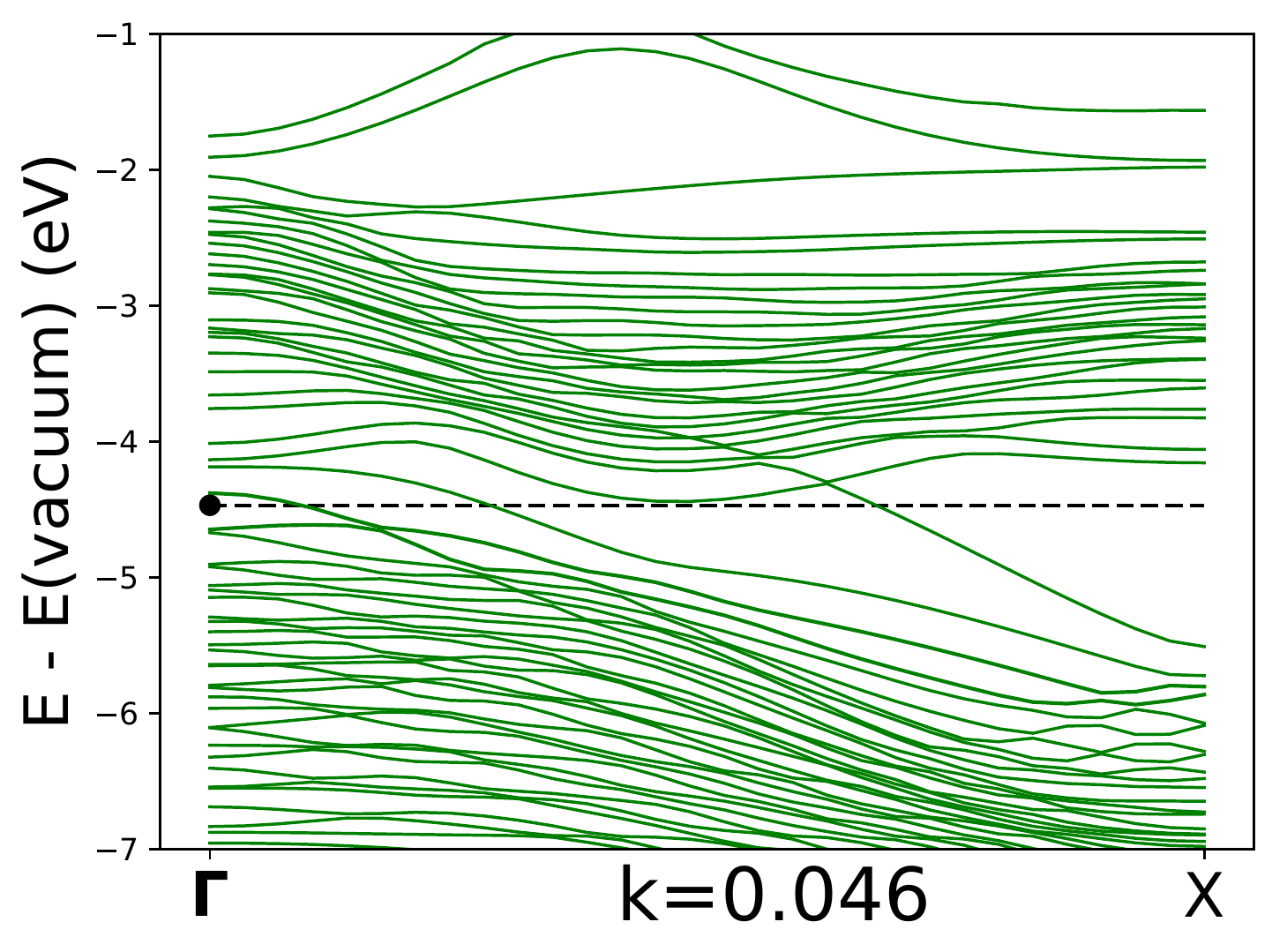}
 	\includegraphics[height=1in, width=1.5in]{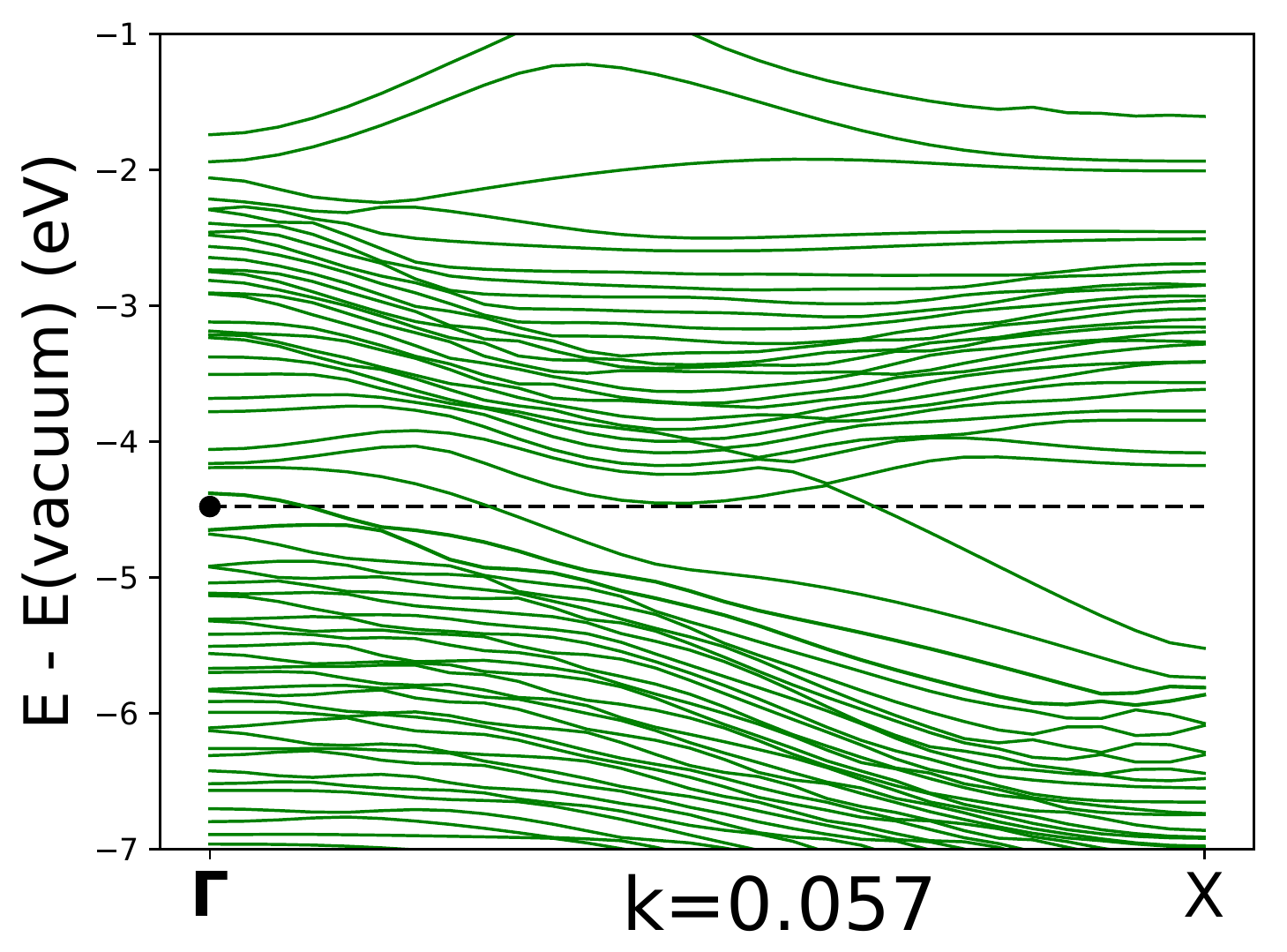}
 	\caption{Band structures with respect to vacuum for group X TMDs; PdTe$_2$ and PtY$_2$ (Y = S, Se, Te). $\kappa$ is the bending curvature ($\AA^{-1}$).}
 	\label{fig:band-X}
 \end{figure}

 \begin{figure}[h!]
 	\renewcommand\thefigure{S9}
 	\includegraphics[scale=0.2]{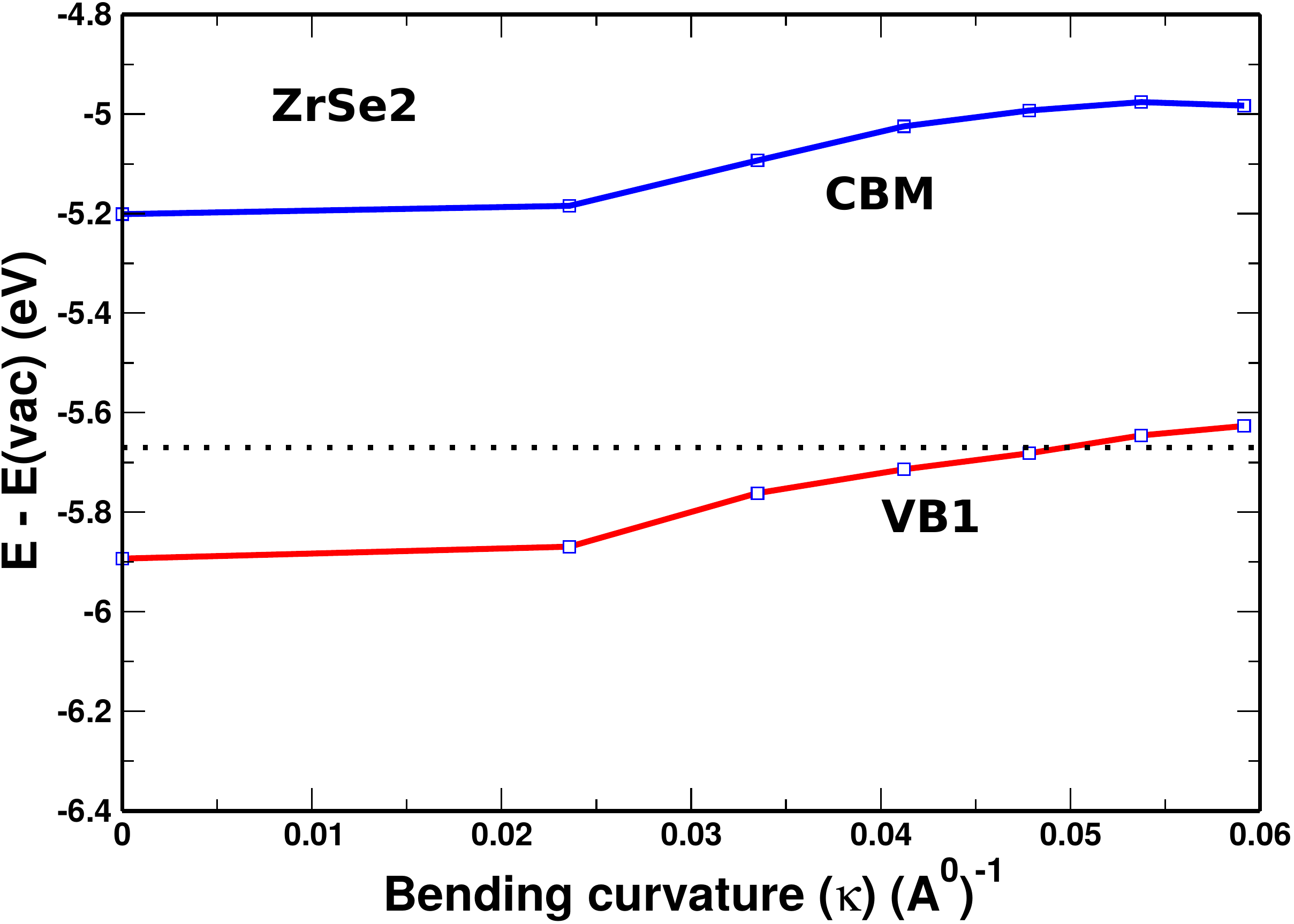}
 	\includegraphics[scale=0.2]{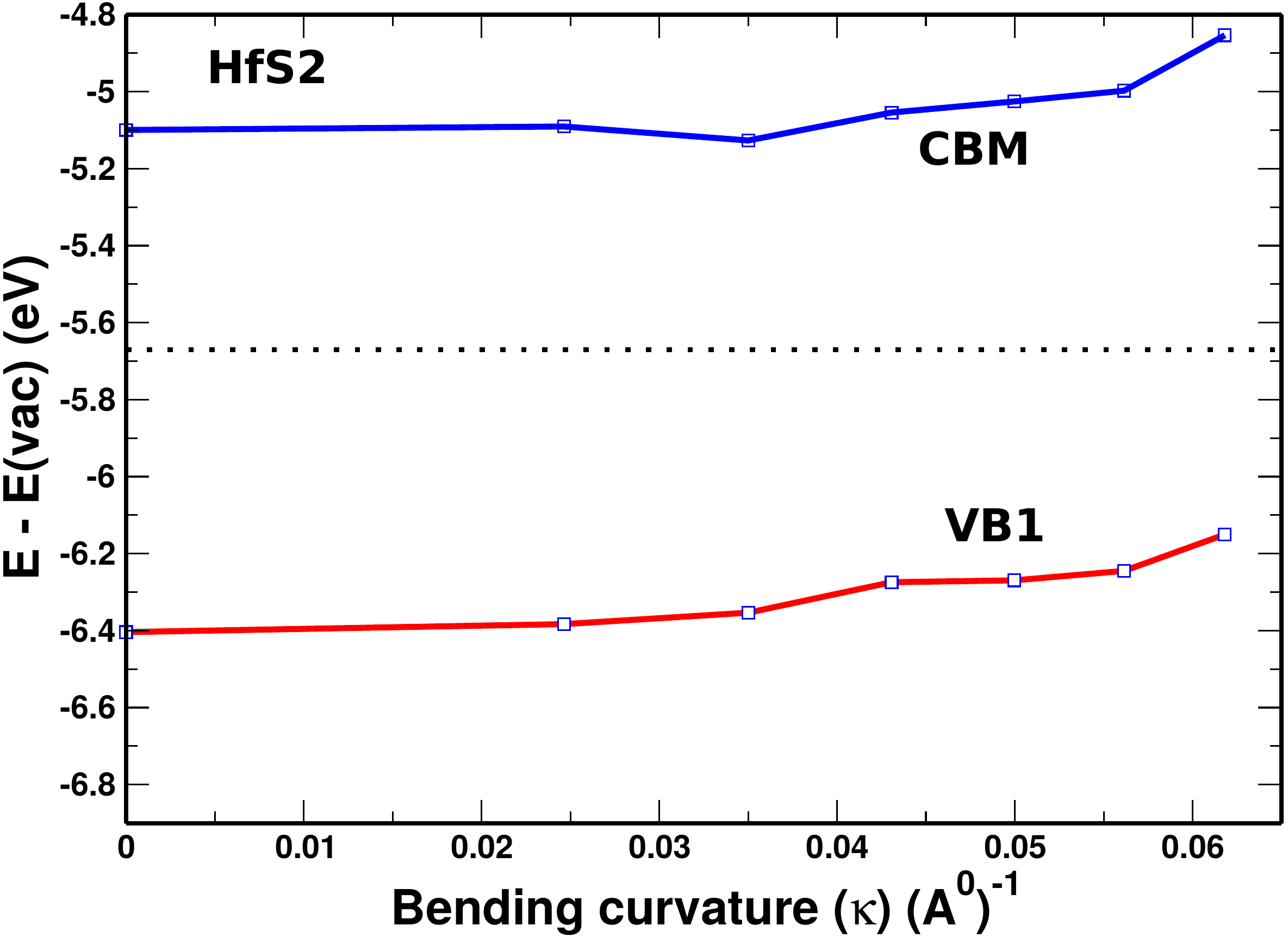}
 	\includegraphics[scale=0.2]{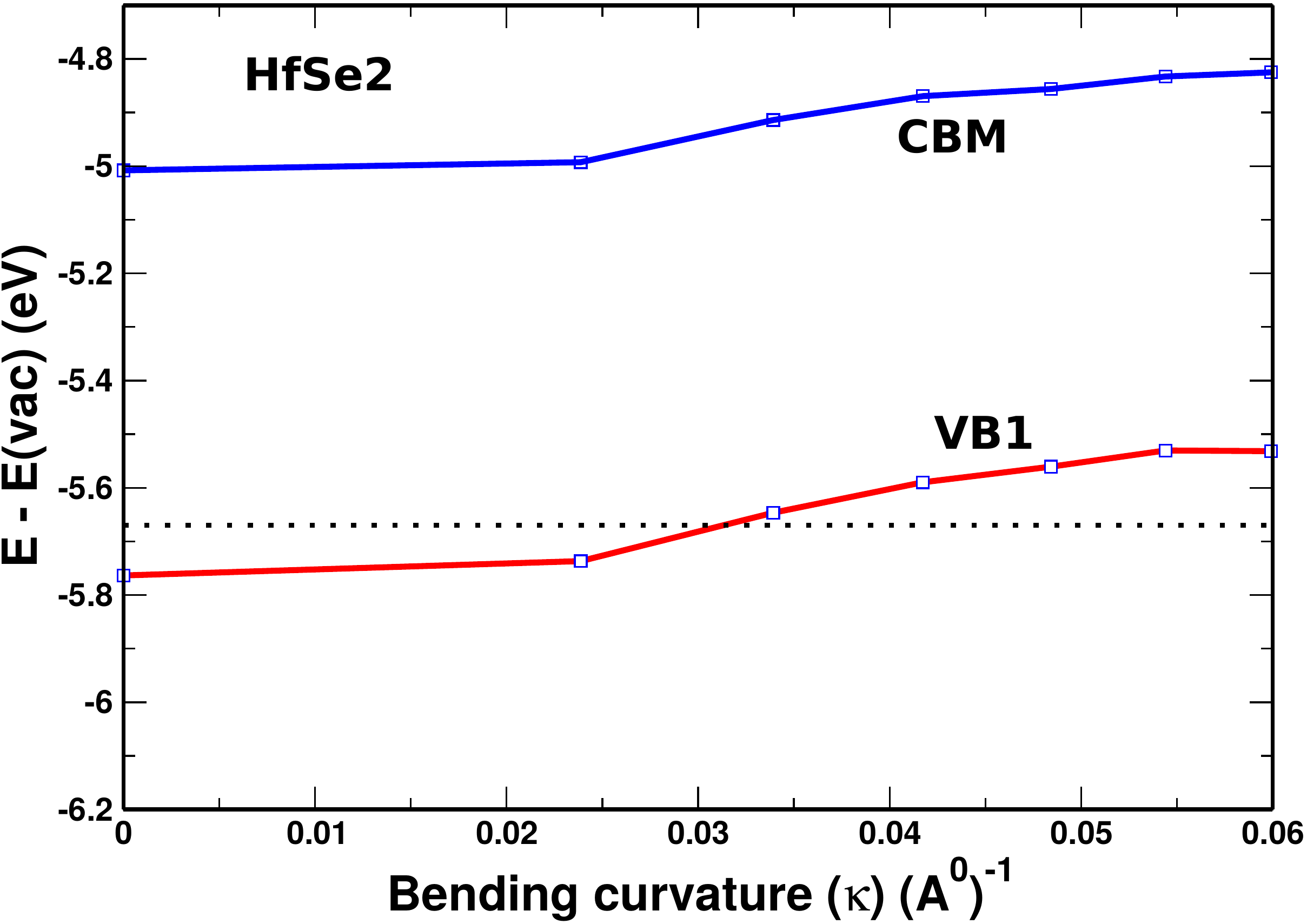}
 	\includegraphics[scale=0.2]{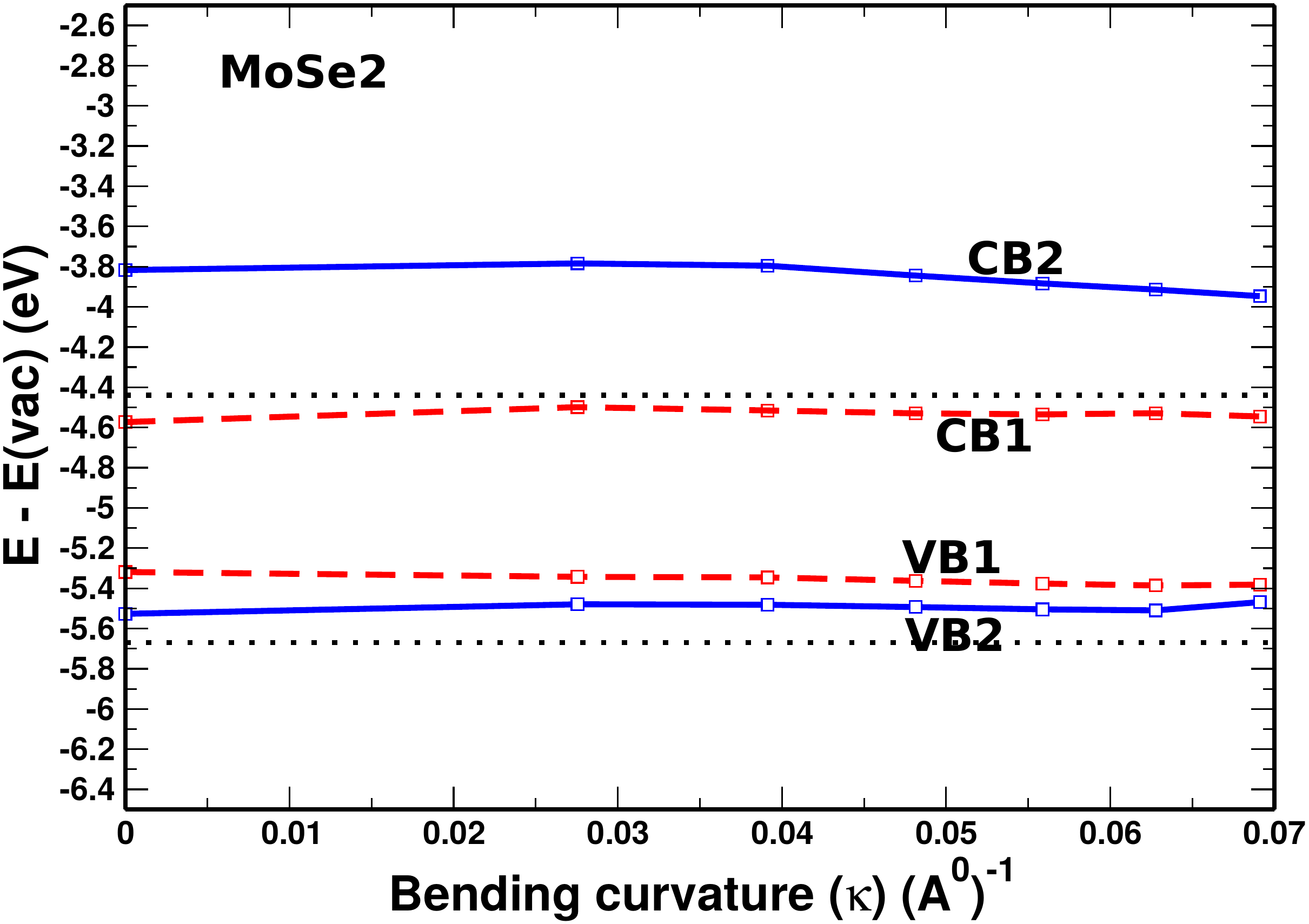}
 	\includegraphics[scale=0.2]{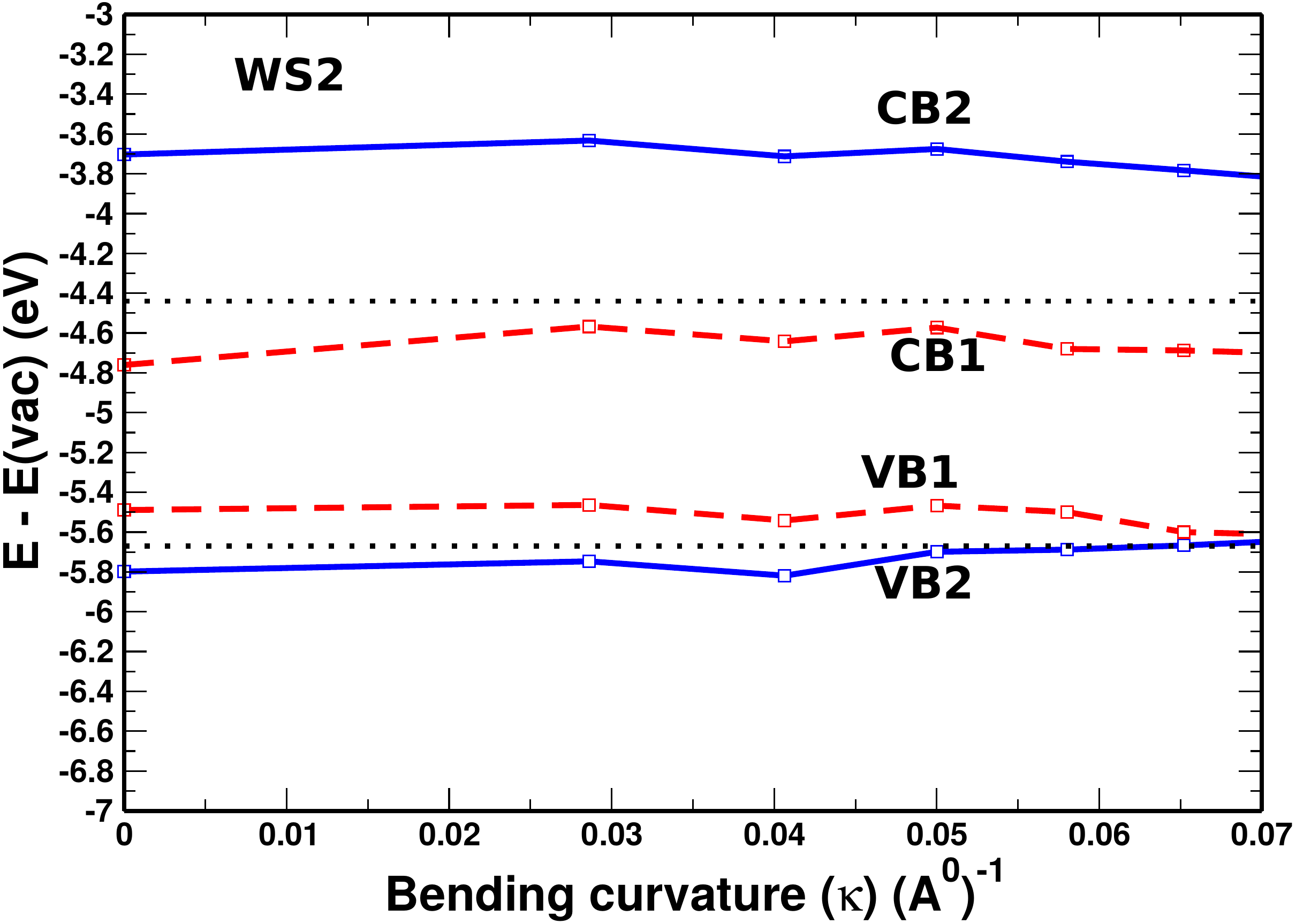}
 	\includegraphics[scale=0.2]{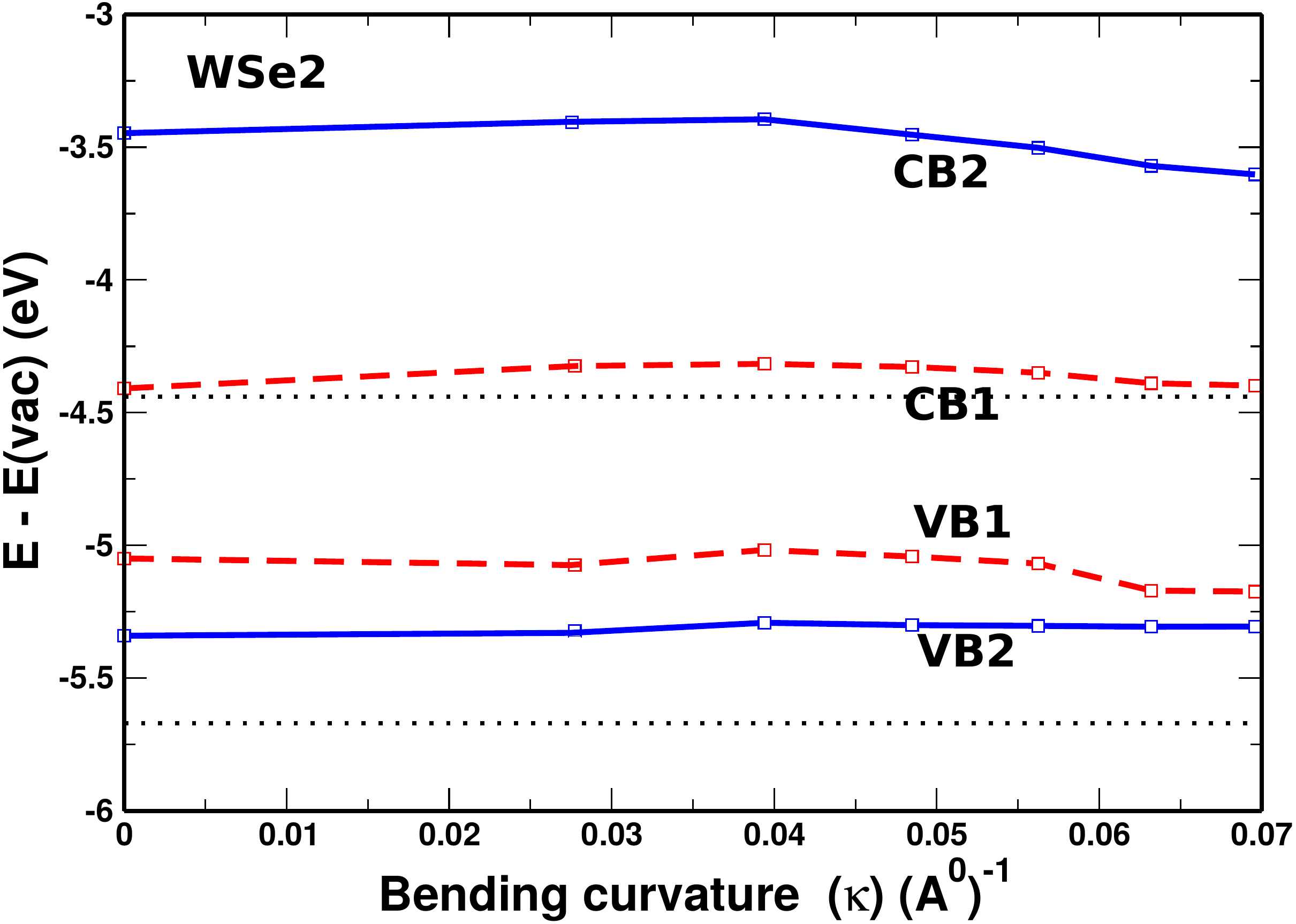}
 	\caption{Band edges with respect to vacuum; CBM, CB1, VB1, CB2, VB2 are the band edges defined in Figure 8 (Main text). The horizontal dotted lines represent water redox potentials: reduction (H$^+$/H2; -4.44 eV), and oxidation (H2O/O2; -5.67 eV).}
 	\label{fig:1T-BE}
 \end{figure}
 
 \begin{figure}[h!]
 	\renewcommand\thefigure{S10}
 	\includegraphics[scale=0.3]{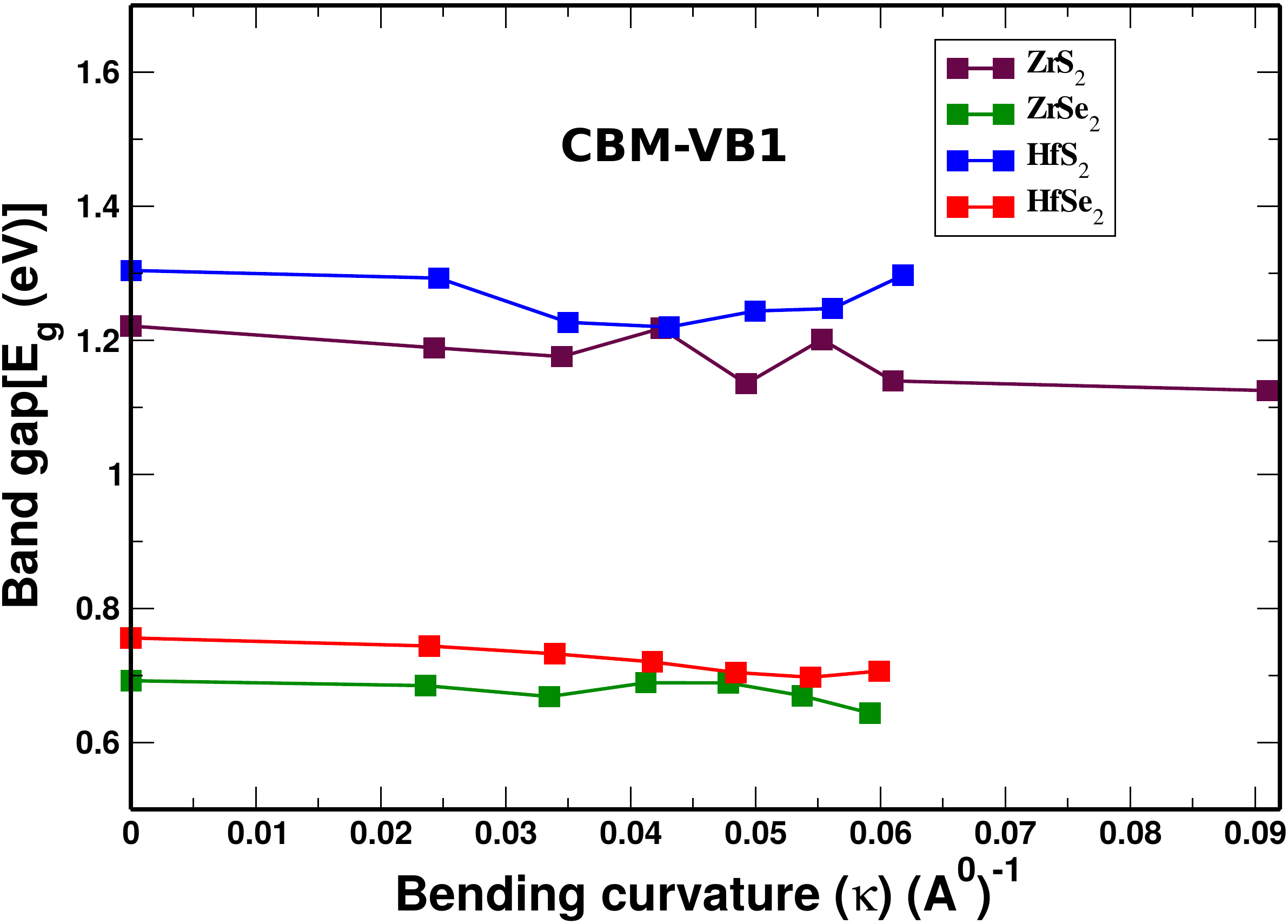}
 	\includegraphics[scale=0.3]{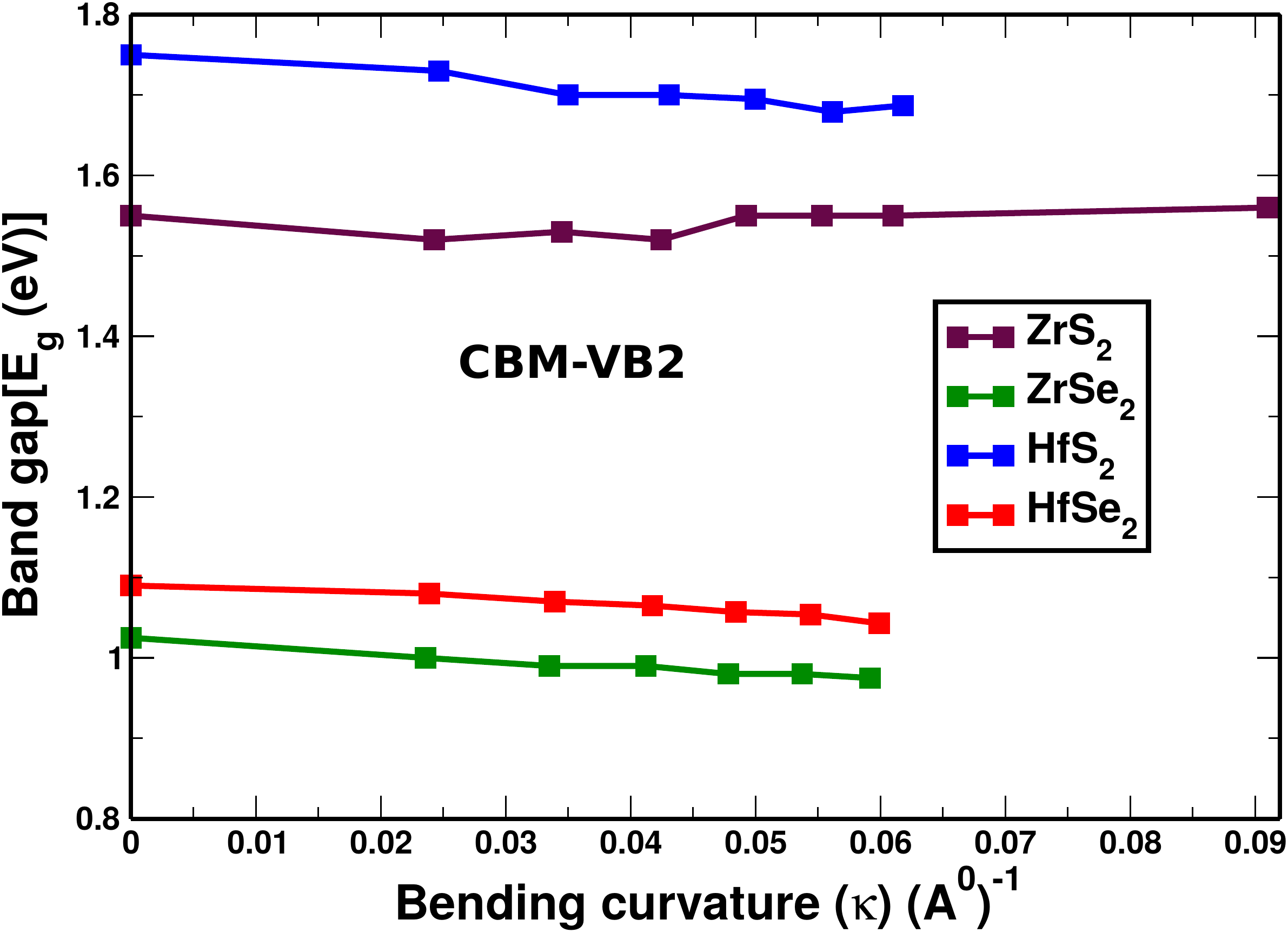}
 	\includegraphics[scale=0.3]{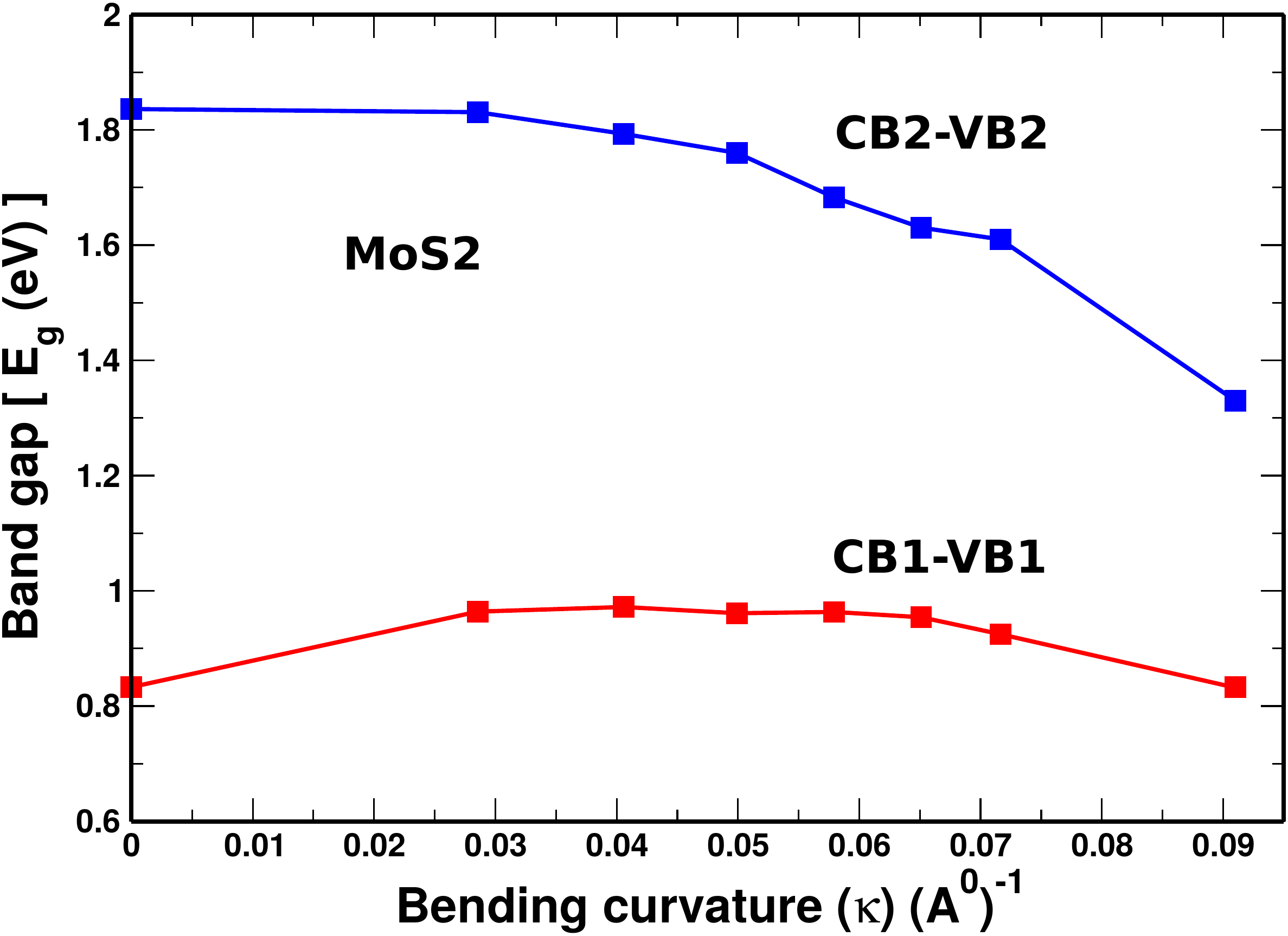}
 	\includegraphics[scale=0.3]{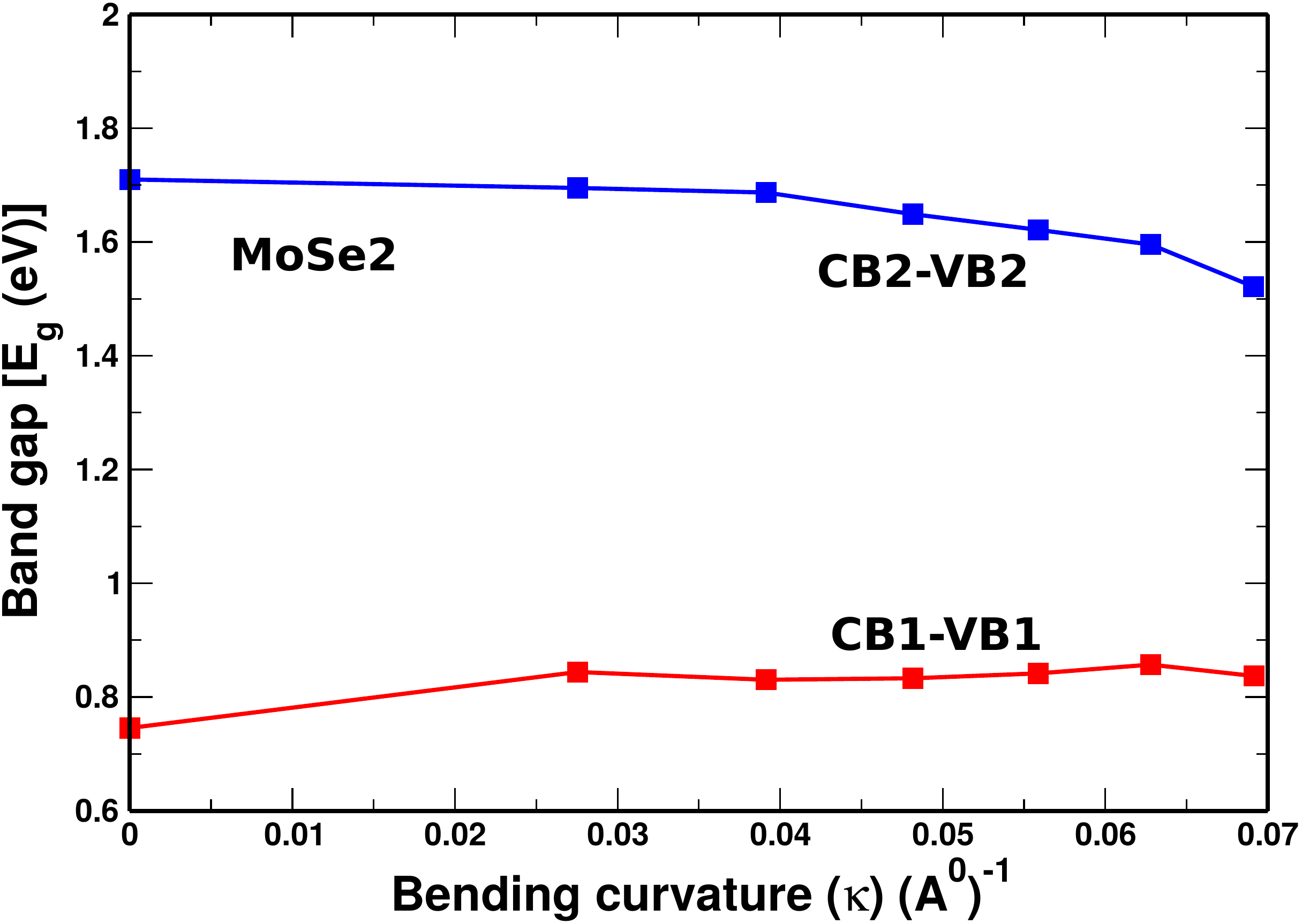}
 	\includegraphics[scale=0.3]{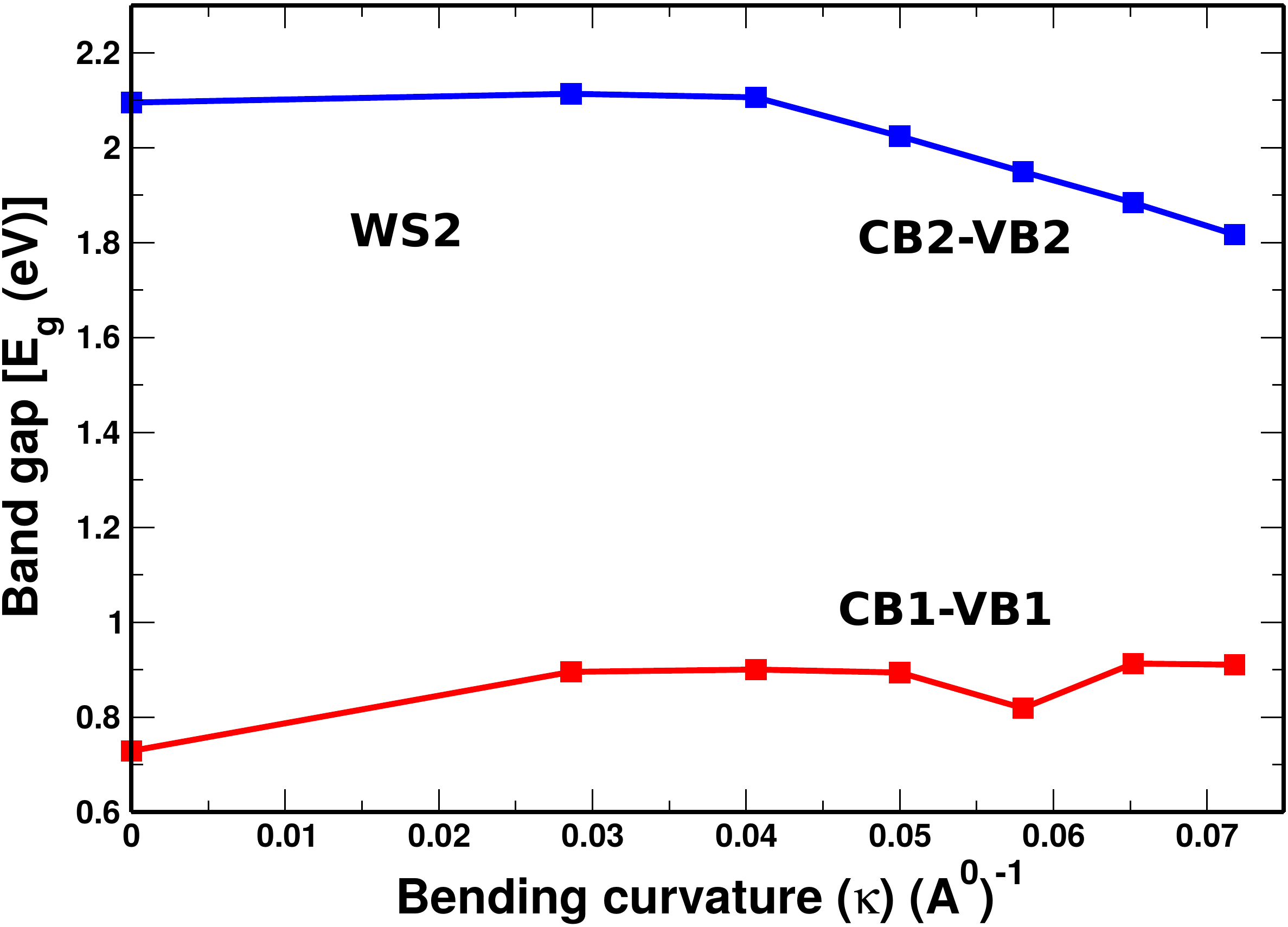}
 	\includegraphics[scale=0.3]{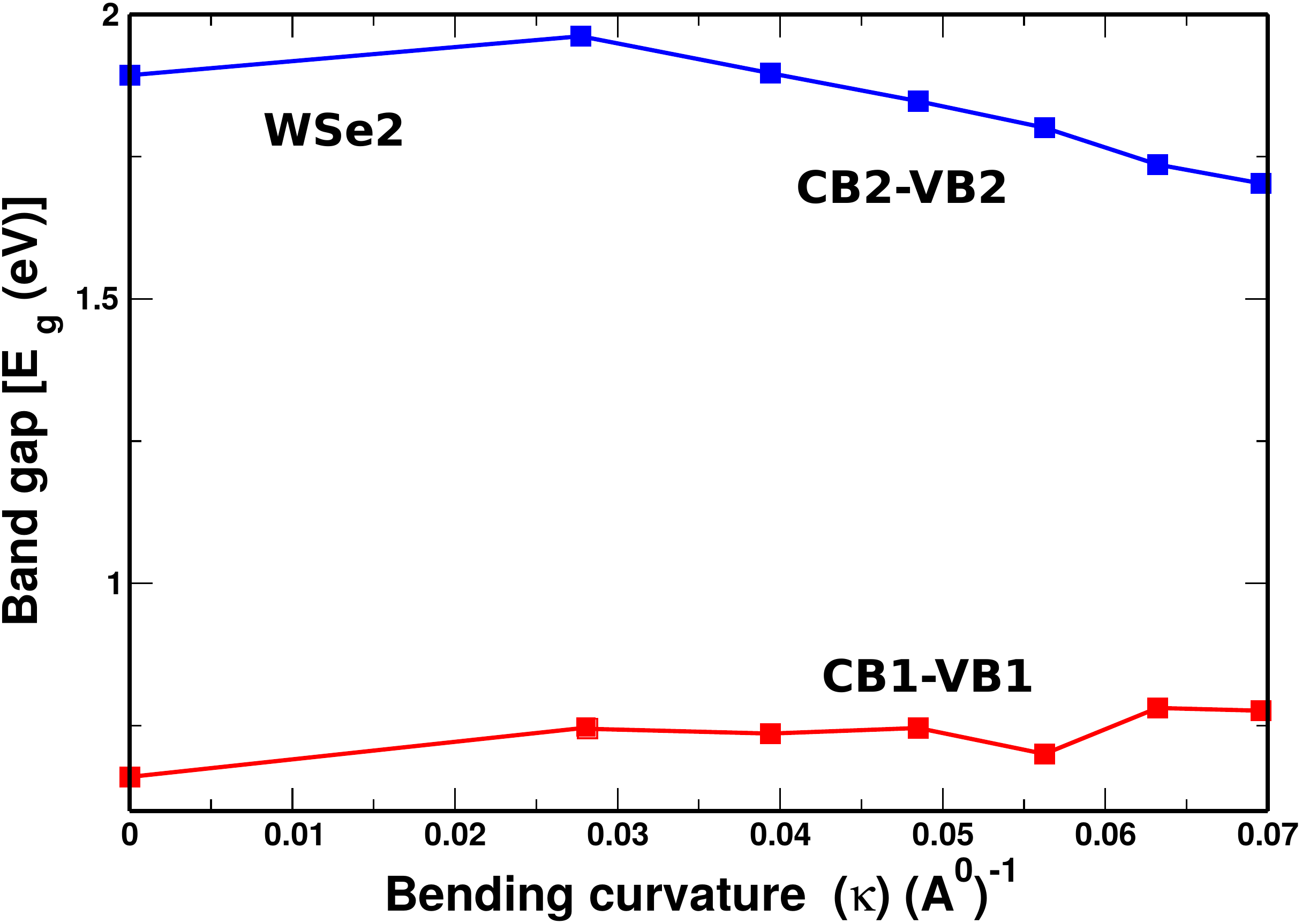}
 	\caption{Band gaps as a function of the bending curvatures. CBM, CB1, VB1, CB2, VB2 are the band edges defined in Figure 8 (Main text). Band gap remains almost constant for 1T semiconductors, while it decreases with the bending curvature for 1H compounds.}
 	\label{fig:1T-BG}
 \end{figure}
 
 \begin{figure}[h!]
 	\renewcommand\thefigure{S11}
 	\includegraphics[scale=0.62]{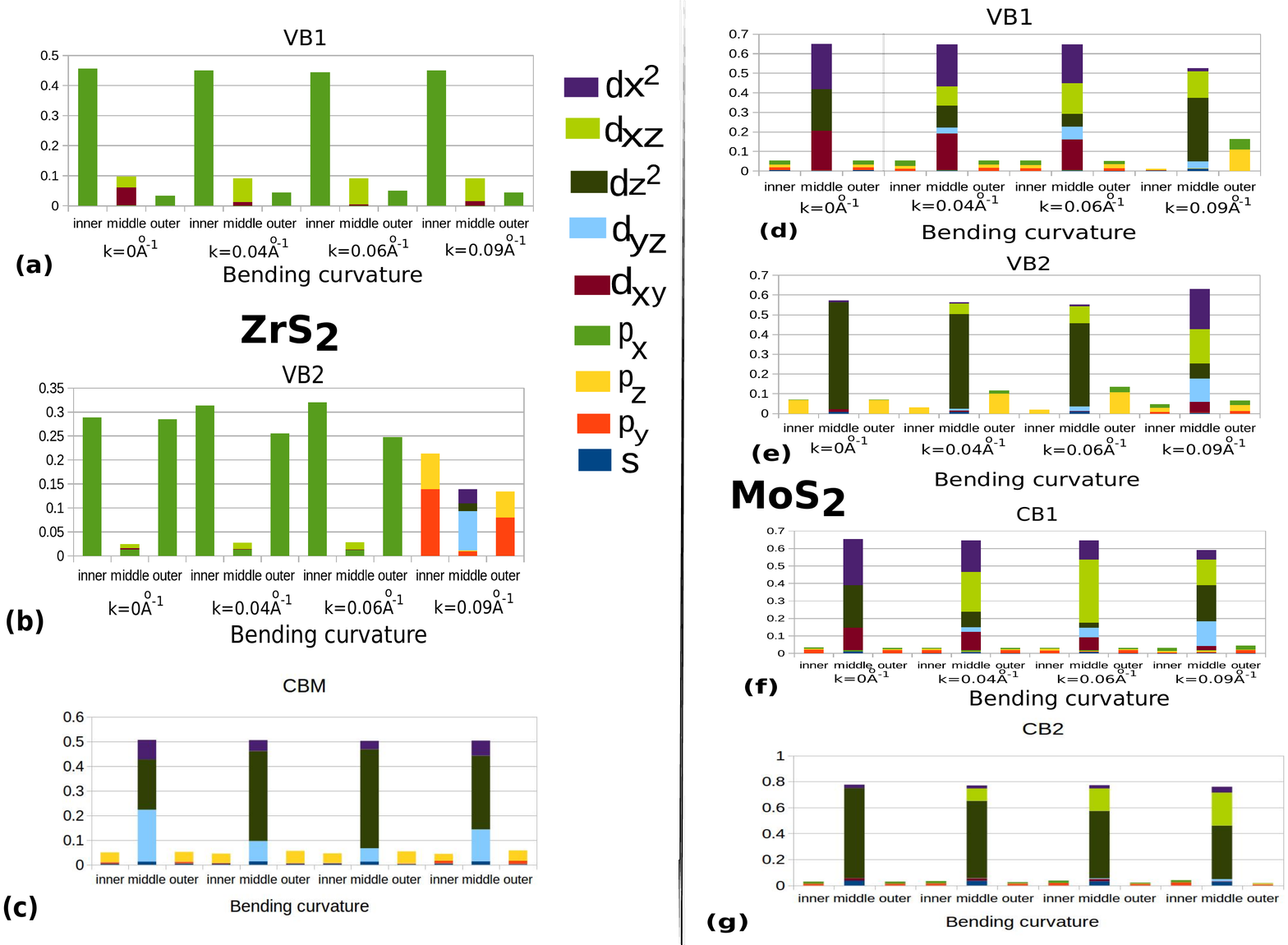}
 	\caption{An angular momentum decomposed wavefunction character of different bands corresponding to a different layer and their variation with the bending curvature; (a) - (c): ZrS$_2$; (d) - (g): MoS$_2$. The inner, middle, outer represent the different layers of the nanoribbon. CBM, CB1, VB1, CB2, VB2 are the band edges defined in Figure 8 (Main text).}
 	\label{fig:wavefunc}
 \end{figure}


\end{document}